\documentclass[12pt]{iopart}
\pdfoutput=1

\newcommand\snowmass{\begin{center}\rule[-0.2in]{4in}{0.01in}\\\rule{4in}{0.01in}\\\vskip 0.1in Submitted to the  Proceedings of the US Community Study\\on the Future of Particle Physics (Snowmass 2021)\\\rule{4in}{0.01in}\\\rule[+0.2in]{4in}{0.01in}\end{center}}

\usepackage[utf8]{inputenc}
\usepackage{graphicx}
\usepackage{textcomp}
\usepackage{amssymb}
\usepackage{amsmath}
\usepackage{cite}
\usepackage{siunitx}
\usepackage{gensymb}
\usepackage{multirow}
\usepackage{makecell}
\usepackage{footmisc}
\usepackage{multicol}
\usepackage{datetime2}
\newcommand{\mydate}{October 11, 2022}
\usepackage{lineno} 

\usepackage[colorlinks,pdfusetitle]{hyperref}
\hypersetup{allcolors=[rgb]{0,0,1}}
\usepackage{cleveref}

\numberwithin{equation}{section}

\begin{document}

\makeatletter
\let\jnl@style=\rm
\def\ref@jnl#1{{\jnl@style#1}}

\def\aj{\ref@jnl{AJ}}                   
\def\actaa{\ref@jnl{Acta Astron.}}      
\def\araa{\ref@jnl{ARA\&A}}             
\def\apj{\ref@jnl{ApJ}}                 
\def\apjl{\ref@jnl{ApJ}}                
\def\apjs{\ref@jnl{ApJS}}               
\def\ao{\ref@jnl{Appl.~Opt.}}           
\def\apss{\ref@jnl{Ap\&SS}}             
\def\aap{\ref@jnl{A\&A}}                
\def\aapr{\ref@jnl{A\&A~Rev.}}          
\def\aaps{\ref@jnl{A\&AS}}              
\def\azh{\ref@jnl{AZh}}                 
\def\baas{\ref@jnl{BAAS}}               
\def\bac{\ref@jnl{Bull. astr. Inst. Czechosl.}}
\def\caa{\ref@jnl{Chinese Astron. Astrophys.}}
\def\cjaa{\ref@jnl{Chinese J. Astron. Astrophys.}}
\def\icarus{\ref@jnl{Icarus}}           
\def\jcap{\ref@jnl{J. Cosmology Astropart. Phys.}}
\def\jrasc{\ref@jnl{JRASC}}             
\def\memras{\ref@jnl{MmRAS}}            
\def\mnras{\ref@jnl{MNRAS}}             
\def\na{\ref@jnl{New A}}                
\def\nar{\ref@jnl{New A Rev.}}          
\def\pra{\ref@jnl{Phys.~Rev.~A}}        
\def\prb{\ref@jnl{Phys.~Rev.~B}}        
\def\prc{\ref@jnl{Phys.~Rev.~C}}        
\def\prd{\ref@jnl{Phys.~Rev.~D}}        
\def\pre{\ref@jnl{Phys.~Rev.~E}}        
\def\prl{\ref@jnl{Phys.~Rev.~Lett.}}    
\def\pasa{\ref@jnl{PASA}}               
\def\pasp{\ref@jnl{PASP}}               
\def\pasj{\ref@jnl{PASJ}}               
\def\rmxaa{\ref@jnl{Rev. Mexicana Astron. Astrofis.}}%
\def\qjras{\ref@jnl{QJRAS}}             
\def\skytel{\ref@jnl{S\&T}}             
\def\solphys{\ref@jnl{Sol.~Phys.}}      
\def\sovast{\ref@jnl{Soviet~Ast.}}      
\def\ssr{\ref@jnl{Space~Sci.~Rev.}}     
\def\zap{\ref@jnl{ZAp}}                 
\def\nat{\ref@jnl{Nature}}              
\def\iaucirc{\ref@jnl{IAU~Circ.}}       
\def\aplett{\ref@jnl{Astrophys.~Lett.}} 
\def\apspr{\ref@jnl{Astrophys.~Space~Phys.~Res.}}
\def\bain{\ref@jnl{Bull.~Astron.~Inst.~Netherlands}} 
\def\fcp{\ref@jnl{Fund.~Cosmic~Phys.}}  
\def\gca{\ref@jnl{Geochim.~Cosmochim.~Acta}}   
\def\grl{\ref@jnl{Geophys.~Res.~Lett.}} 
\def\jcp{\ref@jnl{J.~Chem.~Phys.}}      
\def\jgr{\ref@jnl{J.~Geophys.~Res.}}    
\def\jqsrt{\ref@jnl{J.~Quant.~Spec.~Radiat.~Transf.}}
\def\memsai{\ref@jnl{Mem.~Soc.~Astron.~Italiana}}
\def\nphysa{\ref@jnl{Nucl.~Phys.~A}}   
\def\physrep{\ref@jnl{Phys.~Rep.}}   
\def\physscr{\ref@jnl{Phys.~Scr}}   
\def\planss{\ref@jnl{Planet.~Space~Sci.}}   
\def\procspie{\ref@jnl{Proc.~SPIE}}   

\let\astap=\aap
\let\apjlett=\apjl
\let\apjsupp=\apjs
\let\applopt=\ao
\makeatother

\review{Tau Neutrinos in the Next Decade: from GeV to EeV}

{\renewcommand{\thefootnote}{$\dagger$}
\author{
Roshan Mammen Abraham$^{1}$, 
Jaime Alvarez-Mu\~niz$^{2}$, 
Carlos A.~Arg\"uelles$^{3}$, 
Akitaka Ariga$^{4,5}$, 
Tomoko Ariga$^{6}$, 
Adam Aurisano$^{7}$\footnote{\label{fn:ed}Editor.\\Corresponding email: \href{mailto:pdenton@bnl.gov}{pdenton@bnl.gov}}, 
Dario Autiero$^{8}$, 
Mary Bishai$^{9\color{blue}{\dagger}}$, 
Nilay Bostan$^{10}$, 
Mauricio Bustamante$^{11}$, 
Austin Cummings$^{12}$, 
Valentin Decoene$^{12}$, 
Andr\'e de Gouv\^ea$^{13\color{blue}{\dagger}}$, 
Giovanni De Lellis$^{14,15}$, 
Albert De Roeck$^{16,17\color{blue}{\dagger}}$, 
Peter B.~Denton$^{18\color{blue}{\dagger}}$, 
Antonia Di Crescenzo$^{14,15,16}$, 
Milind V.~Diwan$^{9}$, 
Yasaman Farzan$^{19}$, 
Anatoli Fedynitch$^{20}$, 
Jonathan L.~Feng$^{21}$, 
Laura J.~Fields$^{10}$, 
Alfonso Garcia$^{3,22}$, 
Maria Vittoria Garzelli$^{23}$, 
Julia Gehrlein$^{18\color{blue}{\dagger}}$, 
Christian Glaser$^{24}$, 
Katarzyna Grzelak$^{25}$, 
Steffen Hallmann$^{26}$, 
V Hewes$^{7}$, 
D.~Indumathi$^{27,28}$, 
Ahmed Ismail$^{1}$, 
Sudip Jana$^{29}$, 
Yu Seon Jeong$^{30}$, 
Kevin J.~Kelly$^{31}$, 
Spencer R.~Klein$^{32}$, 
Felix Kling$^{33}$, 
Thomas Kosc$^{34}$, 
Umut Kose$^{16,35}$, 
D.~Jason Koskinen$^{36}$, 
John Krizmanic$^{37}$, 
Jeff Lazar$^{3}$, 
Yichen Li$^{9}$, 
Ivan Martinez-Soler$^{3}$, 
Irina Mocioiu$^{12\color{blue}{\dagger}}$, 
Jiwoo Nam$^{38}$, 
Valentin Niess$^{39}$, 
Nepomuk Otte$^{40}$, 
Sameer Patel$^{41}$, 
Roberto Petti$^{42}$, 
Remy L.~Prechelt$^{43}$, 
Steven Prohira$^{44}$, 
Miriama Rajaoalisoa$^{7}$, 
Mary Hall Reno$^{41}$, 
Ibrahim Safa$^{3}$, 
Carlos Sarasty-Segura$^{7}$, 
R.~Thiru Senthil$^{27,28}$, 
Juliana Stachurska$^{45}$, 
Oleksandr Tomalak$^{46}$, 
Sebastian Trojanowski$^{47,48}$, 
Roger Alexandre Wendell$^{49}$, 
Dawn Williams$^{50}$, 
Stephanie Wissel$^{12\color{blue}{\dagger}}$, 
Barbara Yaeggy$^{7}$, 
Enrique Zas$^{2}$, 
Pavel Zhelnin$^{3}$, and 
Jing-yu Zhu$^{51,52}$}
\address{$^{1}$ Department of Physics, Oklahoma State University, Stillwater, OK, 74078, USA}
\address{$^{2}$ Instituto Galego de F\'\i sica de Altas Enerx\'\i as, Universidade de Santiago de Compostela, Santiago de Compostela, 15782, Spain}
\address{$^{3}$ Department of Physics \& Laboratory for Particle Physics and Cosmology, Harvard University, Cambridge, MA 02138, USA}
\address{$^{4}$ Albert Einstein Center for Fundamental Physics, Laboratory for High Energy Physics, University of Bern, Sidlerstrasse 5, CH-3012 Bern, Switzerland}
\address{$^{5}$ Department of Physics, Chiba University, 1-33 Yayoi-cho Inage-ku, Chiba, 263-8522, Japan}
\address{$^{6}$ Kyushu University, Nishi-ku, Fukuoka, 819-0395, Japan }
\address{$^{7}$ Department of Physics, University of Cincinnati, Cincinnati, Ohio 45221, USA}
\address{$^{8}$ Institut de Physique des 2 Infinis de Lyon, 69622 Villeurbanne, France}
\address{$^{9}$ Physics Department, Brookhaven National Laboratory, Upton, NY 11973, USA}
\address{$^{10}$ Department of Physics, University of Notre Dame, Notre Dame, Indiana 46556, USA}
\address{$^{11}$ Niels Bohr International Academy, Niels Bohr Institute, University of Copenhagen, Denmark}
\address{$^{12}$ Pennsylvania State University, University Park, PA 16802, USA}
\address{$^{13}$ Northwestern University, Department of Physics \& Astronomy, 2145 Sheridan Road, Evanston, IL 60208, USA}
\address{$^{14}$ Dipartimento di Fisica ``E.~Pancini", Università ``Federico II" di Napoli, Italy}
\address{$^{15}$ Istituto Nazionale di Fisica Nucleare, Naples, Italy}
\address{$^{16}$ Experimental Physics Department, CERN, CH-1211 Geneva 23, Switzerland}
\address{$^{17}$ University of Antwerp, Physics Department, Antwerp, Belgium}
\address{$^{18}$ High Energy Theory Group, Physics Department, Brookhaven National Laboratory, Upton, NY 11973, USA}
\address{$^{19}$ School of physics, Institute for Research in Fundamental Sciences (IPM), P.O. Box 19395-5531, Tehran, Iran }
\address{$^{20}$ Institute of Physics, Academia Sinica, Taipei, 11529, Taiwan}
\address{$^{21}$ Department of Physics and Astronomy, University of California, Irvine, CA 92697-4575, USA}
\address{$^{22}$ Instituto de Física Corpuscular (IFIC), Universitat de València (UV), 46980 Paterna, València, Spain}
\address{$^{23}$ Institut fur Theoretische Physik, Universitat Hamburg, Luruper Chaussee 149, D - 22761 Hamburg, Germany}
\address{$^{24}$ Department of Physics and Astronomy, Uppsala University, Uppsala, SE-752 37, Sweden}
\address{$^{25}$ Faculty of Physics, University of Warsaw, 02-093 Warsaw, Poland}
\address{$^{26}$ Deutsches Elektronen-Synchrotron DESY, Platanenallee 6, 15738 Zeuthen, Germany}
\address{$^{27}$ The Institute of Mathematical Sciences, Taramani, Chennai 600113, India}
\address{$^{28}$ Homi Bhabha National Institute, Anushakti Nagar, Mumbai 400094, India}
\address{$^{29}$ Max-Planck-Institut f{\"u}r Kernphysik, Saupfercheckweg 1, 69117 Heidelberg, Germany}
\address{$^{30}$ High Energy Physics Center, Chung-Ang University, Seoul 06974, Korea}
\address{$^{31}$ Theoretical Physics Department, CERN, Esplande des Particules, 1211 Geneva 23, Switzerland}
\address{$^{32}$ Nuclear Science Division, Lawrence Berkeley National Laboratory, Berkeley CA 94720 and Physics Department, University of California, Berkeley CA 94720, USA}
\address{$^{33}$ Deutsches Elektronen-Synchrotron DESY, Notkestr.~85, 22607 Hamburg, Germany}
\address{$^{34}$ University Grenoble Alpes, CNRS, Grenoble INP, LPSC-IN2P3, 38000 Grenoble, France}
\address{$^{35}$ Institute for Particle physics and Astrophysics, ETH Zurich, Switzerland}
\address{$^{36}$ Niels Bohr Institute - University of Copenhagen, Blegdamsvej 17, Copenhagen, Denmark}
\address{$^{37}$ NASA/Goddard Space Flight Center, Laboratory for Astroparticle Physics, Greenbelt, MD20771 USA}
\address{$^{38}$ Department of Physics, Leung Center for Cosmology and Particle Astrophysics (LeCosPA), National Tawian University, Taipei, 10617, Taiwan}
\address{$^{39}$ Clermont~Université, Université~Blaise~Pascal, CNRS/IN2P3, Laboratoire de Physique Corpusculaire, BP~10118, F-63000, Clermont-Ferrand, France}
\address{$^{40}$ Georgia Institute of Technology, Atlanta, GA 30332, USA}
\address{$^{41}$ Department of Physics and Astronomy, University of Iowa, Iowa City, IA 52242, USA}
\address{$^{42}$ University of South Carolina, Columbia, SC 29208, USA}
\address{$^{43}$ University of Hawai'i M\=anoa, Honolulu, HI 96822, USA}
\address{$^{44}$ Department of Physics, Center for Cosmology and AstroParticle Physics (CCAPP), The Ohio State University, Columbus OH, 43210, USA}
\address{$^{45}$ Laboratory for Nuclear Science, Massachusetts Institute of Technology, Cambridge, Massachusetts 02139, USA}
\address{$^{46}$ Theoretical Division, Los Alamos National Laboratory, Los Alamos, NM 87545, USA}
\address{$^{47}$ Astrocent, Nicolaus Copernicus Astronomical Center Polish Academy of Sciences, ul. Rektorska 4, 00-614, Warsaw, Poland}
\address{$^{48}$ National Centre for Nuclear Research, Pasteura 7, 02-093 Warsaw, Poland}
\address{$^{49}$ Department of Physics, Kyoto University, Kyoto 606-8502, Japan}
\address{$^{50}$ Dept.~of Physics and Astronomy, University of Alabama, Tuscaloosa, AL 35487, USA}
\address{$^{51}$ Institut für Astroteilchen Physik, Karlsruher Institut f\"ur Technologie (KIT), Hermann-von-Helmholtz-Platz 1, 76344 Eggenstein-Leopoldshafen, German}
\address{$^{52}$ School of Physics and Astronomy and Tsung-Dao Lee Institute, Shanghai Jiao Tong University, Shanghai 200240, China}

\renewcommand{\thefootnote}{\arabic{footnote}}

\begin{abstract}
Tau neutrinos are the least studied particle in the Standard Model.
This whitepaper discusses the current and expected upcoming status of tau neutrino physics with attention to the broad experimental and theoretical landscape spanning long-baseline, beam-dump, collider, and astrophysical experiments.
This whitepaper was prepared as a part of the NuTau2021 Workshop\footnote{\href{https://www.bnl.gov/nutau2021}{bnl.gov/nutau2021}}.\\[0.1in]
Snowmass topics: NF01, NF03, NF04, NF06, NF08/TF11, NF09, NF10, EF03, EF09, CF07, UF01, IF\\[0.1in]
Preprint numbers: DESY-22-040, LA-UR-21-32255\\[0.1in]
Date: \mydate\\[0.5in]
\snowmass
\end{abstract}

\maketitle


\section*{Executive Summary}
\addcontentsline{toc}{section}{Executive Summary}
The neutrino associated with the tau lepton, the tau neutrino $\nu_\tau$, is generally considered to be the least studied particle.
Due to its low cross section, high tau lepton production threshold, and difficulty in distinguishing it from other neutrino states, the global tau neutrino data set has remained quite low.
Nonetheless, we are currently in a pivotal time for tau neutrino physics from a multitude of directions.
In the last five years, large volume atmospheric neutrino experiments Super-Kamiokande and IceCube have reported definitive detections of tau neutrinos in atmospheric and astrophysical neutrino data.
As these experiments continue to collect data and refine their analyses, a number of other experiments with novel probes of tau neutrinos are expected to come online soon, including the long-baseline accelerator experiment DUNE, forward physics experiments at the LHC, and ultra-high energy neutrino telescopes.

As our experimental efforts to measure tau neutrinos are exploding, the theoretical interest in a deeper investigation of this particle is broad and compelling.
Within the realm of oscillations, detecting atmospheric tau neutrinos provides an important cross check of the oscillation parameters and is one of the most important probes for improving our understanding of the unitarity of the lepton mixing matrix.
There are also many interesting new physics scenarios involving tau neutrinos including sterile neutrinos, non-standard neutrino interactions, neutrino decay, and others.
As experimental progress on tau neutrinos has improved in recent years, a large interest in exploring the new physics scenarios that can be probed by tau neutrinos has increased with it.
In some cases models single out the tau neutrino to satisfy other constraints and in other cases the model does not depend on the flavor of the neutrino but tau neutrinos may be the only means of probing the model. 
For example,  tau neutrinos could be the only neutrino flavor detected at extremely high energies, providing understanding of new parameter space for both standard and beyond standard model scenarios.  
Finally, tau neutrinos play a central role in testing the lepton flavor universality violating hints uncovered in flavor physics experiments.

In order to realize these experimental and theoretical goals, a number of tools need to be developed.
Understanding neutrino cross sections near the tau neutrino charged-current threshold is a notoriously tricky problem.
In addition, tau neutrino propagation through the Earth needs to be handled very carefully, especially for high energy neutrinos.
Finally, tau neutrinos need to be identified in detectors.

In this whitepaper we demonstrate the rich physics case involving tau neutrinos, the exciting experimental landscape that we hope will be fully realized, and describe the software and reconstruction tools required to achieve these goals.
We would also like to draw the reader's attention to the endorsers listed after the conclusions and several new figures in this whitepaper summarizing certain aspects of tau neutrino physics, Figs.~\ref{fig:nu_tau_range}, \ref{fig:NuTauEvents}, \ref{fig:HEnusNewPhysics}, \ref{fig:cross_section_uhe}, \ref{fig:experiments_table}, \ref{fig:UHE_diffuse_sensitivity}, and Table \ref{tab:software_Tau_neutrino_propagation}.
\pagebreak

\tableofcontents\pagebreak

\section{History and Motivation}
\label{sec:history}
The tau neutrino ($\nu_\tau$) is the second most recently discovered and the least studied particle in the Standard Model (SM).
There is a growing theory effort to identify important new physics searches and connections to a broad class of experiments in the tau neutrino sector, see section \ref{sec:theory}.
Despite the difficulty in probing this elusive particle thus far, there are a plethora of experiments across a broad range of energies looking to improve our understanding of the tau neutrino in the coming years, see sections \ref{sec:gev experiments}, \ref{sec:intermediate experiments}, and \ref{sec:he experiments}.
To support this program there are a number of tools to reconstruct, identify, and propagate tau neutrinos that are increasing in sophistication, see section \ref{sec:tools}.
To begin this whitepaper, we show a summary of the physics programs and identification techniques in Fig.~\ref{fig:nu_tau_range} and review the history of the tau neutrino up to today.

\begin{figure}
\centering
\includegraphics[width=\textwidth]{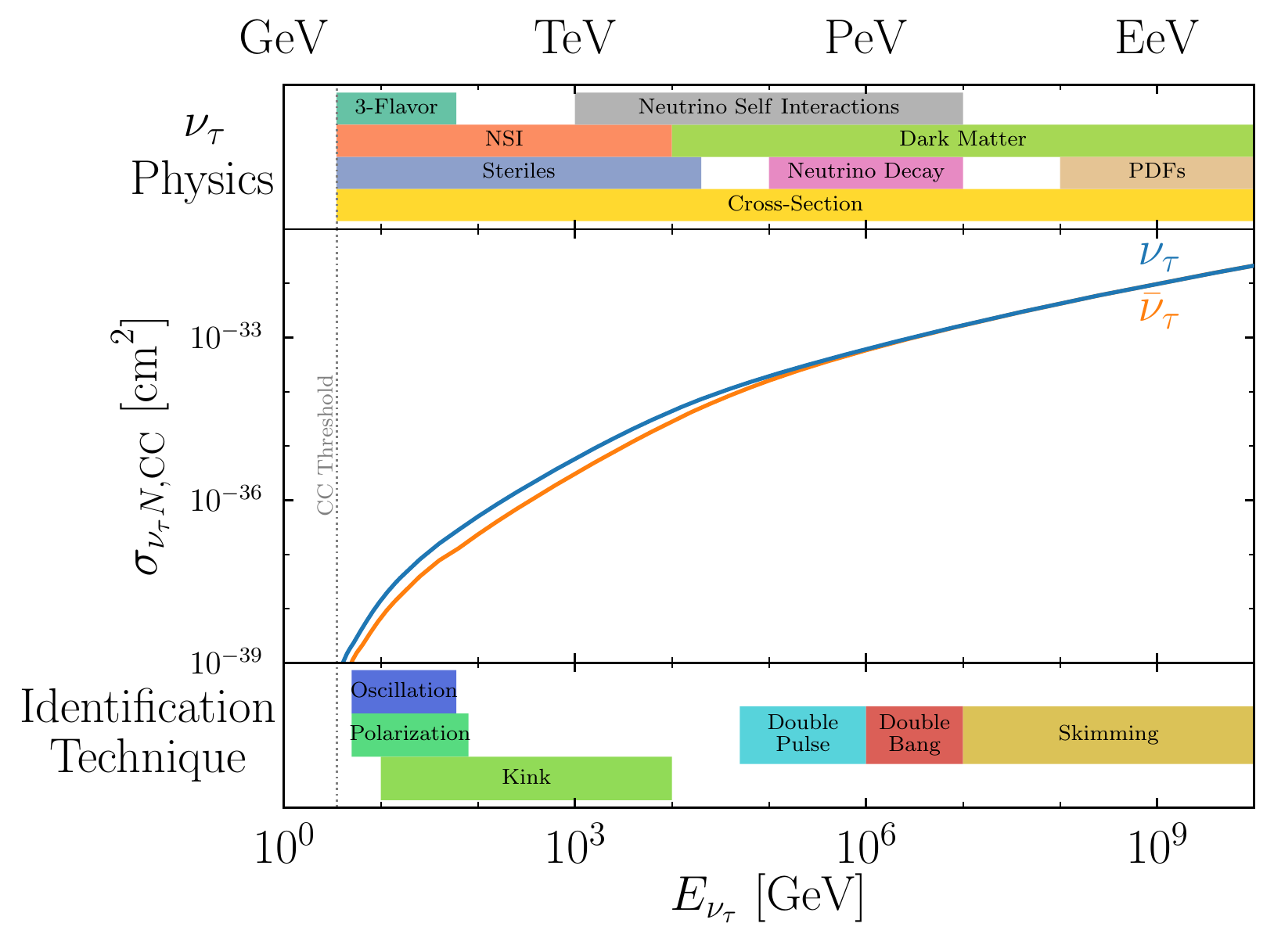}
\caption{\textbf{Top}: An overview of the different physics topics probed by tau neutrinos and the relevant energy scale in tau neutrino energy.
\textbf{Middle}: The tau (anti-)neutrino charged-current cross section with nucleons.
The tau lepton production threshold is denoted on the left.
\textbf{Bottom}: The energy ranges of different tau neutrino detection techniques.
Note that there are additional lower energy probes of tau neutrinos via neutral current scattering at e.g.~SNO combined with atmospherics and unitarity not shown here.
}
\label{fig:nu_tau_range}
\end{figure}

\subsection{Tau Neutrino History}

\subsubsection{Theory}
The existence of a third charged lepton was suggested in 1971 \cite{Tsai:1971vv} and, once it was discovered in 1975 \cite{Perl:1975bf}, it was clear that there should be an associated tau neutrino \cite{Altarelli:1977vp}.
Beyond weak interactions and charged tau lepton physics, tau neutrinos also participate in the phenomenon of neutrino oscillations which provides a significant amount of our knowledge of tau neutrinos.
Two-flavor neutrino oscillations with only electron and muon neutrinos were first discussed somewhat indirectly in 1957 \cite{Pontecorvo:1957cp} and they began to be fleshed out in the 1960s \cite{Maki:1962mu,Pontecorvo:1967fh}.
Three flavor oscillations including tau neutrinos were discussed as early as 1971 \cite{Pontecorvo:1971dc} with further discussions continuing in the late 1970s \cite{Mann:1976mp,Nussinov:1976uw}.
Around the same time it was also realized that neutrino oscillations with three flavors could be a source of CP violation \cite{Wolfenstein:1978uw} -- a topic which remains unresolved.

Given the theoretical landscape at the time, there was an interest in determining how to identify a third neutrino associated with the new heavy charged lepton, a topic that continues to be investigated to today and in coming years.
In the late 1970s, several methods to detect tau neutrinos produced in a hypothetical beam dump experiment were suggested.
One method was to identify a tau neutrino charged current interaction by looking for the signature of tau decaying to muons with significantly more transverse and missing momentum than expected from a muon neutrino interaction \cite{Albright:1978ni}.
Several additional tests were proposed which involved comparing neutral current to charged current ratios and looking for two hadronic showers with missing transverse momentum \cite{Albright:1979vd,Albright:1980ba}.

In the context of atmospheric tau neutrino appearance, it was realized that tau neutrinos could be identified even without individual particle identification \cite{Stanev:1999ki}.
This leverages the tau lepton production threshold, lower cascade energy due to missing energy from tau decays, and the different inelasticity distributions of CC and NC interactions.
This was recently extended to the case where in addition to making no assumptions about any of the oscillation parameters, no assumptions about unitarity were made either, showing that tau neutrino appearance can always be differentiated from electron neutrino appearance \cite{Denton:2021rsa}.

In addition to developing experimental probes of tau neutrinos, it is important to understand the important role they play in many new physics searches. 
Various directions have been explored, such as sterile neutrinos mixing with tau neutrinos \cite{NOMAD:2001xxt,CHORUS:2007wlo,Wendell:2014dka,IceCube:2017ivd,Blennow:2018hto,Dentler:2018sju} and new non-standard neutrino interactions in the tau neutrino sector \cite{Super-Kamiokande:2011dam,Mocioiu:2014gua,Coloma:2017egw,Denton:2018xmq,Coloma:2019mbs,Denton:2020uda,Chatterjee:2020kkm,Denton:2020hop,IceCube:2022ubv}.
In addition, there is an interesting region of viable parameter space for neutrino self interactions in the tau neutrino sector which modifies the evolution of the early universe and may address \cite{Kreisch:2019yzn,Blinov:2019gcj,Creque-Sarbinowski:2020qhz,Esteban:2021tub} the Hubble tension \cite{Kreisch:2019yzn} or relax constraints on inflation models \cite{Barenboim:2019tux}.
Lepton flavor universality seems to be a good symmetry, but we do not have a deeper understanding of why this is so. It is thus crucial to look for new physics hints in every corner of particle physics. Tau neutrino measurements offer a prime place for searches, as current uncertainties are very large and are expected to improve significantly.

Tau neutrinos also have an important role in model building, for example third generation models or models which rely on anomaly free combinations of lepton number present intriguing targets for tau neutrino experiments \cite{Allanach:2018lvl, Babu:2017olk,Altmannshofer:2015mqa,Altmannshofer:2016jzy,Baek:2001kca,Altmannshofer:2016brv,Amaral:2020tga,Amaral:2021rzw}.

\subsubsection{Experimental discoveries}
Using a high resolution emulsion detector and neutrinos produced from an 800 GeV proton beam producing $D_s$ mesons which sometimes decay to tau neutrinos, in 2000 DONuT detected 4 tau neutrino candidate events on a background of 0.34 by looking for kinks in tracks \cite{Kodama:2000mp}.
This detection was the first direct identification of tau neutrinos.
DONuT's final results presented in 2007 include 9 candidate tau neutrino events consistent with the expectation of 10 events \cite{DONuT:2007bsg}.

In 2001 and 2002, SNO reported the detection of a flux of non-electron neutrinos from the Sun.
They compared charged current (CC) events composed entirely of electron neutrinos, elastic scattering (ES) events composed mostly of electron neutrinos with some other flavors, and neutral current (NC) flavor blind events to confirm that electron neutrinos compose only $\sim\frac13$ of the total solar neutrino flux at $E\sim10$ MeV \cite{SNO:2001kpb,SNO:2002hgz}.
These results, combined with Super-Kamiokande's (SuperK) 1998 measurement that muon and tau neutrino mixing is quite high \cite{Kamiokande-II:1992hns}, confirm that a significant fraction of the ES and NC events detected by SNO were tau neutrinos.

After the direct detection of directly produced tau neutrinos by DONuT and the indirect detection of oscillated tau neutrinos by SNO, the next channel to investigate was direct detection of tau neutrinos from oscillations.
OPERA used a beam of muon neutrinos produced at CERN with an average neutrino energy of 17 GeV and an emulsion detector 730~km away in Gran Sasso to look for neutrino events with a kink to identify the tau lepton decay.
In 2010 OPERA reported the first candidate tau neutrino appearance event from oscillations \cite{Agafonova:2010dc} and in 2018 they reported their final results with $6.1\sigma$ evidence for tau neutrino appearance with 10 observed candidate tau neutrino events on a prediction of 6.8 plus 2.0 background events \cite{OPERA:2018nar}.

Atmospheric neutrino experiments have traditionally focused on the muon neutrino disappearance channel, but due to the large atmospheric mixing, nearly all of the muon neutrinos that disappear are tau neutrinos which can be detected if their energies are above the tau lepton threshold.
In 2017 SuperK used a neural network looking at tau leptons decaying hadronically to identify tau neutrino appearance at $4.6\sigma$ with 291 candidate events \cite{Super-Kamiokande:2017edb}.
In 2019 IceCube/DeepCore used a combined fit to muon neutrino disappearance and tau neutrino appearance to identify tau neutrinos in their data set at $3.2\sigma$ with 1804 CC and 556 NC candidate events \cite{IceCube:2019dqi}.
The various detections of tau neutrinos can be parameterized in terms of the cross section normalization over the relevant energy range as shown in Fig.~\ref{fig:normalization comparison}.

\begin{figure}
\centering
\includegraphics[width=0.6\textwidth]{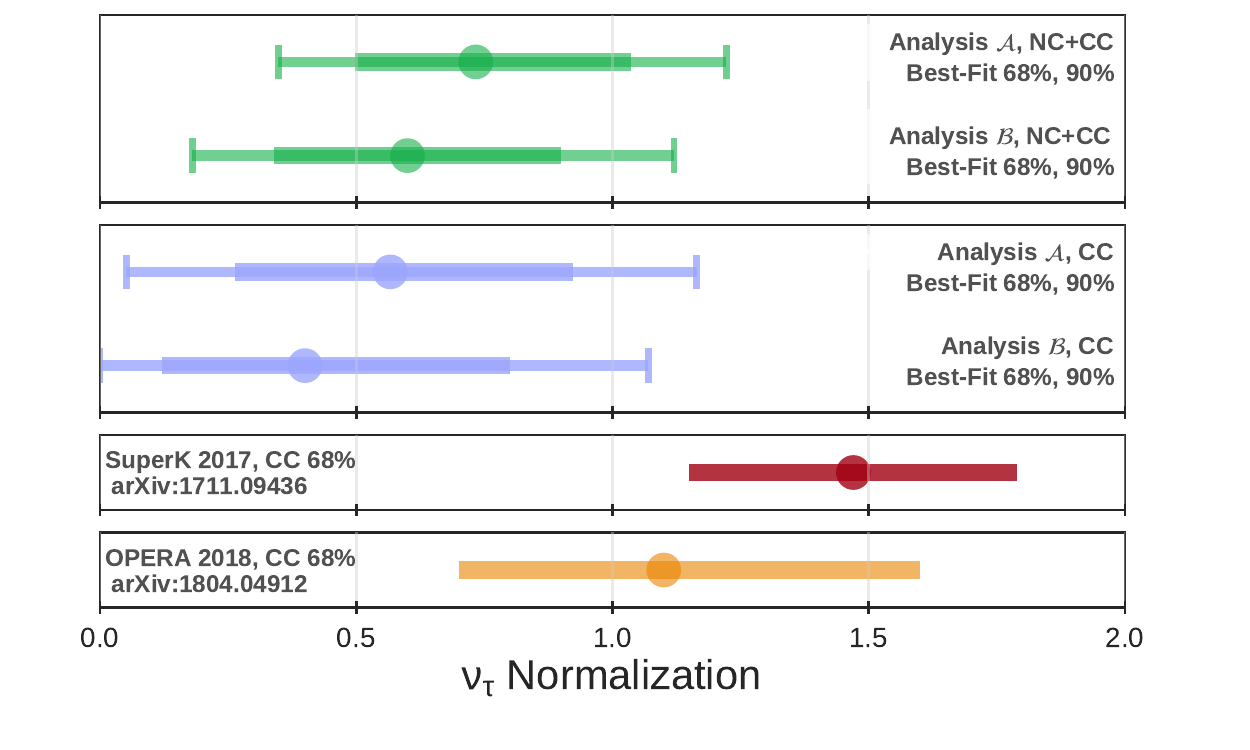}
\caption{The normalization of the weighted average of the tau neutrino cross section compared to the Standard Model expectation.
The top four blue and green lines are from IceCube/DeepCore and contain two different analyses and with/without NC contribution \cite{IceCube:2019dqi}.
The red line is from SuperK \cite{Super-Kamiokande:2017edb} and the orange line is from OPERA \cite{OPERA:2018nar}.
Figure from \cite{IceCube:2019dqi}.}
\label{fig:normalization comparison}
\end{figure}

Recently, IceCube analyzed their high energy starting event data for astrophysical tau neutrino events.
They used two main signatures to differentiate tau neutrinos from electron neutrinos: double cascades \cite{Learned:1994wg} and double pulses \cite{Cowen:2007ny} where the initial hadronic shower can be separated from the hadronic or electron decay of the tau lepton spatially or temporally, respectively.
The double pulse \cite{Meier:2019ypu,Wille:2019pub} and double cascade \cite{IceCube:2020abv} analyses each identified the same two candidate events.
Additional possible channels involving one of the two hadronic showers occurring outside the detector or muonic decays of the tau lepton have thus far evaded detection.
The unfolded tau neutrino flux from these analyses is consistent with other astrophysical flux measurements and a 1:1:1 flavor ratio as expected from lepton flavor universality and terrestrially measured oscillation parameters, albeit with fairly large uncertainties.

The history of reported tau neutrino detections is shown in Fig.~\ref{fig:NuTauEvents} showing the exponential growth in tau neutrino detections over the last two decades.
The cumulative number of detected events has grown at a rate of doubling once every two years and that rate is expected to continue for the foreseeable future.

\begin{figure}
\centering
\includegraphics[width=\textwidth]{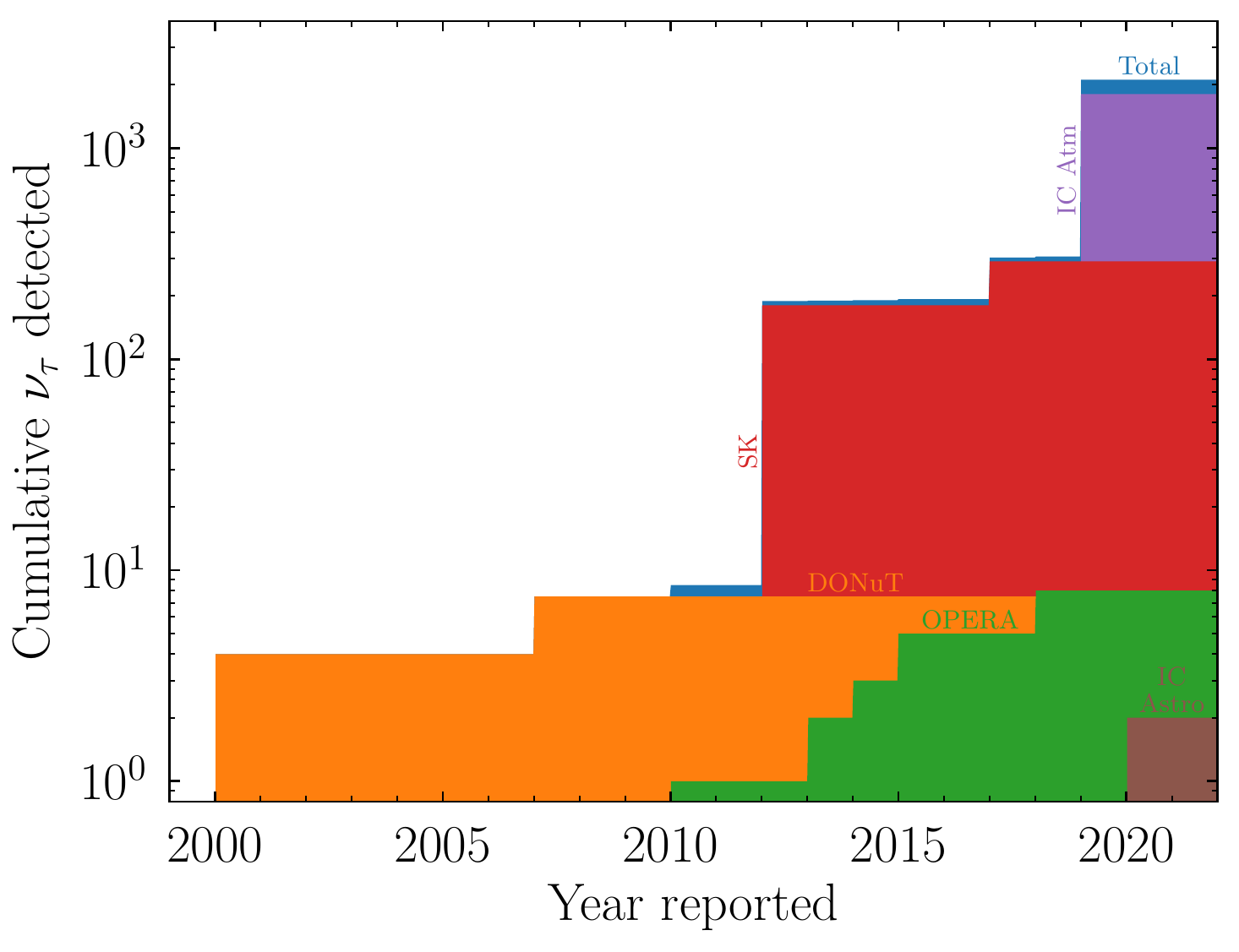}
\caption{The cumulative number of tau neutrinos detected (blue) including contributions from DONuT (orange), OPERA (green), SuperK atmospherics (red), IceCube atmospherics (purple), and IceCube astrophysical (brown).
The doubling rate is about once per two years since four events in 2000.}
\label{fig:NuTauEvents}
\end{figure}

\subsection{Tau Neutrino Motivations}
Given the existing body of literature on tau neutrino theory and the data sets containing tau neutrinos, we believe there is a strong case to significantly expand our efforts to study these particles.
This motivation comes from five main directions.
\begin{enumerate}
\item \textbf{Measure properties of SM particles}: Determining the cross sections and oscillation parameters of each known fermion has been at the center of the particle physics community's efforts for decades; it is time to now turn our efforts to tau neutrinos for which measurements lag behind those of other particles.
\item \textbf{Testing the three flavor picture}: It is necessary to fully explore the oscillation phenomenon and neutrino oscillations provide an excellent place to look for additional instances of new physics.
This requires additional sources of tau neutrinos for oscillations, the necessary detectors and reconstruction tools to identify tau neutrinos, the phenomenology to cast the results in terms of both standard and new physics scenarios, and models to put the new physics scenarios in a broader context.
\item \textbf{Upcoming oscillation experiments}: With the advent of DUNE for long-baseline, Hyper-Kamiokande, IceCube, KM3NeT, and Baikal-GVD for atmospherics, we will have a number of experiments that, while not designed for tau neutrino physics, will be sensitive to tau neutrino physics.
It is essential that the community provides input on how to maximize the secondary physics cases of these experiments.
\item \textbf{Upcoming high energy neutrino experiments}: A large number of experiments designed to detect the neutrino flux in the $E\gtrsim100$ PeV range are currently being proposed and constructed, see \ref{sec:he experiments}.
While the primary motivation of many of these experiments is astrophysics, due to their unique sensitivity to tau neutrinos, it is vital to determine what particle physics can be extracted from them, ideally while still in the planning phase, such that the design can be optimized for maximum physics output.
\item \textbf{Existing anomalies}: A number of channels show moderately significant evidence for a disagreement between second and third generation leptons in $b$-decays \cite{Bifani:2018zmi}, see \ref{sec:flavor}.
In addition, there is an interesting hint of lepton flavor universality violation in electron and muon anomalous magnetic dipole moments \cite{Muong-2:2021vma,Muong-2:2006rrc}.
To fully probe lepton flavor universality, precise measurements of all three generations are required.
Finally, ANITA has reported the detection of several anomalous events in their search for ultra-high energy neutrinos and cosmic rays \cite{ANITA:2016vrp,ANITA:2018sgj,ANITA:2020gmv} that, while observationally consistent with tau neutrinos, are in strong tension with existing limits and Standard Model scenarios for tau neutrinos~\cite{Romero-Wolf:2018zxt,ANITA:2021xxh,Safa:2019ege}.
\end{enumerate}
\pagebreak
\section{Theoretical and Phenomenological Interests}
\label{sec:theory}
There is large theoretical and phenomenological interest in tau neutrinos as their  peculiarities
open the possibility to explore new physics scenarios not accessible with other neutrino flavors.
In fact, tau neutrinos could be related to the solution of various anomalies including the  b-quark anomalies, the Hubble tension which can be alleviated by the introduction of self-interacting neutrinos, and the anomalous ANITA events.
Furthermore, neutrinos of all flavor can provide a window to new physics, therefore it is crucial to test the coupling or mixing of new particles to all neutrino flavors with a special focus on tau neutrinos where the constraints are often the weakest.
While the tau neutrino has been less studied than the other neutrino flavors in the past,  upcoming tau neutrino experiments open the possibility to improve our knowledge in many directions of parameter space.
A thorough test of the tau neutrino sector will shine light on some of the most motivated new physics scenarios which provide a portal between the SM and a new physics sector like  the presence of sterile neutrinos or additional bosons which lead to new neutrino interactions.
Sterile neutrinos could even be related to some of the strongest evidence for physics beyond the SM like the quest for neutrino masses or the matter asymmetry of the Universe which could be explained in the simplest model by the CP violating decays of sterile neutrinos.
Furthermore,  Dark Matter (DM)  could couple to neutrinos thereby connecting two elusive sectors to each other.
In the following we will discuss the current and expected future knowledge of new physics effects in the neutrino sector and identify different experimental goals to probe new physics scenarios.

\subsection{Tau Neutrinos and the Standard Neutrino Paradigm}
\label{sec:3fl_paradigm}
Precision measurements of the oscillation parameters can provide a thorough test of the standard three-flavor neutrino paradigm and yield valuable insights into new physics scenarios like unitarity violation. Therefore, it is crucial to achieve an  improved precision at next-generation accelerator and atmospheric neutrino experiments to $\nu_\mu\to\nu_\tau$ appearance. Together with neutrino scattering experiments and astrophysical neutrino experiments sensitive to tau neutrinos, these independent measurements can serve as crucial consistency checks of the neutrino oscillation paradigm.

\subsubsection{Precision measurement of oscillation parameters}

Given the experimental difficulties of producing and observing the interactions of tau neutrinos, it is no surprise that all current knowledge of neutrino mixing comes nearly completely from measurements involving muon and electron neutrinos.
Within the standard mixing paradigm, due to the assumed unitarity of the leptonic mixing matrix, measurements involving $\nu_e$ and $\nu_\mu$ inform our expectations of $\nu_\mu \to \nu_\tau$ appearance.
Nonetheless, existing and upcoming measurements of $\nu_\mu$ oscillating into $\nu_\tau$ from OPERA~\cite{OPERA:2018nar}, Super-Kamiokande~\cite{Super-Kamiokande:2017edb}, and IceCube~\cite{IceCube:2019dqi} are pushing tau neutrino measurements into the future.

Specifically, for long-baseline and atmospheric neutrino oscillations, current knowledge of neutrino mixing~\cite{Esteban:2020cvm} implies that the amplitude of $\nu_\mu \to \nu_\tau$ appearance is approximated by $4|U_{\mu 3}|^2 |U_{\tau 3}|^2 \approx 0.95$, leading to large probabilities. The current measurements from OPERA, Super-Kamiokande, and IceCube are consistent with this expectation, with relatively large uncertainties.
In this sense the $\nu_\mu\to\nu_\tau$ appearance probability can be determined by measuring the other channels without identifying tau neutrinos, see e.g.~by using unitarity \cite{Martinez-Soler:2021sir}.

The next generation of oscillation experiments, specifically DUNE~\cite{DeGouvea:2019kea,DUNE:2020ypp} and the IceCube Upgrade~\cite{Ishihara:2019aao}, offer improved precision on measuring $\nu_\mu \to \nu_\tau$ appearance. 
Each of the upcoming experiments has the capability to observe $\nu_\tau$ appearance and constrain the associated oscillation parameters, with sensitivity mostly to $\sin^2\theta_{23}$ and $\Delta m_{31}^2$, the atmospheric mixing angle and mass-squared-splitting, respectively. While this sensitivity will be weaker than current (and expected future) measurements, for example by INO \cite{Senthil:2022tmj}, these independent measurements can serve as a consistency check of the three-neutrino paradigm, and constrain, for instance, whether the leptonic mixing matrix is truly unitary~\cite{Parke:2015goa,Hu:2020oba,Ellis:2020hus,Denton:2021mso} (see also the following  subsection).

\subsubsection{Testing the standard neutrino paradigm: unitarity violation}

A crucial test of the standard neutrino paradigm involves probing the unitarity of the leptonic mixing matrix. 
Non-unitary mixing matrices arise in many extensions of the SM like models with neutrinos propagating in extra dimensions \cite{ArkaniHamed:1998vp,ArkaniHamed:1998sj,Bhattacharya:2009nu} and most 
notably in extensions of the SM introducing new, heavy neutrinos potentially connected to neutrino mass generation \cite{Minkowski:1977sc,Schechter:1980gr,Foot:1988aq} or to the solution to the yet unexplained observation of the matter-antimatter asymmetry of the Universe \cite{Fukugita:1986hr}. These scenarios lead to \emph{apparent} low energy unitarity violation (UV) as the mixing matrix of the full theory, including the often kinematically inaccessible sterile neutrinos, is unitary.

While the electron and muon  rows of the leptonic mixing matrix are well constrained due to the large statistics of the electron neutrino disappearances samples from reactor experiments and  muon neutrino disappearance data from long baseline experiments \cite{Ellis:2020ehi,Ellis:2020hus,Hu:2020oba,Forero:2021azc,Agarwalla:2021owd}
the tau row is currently comparably worse constrained and   allows for sizable deviations from unitarity. In order to thoroughly probe the standard neutrino paradigm, all  9 unitarity conditions, following from $UU^\dagger=1$, need to be tested. In fact out of the 9 conditions on the row normalizations and row unitarity triangles, 5 involve tau row matrix elements and all conditions on the column normalization and column unitarity triangles depend on tau row matrix elements.
Therefore it is crucial to improve our knowledge on the tau row  in the future. 

As UV affects weak interactions there are two main ways to test the unitarity of the leptonic mixing matrix: with electroweak precision observables  and  with oscillations. Additionally, if UV arises due to the presence of sterile neutrinos they can be searched for directly at experiments (for a recent review of sterile neutrino bounds across many energy scales see \cite{Bolton:2019pcu}).
In \cite{Antusch:2006vwa,Fernandez-Martinez:2015hxa,Fernandez-Martinez:2016lgt} UV constraints from electroweak precision data have been derived for sterile neutrinos with mass above the weak scale.
A mild preference  between 1 and 2$\sigma$ for non-zero
heavy-active mixing of order $\sim$ 0.03-0.04 has been found in the electron and tau sector \cite{Fernandez-Martinez:2016lgt}. At $2\sigma$ upper bounds are found where the bounds involving tau neutrinos are the least stringent \cite{Fernandez-Martinez:2016lgt}.

Ref.~\cite{Denton:2021mso} investigated the impact of various oscillation and scattering constraints on the tau row. 
In particular, tau row unitarity information arises from atmospheric tau neutrino appearance, astrophysical tau neutrino appearance, and charged current scattering experiments as well as neutral current measurements together with long baseline tau neutrino appearance and atmospheric muon neutrino disappearance data have been used to constrain UV in the benchmark scenario of one kinematically accessible sterile neutrino with averaged out oscillations (this generally  applies to sterile masses $m_N\in [10~\text{eV},15$ MeV]), and kinematically inaccessible steriles with masses above 40 MeV. As it has been shown in \cite{Fong:2016yyh,Blennow:2016jkn,Fong:2017gke} for small sterile mixing angles these two scenarios provide similar constraints up to 4th order in the sterile mixing angles. 
Of importance is also NC data from CEvNS  and at long baseline experiments  \cite{MINOS:2017cae,NOvA:2017geg,NOvA:2021smv,T2K:2019efw,Forero:2021azc} which, even though no tau neutrinos are identified, still provide some constraints on the tau matrix elements.
The relative importance of current and future results on the tau row matrix elements are shown in Fig.~\ref{fig:uti53} for kinematically inaccessible steriles with similar constraint for kinematically accessible but averaged out states.

With the arrival of the next generation of neutrino oscillation experiments which can also be used for direct searches for sterile neutrinos, and more precise electroweak precision data from future colliders \cite{Fan:2014vta,LCCPhysicsWorkingGroup:2019fvj} the standard three-flavor neutrino paradigm can be exhaustively tested.
This will allow for more  insights into new physics scenarios which predict UV and can guide future neutrino research. Most importantly, testing the unitarity of the leptonic mixing matrix allows for one to test one of the most motivated extensions of the SM, the existence of sterile neutrinos.
Therefore it is of utmost importance that planned experiments reach and deliver their expected tau neutrino output and interpret their results in a UV way. To obtain a complete global picture of the constraints on the  unitarity of the leptonic mixing matrix a combination of all constraints from electroweak precision data, from direct searches, and from oscillations needs to be conducted in a statistically sound way in the future.
In addition, as UV often leads to the zero-distance effect, accurate flux predictions independent of near detector measurements are crucial.

\begin{figure}
\centering
\includegraphics[width=\textwidth]{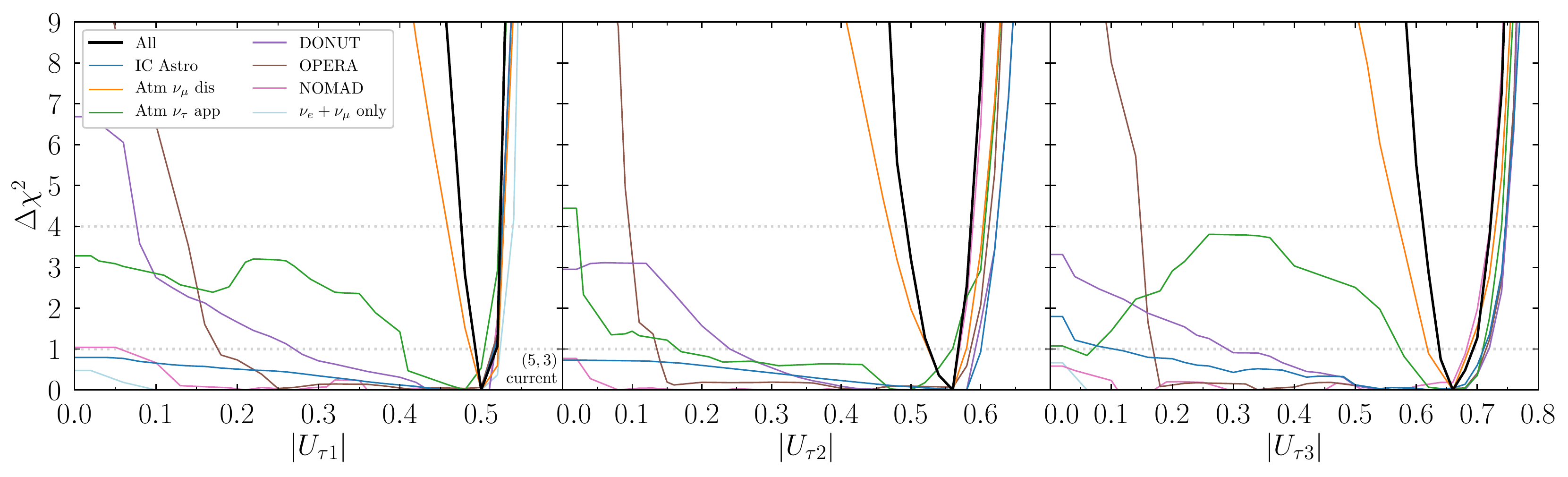}
\includegraphics[width=\textwidth]{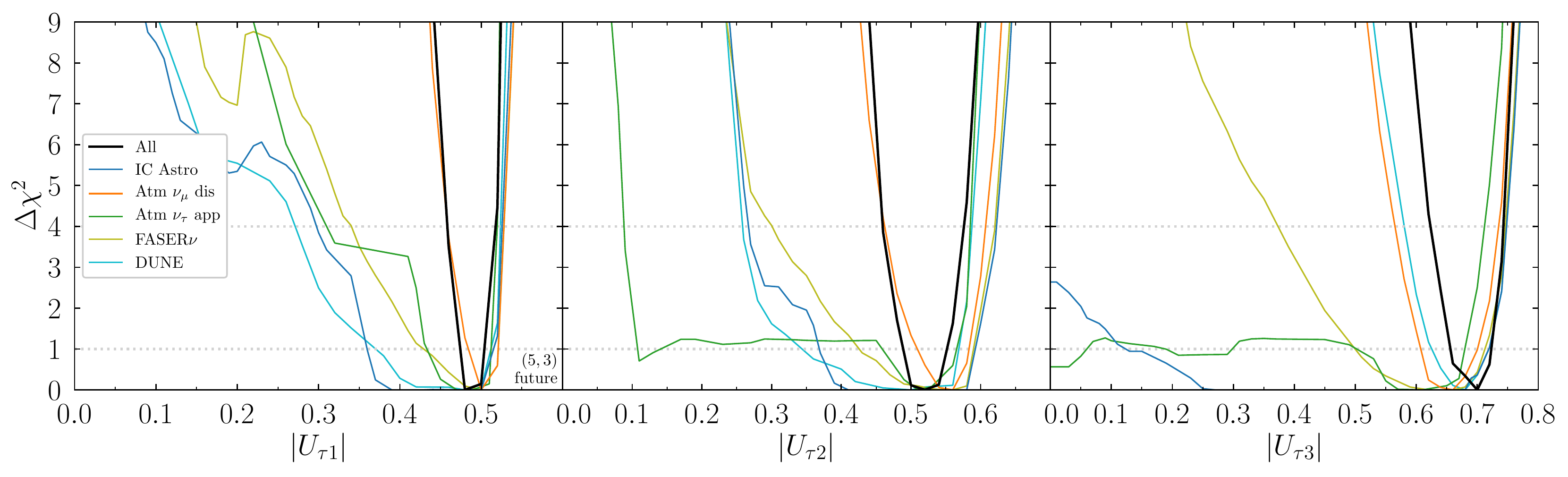}
\caption{The relative information on the individual matrix elements in the tau row from \cite{Denton:2021mso} in the case of kinematically inaccessible steriles using the currently available data in the upper panel, the forecasted data in the lower panel, as well as priors on the electron and muon row elements from \cite{Ellis:2020ehi,Ellis:2020hus}. The different colors represent different data sets which have been included in addition to the constraints from unitarity using priors from the electron and muon row. The black lines include all data sets.}
\label{fig:uti53}
\end{figure}

\subsection{Tau Neutrinos and Sterile Neutrinos}
The neutrino portal, a renormalizable operator which is allowed by all SM gauge and accidental symmetries and just requires the introduction of  sterile neutrinos to the SM, provides one of the best motivated extensions of the SM.  The neutrino portal could be also connected to two of the most pressing open questions of the SM: the quest for the neutrino mass mechanism where one of the best motivated solutions is the seesaw mechanism, and the matter antimatter asymmetry of the Universe via the leptogenesis mechanism.
Depending on the mass scale and active-sterile mixing angles sterile neutrinos have a different phenomenology, therefore we need to use different avenues to probe their existence. As the sterile mixing with tau neutrinos is currently the least constrained active-sterile mixing angle it crucial  to improve these constraints in our search for new physics.

\subsubsection{Light sterile neutrinos}
In  models with one sterile neutrino the 
mixing matrix is a $4\times4$ unitary matrix with 9 parameters\footnote{There are additionally three Majorana phases; one more than in the three-flavor case.
Oscillation experiments are insensitive to the effects of these phases which are suppressed by $(m_\nu/E_\nu)^2\lesssim10^{-15}$.}:
6 mixing angles and 3 oscillation-relevant CP-violating phases.
In addition to the mixing angles $\theta_{12}$, $\theta_{23}$, 
$\theta_{13}$ and one CP-violating phase 
$\delta \equiv \delta_{1}$, the
3+1 model introduces three new mixing angles  
$\theta_{14}$, $\theta_{24}$, $\theta_{34}$ and two phases 
$\delta_{2}$ and $\delta_{3}$.
Furthermore, the presence of a fourth massive neutrino leads to a
new mass-squared difference: $\Delta m^2_{41}$ (or equivalently
$\Delta m^2_{42}$ or $\Delta m^2_{43}$).
There are strong bounds on the mixing of   a sterile neutrino lighter than few 100 MeV with $\nu_\mu$ and $\nu_e$ from various neutrino experiments. However, the bound on the mixing of $\nu_\tau$ with  sterile neutrinos is less constrained. 
In  experiments that study accelerator neutrinos, the
mixing with sterile neutrinos could decrease the number of charged-current
and neutral-current interactions or cause anomalous appearance of 
$\nu_{\tau}$ or $\nu_\mathrm{e}$ events. An
example of the 3+1 oscillation probabilities in the accelerator
neutrino experiments, as a function of $L/E$,
for $\Delta m^2_{41}=6~\mathrm{eV}^2$, $\theta_{14}=0.2$, $\theta_{24}=0.15$, 
$\theta_{34}=0.6$, $\delta_{i}=0$ 
is shown in Fig.~\ref{fig:avg_tau}.
\begin{figure}
\begin{center}
\includegraphics[width=0.85\textwidth,viewport= 12 276 510 522,clip=]{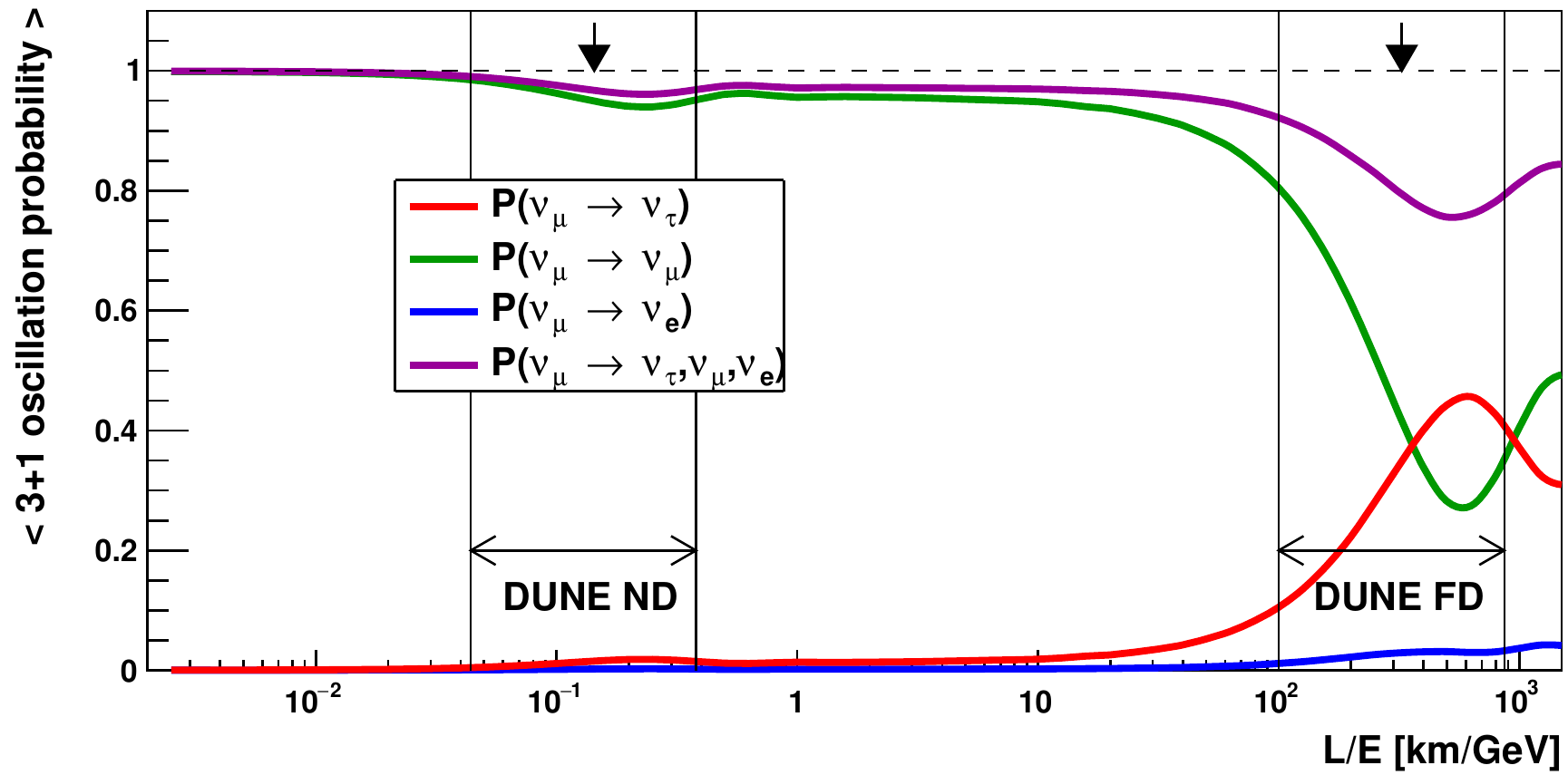}
\caption[Neutrino oscillation probabilities in the model with one sterile
neutrino]
{Neutrino oscillation probabilities in the model with one sterile neutrino
as a function of $L/E$. 
Black arrows indicate $L/E$ values corresponding to the 
maximum of DUNE tau-optimized neutrino flux \cite{DUNE:2020lwj}. 
The vertical lines mark the regions of $L/E$ probed by the near and far 
DUNE detectors. 
Predictions for $\Delta m^2_{41}=6~\mathrm{eV}^2$,
$\theta_{14}=0.2$, $\theta_{24}=0.15$, $\theta_{34}=0.6$
and 
$\delta_{i} = 0$.
Values of $\Delta m^2_{32}$, $\Delta m^2_{21}$,
$\theta_{12}$, $\theta_{13}$ and
$\theta_{23}$ from~\cite{Esteban:2020cvm}. 
} 
\label{fig:avg_tau}
\end{center}
\end{figure}

If there are only three neutrino flavors, near detectors located close
to the source of accelerator neutrinos, measure 
the original neutrino spectrum undistorted by the oscillations.
In such a case, the only source of tau neutrinos are leptonic
decays of the  $\mathrm{D_s}$ mesons produced in the interactions
of protons with the target, $\mathrm{D_s \to \tau+ \nu_{\tau}}$,
followed by the decays $\tau \to \nu_{\tau} + \mathrm{X}$. 

The $\nu_{\tau}$ component of the beam
from prompt $D_s$ decays in CHORUS and NOMAD
was evaluated to be extremely small~\cite{Gonzalez-Garcia:1996mtp,VandeVyver:1996qc}
in comparison to $\nu_{\mu}$ and $\nu_e$ fluxes
and found to be:
\begin{equation}
\frac{N_{\mathrm{CC} \nu_{\tau}}}{N_{\mathrm{CC} \nu_{\mu}}} \sim 3.5-4.9 \cdot 10^{-6}
\end{equation}
or negligible for proton beam of energy 120 GeV.

Therefore, the presence of $\nu_{\tau}$ in the near detectors
would be an interesting signature of sterile neutrinos,
but not for all parameters values.
The oscillation probability at short distances 
from the neutrino source would be modified if $\Delta m^2_{41}$
was not very small ($\Delta m^2_{41} \gtrsim 1~\mathrm{eV}^2$).

In the near detectors the probability of 
$\nu_{\mu} \to \nu_{\tau}$
transition in the model with one sterile neutrino
can be approximated by the formula:
\begin{equation}
\begin{split}
\mathrm{P}_{\nu_{\mu} \to \nu_{\tau}}(L,E) & 
\simeq 4|U_{\mu 4}|^2|U_{\tau 4}|^2
\sin^2 \left ( \frac{\Delta m^2_{41}L}{4E} \right ) \\
 & = \cos^4\theta_{14}\sin^2 2\theta_{24}\sin^2\theta_{34}
\sin^2 \left ( \frac{\Delta m^2_{41}L}{4E} \right ). \\
\label{eq:PSBLmutau2nu}
\end{split}
\end{equation}

It can also be written in the two-flavor form:
\begin{equation}
\mathrm{P}_{\nu_{\mu} \to \nu_{\tau}}(L,E)  
= \sin^22\theta_{\mu\tau}
\sin^2 \left ( \frac{\Delta m^2_{41}L}{4E} \right )
\label{eq:PSBLmutau2nushort}
\end{equation}
where
\[ \sin^22\theta_{\mu\tau} \equiv \cos^4\theta_{14}\sin^22\theta_{24}\sin^2\theta_{34} \]
Thus anomalous appearance of $\nu_{\tau}$ gives access to
the least constrained parameter of 3+1 model, $\theta_{34}$,
but the strength of the method depends
on the value of $\theta_{24}$ parameter.
Limits and sensitivities for the parameters describing
$\nu_{\tau}$ appearance are therefore presented in   
the $\Delta m^2_{41}$ vs $\sin^22\theta_{\mu\tau}$ plane.

Compilation of 90\%~C.L.~limits from the experiments that looked for $\nu_{\tau}$ appearance is presented  in Fig.~\ref{fig:sensitivities}.
Limits from OPERA's $\nu_\tau$ appearance exist, but are quite weak \cite{Meloni:2019pse}.
Limits are compared to 90\%~C.L.~sensitivities obtained with the full MINOS+ simulation and reconstruction, for $\tau \to \mu \nu_{\mu} \nu_{\tau}$ selection.
The influence of reduced systematics and improved signal/background ratio is also demonstrated.
The MINOS+ beam is similar to the DUNE $\nu_{\tau}$ optimized beam, see section \ref{sssec:dune accelerator}.

\begin{figure}
\begin{center}
\includegraphics[width=0.5\textwidth,viewport=0 260 258 522,clip=]
{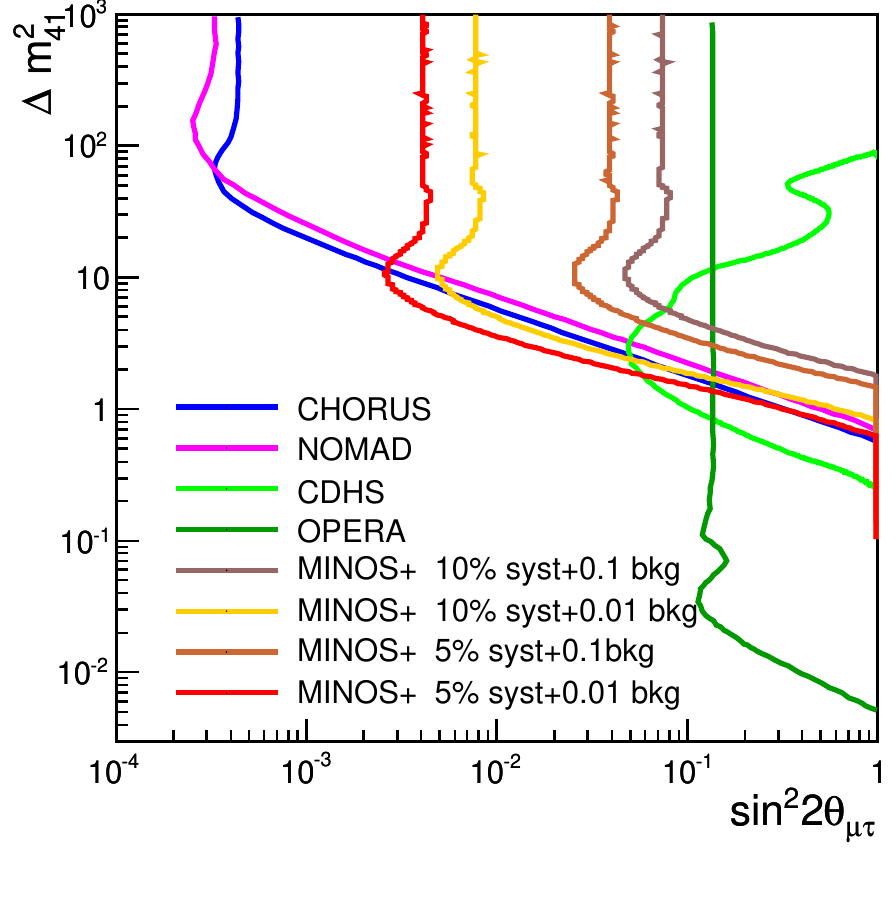}
\caption[Comparison of limits from 
CHORUS, NOMAD, CDHS and OPERA and MINOS+ sensitivities]
{ Compilation of 90\%~C.L.~limits
in the ($\sin^22\theta_{\mu\tau},\Delta m^2_{41}$) plane
from CHORUS~\cite{CHORUS:2007wlo},
NOMAD~\cite{NOMAD:2001xxt}, 
CDHS~\cite{Dydak:1983zq} and OPERA~\cite{OPERA:2015zci} experiments.
Comparison to 90\%~C.L.~MINOS+ sensitivities
for $\tau \to \mu \bar\nu_{\mu} \nu_{\tau}$ selection~\cite{grzelak_katarzyna_2018_1300962,grzelak_katarzyna_2018_WUW}.
Demonstration of the influence of
reduced systematics and improved signal/background ratio. Figure from \cite{grzelak_katarzyna_2018_1300962,grzelak_katarzyna_2018_WUW}. }
\label{fig:sensitivities}
\end{center}
\end{figure}

In addition to searching for steriles with long-baseline experiments atmospheric neutrino experiments provide strong constraints from atmospheric muon neutrino disappearance data. For active-sterile  mass splittings $\Delta m_{41}^2\gtrsim 10~\text{eV}^2$ IceCube \cite{IceCube:2020phf,IceCube:2020tka}, DeepCore \cite{IceCube:2017ivd} and Super-Kamiokande \cite{Super-Kamiokande:2014ndf} constrain the sterile mixing with tau neutrinos due to the presence of the  matter effect whereas for lighter sterile neutrinos only constraints on the mixing with muon neutrinos can be constraints making use of the matter resonance sterile neutrinos experience when crossing the Earth.
Combining the atmospheric neutrino data by IceCube and DeepCore with MINOS/MINOS+ and NO$\nu$A data and exploiting neutral current events measured by SNO, Ref.~\cite{Dentler:2018sju} finds as current bound 
\begin{equation}
|U_{\tau 4}|^2<0.13~(0.17) ~~{\rm at}~~ 90\%~{\rm C.L.} ~(99\%~{\rm C.L.})\,.
\end{equation}

Upcoming  atmospheric experiments like  IceCube-Gen2 \cite{IceCube-Gen2:2020qha} as well as by KM3NeT \cite{KM3NeT:2021uez} and Hyper-Kamiokande \cite{Hyper-Kamiokande:2018ofw} will improve this constraint. 
As the total
MINOS+ statistics corresponds to one-year of DUNE data taking, but with DUNE,
significantly better signal to background ratio is expected.
Therefore, the obtained sensitivities can be treated as lower limit of
what can be achieved in DUNE.

In general, it is found that an excellent a priori understanding of the systematics is crucial for improving accelerator probes of sterile neutrinos due to the non-trivial interplay of the near and far detectors.

\subsubsection{Heavy sterile neutrinos}
\label{sec:HNL}
For heavy neutral leptons (HNL) with masses
around $\sim 1$~GeV, there are tight bounds over the mixing between
the heavy states and the electron and muon neutrinos. In the case of
tau neutrinos, the difficulties in the production and the detection of
$\nu_{\tau}$ makes the bounds weaker. The strongest constraints comes
from DELPHI~\cite{DELPHI:1996qcc}, CHARM~\cite{Orloff:2002de} and
ArgoNeuT~\cite{ArgoNeuT:2021clc}.

Certain unique phenomenological signatures of HNLs exist in some regions of parameter space such as a double bang (DB) signal at IceCube \cite{Coloma:2017ppo,Atkinson:2021rnp}.
The first
bang is created by the a neutral current interaction of $\nu_{\tau}$ inside the
detector. In the same interaction, a HNL is up-scattered and
propagates through the detector. The decay of the sterile neutrino
into charged particles generates the second bang. The small mixing
between active and sterile states allows the propagation of
macroscopic distances~\cite{Coloma:2020lgy} with low initial
energies. The decay length depends on the mixing and the mass of the
heavy state as
\begin{equation}
  L_{{\rm lab}} = c \tau \gamma \beta  \simeq 10~{\rm m}~\left(\frac{10^{-2}}{|U_{\tau 4}|^{2}}\right)~\left(\frac{0.5~{\rm GeV}}{m_N}\right)^{5}~\left(\frac{E_{N}}{1~{\rm GeV}}\right)
  \label{eq:decayL}
\end{equation}
For energies around the GeV scale and neutrino masses of $\sim
0.5$~GeV and mixings on the order of $\sim 10^{-2}$, the decay length
is $\sim 10$~m which is enough to differentiate it from a single cascade provided they both happen in either water Cherenkov or LArTPC detectors.

The small background rate for DB signals allow us to define the
sensitivity region as the values of $|U_{\tau 4}|^2$ and $m_{N}$ with
one background DB event in 10 years. In Fig.~\ref{Fig:Rate}, we show the
sensitivity region using atmospheric neutrinos (left) and the beam
(right). The results indicate that IceCube can probe mixings on the
order of $|U_{\tau 4}|^2\sim 5 \times 10^{-5}$ for $m_{N}\sim
1.8$~GeV, improving the present bounds by one order of magnitude. DUNE
can also provide a complementary sensitivity to the present bounds
using the beam flux. For $m_{N}\sim 1$~GeV, DUNE would be able to
probe masses of $\sim 1$~GeV and mixing up to $|U_{\tau 4}|^2\sim
10^{-3}$. The sensitivity regions are limited by the detector volume
from below. Since the cross section is proportional to the mixing, to
smaller values it is necessary to use larger volumes. That explains why
using atmospheric neutrinos, IceCube shows a stronger sensitivity
compared to DUNE. On the left side, the smaller the mass of the
sterile neutrino the larger the decay length, therefore the region
is limited by detector size. For larger masses, the separation between
the cascade is smaller. The region is limited in the right side by
detector resolution.

\begin{figure}
  \includegraphics[width=.48\textwidth]{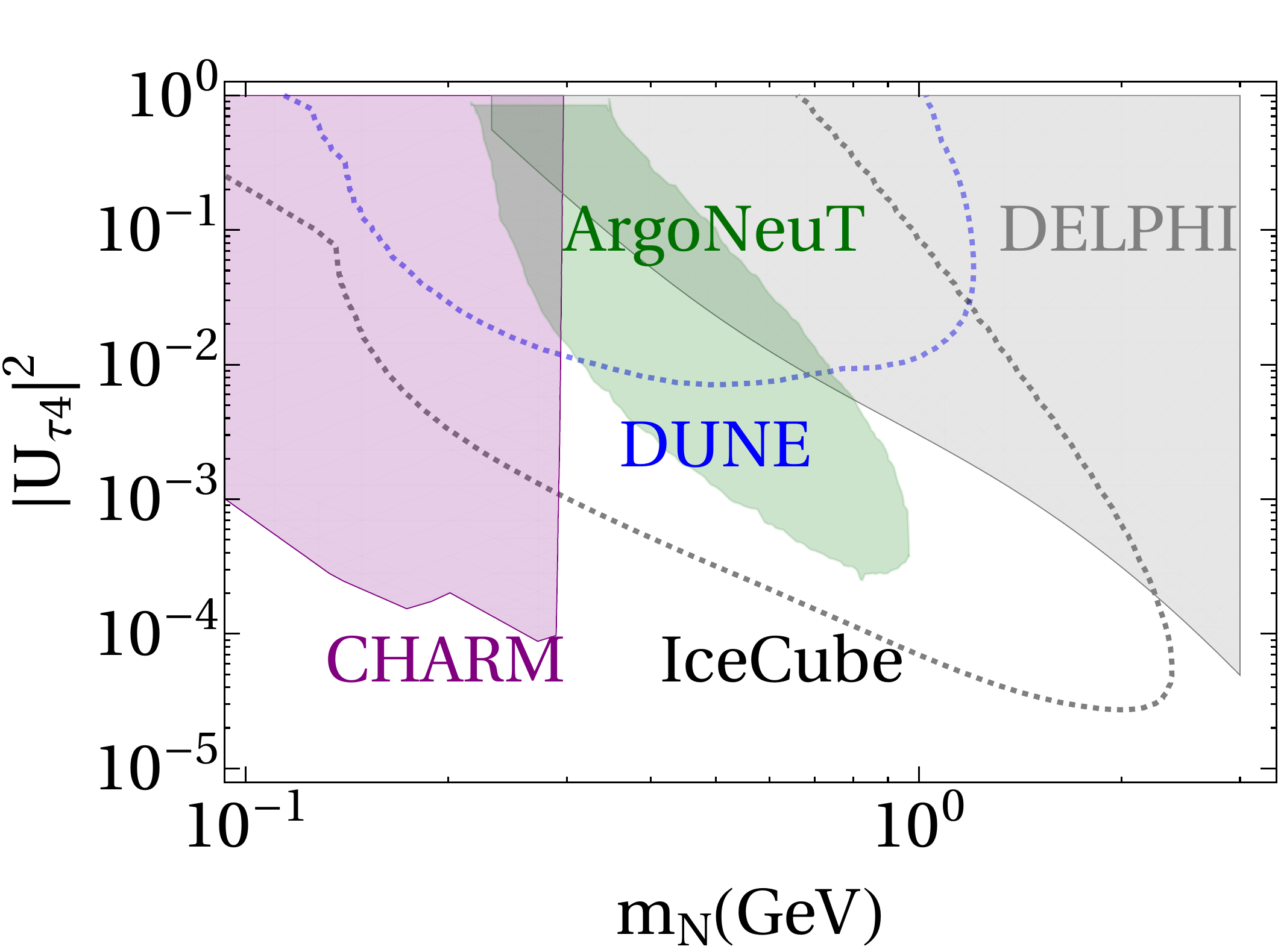}
  \includegraphics[width=.48\textwidth]{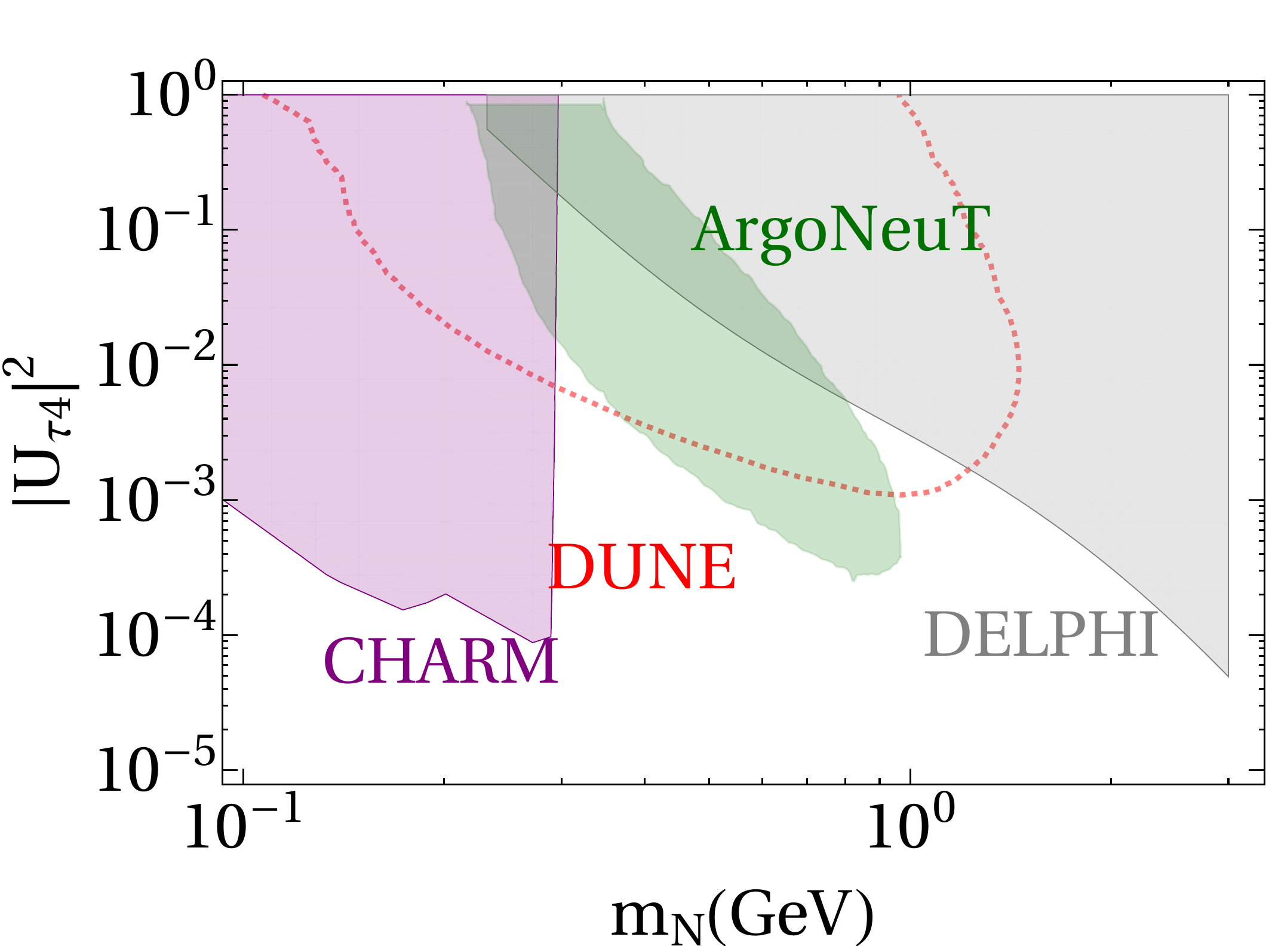}
  \caption{Sensitivity region for IceCube and DUNE using the atmosphere as a source of neutrinos (left) or the beam (right). Each line indicates the region of the parameter space where more than one DB event can be expected. The shaded regions correspond to the exclusion regions by CHARM~\cite{Orloff:2002de}, DELPHI~\cite{DELPHI:1996qcc} and ArgoNeuT~\cite{ArgoNeuT:2021clc}; the CHARM constraints do continue above 300 MeV. Figure based on \cite{Atkinson:2021rnp, Coloma:2017ppo}.}
  \label{Fig:Rate}
\end{figure}

As HNLs have a long decay length for small active-sterile mixing angles (see Eq.~\eqref{eq:decayL}) dedicated long-lived particle detectors can search for $N$ decays, where production can happen in the decays of $D$ mesons, $B$ mesons and $\tau$ leptons. The $N$ decay modes of experimental interest include charged leptons and/or hadrons in the final state. Studies exist of HNLs at current, planned and proposed experiments including NA62~\cite{Drewes:2018gkc}, FASER~\cite{Kling:2018wct}, DUNE~\cite{Berryman:2019dme},  CODEX-b~\cite{Aielli:2019ivi}, MATHUSLA~\cite{Curtin:2018mvb}, and SHiP~\cite{Alekhin:2015byh}.

\subsubsection{Dipole portal}
\label{sec:dipole}
The renormalizable mixing portal is not the only possible interaction between an HNL and the SM. In particular, extensions of the SM that account for neutrino mass generation typically lead to non-zero magnetic moments of neutrinos through loop effects, of the form
\begin{equation}
\mathcal{L} \supset i \bar{N} \gamma^\mu \partial_\mu N - \frac{1}{2} \mu_{\nu_\tau} \bar{\nu}_\tau \sigma_{\mu\nu} N F^{\mu\nu}+\text{H.c.}\,.
\end{equation}
Above the electroweak scale, such a magnetic moment arises from dimension-six terms involving dipole couplings of the third generation lepton doublet to electroweak field strengths. These magnetic moments can in principle be independent of the neutrino mass term generated by the mixing~\cite{Voloshin:1987qy,Barr:1990um,Babu:2020ivd}. Compared to the case of pure $\bar{L} H N$ mixing, neutrino magnetic moments are especially interesting and difficult to constrain because the HNL can only be produced in collisions of energetic neutrinos, rather than in SM decays. The presence of such dipole portal couplings between $\nu_\tau$ and the HNL would lead to distinct signatures in future detectors. These signatures primarily rely on neutrino upscatterings into HNLs, $\nu e^- \to N e^-$, which could be further followed by $N$ decay into a single photon, $N \to \nu_\tau \gamma$. The simultaneous observation of HNL production and decay could lead to double bang events in neutrino telescopes at energies much below that expected for CC $\nu_\tau$ scatterings~\cite{Coloma:2017ppo}. 
Additionally, lighter HNLs can up-scatter on nuclei and nucleons \cite{Schwetz:2020xra}.

Assuming no active-sterile mixing, the production and decay of the HNLs are controlled by the transition magnetic moment.
Fig.~\ref{Fig:RateMM} shows the sensitivity region
of $m_{N}$ and $\mu$ where the number of DB events is
larger than one in $10$~years of data taking. The sensitivity is again
dominated by the measurement of atmospheric neutrinos in IceCube, that
in this case can explore $\mu\sim 5 \times 10^{-10}$ for $m_{N}\sim
1$~GeV, improving the present constraint by three orders of
magnitude. For lower masses, the sensitivity is dominated by low
energy neutrinos, and therefore the difference in size between IceCube
and DUNE becomes less relevant making that DUNE can also improve the
present bounds in at least two orders of magnitude.

\begin{figure}
  \includegraphics[width=.48\textwidth]{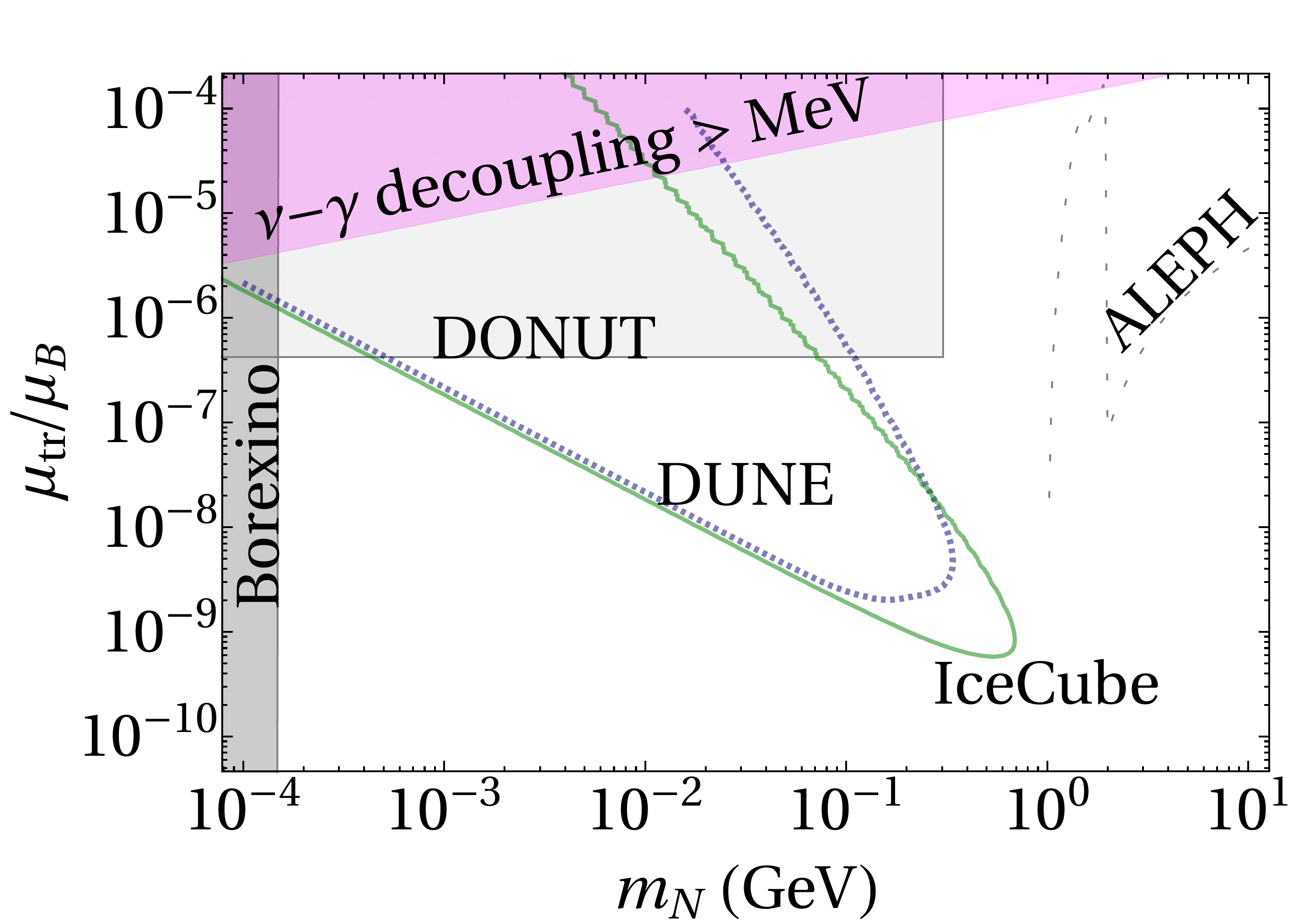}
  \includegraphics[width=.48\textwidth]{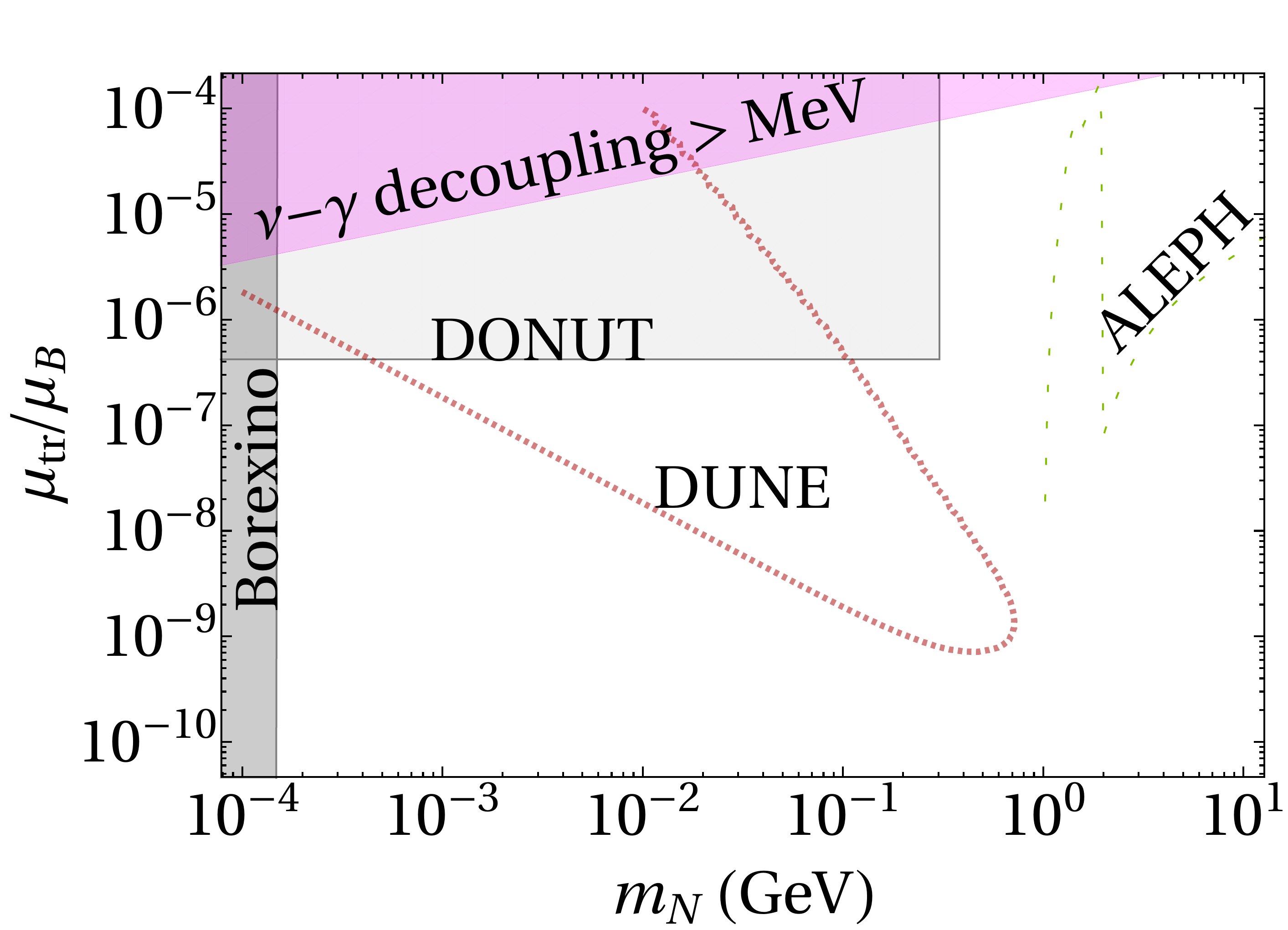}
  \caption{Sensitivity region for IceCube and DUNE in the transition magnetic moment scenario. Each line indicate the parameter space where more than one DB event can be expected. As a neutrino source, we use atmospheric neutrinos (left) and the beam (right). The shaded areas correspond to the regions disfavored by Borexino~\cite{BOREXINO:2014pcl}, DONUT~\cite{DONUT:2001zvi} and ALEPH~\cite{ALEPH:1991qhf}. The purple area is the excluded by the decoupling temperature of the neutrinos in the early Universe~\cite{Coloma:2017ppo}.  Figure based on \cite{Atkinson:2021rnp, Coloma:2017ppo}.}
  \label{Fig:RateMM}
\end{figure}

Further searches for displaced decays of the HNLs could be performed in upcoming intensity frontier searches, e.g., in the proposed SHiP experiment~\cite{Magill:2018jla}. In addition, the dipole portal between $\nu_\tau$ and the HNL can also enhance the neutrino scattering rate off electrons and lead to complementary discovery prospects in the proposed Forward Physics Facility~\cite{Jodlowski:2020vhr,Ismail:2021dyp}. Fig.~\ref{fig:dipoleNMM_tau} compares the bounds obtained on $\mu_{\nu_\tau}$ from $\nu_\tau$ upscattering into HNLs at the FPF with bounds coming from various other experiments.

One can also generate a neutrino magnetic moment without introducing an HNL via the operator $\overline{\nu} \sigma_{\mu \nu} \nu F^{\mu \nu}$, corresponding to a dipole transition between two active neutrino flavors. The corresponding neutrino scattering signal here would closely resemble that from SM $\nu$ neutral current interactions but can be distinguished due to kinematical differences. FLArE-100 can put an upper limit on the $\nu_\tau$ magnetic moment of a few $10^{-8}\mu_B$~\cite{AbariTsaiAbraham} which is an order of magnitude lower than the current bounds from DONUT~\cite{DONUT:2001zvi}. 

At the LHC, the heavy neutrinos can be directly produced from  quark anti-quark pairs and then decay into photons through the dipole portal.
One can constrain $\mu$ by investigating (1) the electron and nuclear recoil energy spectrum, (2) double-bang events (a signal with two visibly
separate cascades) at IceCube, (3) one-single photon signal from the heavy neutrino decay, or (4) a signature similar to neutral-current neutrino events or a single electron event together with the displaced single-photon event.
Most of them are summarized in Fig.~\ref{fig:global-pic-tau}, where $M_4$ denotes the mass of $\nu_4$. 
Constraints  derived from previous experiments are shown in shaded regions while sensitivities based on future
experiments or estimated exclusions (for which no rigorous background/selection efficiency analysis is included) are illustrated with dashed lines except for the DUNE case.
Specifically, the DUNE band denotes the region with
2--20 events/year, corresponding to $95\%$ C.L.~sensitivity over 5 years with 25--2500 background events, and a 100 background events is assumed in the SHiP case. 
See also Refs.~\cite{Atkinson:2021rnp, Ismail:2021dyp, Miranda:2021kre} additional discussions on this topic. 
 
New neutrino interactions can accelerate stellar cooling by the production of exotic
dark particles (if its mass is smaller than the core temperature of the star). For the
observed supernova neutrino burst SN1987A, one can set a limit on ($\mu$) given the
observed neutrino pulses on Earth \cite{Magill:2018jla}. The area enclosed by the cyan
curve in Fig.~\ref{fig:global-pic-tau}  is disfavored by SN1987A, as too many sterile neutrinos
would be produced via the dipole interaction there. Below the curve, the cooling effect is too
weak and above the interaction becomes strong enough so that steriles cannot escape the
collapsing core.  Furthermore, if the sterile is too heavy, the gravitational pull will also
prevent the sterile neutrino from leaving the supernova, leading to a vertical cut-off of
the exclusion curve. 
 From the viewpoint of Big-Bang Nucleosynthesis, the dipole
interaction alters the expansion and cooling rates of the Universe, leading to a corrected
neutron-to-proton ratio and baryon-to-photon ratio \cite{Brdar:2020quo}. The current \(^4{
	\rm He}\) abundance depends on \(M_4\) and \(d_\tau\), resulting in the pink constraint curve in Fig.~\ref{fig:global-pic-tau}. 

It is noticeable that the LEP constraints depend on the UV-completion of the model due to the high energies in consideration.
Above the electroweak scale, the dipole operator allows on-shell Z or W production because of the couplings
to the fields before electroweak symmetry breaking. More information on the model building along this line can be found in Ref.~\cite{Magill:2018jla, Shoemaker:2018vii,Brdar:2020quo}.

\begin{figure}
    \centering
    \includegraphics[width=0.99\textwidth]{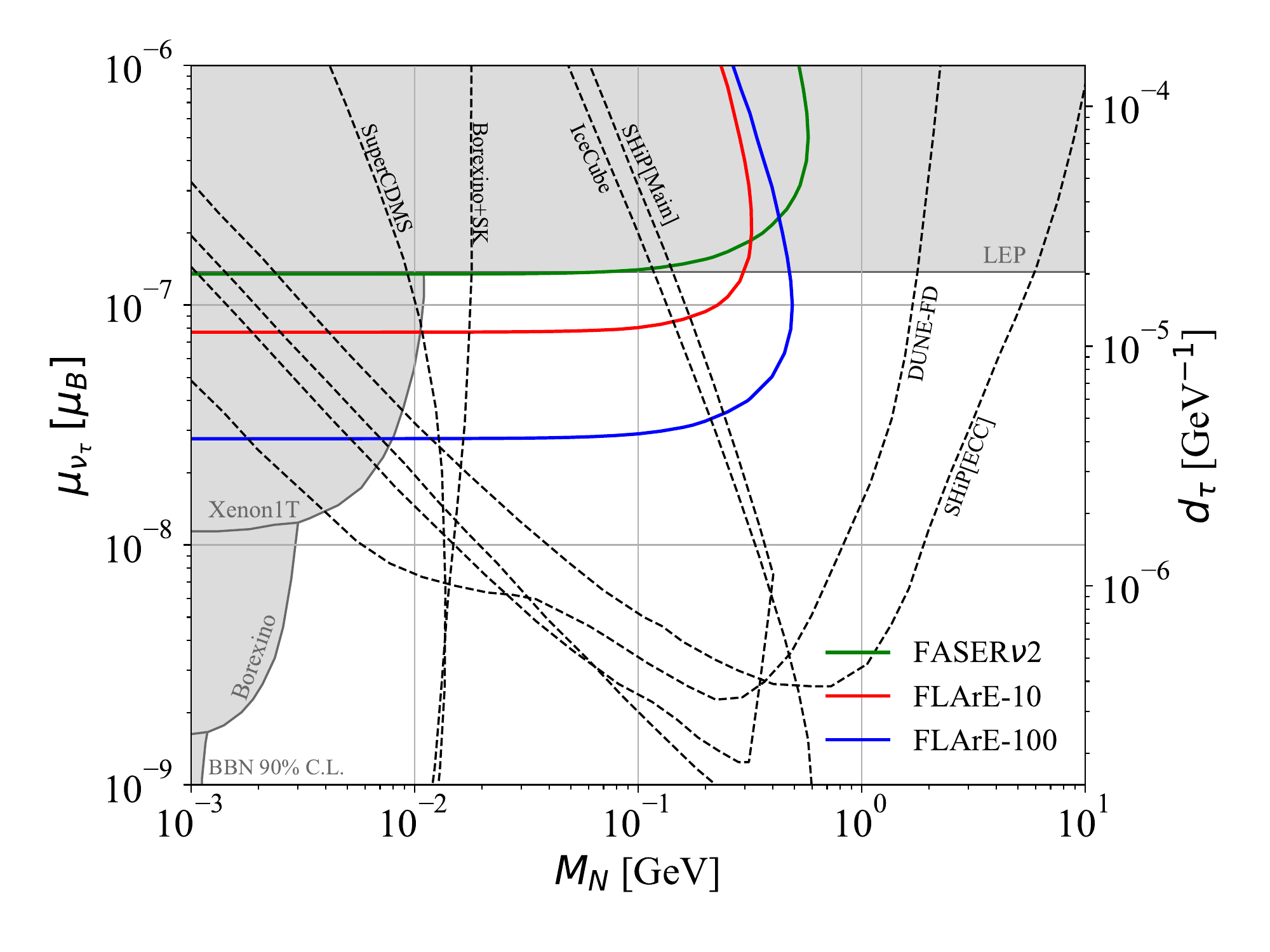}
    \caption{90\% CL Exclusion bounds at FASER$\nu$2, FLArE-10(100) for $\mu_{\nu_\tau}$. Grey shaded region are current constraints coming from the terrestrial experiments and the black dashed lines are projected sensitivities. Figure from \cite{Anchordoqui:2021ghd}.}
    \label{fig:dipoleNMM_tau}
\end{figure}

\begin{figure}[hbt]
	\includegraphics[width=\textwidth]{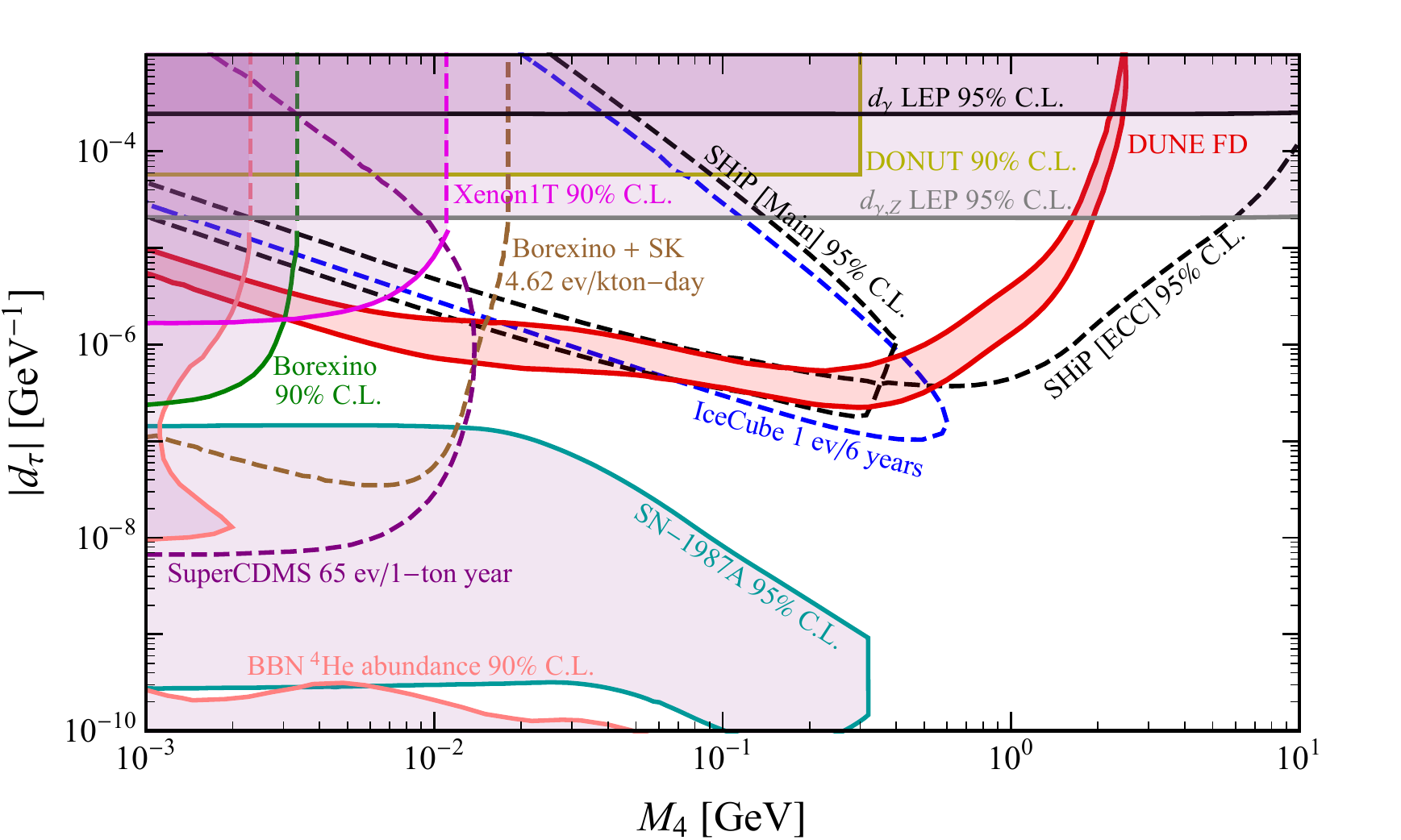}
	\caption{Current constraints (solid curves, shaded regions) and sensitivities (dashed curves, except the DUNE case)  to $|d_{\tau}|=\mu_{\nu_\tau}/2$ with $\mu_{\nu_\tau}$ in units of $\mu_B$
		from previous, on-going or future projects \cite{DONUT:2001zvi,Coloma:2017ppo,Magill:2018jla,Brdar:2020quo,
			Plestid:2020vqf,Schwetz:2020xra}. Figure from \cite{Schwetz:2020xra}.}
	\label{fig:global-pic-tau}
\end{figure}

\subsection{Tau Neutrinos and Other New Physics Scenarios}

In addition to the search for sterile neutrinos the tau sector also allows to probe another portal to new physics: the vector or scalar portals which can induce new neutrino interactions (NSI). It is crucial to test all different flavor structures of these new interactions to obtain a global picture of any deviations from the SM expectations.

Neutrinos might even interact with Dark Matter (DM). The DM couplings to  nucleons, electrons, and weak  bosons  have been searched for with null results. To obtain insights into potential DM-SM connections we need to probe all possible interactions of DM with SM particles including neutrinos across a wide range of DM masses. In particular tau neutrino experiments are ideally suited to probe very heavy DM masses beyond the range of other experiments. 

Finally,  popular neutrino mass models introduce a new decay channel for neutrinos. While there are strong constraints on the neutrino lifetime from a plethora of experiments there are hints for neutrino decay in several data sets involving tau neutrinos.

Future work investigating tau neutrinos as a window to new  physics is required to truly make the best use of upcoming tau data. Neutrino experiments should conduct studies of their sensitivity to BSM-$\nu_\tau$ couplings to identify novel constraints on new physics scenarios.

\subsubsection{Non-standard interactions}
The identification of tau neutrinos can also probe whether SM neutrinos have BSM interactions with matter, typically referred to as non-standard interactions (NSI). In oscillation experiments, measurements of the oscillation probabilities are sensitive to neutral-current NSI between neutrinos and matter while the neutrinos propagate. For earth-based experiments, this corresponds to interactions with electrons, up quarks, and down quarks. The effect is characterized by four-fermion operators,
\begin{equation}\label{eq:LNSI}
\mathcal{L}^{\rm NSI} \supset -2\sqrt{2}G_F \left(\overline\nu_\alpha \gamma_\rho \nu_\beta\right) \left(\epsilon_{\alpha\beta}^{f \bar{f} L} \overline{f}_L \gamma^\rho \tilde{f}_L + \epsilon_{\alpha\beta}^{f \bar{f} R} \overline{f}_R \gamma^\rho \tilde{f}_R\right) + \mathrm{h.c.},
\end{equation}
so that the strength of these new interactions is parameterized relative to the SM weak interactions by $\epsilon_{\alpha\beta}$.

While non-oscillation neutrino experiments may be sensitive to different combinations of these parameters through scattering (see, e.g., Refs.~\cite{Coloma:2017egw,COHERENT:2017ipa,COHERENT:2020iec,Esteban:2018ppq}), oscillations through a particular density of matter are sensitive to the combination of these parameters $\epsilon_{\alpha\beta} = \sum_{f =u,d,e} \epsilon_{\alpha\beta}^f \frac{n_f}{n_e}$ where $\epsilon_{\alpha\beta}^f \equiv \epsilon_{\alpha\beta}^{ffL} + \epsilon_{\alpha\beta}^{ffR}$ (the vector combination of the interactions in Eq.~\ref{eq:LNSI}). Here, $n_f$ is the number density of fermion $f$ in the matter.

The addition of these interactions modifies the matter potential which neutrinos experience while propagating, often in nontrivial ways -- we refer the reader to Ref.~\cite{Proceedings:2019qno} for a review of many of these effects. While tau-related elements $\epsilon_{e\tau}$, $\epsilon_{\mu\tau}$, and $\epsilon_{\tau\tau}$ have prominent effects in oscillations of tau neutrinos (appearance/disappearance of $\nu_\tau$), experiments only sensitive to $\nu_e$ and $\nu_\mu$ neutrinos can still provide strong constraints of these $\nu_\tau$ related parameters. For instance, some of the strongest constraints to date on $\epsilon_{\mu\tau}$ come from measurements of atmospheric $\nu_\mu$ by IceCube~\cite{IceCubeCollaboration:2021euf,IceCube:2022ubv}, and long-baseline measurements of $\nu_\mu \to \nu_\mu$ and $\nu_\mu \to \nu_e$ oscillations at T2K~\cite{T2K:2021xwb} and NOvA~\cite{NOvA:2021nfi} exhibit some preference for non-zero NSI \cite{Denton:2020uda,Chatterjee:2020kkm} that will be tested by future long-baseline experiments of a similar nature.

Measurements of tau neutrinos in the next generation of experiments  using long-baseline neutrinos or atmospheric neutrinos, particularly by the IceCube \cite{IceCube-Gen2:2020qha}, DeepCore, INO \cite{Khatun:2019tad}, KM3NeT \cite{KhanChowdhury:2020xev}, and DUNE~\cite{DUNE:2020ypp} experiments will allow for complementary measurements of these NSI parameters in different oscillation channels. The statistical power of $\nu_\tau$ measurements will likely be limited relative to those of $\nu_\mu$ and $\nu_e$. However, if a new-physics effect is present in one of these oscillation channels and predicted to be nonzero in other channels, then the simultaneous measurement with $\nu_\tau$ observations will provide necessary confirmation. Ref.~\cite{DeGouvea:2019kea} demonstrated the capabilities of DUNE to identify the $\nu_\mu\to\nu_\tau$ appearance signal and to use it to detect nonzero NSI parameters. While the constraints are weaker than what is expected using the DUNE $\nu_\mu \to \nu_e$ and $\nu_\mu \to \nu_\mu$ channels, this complementarity is valuable.

In source and/or detector NSI (CC NSI), explicitly detecting the $\nu_\tau$ in the near detector of an experiment such as DUNE can set the stringent bounds \cite{Giarnetti:2020bmf}.
In addition, lepton flavor violating $\pi^+\to \mu^+ \nu_\tau$ or $\pi^+\to \mu^+ \bar{\nu}_\tau$ decay modes with branching ratios of $O(10^{-3})$, meditated by new heavy scalars, can also be probed at forward experiments such as FASER$\nu$ by looking for a tau neutrino excess \cite{Ansarifard:2021dju,Falkowski:2021bkq}.

\subsubsection{Neutrino decay}

Since neutrinos have mass, they decay via $\nu_j\to\nu_i+\gamma$ \cite{Petcov:1976ff,Marciano:1977wx}, however their lifetimes are far too long to be observed.
If neutrinos couple to a new light or massless particle, then the neutrino decay rate could be significantly enhanced and could be probed in a variety of environments, depending on the parameter space.
The most popular model for neutrino decay involves a light or massless Majoron which is a spin-0 gauge singlet with non-zero lepton number possibly related to neutrino mass generation \cite{Chikashige:1980ui,Gelmini:1980re,Schechter:1981cv,Acker:1991ej}.
Additional models include mirror models \cite{Maalampi:1988vs}, SUSY models \cite{Gabbiani:1990uc,Enqvist:1992ef,Aboubrahim:2013gfa}, left-right symmetric models \cite{Kim:2011ye}, neutrino masses generated by a topological formulation of gravitational anomaly \cite{Dvali:2016uhn}, unparticles \cite{Georgi:2007ek,Zhou:2007zq}, and others.
While these models do not involve any particular focus on tau neutrinos, because of the structure of the lepton mixing matrix and the fact that heavy mass states decay to lighter mass states, there is a phenomenological connection to tau neutrino physics.

Phenomenologically, neutrino decay is often classified into two main categories: invisible and visible decay.
Invisible decay is where the decay products are undetected because either they are sterile neutrinos or they are too low of energy to be detected.
Visible decay involves the detection of the regenerated lower-energy neutrinos.
Neutrino decay constraints are often parameterized in terms of $\tau_i/m_i$ since the absolute neutrino mass scale is unknown, but the neutrino energy is; thus the $m_i$ factor accounts for the Lorentz boost.
Neutrino decay (invisible and visible) has been probed in a wide range of experiments.
Constraints, sensitivity estimates, and degeneracies have been studied in atmospheric and long-baseline accelerator and reactor experiments finding $\tau_3/m_3\gtrsim10^{-10}$ s/eV \cite{Gonzalez-Garcia:2008mgl,Gomes:2014yua,Abrahao:2015rba,Pagliaroli:2016zab,Gago:2017zzy,Coloma:2017zpg,Choubey:2017dyu,Choubey:2017eyg,Choubey:2018cfz,Ascencio-Sosa:2018lbk,deSalas:2018kri,Choubey:2020dhw,Ghoshal:2020hyo,Mohan:2020tbi,Porto-Silva:2020gma}, the solar sector finding $\tau_{1,2}/m_{1,2}\gtrsim10^{-3}$ s/eV \cite{Bahcall:1986gq,Berryman:2014qha,Picoreti:2015ika,Huang:2018nxj,Picoreti:2021yct}, high energy astrophysical neutrinos at IceCube finding $\tau_i/m_i\gtrsim10^{1}$ s/eV \cite{Beacom:2002vi,Beacom:2003zg,Baerwald:2012kc,Bustamante:2016ciw,Rasmussen:2017ert,Denton:2018aml,Abdullahi:2020rge}, galactic supernova finding $\tau/m\gtrsim10^{5}$ s/eV \cite{Kamiokande-II:1987idp,deGouvea:2019goq}, the diffuse supernova neutrino background with an estimated sensitivity of $\tau_i/m_i\sim10^{10}$ s/eV \cite{Ando:2003ie,Fogli:2004gy}, and the cosmic neutrino background with an estimated sensitivity of $\tau_i/m_i\sim10^{15}$ s/eV \cite{Long:2014zva}.
Additionally, measurements of the CMB constrain neutrino decay in a somewhat model dependent fashion to be $\tau_i/m_i\gtrsim10^{11}$ s/eV \cite{Hannestad:2005ex,Escudero:2019gfk,Chacko:2020hmh,Escudero:2020ped} although these bounds may be relaxed by an additional 3-4 orders of magnitude in a separate analysis \cite{Barenboim:2020vrr}.

Hints for neutrino decay exist in long-baseline accelerator data \cite{Gomes:2014yua,Pagliaroli:2016zab,Choubey:2018cfz}, IceCube data \cite{Denton:2018aml,Abdullahi:2020rge}, and cosmological data \cite{Escudero:2019gfk}.
In particular, \cite{Pagliaroli:2016zab} found a mild preference for neutrino decay in OPERA's tau neutrino appearance data.
In addition, \cite{Denton:2018aml,Abdullahi:2020rge} found $\gtrsim3\sigma$ evidence for neutrino decay in IceCube's data by examining the flavor in an energy dependent way; this analysis also predicts a deficit of tau neutrinos at low energies which future observations can test.
These constraints and hints and the potential impact of various new physics scenarios on the high energy astrophysical tau neutrino flux are summarized in Fig.~\ref{fig:neutrino decay}.

\begin{figure}
\centering
\includegraphics[width=0.55\textwidth]{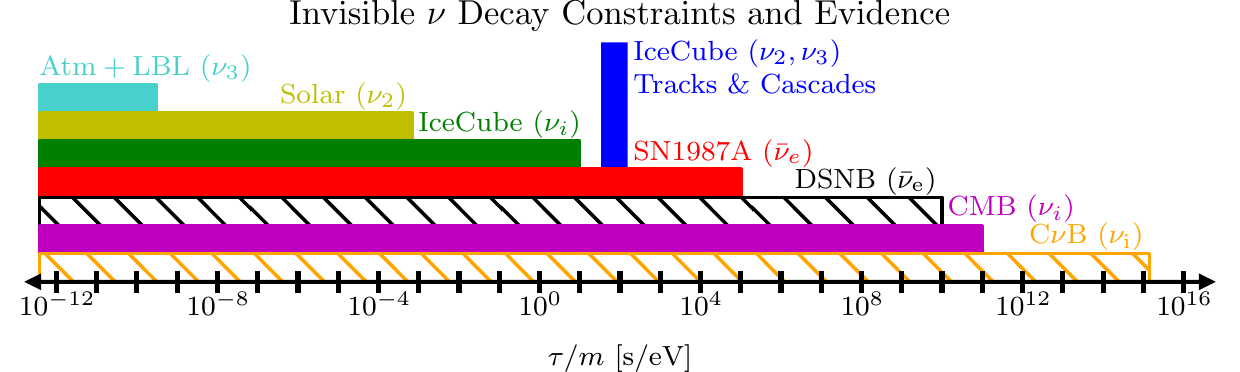}
\includegraphics[width=0.44\textwidth]{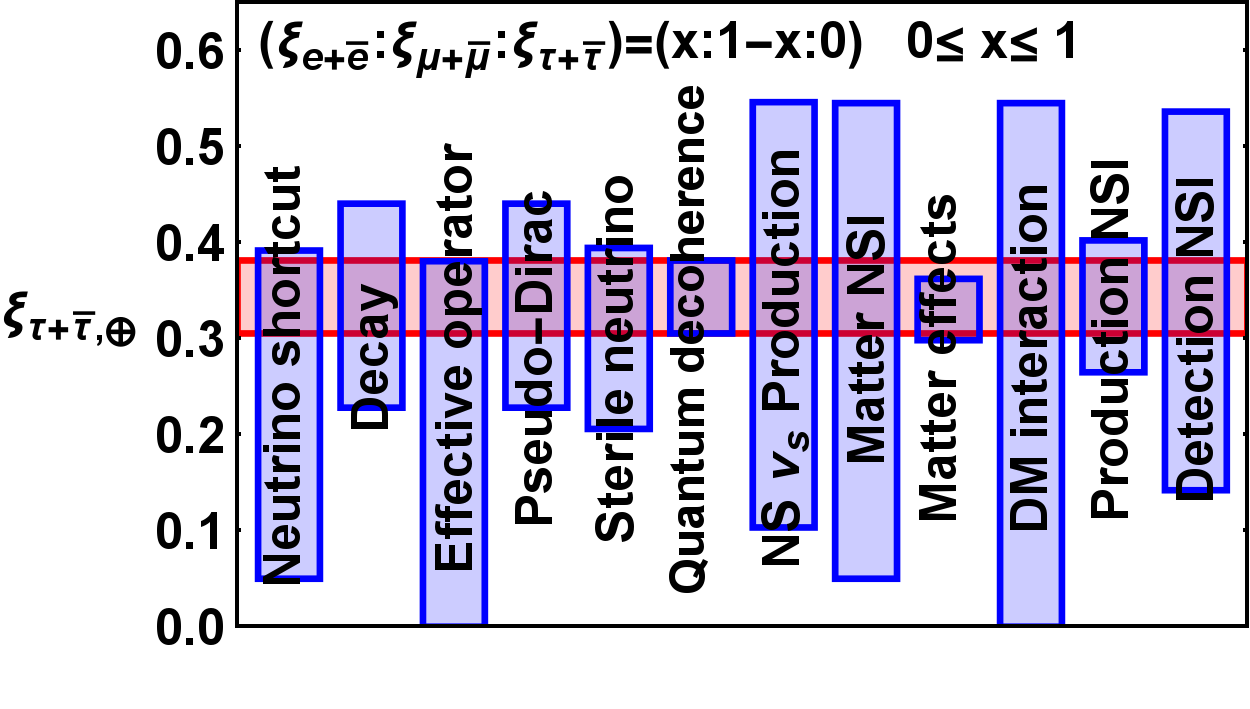}
\caption{\textbf{Left}: Constraints on invisible neutrino decay from a range of experiments.
The blue region represents a hint for neutrino decay \cite{Denton:2018aml,Abdullahi:2020rge} in IceCube data.
Dashed regions represent anticipated sensitivities from future measurements.
Figure adapted from \cite{Abdullahi:2020rge}.
\textbf{Right}: Assuming no tau neutrino are produced at high energy astrophysical sources, $\sim\frac13$ of the flux will be tau neutrinos at the Earth as shown in the red band.
Various BSM scenarios predict deviations from the expected flavor ratio which may also include an energy dependent effect.
Figure from \cite{Rasmussen:2017ert}.}
\label{fig:neutrino decay}
\end{figure}

In general, neutrino decay provides a model independent framework with rich phenomenology that affects neutrino phenomenon in a unique energy and flavor dependent fashion across a broad range of experimental regimes from traditional oscillation experiments to astrophysical experiments to cosmological probes.
In addition, due to the flavor dependence, tau neutrino signals are expected to be enhanced or depleted depending on the region of parameter space providing a target for upcoming experiments to test existing hints and push the constraints farther by increasing their sensitivity to tau neutrinos.

\subsubsection{Tau neutrinos and dark matter}

Probing DM couplings to different SM particles provides insights into connections between DM and the SM and will aid to identify the portal between the SM and the BSM sectors. While the coupling of DM to nucleons, electrons, and weak  bosons is already well constrained for DM masses around the weak scale \cite{Schumann:2019eaa}  it is important to also shed light on  possible DM-neutrino couplings. These two sector share their elusive nature of interacting only weakly with other SM particles however there could be a  possibility of a strong connection between these two sectors 
as the neutrino sector  allows for deviations from the SM expectation, in particular for tau neutrinos. 

In \cite{Blennow:2019fhy}  theoretical models have been derived  that lead to sizable
neutrino-DM interactions. Constraints from DM annihilation into neutrinos at tree-level and loop-induced  DM annihilations into charged leptons as well as loop-induced DM-nucleon interactions together with cosmological constraints which restrict the strength of the DM-neutrino coupling during BBN and CMB apply  to these models. Fig.~\ref{fig:DMnumed} shows the results in a scenario where a DM particle $\chi$ interacts with the SM via the tau neutrino portal and  the DM annihilation processes are mediated by the complex scalar $S$. Cosmological constraints and current constraints from DM annihilation into neutrinos apply and can be improved in the future by Hyper-Kamionde, DUNE and Xenon1T.
In fact, upcoming neutrino experiments which are sensitive to tau neutrinos from DM annihilation  like DUNE, Hyper-Kamiokande 
will  access the parameter space with the correct relic abundance.

\begin{figure}
\centering
\includegraphics[width=0.5\textwidth]{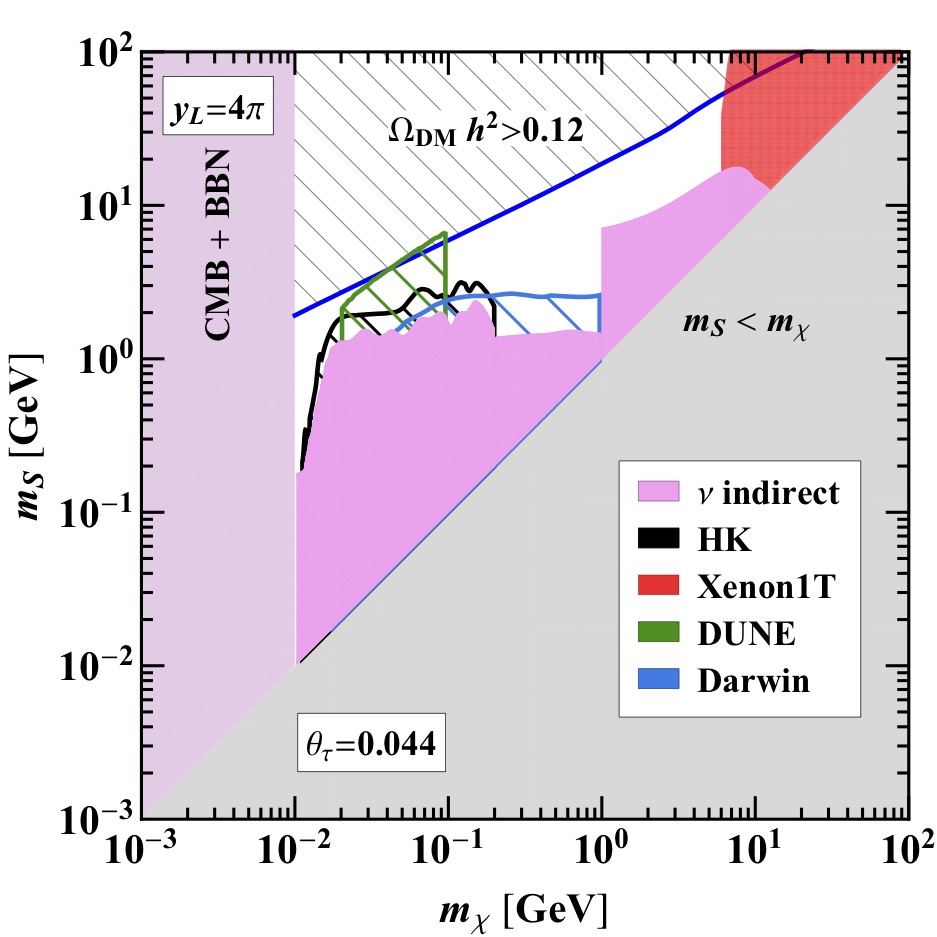}
\caption{Constraints on the DM mass $m_\chi$ and the dark scalar mass $m_S$ which acts as mediator for active-sterile mixing angles
 $\theta_\tau= 0.044,~\theta_{e,\mu}=0$.
Along the blue line the DM relic density matches the observed
value. The colored shaded regions are excluded by different experiments, while the hatched
areas correspond to prospective sensitivities of future experiments. The lower bound $m_\chi<$
10 MeV is set by observations of the CMB and BBN. Figure from  \cite{Blennow:2019fhy}. }
\label{fig:DMnumed}
\end{figure}

Fig.~\ref{fig:DManni} provides a summary of the constraints from DM annihilation into neutrinos. In particular for larger DM masses experiments sensitive to tau neutrinos (see Sec.~\ref{sec:gev experiments}, \ref{sec:intermediate experiments}, \ref{sec:he experiments}) will provide the strongest constraints. 

\begin{figure}
\centering
\includegraphics[width=0.7\textwidth]{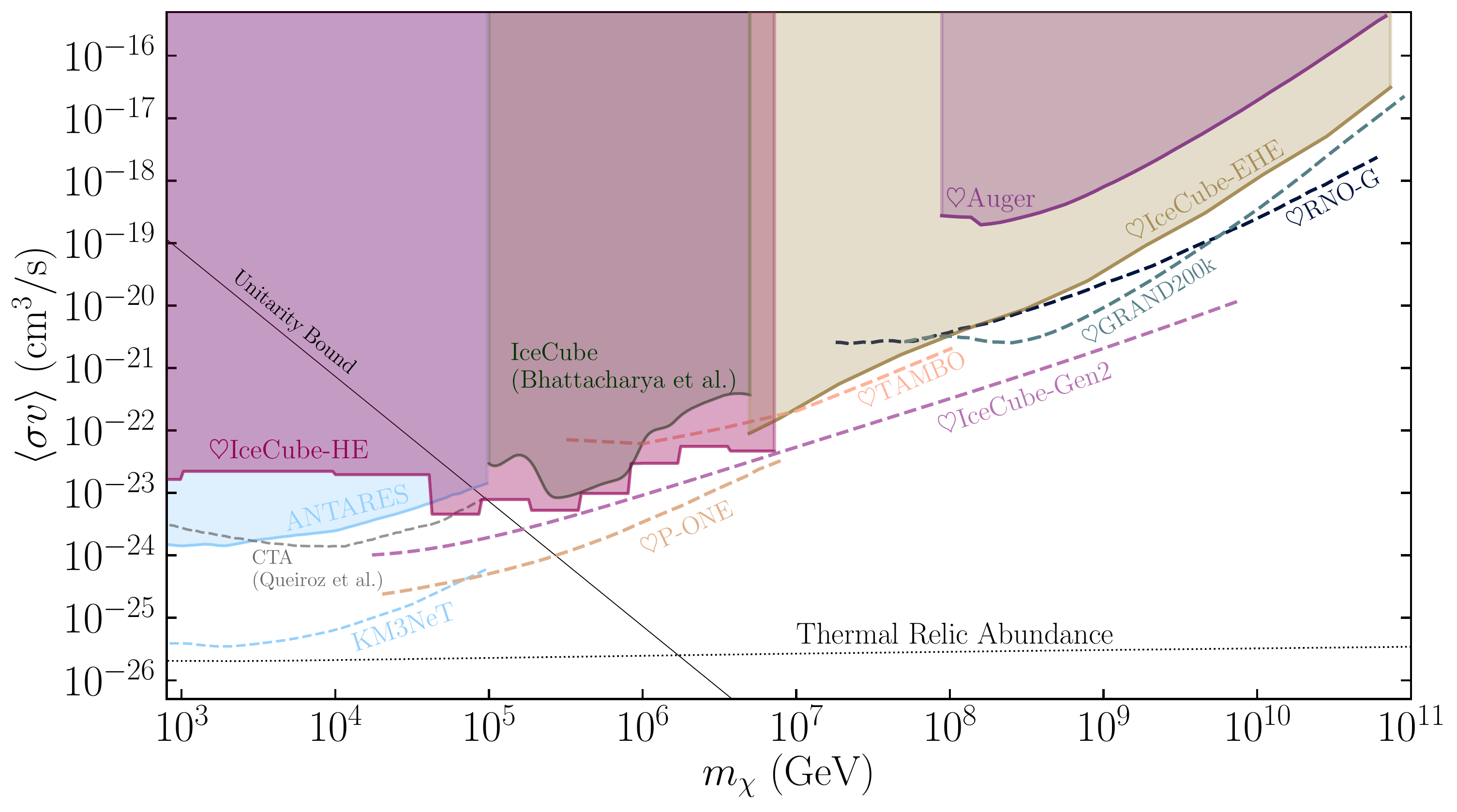}
\caption{The landscape of dark matter annihilation into neutrinos up to $10^{11}$ GeV from  \cite{Arguelles:2019ouk}. Solid
and dashed lines represent 90$\%$ C.L.~limits and sensitivities, respectively. Projected sensitivities assume five years of
data taking for neutrino experiments and 100 hours of observation for CTA. The dotted line corresponds to the value
required to explain the observed abundance via thermal freeze-out. The straight diagonal line, labeled as ``Unitarity
Bound" gives the maximum allowed cross section for a non-composite DM particle. These results assume 100$\%$ of
the dark matter is composed of a given Majorana particle. If instead only a fraction, $f$, is considered these results
should be multiplied by $1/f^2$. In the case of Dirac DM, limits would be scaled up by a factor of two.
All the experimental constraints in this plot are calculated by converting either the detected flux
or the reported upper limit into a conservative upper bound on the DM annihilation cross section. Figure from \cite{Arguelles:2019ouk}.}
\label{fig:DManni}
\end{figure}

In addition to neutrinos from the annihilation of DM these experiments are also sensitive to neutrinos from decaying DM \cite{IceCube:2018tkk,Ng:2020ghe,Guepin:2021qai,Chianese:2021htv}
which have also been proposed as a solution to the ANITA anomaly \cite{Anchordoqui:2018ucj,Heurtier:2019git,Saveliev:2021jtw}.  Dedicated tools have been developed to study the neutrino signature of DM annihilations and decays in the Sun, the Earth, and the Galactic Halo \cite{Blennow:2007tw,Liu:2020ckq} which can be interfaced with  neutrino propagation software (see Sec.~\ref{sec:tools} for an overview) to obtain the expected number of neutrino events at detectors. 

As the tau neutrino data sets increase in the future the potential coupling of DM to tau neutrinos can  be tested as well as  the DM-$\nu_\mu$ or DM-$\nu_e$ couplings. 
So far most of the interest in the DM-$\nu_\tau$ connection was driven by the weaker constraints on tau neutrinos in comparison to  other flavors. In the future models which motivate the 
DM-$\nu_\tau$ couplings from a theoretical point of view should be build to make ideal use of the wealth of future tau neutrino data. These models could for example originate from $L_\mu-L_\tau$ symmetries which are also motivated as a solution the muon anomalies \cite{Altmannshofer:2016jzy,Biswas:2019twf,Amaral:2020tga,Amaral:2021rzw}
(see Sec.~\ref{sec:b_anomalies})). Together with third family models \cite{Allanach:2018lvl} these models strongly mitigate
bounds from DM-electron interactions.
Tau neutrinos can also provide complementary information to constraints from
charged tau leptons assuming that the BSM model couples to the $SU(2)_L$ doublet or can solidify any anomalies which arise in tau physics.
The  tau neutrino data sets spread across many decades of energy which opens the window to test DM masses across a very large range making tau neutrinos a unique probe of DM which cannot be emulated by other experiments. With the rich upcoming tau neutrino data sets as well as future electron and muon neutrino data sets one can also constrain the flavor structure of the DM-neutrino coupling. Furthermore,
the different origin, sources and production mechanisms of neutrinos in terrestrial experiments or from astrophysical sources allow to probe DM-neutrino interactions in different environments.
Finally, these studies don't need to be limited to DM only, also the coupling of tau neutrinos to other BSM particles like long-lived particles  should be studied in neutrino observatories as it has been done in \cite{Arguelles:2019ziu,Kelly:2021mcd}.

\subsubsection{New physics in tau neutrino scattering}

Additional gauge interactions can also affect the properties of tau neutrinos. Dark gauge bosons coupled to the $\nu_\tau$ arise naturally in models with $U(1)_{B-L}$ or $U(1)_{B-3\,L_\tau}$ symmetry. These theories can be made anomaly-free by introducing additional sterile neutrinos. Interestingly, if the new gauge boson is sufficiently light, it can lead to enhanced $\nu_\tau$ production rates that can be probed in the far-forward region of the LHC~\cite{Kling:2020iar}. If the dark gauge boson additionally couples to light dark matter (LDM), further bounds on the model can be obtained from searches for both LDM and BSM neutrino scatterings in the detector~\cite{Batell:2021snh}. This is especially relevant for searches in the Forward Physics Facility (FPF)~\cite{Anchordoqui:2021ghd}, where standard neutrino-induced backgrounds typically correspond to much larger energy depositions and can, therefore, be rejected in the analysis~\cite{Batell:2021aja}.

Finally, models where both HNLs and extra gauge interactions play a role have been intensively studied in the context of the MiniBooNE anomaly~\cite{Bertuzzo:2018itn,Ballett:2018ynz}. In these models, there is a massive dark gauge boson which couples to both SM neutrinos and HNLs. In this case, the upscattering cross section of the active neutrinos to HNLs can be followed by displaced HNL decays into electron-positron pairs, $N \to \nu\,e^+e^-$, via on-shell or off-shell $Z_D$, cf.~Ref.~\cite{Jho:2020jfz} for such discussion for the dominant mixing with the tau neutrinos and the FASER2 experiment. For $m_N<m_{Z_D}/2$, an additional signature can be realized through HNL production in dark gauge boson decays, $Z_D\to NN$, followed by HNL scattering off electrons, $N e\to N e$. This can probe very low values of the $\nu_\tau-N$ mixing angle $U_{\nu N}$ at the FPF~\cite{Jodlowski:2020vhr}.

These models provide a representative sample of theories in which new interactions involving tau neutrinos can be tested with $\nu_\tau$ scattering. In many cases, the relatively high energies of collider and astrophysical facilities allow for probes of $\nu_\tau$ physics at scales above those that can usually be probed with traditional terrestrial neutrino sources. Neutrino scattering thus offers a complementary perspective on BSM physics involving the $\nu_\tau$.

\subsection{Connections of Tau Neutrinos to Other New Physics Phenomena}
\label{sec:b_anomalies}
Tau neutrinos might well be connected to recently observed anomalies in particle physics  which point towards
 lepton universality violation. Even though these anomalies involve charged leptons if new physics mediators couple to the $SU(2)_L$ doublet signatures are also expected in the neutrino sector. Thus this also demonstrates the 
complementarity of neutrino physics to charged lepton physics.

\subsubsection{Tau neutrinos and charged lepton  anomalies}
\label{sec:flavor}

 Neutrinos are part of a $SU(2)_L$ doublet hence any new physics coupling directly to SM neutrinos  automatically implies new physics effects for the charged leptons, however the reverse statement is not true.  Nevertheless, any signal of new physics in the neutrino sector should be  tested in a complementary way with charged leptons probes.

Several results from B-factories 
have shown indications of lepton non-universality  in the decays of
$B$-mesons into semi-leptonic final states.
Global fits \cite{Capdevila:2017bsm,Altmannshofer:2017yso,DAmico:2017mtc,Ciuchini:2017mik,Hiller:2017bzc,Geng:2017svp,Hurth:2017hxg,Alok:2017sui,Alguero:2019ptt,Aebischer:2019mlg,Ciuchini:2019usw} to $b\to s l^+l^-$ data \cite{LHCb:2014cxe,LHCb:2014vgu,LHCb:2015wdu,LHCb:2015svh,CMS:2015bcy,LHCb:2017avl} prefer $e-\mu$ lepton flavor violating  new physics over the SM interpretation.
Furthermore, data from $b\to c l\nu$ transitions points towards $\tau-\mu$ flavor violation \cite{BaBar:2012obs,BaBar:2013mob,LHCb:2015gmp,LHCb:2017rln,LHCb:2017smo}, and may even point to the existence of new tau flavored right handed neutrinos \cite{Greljo:2018ogz,Asadi:2018wea}.
Finally, the recent confirmation of a discrepancy between the measured and predicted value of the anomalous magnetic moment of the muon \cite{Muong-2:2021vma,Muong-2:2006rrc,Muong-2:2021ojo, Aoyama:2020ynm} requires new physics in the muon sector. As any short-range contribution to this observable scales with the lepton’s mass,
this observable can be considered a probe of lepton non-universality. To probe lepton non-universality further a more precise  measurement \cite{DELPHI:2003nah} of the anomalous magnetic moment of the tau  is desirable for which some ideas have been put forward recently \cite{delAguila:1991rm,Beresford:2019gww,Dyndal:2020yen,Crivellin:2021spu,Burmasov:2022gnl}.

New physics models which aim to provide an explanation to these anomalies have often also implications for the neutrino sector. A popular explanation relies on extending the SM gauge group by gauging an anomaly free combination of individual lepton number like $L_\mu-L_\tau$ \cite{Altmannshofer:2015mqa,Altmannshofer:2016jzy,Baek:2001kca,Altmannshofer:2016brv,Amaral:2020tga,Amaral:2021rzw}. Alternative explanations which rely on  models which introduce new gauge symmetries for the third generation only \cite{Allanach:2018lvl, Babu:2017olk} have also been proposed.
In fact, due to strong constraints on couplings to electrons such anomaly free gauge symmetries almost  inevitably involve couplings to taus \cite{Greljo:2021npi}.
These models predict signatures in the neutrino sector like  neutrino tridents \cite{Altmannshofer:2014pba}, effects in neutrino-electron scattering and coherent neutrino-nucleus scattering  \cite{Kaneta:2016uyt}, and impacts on astrophysical or cosmological neutrino observables \cite{Escudero:2019gzq, Kamada:2015era,Bauer:2018onh}.
Much of the phenomenological studies focus on the impact on the muon sector however the wealth of current and upcoming tau neutrino data sets will provide valuable additional insights into the solution to the experimental anomalies and will be crucial in distinguishing different scenarios.

The different masses of the SM leptons  already demonstrate that the SM is lepton non-universal; the 
observed anomalies provide hints that there is lepton non-universality even in the absence of Yukawa couplings.
However any departure from
lepton universality is necessarily associated with the violation of lepton flavor conservation \cite{Glashow:2014iga} although this relation does not hold in a 
 class of new physics scenarios with minimal flavor violation  \cite{Alonso:2015sja}.
 The mixing of different neutrino flavors demonstrates that lepton flavor is not a conserved quantity for neutral leptons\footnote{Up to now there are only upper limits on the violation of total lepton number from searches for neutrinoless double beta decay \cite{Vergados:2016hso, DellOro:2016tmg, Dolinski:2019nrj,Agostini:2022zub}.}, whereas we only have bounds on lepton flavor violation with charged leptons \cite{Zyla:2020zbs}. Charged lepton flavor violation is extremely suppressed in the SM such that  any evidence would provide indisputable evidence of BSM physics. In fact, considering the SM as an effective field theory, charged lepton flavor violation can arise from dimension-6 operators 
and provides therefore a deep look into high scale physics.
The current best limits on charged lepton flavor violation come from the muon sector \cite{Zyla:2020zbs} yet the flavor structure of new physics is unknown therefore it is imperative to also probe charged lepton flavor violation  using other flavors
to obtain a set of measurements which also allows to distinguish among the various models.
Additionally, in many models the
predicted tau lepton flavor violating branching ratios are several orders of magnitude larger than those of the muons \cite{Mihara:2013zna}. B-factories \cite{Belle-II:2018jsg,LHCb:2013fsr} and  tau-charm factories
\cite{BESIII:2020nme,Lyu:2021tlb,Charm-TauFactory:2013cnj} produce an abundance of tau leptons and will therefore drive future tau lepton studies.  

Belonging to the heaviest fermion generation the tau lepton can also provide valuable insights into the flavor puzzle. In fact, one naively expects the heavier fermions to be more sensitive to whatever
dynamics is responsible for the observed hierarchies in the fermion masses. 
Therefore thorough probes of the tau lepton together with exhaustive tests of the underlying paradigms of neutrino oscillations and mechanisms for neutrino mass generation  are essential to unveil the solution to the flavor puzzle.

Tau neutrinos and its accompanying charged lepton offer tremendous potential to understand open questions of the SM. In the future we need to make extensive use of the complementarity of these probes to learn more about the tau sector in the SM and beyond.

\subsection{Looking for New Phenomena with HE and UHE Tau Neutrinos}
\label{HE}
High-energy and ultra high energy neutrinos can open a window to new physics, in fact tau neutrinos are special as even though they are not produced in astrophysical sources 
due to oscillations we expect a sizable UHE tau neutrino flux at Earth. 
In fact, the ANITA experiment observed anomalous UHE $\nu_\tau$ events which have not been conclusively explained.
In addition, HE and UHE neutrinos allow to probe the self interacting neutrino solution to the tension in the  measurements of the Hubble constant. However strong constraints apply for this solution involving $\nu_e$ and $\nu_\mu$ whereas currently self-interacting $\nu_\tau$ are still allowed. The next generation of HE and UHE tau neutrino experiments will be fundamental in probing the solutions to these anomalies in particle physics and cosmology.

\subsubsection{Astrophysical neutrinos}
\label{sec:anita}

High energy neutrinos can be produced via scattering of energetic protons off protons or photon gas in a galactic halo. They can  also be produced by the scattering of ultra high energy cosmic rays ({\it i.e.,} protons)
off the background CMB photons. The scattering can abundantly yield charged pions which in turn decay into leptons, producing a flux of high energy neutrinos with flavor ratios: $F_{\nu_e}^0:F_{\nu_\mu}^0:F_{\nu_\tau}^0\simeq 1:2:0$ at the source. The $\nu_\tau$ component at the source will be negligible. The main contribution to the $\nu_\tau$ flux can come from the decay of the charm or beauty quark whose  productions in the $pp$ or  $\gamma p$ collisions
are suppressed. As is well-known, the oscillation of neutrinos \textit{en route} from the source to the Earth will convert the flavor ratio from $F_{\nu_e}^0:F_{\nu_\mu}^0:F_{\nu_\tau}^0\simeq 1:2:0$ to $F_{\nu_e}^\oplus:F_{\nu_\mu}^\oplus:F_{\nu_\tau}^\oplus\simeq 1:1:1$. Notice that despite $F^0_{\nu_\tau}/F^0_{\nu_e},F^0_{\nu_\tau}/F^0_{\nu_\mu}\ll 1$ at source, the $\nu_\tau$ flux at the detector will be comparable to $F_{\nu_e}^\oplus\sim F_{\nu_\mu}^\oplus$. Indeed, it has been shown that even with an arbitrary flavor composition at the source, the flavor composition at the Earth will be democratic  $F_{\nu_e}^\oplus\sim F_{\nu_\mu}^\oplus\sim F_{\nu_\tau}^\oplus$  \cite{Bustamante:2015waa}.  To be more precise, only the flavor ratios shown with a blue butterfly-shaped region  in Fig.~\ref{fig:astro_ratios}  can be covered. More intriguingly, as long as the flux reaching the Earth is an incoherent combination of neutrino mass eigenstates, we still expect $F_{\nu_e}^\oplus\sim F_{\nu_\mu}^\oplus\sim F_{\nu_\tau}^\oplus$ as demonstrated in Fig.~\ref{fig:astro_ratios}. This implies even if well-studied new physics such as neutrino decay, quantum decoherence or mixing with sterile neutrino is invoked, the neutrino flavor ratio cannot move out of the blue region in Fig.~\ref{fig:astro_ratios}. In particular, $F_{\nu_\tau}^\oplus$ will remain sizable.

\begin{figure}
\centering
\includegraphics[width=0.4\textwidth]{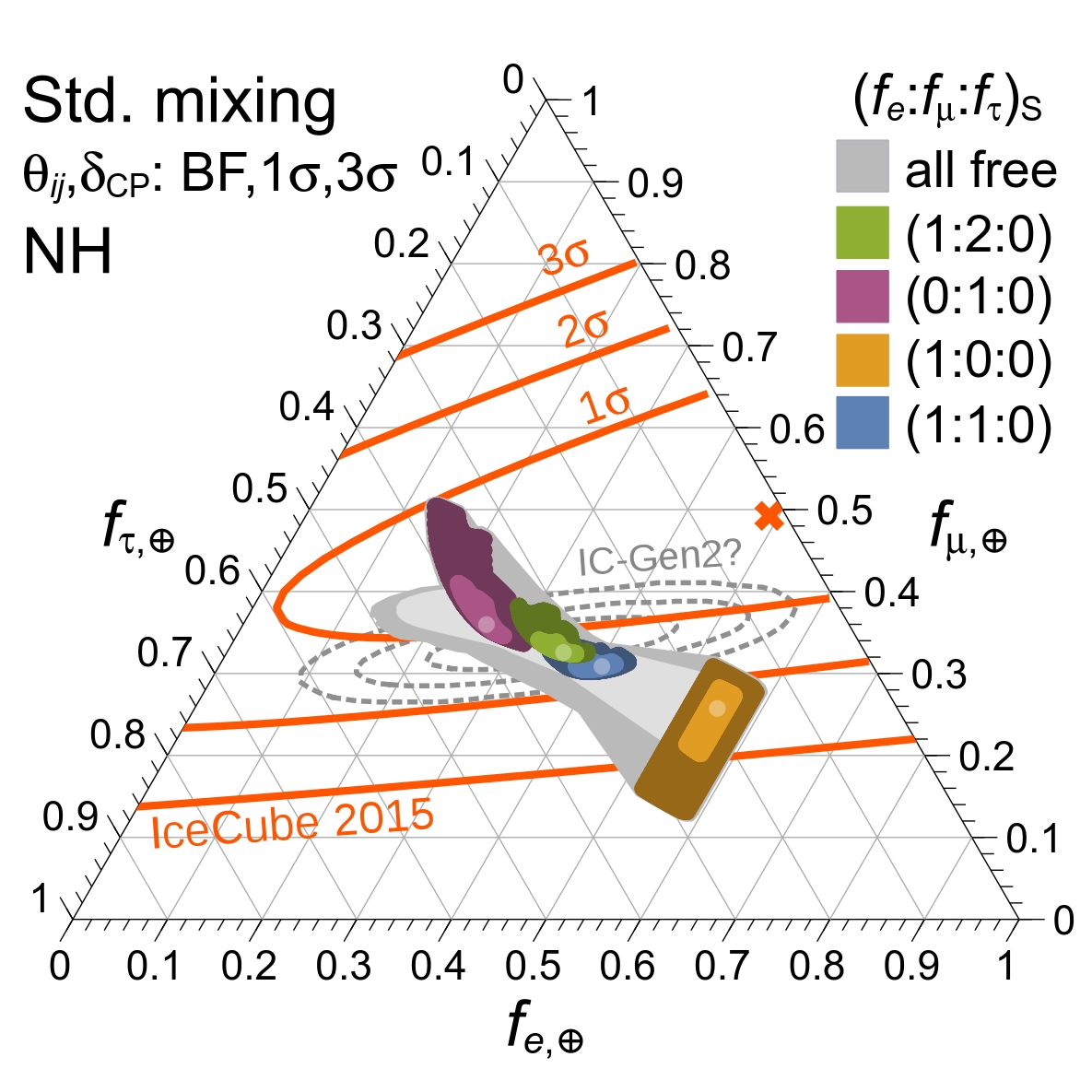}\hspace{0.5cm}
\includegraphics[width=0.4\textwidth]{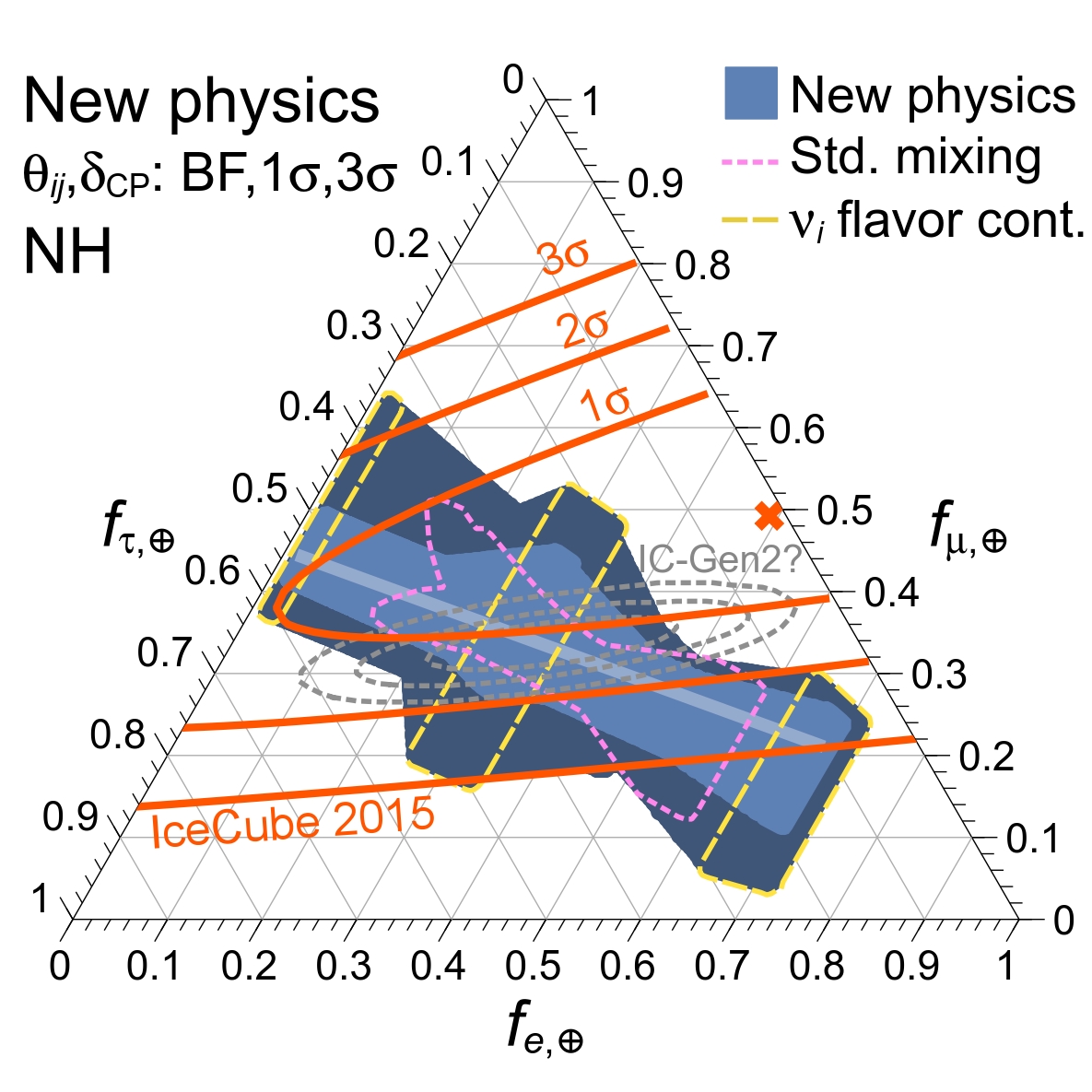}
\caption{Left: Allowed flavor ratios at Earth for different choices of
source ratios, assuming standard mixing. Projected 1$\sigma$, 2$\sigma$,
and 3$\sigma$ exclusion curves from IceCube-Gen2 are included for
comparison (gray, dotted). Right: Allowed flavor ratios at Earth in a general class of
new-physics models; the entire triangle can be covered from general new physics scenarios. These produce linear combinations of
the flavor content of $\nu_3,~\nu_2$ and $\nu_1$, shown as yellow (dashed)
curves, from left to right. The standard mixing 3$\sigma$ region
from the left is shown as a magenta (dotted) curve. Figures from  \cite{Bustamante:2015waa}. }
\label{fig:astro_ratios}
\end{figure}

As shown in Ref.~\cite{Farzan:2018pnk,deSalas:2016svi}, a coupling between ultralight dark matter and neutrinos breaks this rule, such that the original flavor ratio with $F_{\nu_\tau}\ll F_{\nu_e},F_{\nu_\mu}$ can be maintained up to the Earth for neutrino flux  produced in a source immersed in dark matter halo.
A current-current form of interaction between neutrinos and ultralight dark matter, $\phi$ of form
$$(\phi^*\partial_\mu \phi-\phi\partial_\mu \phi^*)(\bar{\nu}_\alpha \gamma^\mu \nu_\alpha)$$
induces a constant (without time modulation) effective mass of form
$m_{eff}^\alpha \nu_\alpha^\dagger\nu_\alpha$ for neutrinos in the background of dark matter. Such a current-current interaction can come by gauging  the flavor symmetry $L_\alpha -L_\beta$ and by assigning flavor to $\phi$.  As long as $|m_{eff}^\alpha-m_{eff}^\beta| \gg \Delta m^2_{atm}/E_\nu$, the mass eigenstates will correspond to flavor eigenstates and oscillation
of the flavors cease to happen. Consider a neutrino flux produced in a source located inside Dark Matter  (DM) halo such that at production the dark matter induced effective mass dominates over $\Delta m^2_{atm}/E_\nu$. Since the variation of dark matter density along the route of the flux is smooth, the flavor evolution  will be adiabatic and as a result, the initial flavor ratio will be maintained.\footnote{Notice that the dark matter effects maintain the coherence of neutrino mass eigenstates up to the Earth so the general remark made in \cite{Bustamante:2015waa} regarding new physics effects on flavor ratio does not apply for this scenario.}  The two $\nu_\tau$ events registered by IceCube \cite{IceCube:2020abv} put a strong bound on possible coupling between ultralight dark matter and neutrinos, implying that for PeV neutrinos, dark matter induced mass cannot be significant \cite{Farzan:2021gbx,Farzan:2021slf}. However, for 
yet higher energy neutrinos, the dark matter effects inside dark matter halo can dominate ({\it i.e.,} for $E_\nu\stackrel{>}{\sim}$
EeV inside the DM halo, $|m_{eff}^\alpha-m_{eff}^\beta|\gg \Delta m^2_{atm}/E_\nu$). However, if neutrinos are produced outside the halo, the vacuum term will dominate and the coherence of the mass eigenstates reaching the halo of the Milky Way will be lost. That is for cosmogenic neutrinos, we expect the canonic flavor ratio of
$F_{\nu_e}^\oplus:F_{\nu_\mu}^\oplus:F_{\nu_\tau}^\oplus\simeq 1:1:1$. If the future searches confirm a flux of ultrahigh energy neutrinos with $F_{\nu_e}\sim F_{\nu_\mu}\gg F_{\nu_\tau}$, we may conclude (i) They originate from a source inside a DM halo so they cannot be cosmogenic; (ii) There is a current-current form of interaction between ultralight DM background and neutrinos which maintains the coherence of mass eigenstates comprising flavor eigenstates and consequently preserves the original flavor ratios of neutrinos \cite{Farzan:2021gbx,Farzan:2021slf}.

For neutrinos of energy of $O(\text{EeV})$, the attenuation 
length in Earth is of order of $\mathcal{O}(100)$ km.  The electron and muon neutrinos, traveling inside the Earth produce electron or muon which can be absorbed. The tau neutrinos produce charged taus which decay back to lower energy neutrinos and charged leptons. This is called regeneration mechanism. The EeV tau neutrino skimming through the Earth ({\it i.e.,} traversing chords with lengths of order of or smaller than the attenuation length) gives rise to extensive air shower signal which can be detected by various upcoming detectors such as GRAND and POEMMA \cite{GRAND:2018iaj,POEMMA:2020ykm}. On the other hand, the EeV $\nu_\tau$ flux passing through longer chords inside the Earth can produce a flux of PeV neutrino flux through regeneration. The upper bounds from IceCube on the PeV neutrino flux constrain the $\nu_\tau$ flux of $\sim $EeV energy from both continuous or transient sources \cite{Barbano:2019yrk,Safa:2019ege}.

In 2006 and 2014, the ANITA radio telescope flying over Antarctica 
registered two events that looked like $\nu_\tau$ of $\sim 0.6$ EeV, emerging from deep down the Earth after crossing chords of sizes 5800~km and 7300~km, respectively 
\cite{ANITA:2016vrp,ANITA:2018sgj,ANITA:2020gmv}. Considering that the Earth is opaque for neutrinos of such high energy, the observation defies an explanation within the SM. To explain the two events, various beyond SM scenarios have been developed, in particular, Ref.~\cite{Cherry:2018rxj} suggests a $3+1$ scheme with a sterile neutrino mixed with $\nu_\tau$ as a solution.
Refs.~\cite{Farzan:2021gbx,Farzan:2021slf} revisit this scenario,
taking into account that (1) when neutrino mass eigenstates (including $\nu_4$) with energies of EeV scale cross the Earth, their active components are eliminated and unless active sterile oscillation takes place across the Earth, they emerge as pure sterile neutrinos;  (2) If $\nu_s$ only mixes with $\nu_\tau$, the $\nu_4$ flux arriving at the Earth cannot be intensive because the $\nu_\tau$ production at source is negligible.
That is $F_{\nu_e}^0:F_{\nu_\mu}^0:F_{\nu_\tau}^0:F_{\nu_s}^0=1:2:0:0$ will lead to $F_{\nu_1}:F_{\nu_2}:F_{\nu_3}:F_{\nu_4}=1:1:1:0$. To have a significant $\nu_4$ flux at Earth, the sterile neutrino should also mix with $\nu_e$ or $\nu_\mu$. On the other hand, in order for 
$\nu_4$ entering the Earth to emerge from the other side as $\nu_\tau$, the sterile active oscillation length should be of order of the chord size, $L$, which requires $m_{\nu_s}\sim \sqrt{E_\nu/L}\sim$few 100 eV for $E_\nu\sim$EeV. This has been shown in Fig.~\ref{fig:osci_chord}.

Refs.~\cite{Farzan:2021gbx,Farzan:2021slf} show that by saturating the bounds on the mixings with $\nu_\tau$ and $\nu_e$ and taking an EeV neutrino flux arriving at the Earth that saturates the IceCube bound from regeneration \cite{Barbano:2019yrk}, it will be  possible for  future experiments such as PUEO or POEMMA to detect  a flux of $\nu_\tau$ emerging from deep down the Earth with chord sizes far exceeding the attenuation. However, the flux would still be too small to account for the anomalous ANITA events.  If such future experiments register an EeV $\nu_\tau$ flux from deep down the Earth, we will expect accompanying PeV neutrino flux to be detected by neutrino telescopes as well as signatures of $3+1$ scheme with a sterile neutrino of few 100 eV mass and large mixing which can be tested {\it e.g.,} by scrutinizing the unitarity of the $3\times 3$ PMNS mixing submatrix.

\begin{figure}
\centering
\includegraphics[width=0.45\textwidth]{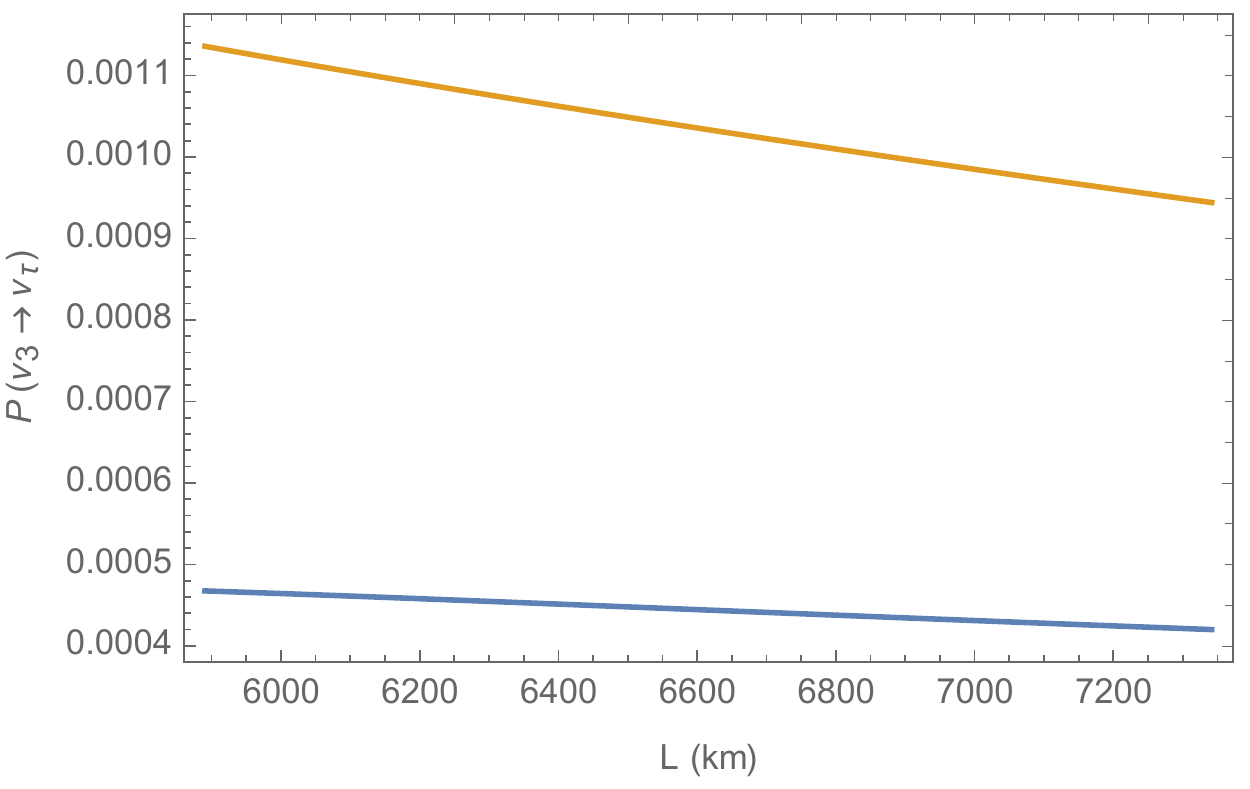}\hspace{0.5cm}
\includegraphics[width=0.45\textwidth]{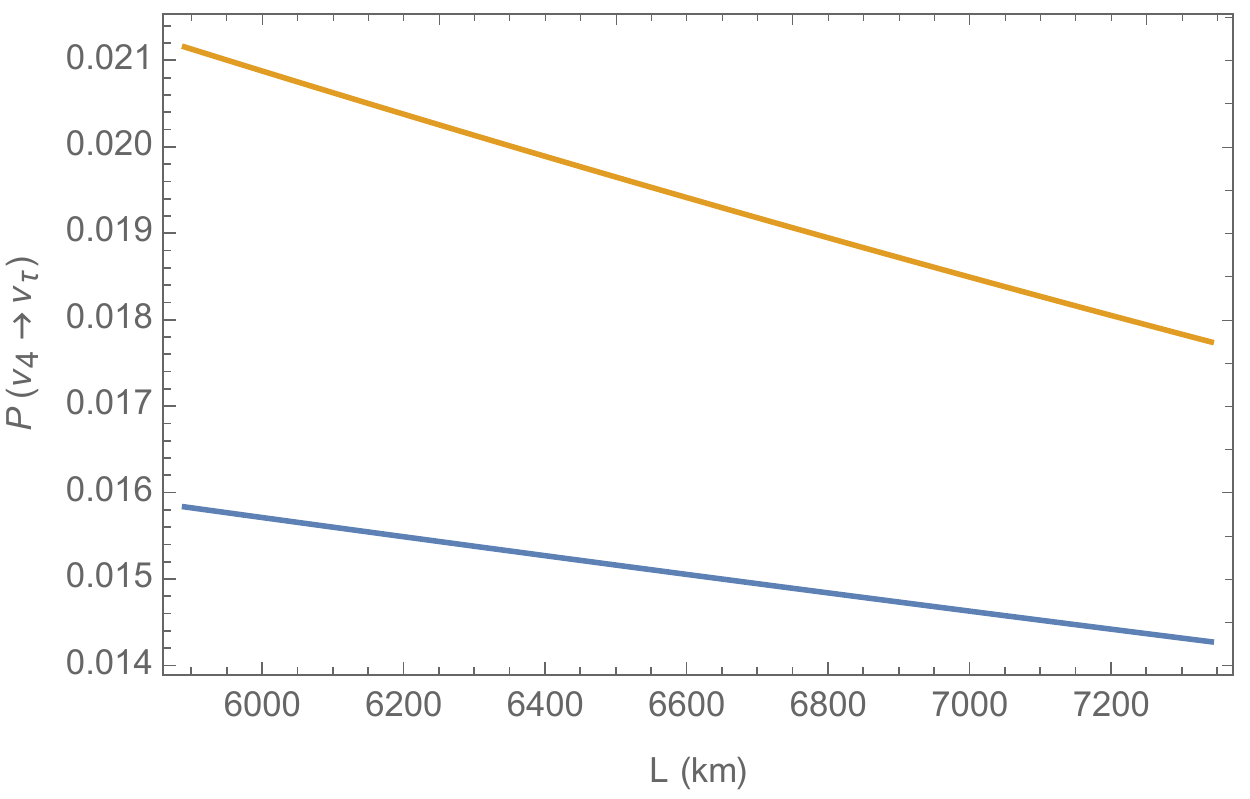}
\caption{ Probability of conversion
of $\nu_3$ or $\nu_4$
arriving at the Earth into $\nu_\tau$
after traversing a chord of size $L$ in the mantle. We have taken $E_\nu$=EeV, the neutrino-matter scattering rate $\Gamma =
0.003~\text{km}^{-1}$ (corresponding to $\rho$ = 4.5 $\text{gr/cm}^3$ and cross section of $1.1\times 10^{-32}~\text{cm}^2$ taken from \cite{Safa:2019ege}),
$|U_{\tau 4}|^2 $ = 0.1 and $\Delta M^2/2E_\nu$ = 0.35$\Gamma$ which corresponds to $\sqrt{\Delta M^2}$= 643 eV. The orange and blue
lines respectively correspond to the antineutrino and neutrino modes. Figure from \cite{Farzan:2021gbx}. }
\label{fig:osci_chord}
\end{figure}

\subsubsection{Neutrino self-interactions}

Exploring $\nu_\tau$ properties is challenging, as they are difficult to directly produce in the laboratory. A simple example is the quest for neutrino self-interactions ($\nu$SI), where the $\nu_\tau$ sector remains mostly unexplored. Indeed, strong neutrino self-interactions that impact our understanding of the Early Universe (see Refs.~\cite{Blinov:2019gcj,Cyr-Racine:2013jua, Archidiacono:2013dua, Lancaster:2017ksf, Oldengott:2017fhy, Huang:2021dba}) are only allowed in the $\nu_\tau$ sector.

Figure \ref{fig:bounds} illustrates this: while $\nu$SI constraints are strong for $\nu_e$, they are incomplete for $\nu_\mu$ and nearly nonexistent for $\nu_\tau$. The present constraints (at $2\sigma$) are shown in shaded contours~\cite{Brdar:2020nbj,Blinov:2019gcj, Blum:2014ewa, Kelly:2019wow,Esteban:2021tub}. The dashed purple line indicates the interaction strength below which cosmological neutrinos free-stream as expected at times relevant to observational cosmology. As an example of $\nu$SI that is presently allowed, the hatched region indicates the Moderately Interacting Neutrino (MI$\nu$) solution~\cite{Kreisch:2019yzn, Blinov:2019gcj}, which has been argued to affect cosmological parameter extraction from CMB data, especially the Hubble constant $H_0$ and the amplitude parameter $\sigma_8$.  Even if the MI$\nu$ solution fades away once more data is accumulated (there seem to be indications for that~\cite{Cyr-Racine:2013jua, Archidiacono:2013dua, Lancaster:2017ksf, Oldengott:2017fhy, Huang:2021dba}), it will remain important to probe the full parameter space above the purple dashed line.

\begin{figure}[hbtp]
  \centering
  \includegraphics[width=\textwidth]{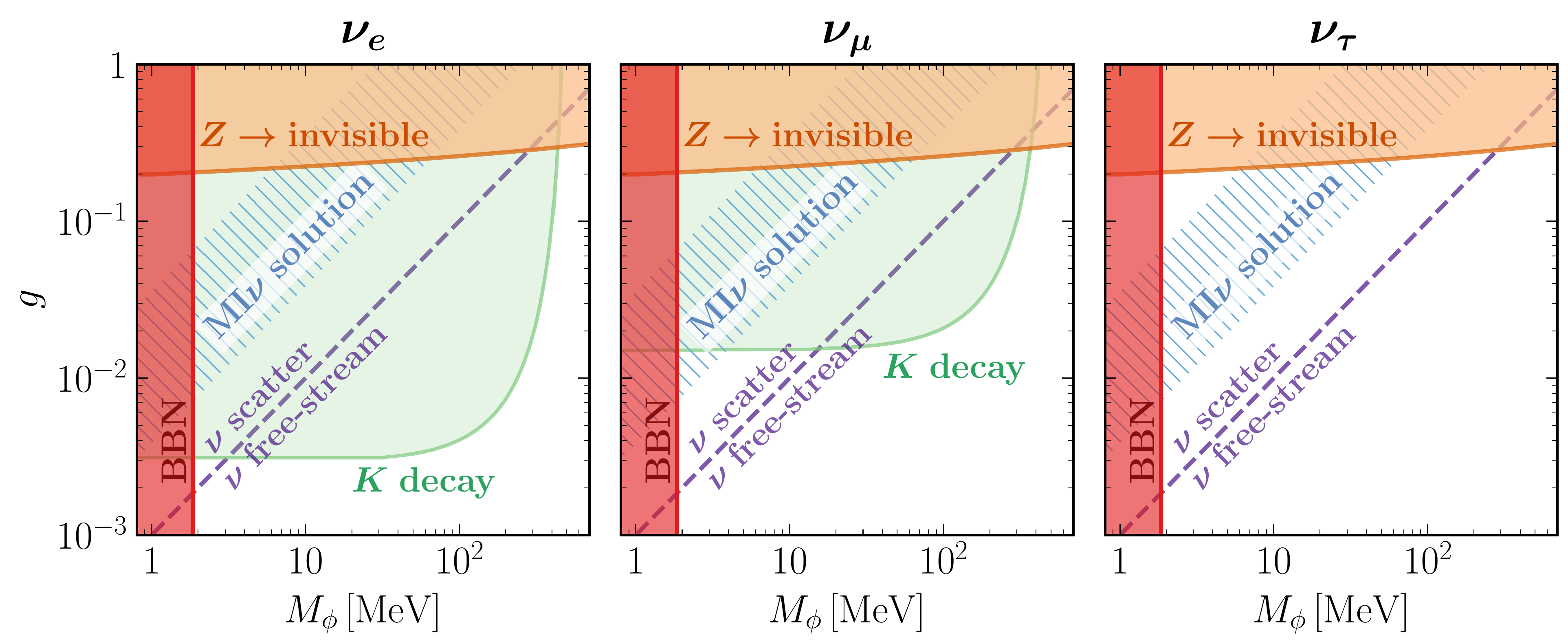}

  \caption{Present constraints on scalar neutrino self-interactions, with
      coupling strength $g$ and mediator mass $M_\phi$, for each of the three neutrino flavors.
      The hatched region is the ``Moderately Interacting
      neutrino'' (MI$\nu$) solution~\cite{Kreisch:2019yzn}, argued to affect CMB observables.
      The dashed purple line is the interaction strength below which cosmic
      neutrinos free-stream as expected at cosmologically relevant times. Above this line, our understanding of the early Universe would be affected. \emph{As shown, $\nu_\tau$ self-interactions are the least explored, leaving room for
      significant cosmological neutrino effects.} Figure from Ref.~\cite{Esteban:2021tub}.}
      \label{fig:bounds}
\end{figure}

HE and UHE neutrinos are a unique window to make progress and explore $\nu_\tau$ self-interactions. The first insight is that, because of flavor mixing, astrophysical neutrinos must always contain a large $\nu_\tau$ component. And, if there are strong $\nu$SI in the $\nu_\tau$ sector, these will affect their propagation. The basic physics is as follows~\cite{Kolb:1987qy, Hooper:2007jr, Ng:2014pca, Ioka:2014kca}: en route to Earth, high-energy astrophysical neutrinos may scatter with neutrinos in the C$\nu$B. As a consequence, high-energy neutrinos are absorbed and lower-energy neutrinos are regenerated.  This leads to unique dips and bumps in the astrophysical neutrino spectrum.

Furthermore, $\nu$SI in astrophysical neutrinos can be resonantly enhanced. This happens when the center-of-mass energy of the interaction equals the mediator mass. For $M_\phi \gtrsim 1 \, \mathrm{MeV}$ (see Fig.~\ref{fig:bounds}), $\nu$SI are resonant for neutrino energies $\gtrsim 10^5 \, \mathrm{GeV}$. That is, astrophysical HE and UHE neutrinos have precisely the energies where they can probe unexplored, cosmologically relevant $\nu$SI.

And, whereas present IceCube data does not have enough statistical power for this, future observatories will open the era of precision exploration of $\nu_\tau$ self-interactions. Keys to this will be their high-energies, large statistics, and relatively good energy resolution.

Figure \ref{fig:contours} shows the power of future observations. IceCube-Gen2 optical~\cite{IceCube-Gen2:2020qha} will have superb sensitivity, covering a huge range of cosmologically relevant $\nu$SI parameters. Its sensitivity decreases for $M_\phi \gtrsim 20 \, \mathrm{MeV}$, the mediator masses above which the $\nu$SI spectral features are at energies $E_\mathrm{dep} \gtrsim 10^7 \, \mathrm{GeV}$, higher than the ones that IceCube-Gen2 optical will be sensitive to. The remaining allowed region could be explored with higher-energy observatories~\cite{GRAND:2015uko, ARA:2019wcf, IceCube-Gen2:2020qha, Abarr:2020bjd, POEMMA:2020ykm,  Prohira:2021vvn, Fiorillo:2020jvy, Fiorillo:2020zzj}.

\begin{figure}
\centering
\includegraphics[width=0.5\textwidth]{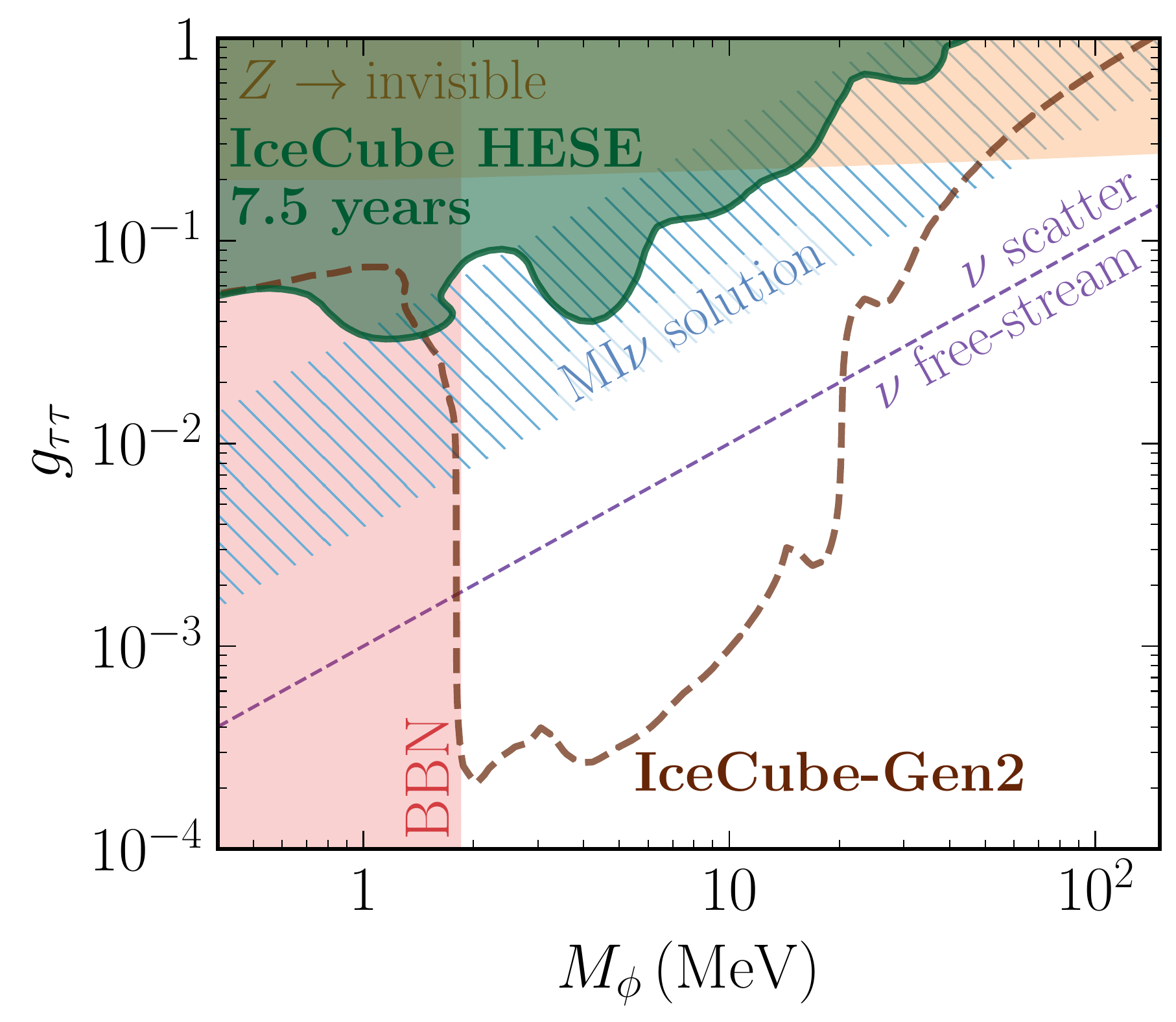}
\caption{Present and future sensitivity to $\nu_\tau$
  self-interactions, along with present bounds and cosmologically relevant
  regions (c.f.~Fig.~\ref{fig:bounds}). The dark green region, including part of the MI$\nu$ region, is excluded
  by IceCube data, and the dashed brown line shows the IceCube-Gen2 optical sensitivity (2$\sigma$). 
  \emph{IceCube-Gen2 will be sensitive to
  a large parameter space where neutrinos have a non-trivial cosmological behavior}. The sensitivity to other flavors is comparable. Figure from Ref.~\cite{Esteban:2021tub}.}
\label{fig:contours}
\end{figure}

Next generation astrophysical neutrino observations will open many additional windows. As mentioned above, UHE neutrinos will explore the higher-mass end of the parameter space. Other key observable is the flavor composition (see, e.g., Ref.~\cite{Song:2020nfh}), different for different $\nu$SI models. Another is exploiting point sources: the main purpose of next-generation neutrino astronomy is to resolve individual neutrino  sources~\cite{IceCube-Gen2:2020qha}; any detection would be highly valuable to explore $\nu$SI~\cite{Kelly:2018tyg}. The reason is that nearby sources should not be affected by $\nu$SI and could provide a better understanding of the high-energy astrophysical neutrino spectrum. The appearance of spectral signatures in far but not near sources would be a smoking gun for $\nu$SI. In addition, the scattering of neutrinos en route to the Earth could introduce measurable time delays~\cite{Ng:2014pca}. Finally, hints for $\nu$SI could leave signatures in future precise cosmological observables~\cite{EUCLID:2011zbd, Maartens:2015mra, DESI:2016fyo, CMB-S4:2016ple, LSSTDarkEnergyScience:2018jkl}.

\pagebreak
\section{Experimental Probes at the GeV Scale}
\label{sec:gev experiments}
\subsection{Short-baseline Accelerator}

\subsubsection{NOMAD}
\label{sec:nomad}
The Neutrino Oscillation MAgnetic Detector (NOMAD) experiment was designed to search for $\nu_\tau$ appearance from neutrino oscillations in the CERN wide-band neutrino beam produced by the 450 GeV proton synchrotron \cite{NOMAD:1997pcg}. 
The average neutrino flight path to NOMAD was 625 m, the detector being 836 m downstream of the beryllium target for the primary protons.
The relative composition of CC events in NOMAD was $\nu_\mu$ CC : $\bar \nu_\mu$ CC : $\nu_e$ CC : $\bar \nu_e$ CC = 
1.00 : 0.025 : 0.015 : 0.0015, with average neutrino energies of 45.4, 40.8, 57.5, and 51.5 GeV, respectively.
The prompt $\nu_\tau$ component was negligible.
The experiment collected $5 \times 10^{19}$ protons on target from 1995 to 1998, corresponding to 
about $1.4 \times 10^6$ $\nu_\mu$ CC interactions in the NOMAD fiducial volume. 

\begin{figure}[tb]
\centering
\includegraphics[width=.55\linewidth]{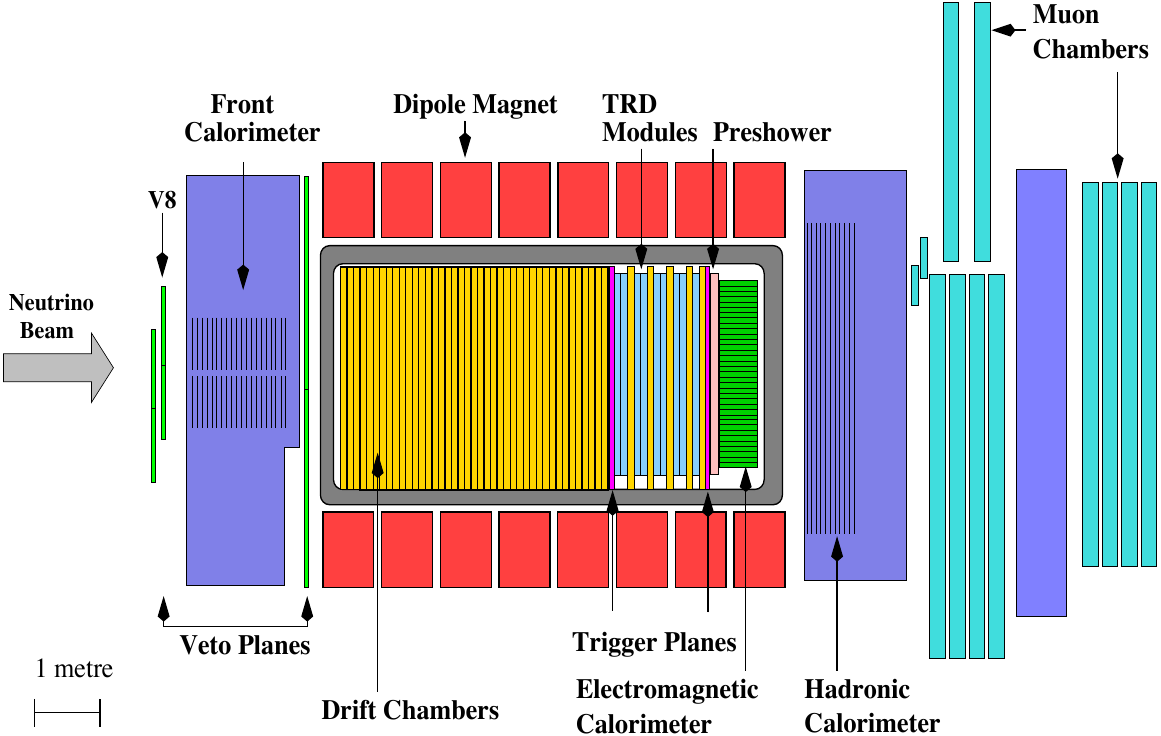}
\caption{\label{NOMAD-detector} Side view of the NOMAD detector from \cite{NOMAD:1997pcg}.} 
\end{figure}

The search for a signal from $\nu_\tau$ appearance in NOMAD relied on the identification of $\nu_\tau$ CC interactions using kinematic criteria~\cite{Albright:1978ni}. 
The spatial resolution of the detector did not resolve the $\tau$ decay vertex from the $\nu_\tau$ CC interaction.
The identification of $\nu_\tau$ CC 
events was thus achieved by exploiting the kinematic constraints which can be constructed from a precise measurement of the four-momenta of 
all visible final-state particles. This requires a detector with good momentum, angular, and energy resolutions, as well as sophisticated analysis schemes. 
In particular, NOMAD was designed to achieve an accurate reconstruction of the event kinematics in a plane transverse to the beam direction to detect both the direction and magnitude of the transverse kinematic imbalance~\cite{NOMAD:2001xxt} and was the first experiment to do so.

The NOMAD detector (Fig.~\ref{NOMAD-detector}) can be considered a low density spectrometer \cite{NOMAD:1997pcg}.
The low-density design was essential to achieve an accurate reconstruction of the transverse plane kinematics and was based on drift chambers (DCH) acting both as a high resolution tracker and as an active neutrino target, 
with an average density of about 0.1 g/cm$^3$ and a total thickness of about one radiation length. 
The fiducial mass of the NOMAD DCH-target was about 2.7 tons and was composed primarily of carbon.  
Downstream of the DCH, there was a transition radiation detector (TRD), 
followed by a preshower (PRS) and a lead-glass electromagnetic calorimeter (ECAL). 
The ensemble of DCH, TRD, and PRS/ECAL was placed within a dipole magnet providing a 0.4 T
magnetic field. Downstream of the magnet was a hadron calorimeter, 
followed by two muon stations, each comprising large area drift chambers and separated by an iron filter. 
The charged tracks in the DCH were measured with a momentum (p) resolution of $\sigma (p) / p = 0.05/ \sqrt{L} \oplus 0.008 p / \sqrt{L^5}$ 
(p in GeV/c and L in meters) with unambiguous charge separation in the energy range of interest. 
The energy deposition from $e$ and $\gamma$ were measured in
ECAL with an energy resolution of $\sigma (E) / E = 1.04\% + 3.22\% / \sqrt{E}$ (E in GeV).

For each of the $\tau$ decay channels, background events can be divided into two categories with different kinematic configurations. 
In $\nu_\mu$ and $\nu_e$ CC interactions the leading lepton is typically well-isolated and balances the momentum of the remaining hadronic 
system in the transverse plane. Conversely, in NC interactions all visible particles are part of the hadron jet, and a large missing transverse momentum 
associated with the escaping neutrino is almost opposite to the direction of the hadronic system. 
The signal from $\tau$ decays has intermediate properties between these two extremes: 
the $\tau$ decay neutrino(s) introduce a modest missing transverse momentum and the non-collinearity of the $\tau$ and its visible decay products 
can reduce their isolation. Consequently, in $\nu_\tau$ CC events the transverse component of the total visible momentum and the variables describing 
the visible decay products have different absolute values and different correlations with the remaining hadronic system than in $\nu_\mu (\nu_e)$ 
CC and NC interactions.
In order to optimize separately the rejection of each of the two opposite background sources, 
NOMAD implemented an event classification based on the use of two distinct likelihood functions exploiting all the degrees of freedom of the event kinematics (and their correlations)~\cite{NOMAD:2001xxt}.

The NOMAD experiment found no evidence for $\nu_\tau$ appearance, setting limits on oscillation parameters $\sin^2 \theta_{\mu \tau} < 3.3 \times 10^{-4}$ at large $\Delta m^2$ 
and $\Delta m^2 < 0.7$ eV$^2$/c$^4$ at $\sin^2 \theta_{\mu \tau} = 1$ at 90 \% confidence level~\cite{NOMAD:2001xxt}. This value still represents the most stringent limit 
available on $\nu_\tau$ appearance at large $\Delta m^2$. The NOMAD sensitivity was not limited by backgrounds, but was essentially defined 
by the available statistics.
The NOMAD $\nu_\tau$ search developed the analysis of transverse kinematic imbalance into a mature technique, 
which was subsequently applied in other measurements.

\subsubsection{DUNE near detector}

The Deep Underground Neutrino Experiment (DUNE) is a long baseline experiment that aims to measure the unknown parameters of the PMNS matrix and at the same time, seeks to search for beyond the Standard Model physics~\cite{DUNE:2020ypp, DUNE:2020lwj, DUNE:2020txw}. DUNE uses a neutrino beam provided by the Long Baseline Neutrino Facility (LBNF), which is located at the Fermi National Accelerator Laboratory in Batavia, Illinois. The current nominal neutrino beam design for DUNE is obtained using a 1.2 MW proton beam, with an energy of 120 GeV, that impinges on a cylindrical graphite target with a diameter of 16 mm and a length of 2.2 m. This proton-target collision produces hadrons which are focused by three 300 kA magnetic horns. Those hadrons travel along a 194 m decay pipe before decaying into $\nu_\mu$ (or $\bar{\nu}_{\mu}$, according to the polarity of the hadrons), with a small contamination of $\nu_e$ ($\bar{\nu}_e$). The resulting $\nu_\mu$ energy spectrum peaks at around 2.5 GeV (close to the first oscillation maximum) with most neutrinos having an energy of 0-6 GeV, as shown in figure \ref{tauoptim}. This low energy neutrino beam configuration is optimized for the search for CP-violation. However, the beamline could potentially be modified to generate a higher energy $\nu_\mu$ beam. This would be done by replacing the three focusing horns with two parabolic horns, separated by a distance of 17.5 m, and by modifying the target to be 1.5 m long and 10 mm wide, positioned 2 m from the first horn  (more details can be found in \cite{DUNE:2020mra}). With this alternate beamline configuration, the $\nu_\mu$ beam peaks at around 5 GeV (Figure \ref{tauoptim} in red),  where a significant portion of the $\nu_\mu$ have an energy above the 3.4 GeV kinematic threshold for $\tau$ leptons production.

\begin{figure}
\centering
\includegraphics[width=.55\linewidth]{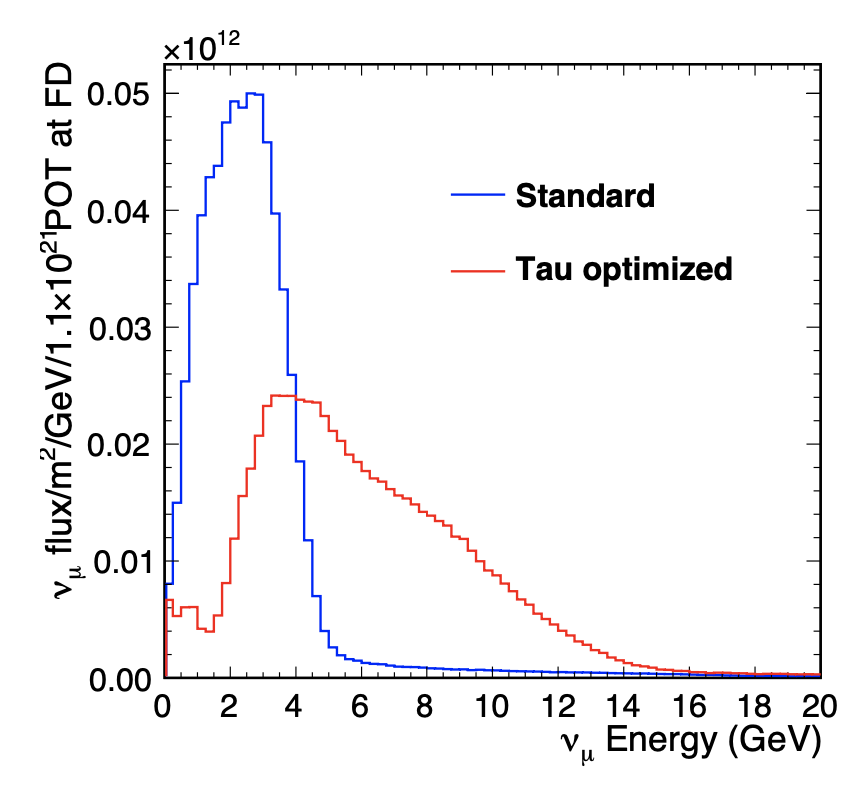}
\caption{\label{tauoptim} Comparison between the standard CP-optimized and a potential tau-optimized neutrino fluxes for the DUNE experiment~\cite{DUNE:2020ypp}.}
\end{figure}


The DUNE Near Detector (ND)~\cite{DUNE:2021tad} will consist of a collection of detectors located 574 m downstream of the neutrino source designed to measure with high precision the spectrum and flavor composition of the initial neutrino beam. The current design of the DUNE Near detector has three components : a Liquid Argon TPC (LArTPC) named ND-LAr, a gaseous argon TPC (ND-GAr) and the System for on-Axis Neutrino Detection (SAND). The ND-LAr has an active volume of $3 \times 4 \times 7$ $m^3$, where it is $5$ meters long in the beam direction, $3$ meters tall and $7$ meters transverse to the beam, with an active mass of $147$ t . It is composed of several argon cube modules where each module is $1 \times 1 \times 3.5$ $m^3$. The ND-GAr is based on a High Pressure gaseous Argon TPC (HPgArTPC). The HPgArTPC is held at a $10$ atm pressure and has an active mass of $1.8$ tons. It is surrounded by an electromagnetic calorimeter called the Barrel ECAL designed to detect photons and neutrons, and provide the initial timing for interactions inside the HPgArTPC. Located at the end of the detector is SAND, which monitors the on-axis beam~\cite{DUNE:2020mra}.

The SAND detector is expected to provide an excellent sensitivity to $\nu_\tau$ searches since it includes a low-density integrated target-tracking system based on the NOMAD concept described in Sec.~\ref{sec:nomad}. A Straw Tube Tracker (STT) is alternating thin passive layers -- typically 1-2\% of radiation length -- of polypropylene (CH$_2$) and graphite (C) with straw layers with negligible mass (about 3\% of the total mass). The average density can be tuned in the range 
$0.005 \leq \rho \leq 0.18$ g/cm$^3$ and the total thickness of the STT is comparable to one radiation length. The default fiducial mass is about 5 tons. 
The STT has more than one order of magnitude higher track sampling compared to NOMAD, resulting in improved resolutions and acceptance. 
The STT is surrounded by a 4$\pi$ hermetic electromagnetic calorimeter, both located inside a 0.6 T solenoidal magnetic field. 
The high momentum, angular, and energy resolution of SAND and its 4$\pi$ acceptance for final state particles allow an accurate reconstruction of 
transverse plane kinematics, which is essential for the $\nu_\tau$ detection and can compensate the relatively small fiducial mass. 

The primary purpose of the ND is to reduce systematic uncertainties for long-baseline oscillation analyses. The LArTPC technology offers millimeter-scale spatial resolution, excellent particle tracking performance, and accurate calorimetric information. However, these properties, along with its exposure to the intense neutrino flux from LBNF, and its short baseline of 547 m, make the DUNE ND an excellent candidate for probing anomalous short-baseline $\nu_{\tau}$ appearance which may occur either through zero-baseline oscillations due to non-unitarity of the PMNS matrix, or through sterile-driven oscillations involving sterile states with masses in the $\sim$1-10~eV range.

For $\nu_\tau$ CC interactions, the produced $\tau$-lepton's short lifetime makes it impossible to detect it directly in the DUNE ND. Instead, the $\tau$ is reconstructed from its decay product. However, the $\tau$-lepton can decay many ways including leptonically which mimic the signature of $\nu_e$ CC and $\nu_{\mu}$ CC interactions and hadronically which mimic NC interactions.  It has previously been shown that transverse plane kinematic differences are useful for distinguishing $\nu_{\tau}$ CC signal from backgrounds~\cite{Albright:1978ni}.  Therefore, a variety of variables characterizing the kinematics of the visible $\tau$ decay system and the hadronic recoil system produced by the nucleus were combined using a Boosted Decision Tree (BDT) trained to separate signal and background.  Currently, separate selectors have been developed focusing on three $\tau$-lepton decay channels: $\tau \rightarrow e \nu_{\tau} \bar{\nu}_{e}$, $\tau \rightarrow \mu \nu_{\tau} \bar{\nu}_{\mu}$, and $\tau \rightarrow \rho \nu_{\tau}$.  In the $e$ and $\mu$ channels, we assume that the lepton is identifiable to construct the input kinematic variables.  In the $\rho$ channel, we consider all $\pi^{\pm}\pi^{0}$ pairs as $\rho$ candidates and score them using a $\rho$ selection BDT.  The $\rho$ candidate with the highest BDT score is used to construct the input kinematic variables for the $\nu_{\tau}$ selector. 

Using these selectors, it is possible to determine the sensitivity of the DUNE ND to anomalous $\nu_{\tau}$ appearance based on event counting.  In long-baseline experiments, background systematic uncertainties are usually constrained by near detector measurements.  Since this analysis only uses the ND, it is not possible to constrain flux and cross section uncertainties.  Therefore, we choose to require high BDT scores in order to select a nearly background-free sample.  For the leptonic channels, it is possible to select signal with $\sim$20\% efficiency with negligible background, and for the $\rho$ channel, it is possible to select $\sim$40\% of signal events containing a well-identified $\rho$ candidate with negligible background.  

The combined sensitivity for one year of running ($1.1 \times 10^{21}$ POT) using the tau-optimized beam, a fiducial mass of the ND-LAr detector of 67 t , and 10\% systematic uncertainties is shown in Fig.~\ref{fig:dune_nd_combined_sensitivity}.  Due to the choice of requiring high BDT scores, the effect of systematic uncertainties on the backgrounds is low, and the resulting sensitivities are statistically limited. The black curve shows the expected sensitivity assuming perfect reconstruction, and the gray curve shows the expected sensitivity assuming reconstruction performance similar to other LArTPC experiments~\cite{Furmanski:2020smg}.  These are shown along with limits on short-baseline $\nu_{\tau}$ appearance from other experiments.  

\begin{figure}
    \centering
    \includegraphics[width=0.7\textwidth]{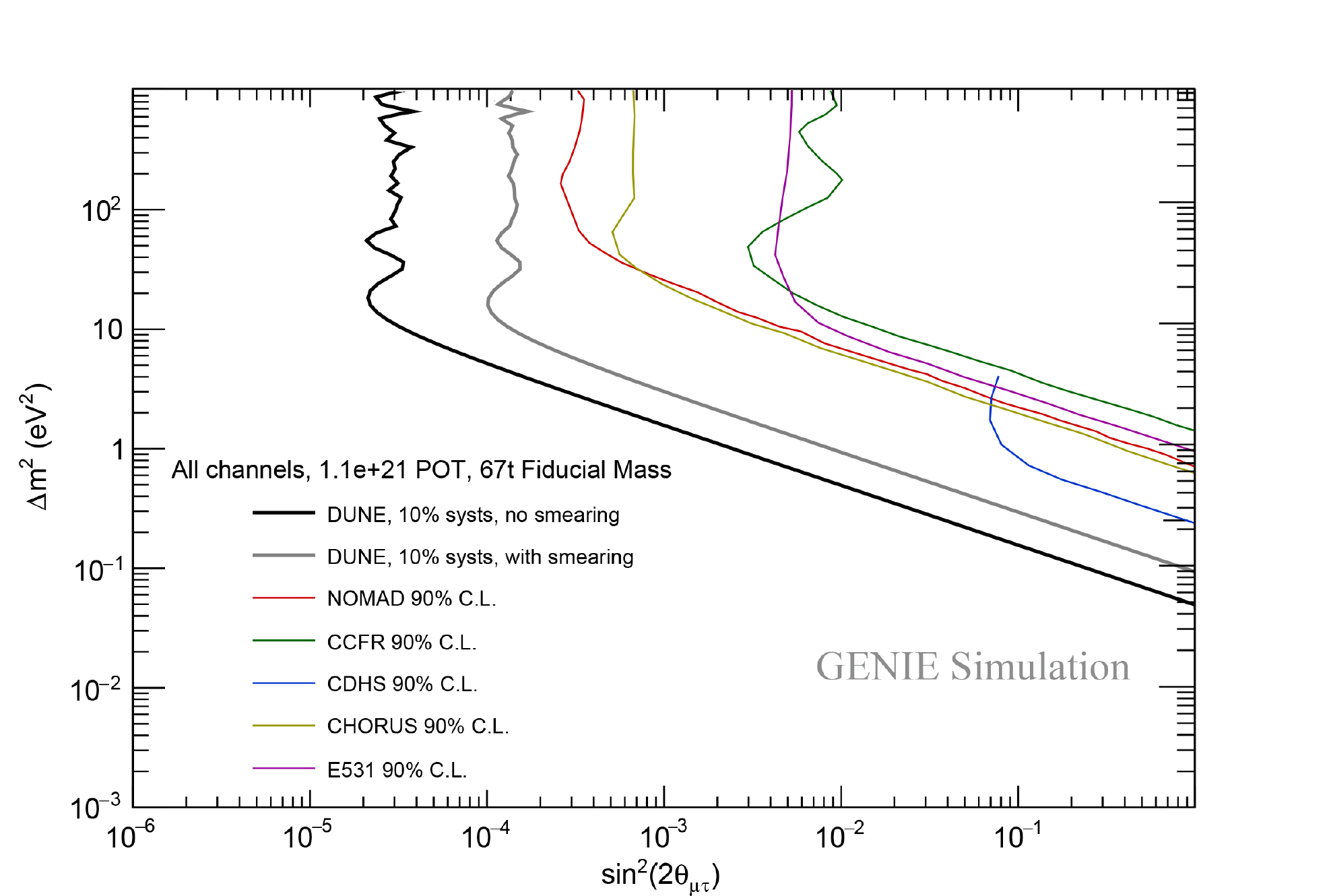}
    \caption{The combined sensitivity from the $e$, $\mu$, and $\rho$ channels. This sensitivity is calculated assuming $1.1\times 10^{21}$ protons-on-target exposure to the tau-optimized beam, a fiducial mass of 67 tons, and 10\% systematic uncertainty.}
    \label{fig:dune_nd_combined_sensitivity}
\end{figure}

\subsection{Fixed Target Experiments}
\label{sec:fixed}

\subsubsection{The NA65 experiment at CERN}

Lepton Flavor Universality is one of the basic assumption in the Standard Model, and has been tested in leptonic decays with high precision. However, it has not been
meaningfully tested in neutrino scattering. Moreover, recent measurements made  e.g. by the 
LHCb experiment show interesting hints of potential Flavor Universality Violation~\cite{LHCb:2017avl,LHCb:2021lvy}.
To date, the tau neutrino was  observed directly by only a few experiments. Among these, only the DONUT experiment~\cite{DONuT:2007bsg} studied tau neutrinos that were ``directly'' produced by the accelerator. Other experiments
made use of  ``oscillated'' neutrinos, which can be affected by secondary effects in the transportation through the Earth. For the test of tau neutrino nature, a precision study of tau neutrinos with a controlled beam is necessary.
Figure \ref{fig:dstau_cross-sec-nu} shows the result of DONuT, compared with other measurements in the deep inelastic scattering (DIS) regime. DONUT's cross section result is strongly affected by the systematic uncertainty of tau neutrino production. In particular, the uncertainty of $D_s$ double differential production cross section in proton interactions leads to an uncertainty of about 50\% with respect to the tau neutrino cross section value. The statistical uncertainty is smaller and amounts 33\%. Therefore, the knowledge of $D_s$ production is vital for high precision tau neutrino studies. The NA65/DsTau experiment~\cite{DsTau:2019wjb} is designed and approved to study tau neutrino production using  400 GeV protons from the CERN-SPS.

\begin{figure}[htbp]
\centering
\includegraphics[width=0.7\textwidth]{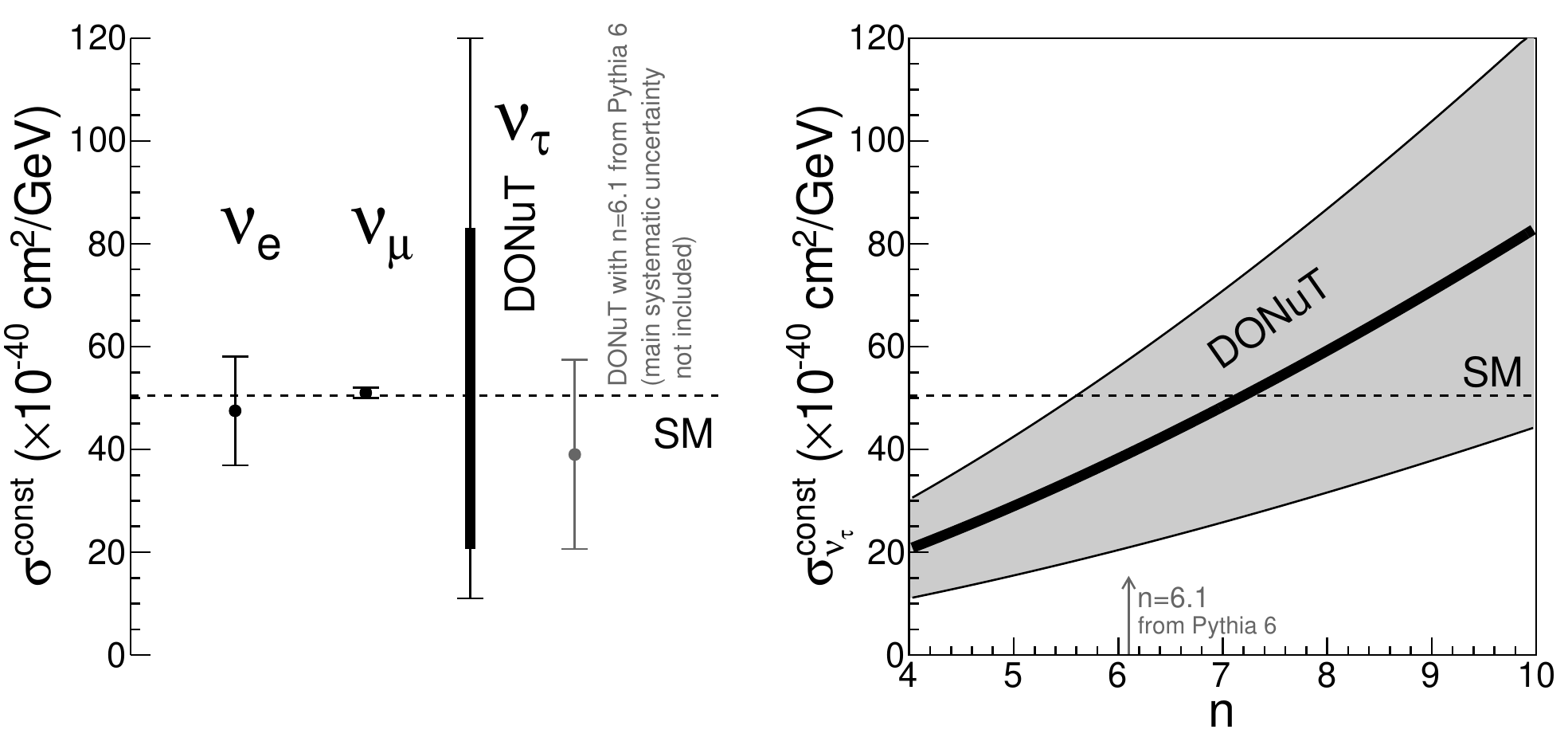}
\caption{Left: $\nu$, $\overline{\nu}$ averaged energy independent cross section of the three neutrino flavors \cite{DsTau:2019wjb}.
Right: The cross section result by DONUT, as a function of unknown kinematical parameter $n$ of the $D_s$ double differential production\cite{DONuT:2007bsg}.
}
\label{fig:dstau_cross-sec-nu}
\end{figure}

The main concept of the artificial tau neutrino beam is to exploit the leptonic decay of $D_s$ mesons produced in proton-nucleus interactions, and the subsequent tau decay, namely $D_s\rightarrow \tau\nu_\tau$ and $\tau\rightarrow \nu_\tau X$. 
Both decays take place at a distance scale of millimeters, and the kink angle at the $D_s\rightarrow \tau$ decay is expected to be very small, at the level of a few mrad. Therefore, such a measurement requires a very precise tracking detector. NA65/DsTau profits from the high spatial resolution of the emulsion detector, 50~nm position resolution and 0.35~mrad angular resolution. The $D_s \rightarrow \tau \rightarrow X$ decays can be identified by the double-kink topology. 

The detector set-up is shown in Figure \ref{fig:DsTauModule}. It consists of 10 target units, each made of 500-$\mu m$ thick tungsten plate, followed by 10 emulsion films interleaved with 200-$\mu m$ thick plastic sheets. This set-up includes a proton target, a decay volume for charmed hadrons and tau leptons as well as a high-precision tracking detector. The target units are followed by three additional tungsten plates for momentum measurement based on  multiple Coulomb scattering. In addition five additional emulsion films placed upstream of the module, act as a trigger for incoming protons. A single module containing 130 emulsion films has a transverse area of 25 cm $\times$ 20 cm$^2$ and 8 cm thick. A schematic view of the experimental setup is shown in Figure~\ref{fig:DsTauModule}. About 60 modules will be exposed to 400 GeV protons from the CERN-SPS. This will yield 200 million proton interactions, one half on tungsten and the other half on molybdenum nuclei, and 1000 $D_s\rightarrow\tau\rightarrow X$ decays will be collected. The $D_s$ double differential production cross section will be obtained using this sample, and tau neutrino production will then be constrained at the precision of 10\%.

\begin{figure}[htbp]
\begin{center}
 \includegraphics[width=0.44\textwidth, trim = 0cm 6.35cm 18.6cm 0, clip]{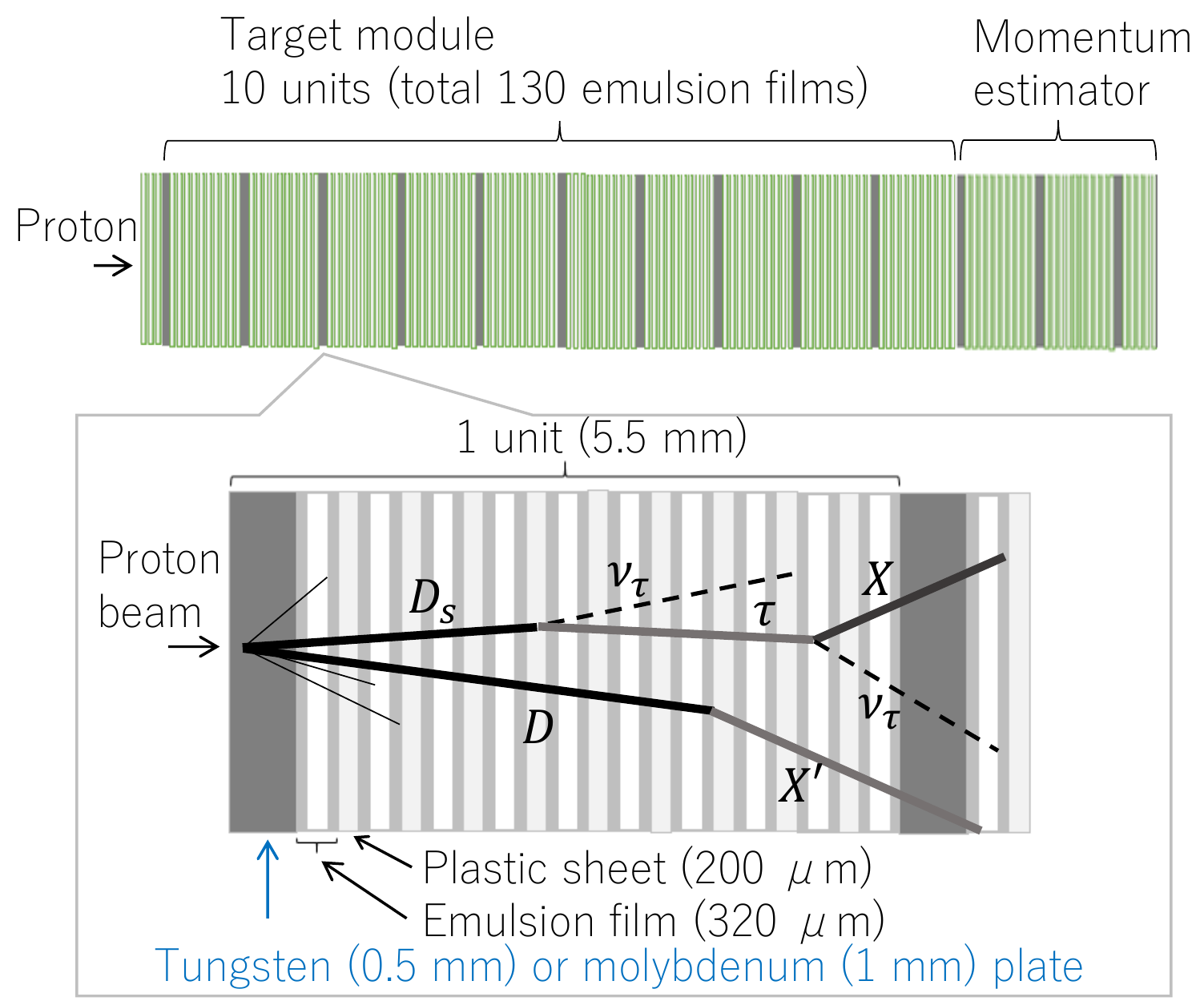}
 \includegraphics[width=0.55\textwidth]{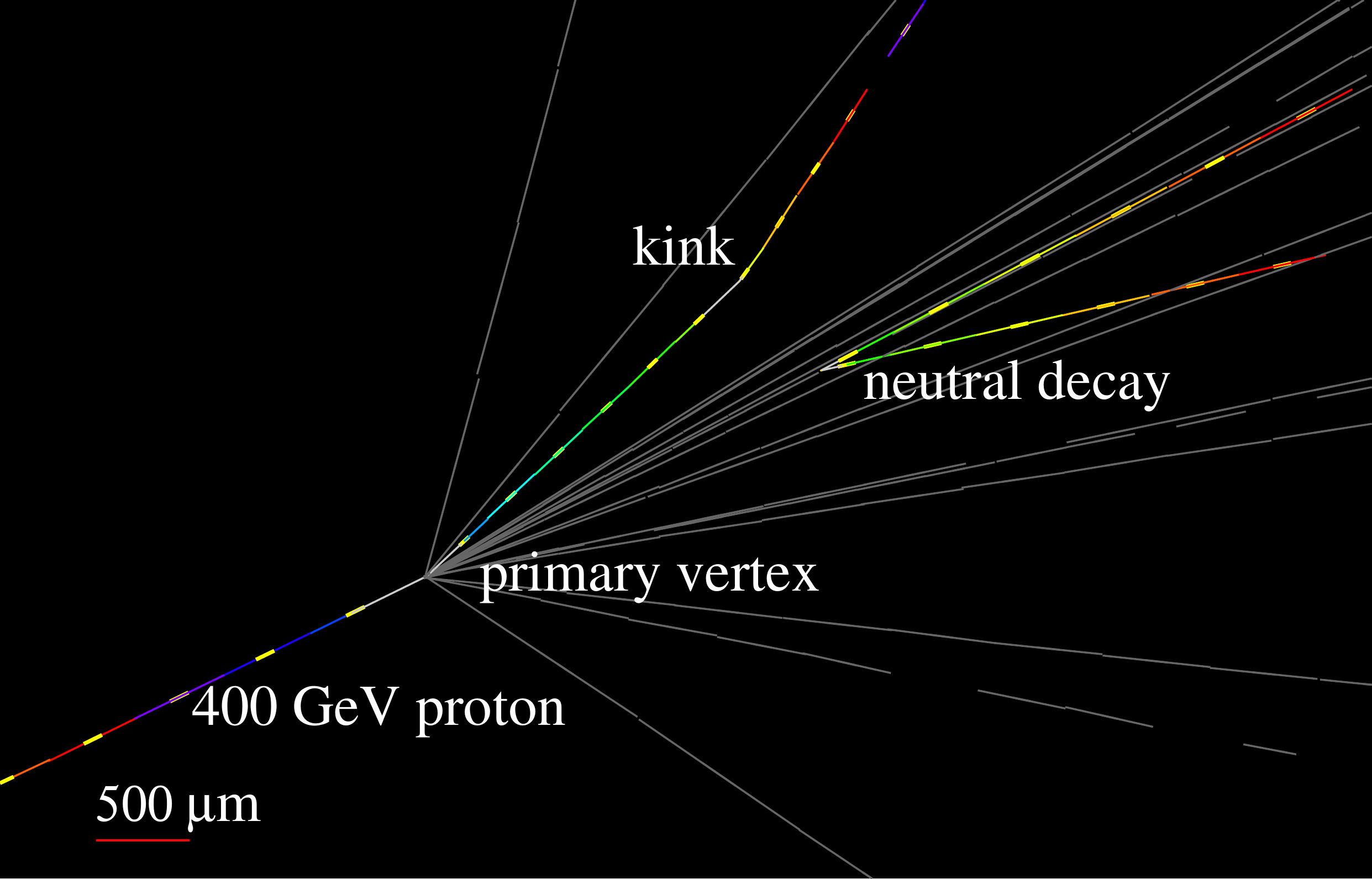}
\caption{Schematic view of the module structure. Each tungsten target plate followed by 10 emulsion films alternated by 9 plastic sheets acting as a tracker and decay volume of 4.8 mm. The sensitive layers of emulsion detectors are indicated by green color. This basic structure is repeated 10 times, and then followed by a lead-emulsion ECC structure for momentum measurement of the daughter particles. A double charm candidate event with a neutral 2-prong (vee) and a charged 1-prong (kink) topology (tilted view). Figure from \cite{DsTau:2019wjb}.}
\label{fig:DsTauModule}
\end{center}
\end{figure}

In addition to $\nu_\tau$ production, DsTau is expecting to collect a large sample of charmed particle pairs, at the level of 10$^5$. The analysis of such event would provide by-products, such as measurement of the forward charm production, the intrinsic charm content in proton, the interaction cross section of charm hadrons, etc.

Two test runs were performed at the CERN SPS H2 and H4 beam lines in 2016 and 2017 in order to test the various steps of the experiment from emulsion production to data analysis and to improve the detector structure and exposure scheme. A pilot run was conducted in 2018, collecting about 15 million interactions in tungsten.  
The obtained results provide a proof of feasibility for the full-scale physics runs in 2021 and 2022. 
DsTau collaboration performed its first physics run in September-October 2021 at the CERN SPS H2 beamline, collecting additional $\sim$ 30 million events. Another data taking will be performed in the future to fulfill the nominal total number of interactions of 200 million events

\subsubsection{The SHiP fixed target experimental proposal at CERN}
The  proposed Search for Hidden Particles (SHiP) beam-dump experiment \cite{Anelli:2007512} at the CERN SPS accelerator is designed to both search for feebly interacting GeV-scale particles and to perform measurements in neutrino physics. The experiment is optimised 
to make measurements on tau neutrinos and on neutrino-induced charm production by all three species of neutrinos.

The setup consists of a high-density
proton target located in a target bunker~\cite{PhysRevAccelBeams.22.123001,PhysRevAccelBeams.22.113001,Kershaw:2018pyb}, followed by a hadron stopper and a muon shield~\cite{Akmete:2017bpl}. The target is made of blocks of a titanium-zirconium doped molybdenum alloy (TZM), followed by blocks of pure tungsten. The total target depth is twelve
interaction lengths over 1.4\,m. The high atomic numbers and masses of the target material maximizes the production of charm and beauty as sources of Hidden Sector particles and neutrinos.

The SHiP detector consists of two complementary apparatuses, the Scattering and Neutrino Detector (SND) and the Hidden Sector Decay Spectrometer. 
SND aims at performing measurements with neutrinos and to search for light dark matter (LDM). 
The detection of tau neutrinos, and the aimed  observation for the first time of tau anti-neutrinos, pose a challenge for the design of the detector, which has to fulfill conflicting requirements: a large target mass to collect enough statistics, an extremely high spatial accuracy to observe the short-lived tau lepton and a magnetic field to disentangle neutrinos from anti-neutrinos.
The optimization of the SND layout also had to take into account constraints given by the location and the muon flux.
The overall layout, as implemented in simulation, is shown in Fig.~\ref{fig:spectro_layout}.

\begin{figure*}
\begin{center}
\includegraphics[width=0.7\linewidth]{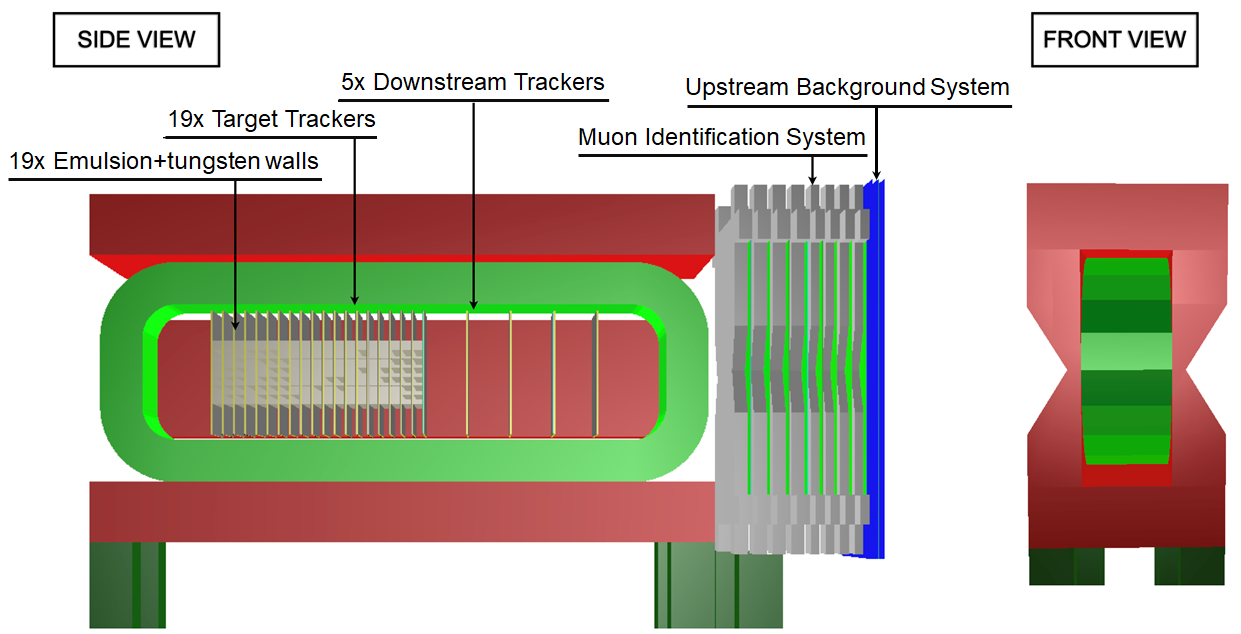}
\caption{Schematic layout of the Scattering and Neutrino Detector (SND)
~\cite{SHIP:2021tpn}.
}
\label{fig:spectro_layout}
\end{center}
\end{figure*}

SND consists of a $\sim$7\,m long  detector inserted in a magnet \cite{SHiP:2019qbi} providing a 1.2\,T horizontal magnetic field, followed by a muon identification system.  The magnet hosts the emulsion target, interleaved with target tracker planes, and a downstream tracker.
The emulsion target has a modular structure: the unit cell consists of an emulsion cloud chamber (ECC) made of tungsten plates interleaved with nuclear emulsion films, followed by a compact emulsion spectrometer (CES) for the momentum and charge sign measurement of particles produced in neutrino interactions. The ECC bricks are arranged in walls alternated with target tracker planes, providing the time stamp of the interactions occurring in the target. The downstream tracker is made of three target tracker planes separated by $\sim$50\,cm air gaps. It is used to measure the charge and momentum of muons exiting the target region, thus extending significantly the detectable momentum range of the CES. The downstream tracker planes also help to connect the tracks in the emulsion films with the downstream SND muon identification system. 
The SND muon identification system is made of a sequence of iron filters and resistive plate chamber (RPC) planes, totaling about two meters in length. 

The emulsion target is in the current baseline made of 19 emulsion brick walls and 19 target tracker planes. The walls are divided in 2$\times$2 cells, each with a transverse size of 40$\times$40 cm$^2$, containing ECC and a CES as illustrated in Fig.~\ref{fig:target_layout}.

\begin{figure}
\begin{center}
\includegraphics[width=0.5\linewidth]{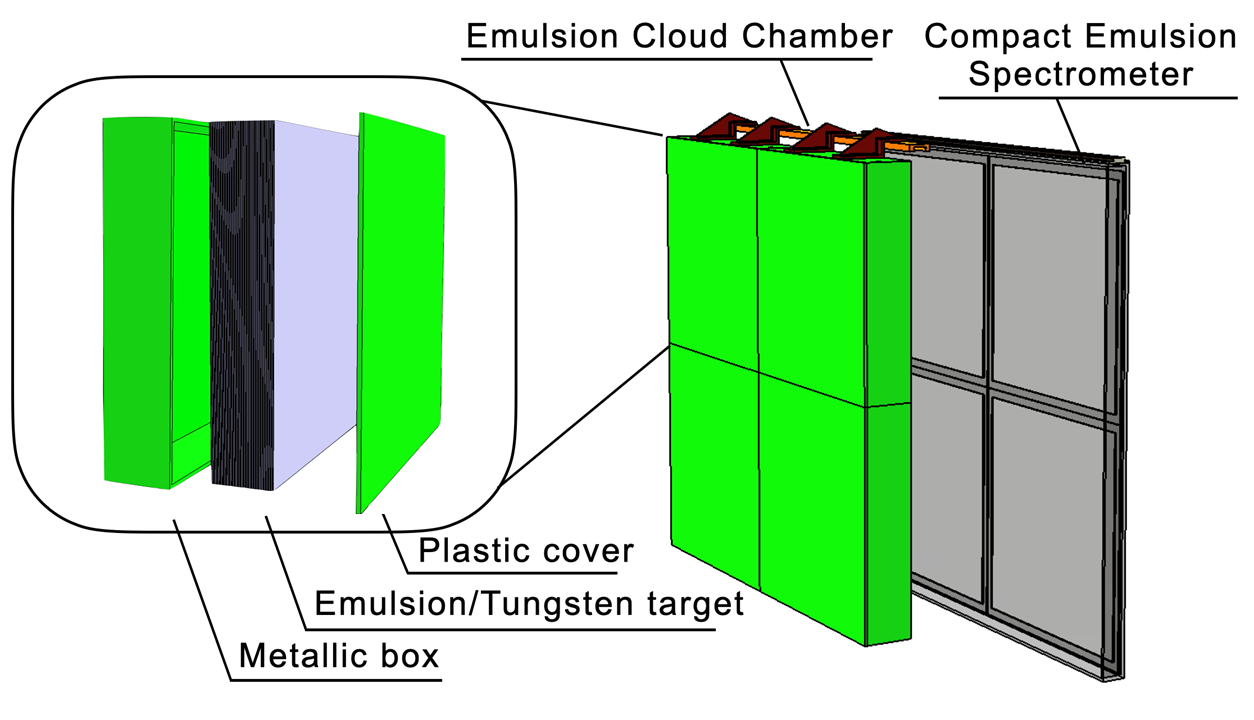}
\caption{Layout of the emulsion target and closeup view of one emulsion brick wall of four cells, each containing an ECC and a CES
~\cite{SHIP:2021tpn}.
}
\label{fig:target_layout}
\end{center}
\end{figure}

The ECC technology makes use of nuclear emulsion  films interleaved with passive absorber layers to build up a tracking device 
with sub-micrometric position and milliradian angular resolution, as demonstrated by the OPERA experiment~\cite{Acquafredda:2009zz}. 
The technique allows detecting tau leptons~\cite{OPERA:2018nar} and charmed hadrons~\cite{Agafonova:2014khd} by disentangling their production and decay vertices.  The high spatial resolution of the nuclear emulsion allows measuring the momentum of charged particles through the detection of multiple Coulomb
scattering in the passive material~\cite{OPERA:2011aa}, and identifying electrons by observing electromagnetic showers in the brick~\cite{Agafonova:2018dkb}.
An ECC brick is made of 36 emulsion films with a transverse size of  40$\times$40 cm$^2$, interleaved with 1\,mm thick tungsten layers. Tungsten has been chosen in place of lead as in the OPERA experiment  for its higher density and for its shorter radiation length and smaller Moli\`ere radius in order to improve the electromagnetic-shower containment. The resulting brick has a total thickness of $\sim$5 cm, corresponding to $\sim$10\,$X_0$, and a total weight of $\sim$100\,kg. The overall target weight with 19 walls of 2$\times$2 bricks is about 8\,tonnes. 
The CES modules aim at measuring the electric charge of hadrons produced in tau lepton decays, thus providing the unique feature of disentangling $\nu_\tau$ and $\overline{\nu}_\tau$ CC interactions also in their hadronic decay channels. It complements the use of the ECC in the momentum measurement for hadrons and soft muons which are emitted at large angles and which do not reach the downstream tracker. The basic structure of the CES is made of three emulsion films interleaved by two layers of low density material. 

The baseline technology for the SND tracker systems consists of a scintillating-fibre tracker (SciFi). 
The  SND muon identification system is designed to identify with high efficiency the muons produced in neutrino interactions and $\tau$ decays occurring in the emulsion target. 
The system consists of eight hadron filters of iron interleaved with tracking planes instrumented with RPCs.

The nuclear emulsion technology combined with the information provided by the SND muon identification system makes it possible to identify the three different neutrino \linebreak flavors in the SND detector.
The neutrino flavor is determined through the flavor of the primary charged lepton produced in neutrino CC interactions. The muon identification is also used to distinguish between muons and hadrons produced in the $\tau$ decay and, therefore, to identify the $\tau$ decay channel. In addition, tracking in the SND magnetic spectrometer will allow for the first time to distinguish between $\nu_\tau$ and $\overline{\nu}_\tau$ by measuring
the charge of $\tau$ decay products. The charge sign of hadrons and muons is measured by the CES, the downstream tracker, and by the muon identification system. 
The neutrino fluxes produced in the beam dump have been estimated with simulations, including the contribution from cascade production in the target. 
The number of CC deep inelastic scattering (DIS) interactions in the neutrino target is evaluated by convoluting the generated neutrino spectrum with the cross-section provided by the \textsc{GENIE}~\cite{Andreopoulos:2009rq} Monte Carlo generator. The expected number of CC DIS in the target of the SND detector is reported in the first column of Table~\ref{tab:neu_yield}. 

\begin{table}
\begin{center}
\begin{tabular}{c   c c }
&   CC DIS  & CC DIS  \\
&  interactions   & w. charm prod.\\
 \hline\hline
 $N_{\nu_e}$ & $8.6 \times 10^{5}$ & 5.1 $\times$10$^4$ \\
 $N_{\nu_\mu}$ & $2.4 \times 10^{6}$ &  1.1 $\times$10$^5$  \\
 $N_{\nu_\tau}$ & $2.8 \times 10^{4}$  & 1.5 $\times$10$^3$\\
 $N_{\overline{\nu}_e}$ & $1.9 \times 10^{5}$ & 9.8 $\times$10$^3$ \\
 $N_{\overline{\nu}_\mu}$ & $5.5 \times 10^{5}$ & 2.2 $\times$10$^4$ \\
 $N_{\overline{\nu}_\tau}$  & $1.9 \times 10^{4}$  & 1.1 $\times$10$^3$\\
 \hline
\end{tabular}
  \caption{ \label{tab:neu_yield} Expected CC DIS interactions in the SND assuming $2\times10^{20}$ protons on target.
  From \cite{SHIP:2021tpn}.}
\end{center}
\end{table}

With $2\times 10^{20}$ protons on target, more than $\sim$2$\times10^5$ neutrino-induced charmed hadrons are expected, as reported in the second column of Table~\ref{tab:neu_yield}. The total charm yield exceeds the samples available in previous experiments by more than one order of magnitude.
 
\textbf{Sensitivity to $F_4$ and $F_5$} The CC $\nu_\tau$ $(\overline{\nu}_\tau)$ differential cross-section is represented by a standard set of five structure functions:  
\begin{multline}
\frac{d^2\sigma^{\nu(\overline{\nu})}}{dxdy} = \frac{G_F^2 M E_\nu}{\pi (1+Q^2/M^2_W)^2}
 \biggl(
  (y^2x + \frac{m_\tau^2y}{2E_\nu M})F_1 + 
 \left[ (1-\frac{m_\tau^2}{4E_\nu^2}) - (1+\frac{Mx}{2E_\nu}) \right] F_2 \\
 \pm  \left[ xy(1-\frac{y}{2}) - \frac{m_\tau^2y}{4E_\nu M}\right] F_3 +
 \frac{m_\tau^2(m_\tau^2+Q^2)}{4E_\nu^2M^2x} F_4 -
 \frac{m_\tau^2}{E_\nu M} F_5  
  \biggr)
 \end{multline}
where $\{x, y, Q^2\}$ are the standard DIS kinematic variables related through $Q^2 = 2M_NE_\nu x y$.\\
The structure functions $F_4$ and $F_5$, pointed out by Albright and Jarlskog in Ref.~\cite{Albright:1974ts}, are neglected in muon neutrino interactions because of a suppression factor depending on the
square of the charged lepton mass divided by the nucleon mass times neutrino energy. Given the higher mass value of the $\tau$ lepton, $F_4$ and $F_5$ structure functions contribute, instead, to the tau neutrino cross section.
Thus one could interpret a measurement of the $\nu_\tau$ cross section, in combination with the often well-measured $\nu_\mu$ cross section, as a check on the PDF's and our understanding of $F_4$ and $F_5$.
At leading order, in the limit of massless quarks and target hadrons, $F_4=0$ and $2xF_5=F_2$, 
where $x$ is the Bjorken-$x$ variable (Albright-Jarlskog relations). Calculations at NLO show that $F_4$ is lower than 1\% of $F_5$ \cite{Reno:2006hj}.\\
With the statistics of tau neutrino interactions collected in five years run, the SHiP experiment will have 
the capability of being sensitive to $F_4$ and $F_5$.
In addition, IceCube/DeepCore \cite{Koskinen:NuTau21} is investigating the impact of the structure functions and future forward physics facilities at the LHC may have sensitivity to them as well.
The hypothesis of $F_4 = F_5 = 0$ would result in an increase of the $\nu_\tau$ and $\overline{\nu}_\tau$ CC DIS cross sections and consequently, of the number of expected $\nu_\tau$ and $\overline{\nu}_\tau$ interactions.
\begin{figure}
\begin{center}
\includegraphics[scale=0.4]{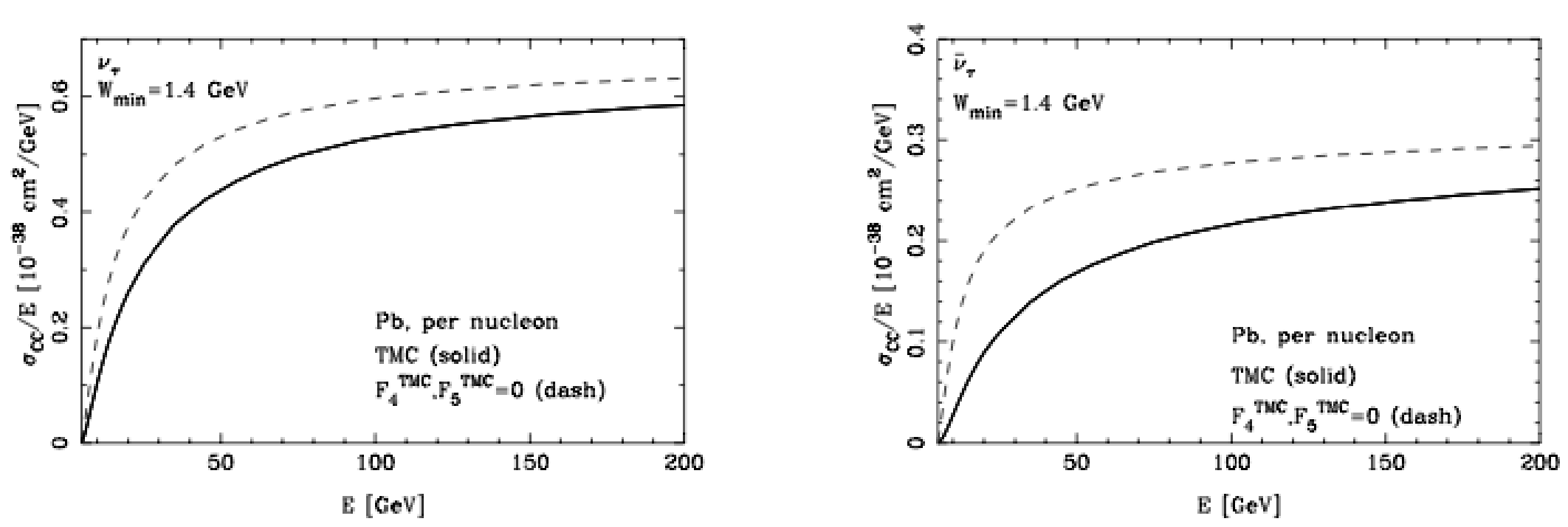}
\end{center}
\caption{
$\nu_\tau$ (right) and $\overline{\nu}_\tau$ (left) CC DIS cross-section  predicted by the SM (solid) and in the $F_4=F_5=0$ hypothesis (dashed)~\cite{Anelli:2007512}.
}
\label{fig:xsec_f5}
\end{figure}\\
Figure \ref{fig:xsec_f5} shows that the difference between the cross sections in the $F_4=F_5=0$ hypothesis  and the SM one is larger for lower neutrino energies.  This behavior reflects in the energy dependence of the variable $r$, defined as the ratio between the cross section in the two hypotheses: it is higher for lower neutrino energies, where the discrepancy of the two curves is larger, and decreases, tending to one, for higher energies, where the contribution of $F_4$ and $F_5$ becomes negligible.\\ 
The ratio $r$ is reported for $\overline{\nu}_\tau$ in the left plot of Fig.~\ref{fig:f4f5}. 
To have evidence of a non-zero value of $F_4$ and $F_5$, the ratio $r$ is required to be larger than $3\sigma$,  being $\sigma$ the uncertainty on the incoming neutrino flux, amounting to 20\%. 
This condition is satisfied for $E_{\overline{\nu}_\tau}<38$ GeV, where we expect to observe about 3600  $\overline{\nu}_\tau$ interactions.\\
The ratio $r$ was estimated also for the sum of $\nu_\tau$ and $\overline{\nu}_\tau$. The right plot of Fig.~\ref{fig:f4f5} shows that in this case $r>3\sigma$ for neutrino energies below 20 GeV, where the number of observed $\nu_\tau+\overline{\nu}_\tau$ interactions, not requiring the leptonic number to be measured, is expected to be about 4900.

\begin{figure}
\begin{minipage}{14pc}
\includegraphics[scale=0.45]{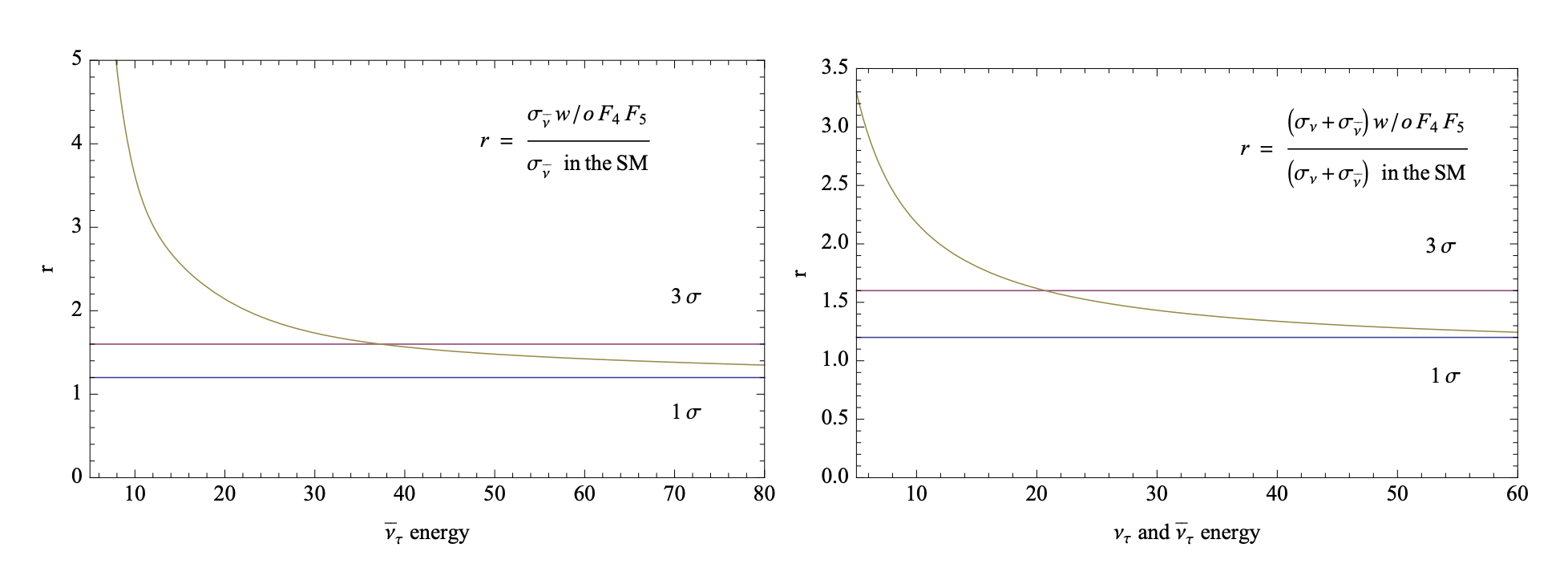}
\end{minipage}
\caption{Energy dependence of the ratio $r$ between the DIS cross-section in the $F_4=F_5=0$ hypothesis and in the SM hypothesis, for $\overline{\nu}_\tau$ (left) and for the sum of $\nu_\tau$ and $\overline{\nu}_\tau$ (right)
~\cite{Anelli:2007512}.}
\label{fig:f4f5}
\end{figure}

\textbf{$\boldmath{\nu_\tau}$ magnetic moment}
The presence of a non-zero magnetic moment adds an extra component to the elastic cross-section
for the process $\nu e^- \to \nu e^-$ that in the SM involves only the neutral current, except for $\nu_e$ (see also sec.~\ref{sec:dipole}). Therefore,
an increase of the measured cross-section can prove the hypothesis of
anomalous magnetic moment. So far, a non-zero magnetic moment for $\nu_e$ and $\nu_\mu$ has been excluded down to $\mu_{\nu_e}<1.9\times10^{-11}\mu_B$ and to $\mu_{\nu_\mu}<6.9\times10^{-10}\mu_B$~\cite{PDG:2018}, respectively. 
With the expected statistics of $\nu_\tau$ CC interactions in the SND,  SHiP  can significantly constrain  the $\nu_\tau$ magnetic moment. 

In the elastic scattering of a neutrino on an electron, the scattering angle of the outgoing electron with respect to the direction of the incoming neutrino is limited, in the laboratory frame, by kinematic constraints~\cite{Radel:1993sw}:
\begin{displaymath}
\theta^2_{\nu-e} < \frac{2m_e}{E_e}.
\end{displaymath}
Therefore, for $E_e>1$ GeV, $\theta_{\nu-e}$ is smaller than 30\,mrad. This can help suppressing the
background from events with the same topology. 

The main background sources for this analysis are: $(i)$ neutrino elastic scattering (ES) with electrons of the detector target, $(ii)$ electron neutrino and anti-neutrino quasi elastic scattering (QE) with nucleons of the target
with non detected outgoing nucleons, $(iii)$ CC deep-inelastic interactions (DIS) of electron neutrinos and anti-neutrinos with nucleons in the detector target with no revealed hadrons in the final state, and $(iv)$ electron neutrino and anti-neutrino resonant processes. In order to take into account the uncertainty on the neutrino interaction position in the
detector target, a smearing of the electron angle by 1\,mrad was introduced although the angular uncertainty is dominated by the measurement accuracy. The GENIE generator was used to estimate the number of expected background events surviving the following selection criteria: only the electron reconstructed  in the final state, $E_e>1$ GeV,  $\theta_{\nu-e} < $ 30\,mrad.
The overall contribution from different background sources amounts to $14\times10^3$ events, mainly  from QE processes. 

Denoting with $\Phi_{\nu_\tau}$ the incoming tau neutrino flux on the detector, with $N$ the number
of nucleons in the neutrino target,  with $\sigma^\mu$ the contribution of the non-zero magnetic moment to the cross-section and with $\mu_B = 5.8\times 10^5 eV T^{-1}$ the Bohr magneton, the number of expected
events for a magnetic moment $\mu_\nu$ is given by:
\begin{equation}
n_{evt} = \frac{\mu_\nu^2}{\mu_B^2} \times \int {\Phi_{\nu_\tau} \sigma^\mu N dE}  = 4.3\times 10^{15} \frac{\mu_\nu^2}{\mu_B^2} 
\end{equation}
Assuming a 5\% systematic uncertainty, the evidence for a tau neutrino anomalous magnetic moment with a significance of 3$\sigma$ requires the observation of an excess of
about 540 events over the background. Hence, a region
down to a magnetic moment $1.3\times10^{-7} \mu_B$ could be explored.

\subsection{Long-baseline Accelerator}

\subsubsection{OPERA}

Several experiments have proven that neutrinos oscillate and thus they are massive particles. 
In the atmospheric sector, Super-Kamiokande provided the first compelling evidence for $\nu_\mu$ oscillations in 1998 as an explanation of the $\nu_\mu$ deficit~\cite{Fukuda:1998mi}.
This result was interpreted as $\nu_\mu$ disappearance due to $\nu_\mu\to\nu_\tau$ oscillations, although at that time the $\nu_\tau$ had not yet been directly observed.
The result was confirmed with neutrino beams produced at accelerators by the K2K~\cite{Ahn:2006zza} and MINOS~\cite{Michael:2006rx} experiments. The appearance of $\nu_\tau$ from $\nu_\mu\to\nu_\tau$ transitions was required for the complete confirmation of the three-flavor neutrino oscillation picture in the so-called atmospheric sector.

The OPERA experiment was designed in the late 1990s to conclusively prove $\nu_\mu \to \nu_\tau$ oscillations in appearance mode \cite{DeLellis:2021ovm}. The detector was located in the underground Gran Sasso Laboratory and consisted of an emulsion/lead target complemented by electronic detectors.
It was exposed to the CNGS (CERN Neutrinos to Gran Sasso) beam from 2008 to 2012.
The beam was an almost pure $\nu_\mu$ beam with a baseline of 730 km and the OPERA detector collected a total of $1.8\cdot 10^{20}$ protons on target. OPERA was unique at the time in its capability of detecting all three neutrino flavors.

The $\nu_\tau$ interactions were identified through the observation of the decay of the $\tau$ lepton produced in $\nu_\tau$ CC interactions. Once a vertex had been identified in the scanned volume, a decay search procedure was applied \cite{Agafonova:2014khd} looking for all $\tau$ decay modes including leptonic and hadronic modes.
Two different methods for a decay topology were used: the observation of an impact parameter of the track larger than 10 $\mu$m with respect to the reconstructed vertex and the observation of a kink along a track coming out of the vertex. Due to the high spatial resolution of nuclear emulsions, both the impact parameter and the kink angle could be measured with high accuracy.
In addition, a kinematical analysis was used to suppress background sources showing a similar topology.

The analysis of the first two runs in 2008 and 2009 was performed without any kinematical pre-selection of the events, to avoid any bias before demonstrating a full understanding of the data. 
In this sample which contained $30\%$ of the overall number of p.o.t., one event was recognised as a $\nu_\tau$ candidate decaying to a $\rho$~\cite{OPERA:1tau}. 
Once the agreement was demonstrated~\cite{Agafonova:2011zz},  the scanning strategy was optimised to speed up the analysis:   events without any muon in the final state as well as those with muon momentum $P_\mu < 15$ GeV/c were selected, thus significantly reducing the charm background while marginally affecting signal efficiency. 

Moreover, the development of new automated scanning systems \cite{Fukuda:2014vda, Alexandrov:2015kzs} allowed the detection of large-angle nuclear fragments. These systems showed
a detection efficiency of $95\%$ or higher to recognize large-angle minimum ionizing particles ($1.0 \le |\tan\theta| \le 3.5$), thus rejecting background topologies more efficiently.

The processes mimicking the $\tau$ decay are: i) the decay of charmed particles produced in $\nu_{\mu}$~CC interactions; ii) re-interaction of hadrons from $\nu_{\mu}$ events in lead; iii) the large-angle scattering (LAS) of muons produced in $\nu_{\mu}$~CC interactions. Processes i) and $\nu_{\mu}$~CC in ii) represented a background source only when the $\mu^-$ at the primary vertex was not identified. 

\textbf{Charmed Particles Production}

The sample of charmed hadron particles produced in muon neutrino interactions was used as a  control sample to cross-check the topological $\tau$ detection efficiency, due to the very similar decay patterns. The decay search procedure was applied to search for charmed hadrons: 50~charm decay candidate events were observed in a sample of located $\nu_\mu$ CC interactions from 2008-2010 runs, while $54\pm 4$ were expected from MC simulation~\cite{Agafonova:2014khd}. This result proved that the detector performance and the analysis chain applied to neutrino events were well reproduced by the MC simulation and thus validated the methods for $\nu_\tau$ appearance detection. 

\textbf{Hadronic Interactions}

To study topological and kinematical characteristics of hadron interactions, a lead-emulsion target was exposed to 2, 4 and 10 GeV/c hadron beams. A total of 318 hadron interactions were found and reconstructed by following 60 m $\pi^{-}$ tracks in the brick, together with secondary charged particle tracks from interaction vertices. The multiplicity of charged particles and the emission angle of each secondary particle were measured. Their distributions were found to be in good agreement with the FLUKA Monte Carlo simulation. 
This result was used to evaluate the  background from hadronic interaction in the $\tau$ lepton search. The detailed description of this study can be found in a dedicated paper \cite{Ishida:2014qga}. 

\textbf{Large angle scattering}

Large angle scattering (LAS) of muons was a background source for $\tau\to\mu$ decay channel. Simulation for LAS was performed using the GEANT4 package. About 1.1~billion incident $\mu^{-}$ were generated with a flat momentum distribution between 1~and 15~GeV/c, with orthogonal incidence on the lead-film double cell. The simulation exploited a dedicated treatment of the form factors tuned for the description of the  available data. Results for $\theta_{kink}$ and $p^{T}_{2ry}$ were validated thanks to a detailed comparison with experimental data from the literature.  
The detailed description of the study of the LAS of muons can be found in a dedicated paper \cite{Longhin:2015dsa}.

\textbf{Discovery of $\boldmath{\nu_\tau}$ oscillations in the CNGS beam}

In the sample analyzed until 2015, corresponding to 5408 located neutrino interactions with the decay search performed, five $\nu_\tau$ candidates survived the stringent kinematical cuts~\cite{OPERA:2tau}: three in the $\tau\to 1h$ decay channel~\cite{OPERA:1tau, OPERA:4tau,OPERA:5tau}, one in the $\tau\to 3h$~\cite{OPERA:2tau} and one in the $\tau \to \mu$~\cite{OPERA:3tau} decay channel. 
The expected background was estimated to be ($0.25\pm0.05$) events. 

The observation of five candidates with such a low background level translated into a probability of the background only hypothesis of $1.1 \times 10^{-7}$, equivalent to a significance of $5.1\sigma$ in the observation, thus reporting the discovery in appearance mode of muon to tau neutrino oscillations~\cite{OPERA:5tau}.

\textbf{Final results on $\boldmath{\nu_\tau}$ oscillations}

After these results, the Collaboration set a new goal: reduce as much as possible the statistical uncertainty,  estimate the oscillation parameters for the first time in appearance mode and constraint the oscillation parameters through a global fit of all available data, including electron neutrinos. 
Given the successful description of the data achieved with the Monte Carlo simulation, established with different control samples, a new analysis strategy was developed, fully exploiting the topological and kinematic features of $\nu_\tau$ events with a multivariate approach.
The newly developed selection was applied to the final data sample, corresponding to 5603 fully reconstructed neutrino interactions. 
Details about the new selection method are reported in a dedicated paper~\cite{OPERA:final}. The total expected signal was estimated to be $(6.8\pm 1.4)$ events, with $(2.0\pm0.4)$ background events. Ten events survived the new topological and kinematic cuts. The additional five $\nu_\tau$ candidates were found in the hadronic decay channel: three 1-prong and two  3-prong decays.

An analysis with the Boosted Decision Tree (BDT) multivariate analysis was implemented, exploiting all the characteristics of the interactions, as reported in a dedicated paper~\cite{OPERA:final} and the results are shown in Fig.~\ref{fig:BDT} for each $\tau$ decay channel. 
\begin{figure}[h]
\centering
\includegraphics[width=0.7\linewidth]{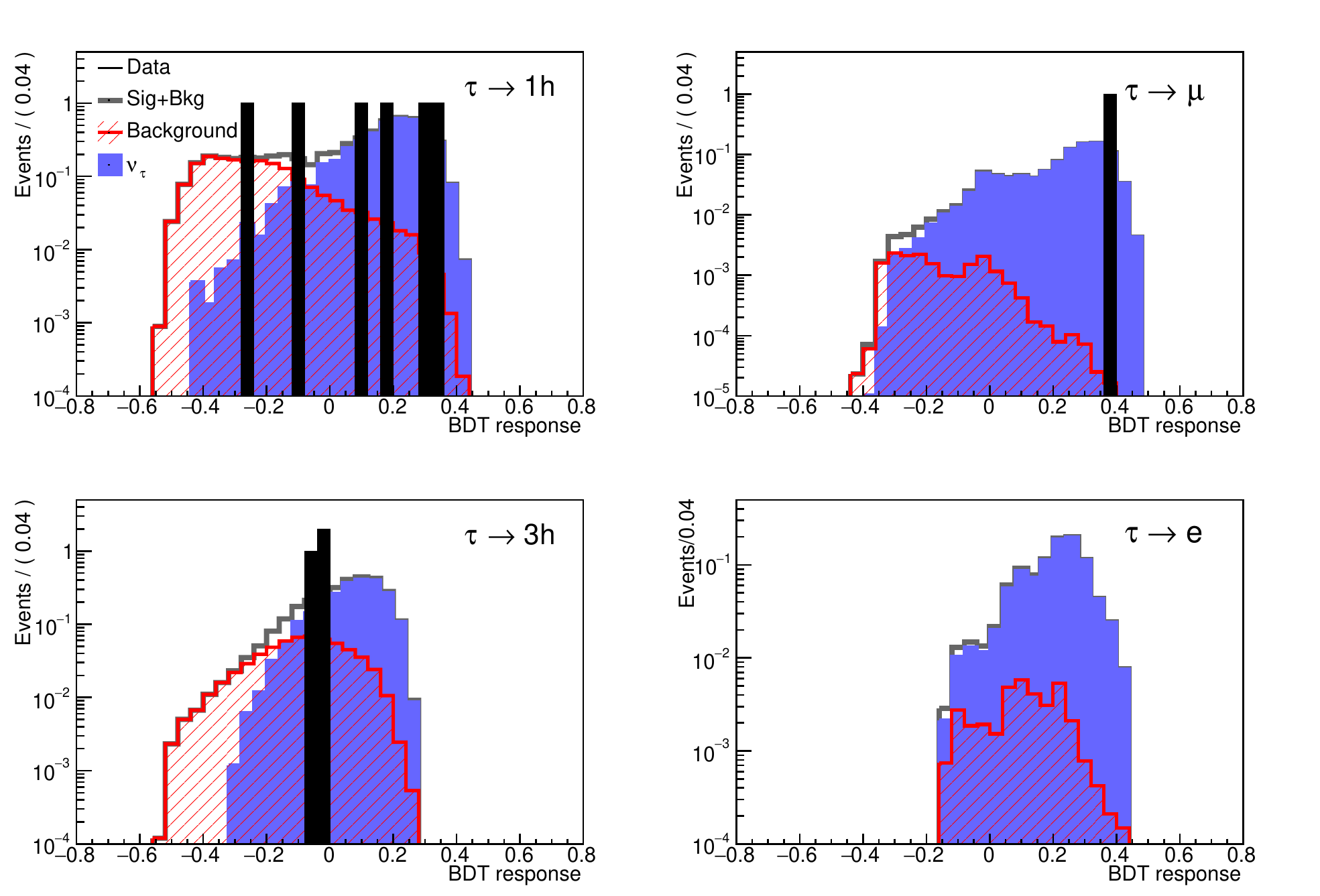}
\caption{BDT response for each $\tau$ decay channel~\cite{OPERA:final}.}\label{fig:BDT}
 \end{figure}
 
 The statistical analysis used to evaluate the significance for the $\nu_\tau$ appearance was based on an extended likelihood constructed as the product of a probability density function given by the BDT response, a Poisson probability term which takes into account the number of observed events and the expected background in each decay channel, and a Gaussian term which accounts for systematics.
 The null hypothesis was excluded with the improved significance of 6.1~$\sigma$~\cite{OPERA:final}, corresponding to a background fluctuation probability of
$4 \times 10^{-10}$.

The number of observed $\nu_\tau$ candidates after background subtraction is a function of the product of $\nu_\tau$ CC cross-section ($\sigma_{\nu_\tau}^{CC}$) and the oscillation parameter $\Delta m^2_{23}$.

The squared mass difference was measured for the first time in appearance mode, $|\Delta m^2_{23}| =  {(2.7^{+0.7}_{-0.6})\cdot 10^{-3}\textnormal{eV}^{2}}$, assuming ${\sin^2 2\theta_{23}=1}$ and the Standard Model $\nu_{\tau}$ cross-section. The result is consistent with the measurements performed in disappearance mode by other experiments and with the Particle Data Group  fit~\cite{PDG:2016}.

The first measurement, with a negligible contamination from $\bar{\nu}_\tau$, of the $\nu_\tau$~CC cross-section was also obtained: on the OPERA lead target it is equal to ${(5.1^{+2.4}_{-2.0}) \cdot 10^{-36} \textnormal{cm}^2}$, assuming ${|\Delta m^2_{23}|=2.50\cdot 10^{-3}\ \textnormal{eV}^{2}}$.

In the muonic channel, neutrinos could be distinguished from anti-neutrinos by the charge of the produced muon.
For the $\tau\to\mu$ candidate, the muon charge was determined as negative at 5.6~$\sigma$ confidence level. Performing a dedicated BDT analysis, which included also the background from $2\%$ $\bar\nu_\mu$ beam contamination, the first direct evidence for the leptonic number of $\tau$ neutrinos with a significance of $3.7\sigma$ was obtained~\cite{OPERA:final}.

\textbf{Combined oscillation analysis}	

The final samples of $\nu_\tau$ and $\nu_e$ CC interactions were used, for the first time, in a combined analysis to constrain the oscillation parameters both in the standard three-flavor model and in the extended $3+1$ one.

The visible neutrino energy defined in~\cite{OPERA:final} and the reconstructed neutrino energy defined in~\cite{Agafonova:2018dkb} were used as observables for the $\nu_{\tau}$ and $\nu_e$ samples, respectively. The statistical analysis of the data was based on a maximum-likelihood joint fit across the two samples. 
The data are compatible with the three-flavor neutrino model and constraints on $\theta_{23}$ and $\theta_{13}$ were derived jointly for the first time exploiting tau and electron neutrino appearance channels.
The best fit value and the $1\sigma$ confidence interval for the $\theta_{23}$ parameter were found to be $\theta_{23}=0.78^{+0.32}_{-0,31}$ rad, while the $1\sigma$ confidence interval for $\theta_{13}$ was found to be $[0,0.20]$ rad. 

In addition, $\nu_{\tau}$ and $\nu_e$ appearance channels were combined for the first time to constrain parameters of the $3+1$ sterile mixing model. For $\Delta m^2_{41} > 0.1 eV^2$, upper limits on $\sin^2  2 \theta_{\mu \tau}$ and $\sin^2  2 \theta_{\mu e}$ were set to 0.10 and 0.019 for NH and IH. The results on the exclusion region for the $3+1$ sterile mixing model are reported in Fig.~\ref{fig:combined_exclusion}.
The MiniBooNE best-fit~\cite{Aguilar-Arevalo:2018gpe} values ($\Delta m^2_{41} =  0.041 eV^2$, $\sin^2  2 \theta_{\mu e} =  0.92$) were excluded with 3.3$\sigma$ significance.

\begin{figure}[tbh]
\centerline{\includegraphics[width=0.8\textwidth]{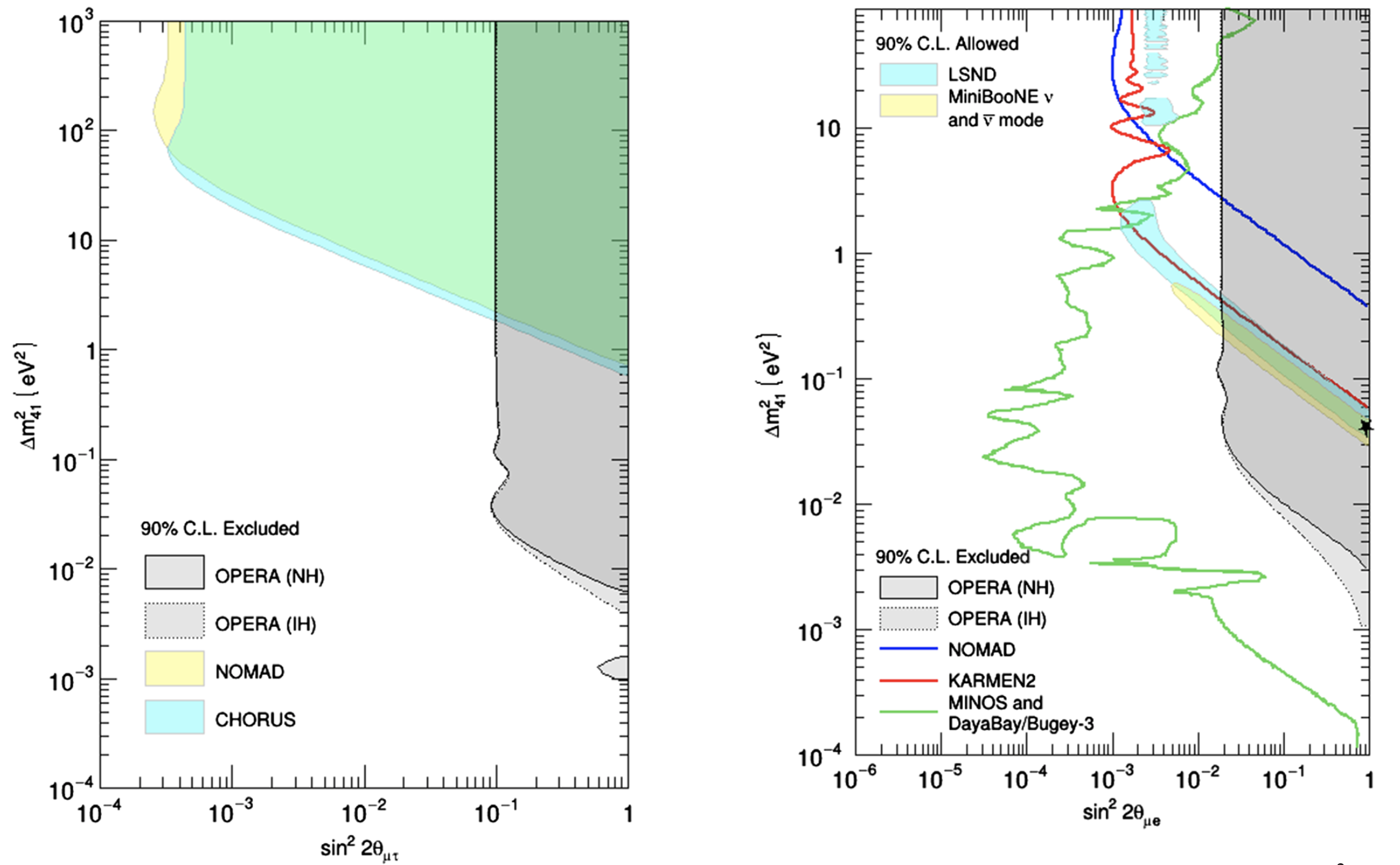}}
\vspace*{8pt}
\caption{Left: The $90\%$~C.L.~exclusion region in the $\Delta m^2_{41}$ and $\sin^2 2\theta_{\mu \tau}$ parameter space for the normal (solid line) and inverted (dashed line) hierarchy of the three standard neutrino masses. The exclusion regions by NOMAD~\cite{NOMAD:2001xxt} and CHORUS~\cite{Eskut:2007rn} are also shown. Right: the $90\%$~C.L.~exclusion region in the $\Delta m^2_{41}$ and $\sin^2 2\theta_{\mu e}$ plane is shown for the normal (solid line) and inverted (dashed line) hierarchy of the three standard neutrino masses. The plot also reports the $90\%$~C.L.~allowed region obtained by LSND~\cite{Louis:2009zza} (cyan) and MiniBooNE combining $\nu$ and $\bar{\nu}$ mode~\cite{Aguilar-Arevalo:2018gpe} (yellow). The blue and red lines represent the $90\%$~C.L.~exclusion regions obtained in appearance mode by NOMAD~\cite{Astier:2003gs} and KARMEN2~\cite{Armbruster:2002mp}, respectively. The $90\%$~C.L.~exclusion region obtained in disappearance mode by the MINOS and DayaBay/Bugey-3 joint analysis~\cite{Adamson:2016jku} is shown as green line. The black star ($\star$) corresponds to the MiniBooNE best-fit values for the combined analysis of $\nu$ and $\bar{\nu}$ data~\cite{OPERA:final}.
\protect\label{fig:combined_exclusion}}
\end{figure}

\subsubsection{DUNE}
\label{sssec:dune accelerator}

The DUNE Far Detector (FD) reference design consists of four 10 kt fiducial mass LArTPC modules located at a baseline of 1285 km from the LBNF neutrino source at the 4850 km level of the Sanford Underground Research Facility in Lead, South Dakota.  The long-baseline, large detector mass, and intense beam will allow DUNE to measure all three flavor oscillation parameters in a single experiment.  While DUNE is optimized to measure $\nu_e$ appearance in a $\nu_{\mu}$ beam the broadband beam and long baseline lead to significant $\nu_{\tau}$ appearance above the kinematic threshold to produce a $\tau$-lepton.  Due to this, DUNE is the only upcoming neutrino experiment expected to be able to collect a sample of oscillated $\nu_{\tau}$ CC beam events.   

With the excellent spatial and energy resolution of LArTPC technology, DUNE is expected to be able to select and reconstruct $\nu_{\tau}$ CC interactions well.  The energy threshold for detecting $p$, $\pi^{\pm}$, and $\gamma$ is $\mathcal{O}(10 MeV)$, and through a combination of energy deposition per unit length, range, multiple scattering, and topology, it will be possible to distinguish between electrons, photons, muons, and pions with nearly 100\% efficiency at the typical energies of beam produced $\nu_{\tau}$ CC interactions.  In Ref.~\cite{Machado:2020yxl}, a truth-level study of $\nu_{\tau}$ CC interactions where the $\tau$-lepton decayed hadronically suggests that relatively simple kinematic requirements on events containing at least one $\pi^{\pm}$ could confirm $\nu_{\tau}$ CC appearance with a significance of 3.1$\sigma$ in one year of running in the CP-optimized beam mode or 7.9$\sigma$ in one year of running in the tau-optimized beam mode, assuming 1.2 MW beam power and 40 kt fiducial mass.  This selection corresponds to $\sim$60\% signal efficiency and $\sim$80\% NC background rejection efficiency.  A separate study of exclusive final states using transverse kinematic variables showed that it is possible to select $\nu_{\tau}$ CC interactions where the $\tau$-lepton produces either a single $\pi$, $e$, or $\rho$~\cite{Kosc:2021huh}.  As shown in Fig.~\ref{fig:DUNE_Kosc_ROC}, these exclusive final states can be selected with a signal efficiency of $\sim$20\% while rejecting nearly 100\% of the background. Finally, in a study of atmospheric $\nu_{\tau}$ CC interactions in a large LArTPC, it was found that simple kinematic requirements could select signal with 30\% efficiency while selecting NC background with only 0.5\% efficiency~\cite{Conrad:2010mh}.  We assume this performance in the following studies.

\begin{figure}
    \centering
    \includegraphics[width=0.6\textwidth]{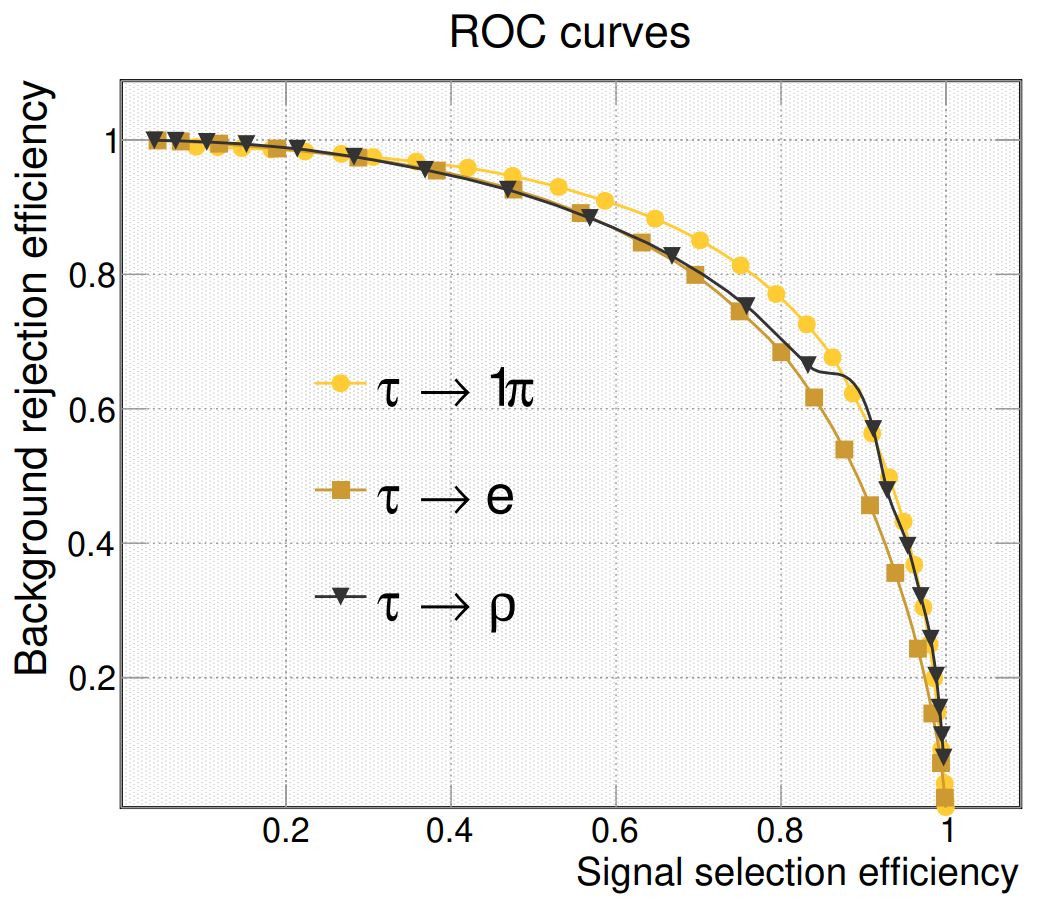}
    \caption{Background rejection vs. signal efficiency for selecting $\nu_{\tau}$ CC interactions where the $\tau$-lepton decays involve an electron, single $\pi^{\pm}$ or a $\rho$ using the transverse kinematic variable technique. Figure reproduced from Ref.~\cite{Kosc:2021huh}.}
    \label{fig:DUNE_Kosc_ROC}
\end{figure}

Using the predicted LBNF fluxes for the CP-optimized and the tau-optimized beams and cross-section predictions from GENIE 2.12.2~\cite{Andreopoulos:2009rq}, we estimate that DUNE will record $\sim$130 $\nu_{\tau}$ CC interactions per year in CP-optimized neutrino mode, $\sim$30 $\bar{\nu}_{\tau}$ CC interactions per year in CP-optimized antineutrino mode, and $\sim$800 $\nu_{\tau}$ CC interactions per year in tau-optimized neutrino mode.  Using two configurations (3.5 years each in CP-optimized neutrino and antineutrino modes or 3 years each in CP-optimized neutrino and antineutrino modes and one year in tau-optimized neutrino mode, Ref.~\cite{DeGouvea:2019kea} studied the impact of including a selected $\nu_{\tau}$ sample on three-flavor mixing, a non-unitary PMNS matrix, light sterile neutrinos, and NC NSI, see e.g.~\cite{Giarnetti:2021wur}. 

As shown in Fig.~\ref{fig:DUNEBeamSensitivity}, it is possible to simultaneously constrain $\Delta m^2_{31}$ and $\sin^2 2\theta_{23}$ using three independent samples, $\nu_{\mu}$ disappearance, $\nu_e$ appearance, and $\nu_{\tau}$ appearance.

Although the $\nu_{\tau}$ sample provides weaker constraints than the others, the three constraints are not independent if the PMNS matrix is unitary.   Using all three channels should allow DUNE to constrain $|U_{e3}|^2 + |U_{\mu 3}|^2 + |U_{\tau 3}|$ to 6\% in a model-independent way.  If instead we parameterize non-unitarity effects by multiplying a unitary mixing matrix by a lower diagonal matrix to model the effects of a heavy sterile neutrino state, we find that adding a $\nu_{\tau}$ appearance sample improves the constraint on the non-unitarity parameters $\alpha_{32}$ and $\alpha_{33}$ compared to using $\nu_{\mu}$ disappearance alone.  Similarly, the $3+3+1$ scenario further improves the physics reach over the $3.5+3.5$ scenario.

\begin{figure}
    \centering
    \includegraphics[width=0.45\textwidth]{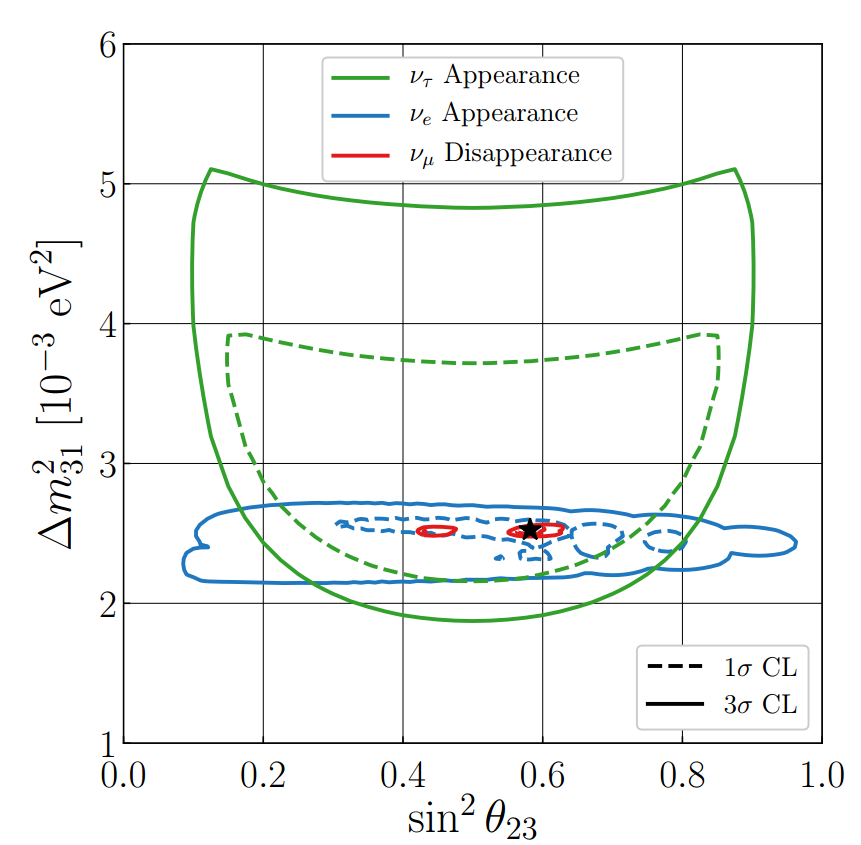}
    \includegraphics[width=0.45\textwidth]{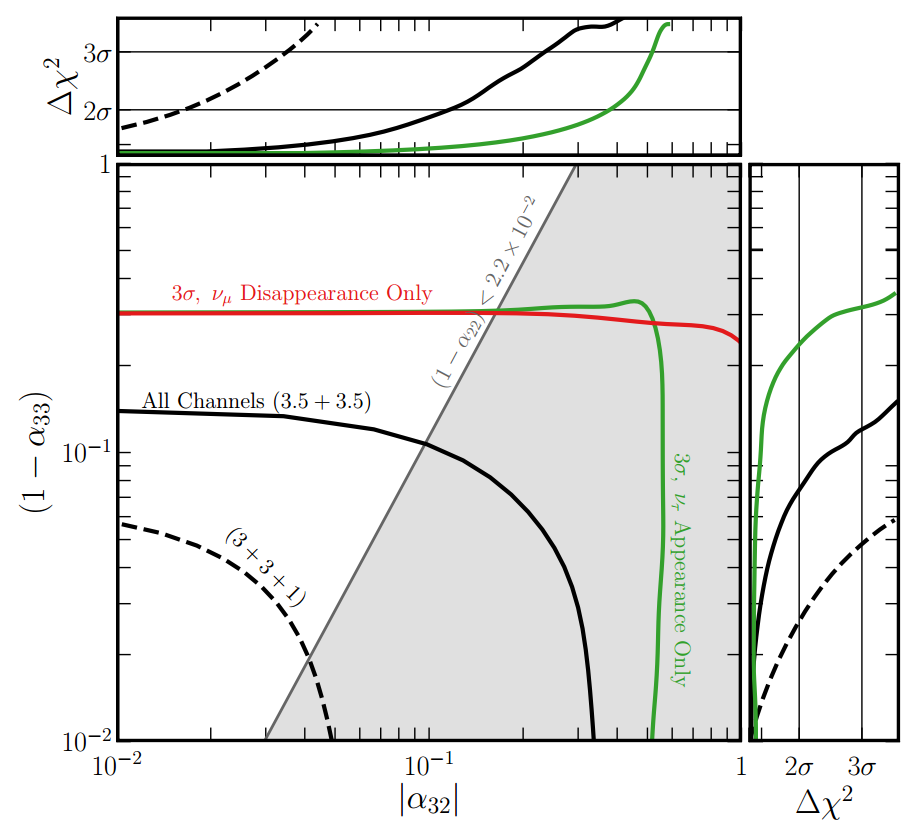}
    \caption{Left: Sensitivity for $\Delta m^2_{31}$ vs. $\sin^2 \theta_{23}$ using $\nu_e$ appearance (blue), $\nu_{\mu}$ disappearance (red), and $\nu_{\tau}$ appearance (green) with 3.5 years each of the CP-optimized beam in neutrino and antineutrino modes. Right: Sensitivity for the non-unitarity parameters $(1 - \alpha_{33}$ vs $|\alpha_{32}|$. The red line shows the sensitivity if only the $\nu_{\mu}$ disappearance channel were used, and the green line shows the sensitivity if only the $\nu_{\tau}$ appearance channel were used.  The solid black line shows the improvement of the sensitivity if all three channels were used.  The dashed black line shows the improvement if one year of CP-optimized beam were replaced with a year of tau-optimized beam. Figures reproduced from Ref.~\cite{DeGouvea:2019kea}.}
    \label{fig:DUNEBeamSensitivity}
\end{figure}

In addition to oscillation physics, the large $\nu_{\tau}$ CC sample that could be collected using the tau-optimized beam configuration would make it possible to measure not only the normalization of the $\nu_{\tau}$ CC cross section, but also differential cross sections.  This may allow DUNE to constrain currently unmeasured parameters like the $F_4$ and $F_5$ structure functions and the pseudoscalar form factor.  

\subsection{Atmospheric}
Due to their massive size and ability to detect unprecedented amounts of atmospheric neutrinos over energies and baselines which have oscillation signatures, neutrino telescopes have a unique role in understanding the fundamental nature of tau neutrinos.
This is particularly important in the case of studying $\nu_\tau$ which are kinematically forbidden from CC interactions at energies below $\sim$3.5 GeV, and have a cross section suppression compared to both $\nu_e$ and $\nu_\mu$ at energies up to $\sim1$ TeV.

The key feature for neutrino telescopes and atmospheric oscillations measurements is their large size coupled with oscillation baselines up to the Earth diameter of 12\,750\,km, which produce an oscillation maximum for $\nu_\tau$ appearance at $\sim25$\,GeV at much higher energies than probable in accelerator neutrino experiments.
The effective volumes of current atmospheric detectors detectors range from tens of kilotons (Super-Kamiokande) and report $\sim338$ CC $\nu_\tau$ events \cite{Super-Kamiokande:2017edb} to megatons (IceCube/DeepCore\cite{IceCube:2011ucd}) with $\sim$1800 CC $\nu_\tau$ \cite{IceCube:2019dqi}.

The impressive statistics for collected $\nu_\tau$ events are experimentally offset by the reality that identifying a $\nu_\tau$ event faces numerous challenges that are handled in different ways with different detector designs.

\subsubsection{Super-Kamiokande/Hyper-Kamiokande}

The existing and on-going atmospheric $\nu_\tau$ appearance analyses at SuperK currently adopt the OPERA convention \cite{OPERA:2018nar} of reporting $\nu_\tau$-appearance via ``a scale factor on the number of events expected by the model of neutrino interactions'' known as the $\nu_\tau$-normalization ($N_{\nu_\tau}$), where $N_{\nu_\tau}=0$ is for no $\nu_\tau$-appearance, $N_{\nu_\tau}=1$ is consistent with $3\times3$ PMNS unitarity and no new physics, and $N_{\nu_\tau}>1$ would be the observation of more $\nu_\tau$ than what could be expected via conventional neutrino oscillations.
The SuperK results are shown in Fig.~\ref{fig:normalization comparison} and, given that they include 15 years of data, are not likely to improve significantly in coming years.
SuperK leverages polarization information in the $\tau$ decay, as well as oscillation features and $\tau$ threshold effects, to determine the $\nu_\tau$ signature; for more see Sec.~\ref{sec:tools water and ice low energy} for more.
With HyperK sensitivity to the 5\% level is anticipated with 5.6 Mton$\cdot$year exposure corresponding to 15 years with both tanks although an in depth analysis of the sensitivity does not yet exist \cite{Hyper-Kamiokande:2018ofw}.

\subsubsection{IceCube/DeepCore}
DeepCore is a dense infill inside the larger IceCube detector.
The detector is a 3D array of digital optical modules containing PMTs distributed through the Antarctic ice.
Due to the fact that the charged $\tau$ lepton from a $\mathcal{O}(20)$\,GeV charged current $\nu_\tau$ interaction will decay over distances of $\ll1$\,m, observing the track is not possible in DeepCore.
In addition, the fact that the $\tau$ decay includes one or two neutrinos means that there will be significant missing energy which also makes the events harder to detect due to the steep atmospheric spectrum.
In a 3D detector essentially two event topologies can be identified: tracks which involve muons from $\nu_\mu$ CC events or $\nu_\tau$ CC events where the $\tau$ decays to a $\mu$, and cascades from all other neutrino interaction categories.
This makes it experimentally impossible to identify on an event-by-event basis, a $\nu_\tau$ interaction in the atmospheric energy range.
Nonetheless, it is possible to measure the appearance of $\nu_\mu \rightarrow \nu_\tau$ oscillations by observing cascade-like events which are consistent with the expectation from $\nu_\tau$-appearance combined with known properties of $\tau$ leptons \cite{Stanev:1999ki,Giordano:2010pr, Denton:2021rsa}, see Fig.~\ref{fig:AppearanceBump}.

\begin{figure}
\centering
\includegraphics[trim={0cm 0cm 0cm 0cm}, clip, scale=1]{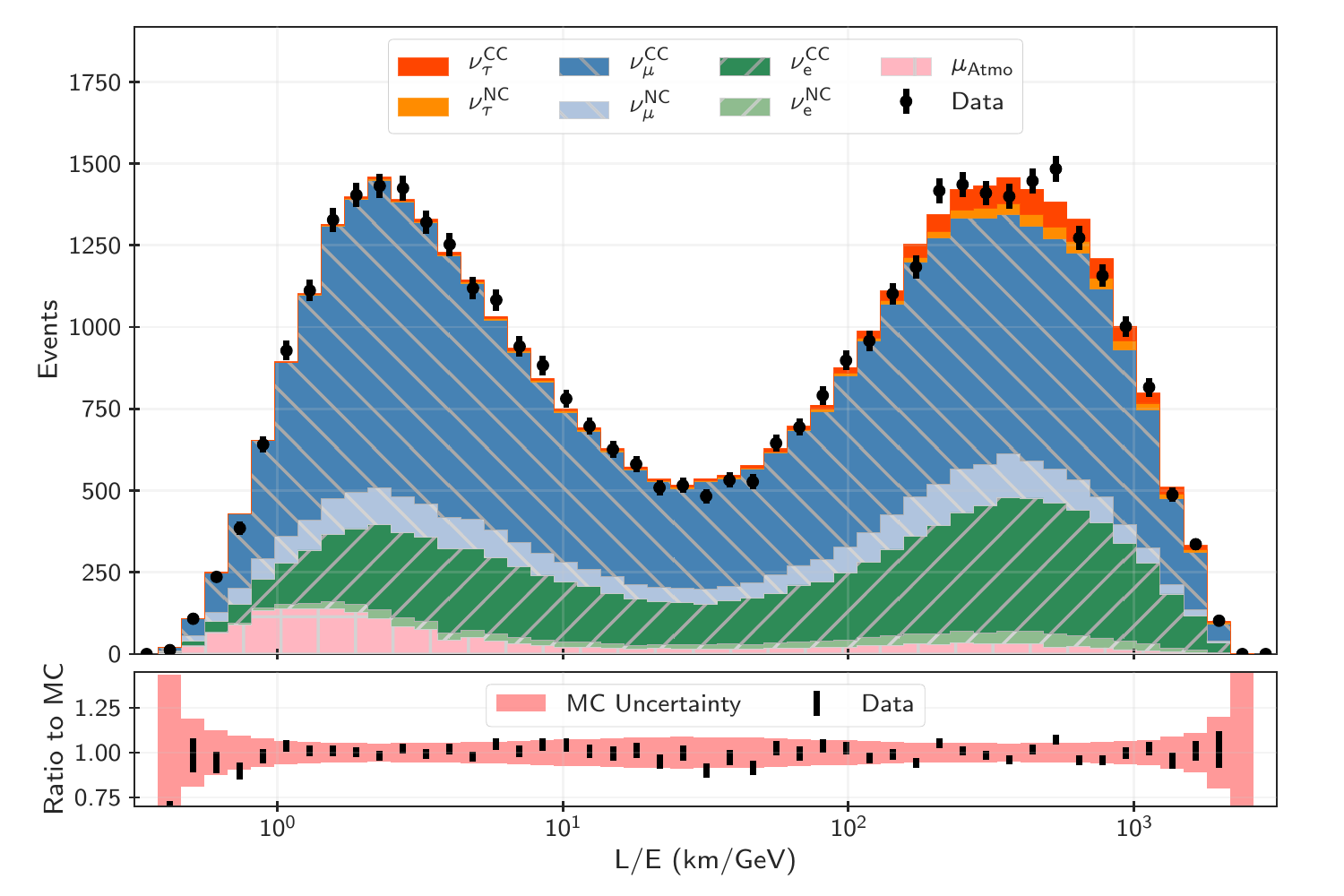}
\caption{Monte Carlo estimates for the various particle types present in the 3-year oscillated neutrino sample from IceCube \cite{IceCube:2019dqi}.
The presence of oscillated $\nu_\tau$ can be seen in red/orange and are needed in order to explain the observed data.}
\label{fig:AppearanceBump}
\end{figure}

The current experimental results are shown in Fig.~\ref{fig:normalization comparison}, where the uncertainty on $N_{\nu_\tau}$ is at the level of 25\%-30\% for SuperK and IceCube/DeepCore, and slightly larger for OPERA.
None of the results show any significant disagreement with a $N_{\nu_\tau}=1$.
DeepCore is currently analyzing an 8-year data sample with an improved event selection that is expected to contain $\sim$10k CC $\nu_\tau$ events, which will improve upon the 1.8k CC $\nu_\tau$ collected for the 3-year sample and potentially result in a sub-15\% uncertainty on $N_{\nu_\tau}$. 

Pioneered by the KM3NeT consortium, both the IceCube Upgrade and Hyper-Kamiokande will deploy multi-PMT photosensors which provide improved granularity for particle identification and event reconstruction.
With these new photosensors, the goal of developing event reconstruction algorithms using new machine learning techniques \cite{KM3NeT:2020zod,IceCube:2021dvc} is emerging as a feature of overlap between experimental efforts.
With the upgrade, for which $\nu_\tau$ appearance is a primary goal, IceCube/DeepCore can reach 10\% precision on $N_{\nu_\tau}$ in one year and 6\% precision in three years not including several features including improved calibration applied to the existing 10 years of data \cite{Ishihara:2019aao}.

There is also an increased push to use atmospheric neutrino oscillations to fit individual elements of the PMNS matrix in addition to fitting the atmospheric mixing parameters ($\Delta m^2_{32}$ and $\theta_{23}$) and the tau neutrino normalization.
It is crucial that experiments report these results in a general way so that they can be combined with other experiments to fully test the three flavor oscillation hypothesis and to test for new physics scenarios that affect oscillations in the tau neutrino sector.

\subsubsection{KM3NeT/ORCA}
KM3NeT/ORCA is optimized to determine the neutrino mass ordering in the few-GeV energy region.
With an instrumented mass of $\sim$7 Mton KM3NeT/ORCA will collect more than 3000 oscillated $\nu_\tau$ CC interactions per year at analysis-level \cite{KM3NeT:2021ozk} and thus overcome the statistical limitations of current $\nu_\tau$ appearance measurements.

By fitting the appeared $\nu_\tau$ contribution to the oscillated atmospheric flux on a statistical basis using 3D binned data (direction, energy, event type), the $\nu_\tau$ normalization can be constrained to within $\pm$20\% at $3\sigma$-level ($\pm$7\% at $1\sigma$-level) within 3 years with the full detector \cite{KM3NeT:2021ozk}. Given the high statistics and typical $\nu_\tau$ energies well above threshold ($\sim$25\,GeV for neutrinos that have traversed Earth), KM3NeT/ORCA will already be sensitive to measure $\nu_\tau$ appearance at an early stage of construction \cite{Hallmann:2020pip}.

KM3NeT/ORCA has recently performed its first oscillation analysis in $\nu_\mu$ disappearance mode using one year of data \cite{KM3NeT:2021hkj} recorded with only the first six deployed Detection Units out of a planned 115.
The analysis demonstrates that the instrument is able to collect a clean neutrino sample and measure the atmospheric oscillation parameters.
The analysis confirms the presence of neutrino oscillations in the atmospheric flux with a $5.9\sigma$ CL preference over non-oscillation.

\subsubsection{ICAL at INO}
The iron calorimeter (ICAL) at the India Neutrino Observatory (INO) \cite{ICAL:2015stm} is a proposed magnetized iron calorimetric detector to study atmospheric neutrino oscillations.
Because neutrino telescopes analyzing $\nu_\tau$ from neutrino oscillations are dependent on the characterization of the initial unoscillated atmospheric flux of $\nu_\mu$ and $\nu_e$, improved modeling and treatment of systematic uncertainties related to the atmospheric neutrino flux will be a priority for neutrino physicists.
Prevalent models used for the atmospheric neutrino flux have historically come from M.\@ Honda and collaborators \cite{Honda:2015fha,Honda:2011nf}.
The `Honda' models included the impact of the geomagnetic field at neutrino energies $< 10$\,GeV where geomagnetic deflection can introduce a reasonable impact on the overall flux.
As larger neutrino telescopes such as IceCube and KM3NeT introduce detector extensions which can reconstruct and analyze neutrinos at $\mathcal{O}(1)$\,GeV energies the impact of geomagnetic deflection becomes more relevant.

This is an area where the ICAL detector can provide crucial measurements not only for understanding the atmospheric flux, but also for testing new physics scenarios that involve $\nu_\tau$ and $\bar\nu_\tau$ behaving differently.

\subsubsection{DUNE far detector}

The LBNF beam in both the CP-optimized and tau-optimized configuration will produce large numbers of $\nu_{\tau}$ CC interactions in the DUNE FD, but due to the 1300 km baseline of the DUNE FD, the first atmospheric oscillation maximum occurs below the kinematic threshold for creating a $\tau$-lepton.  Because of this, the effect of $\Delta m^2_{31}$ and $\sin^2 \theta_{23}$ on the $\nu_{\tau}$ appearance probability is partially degenerate.  To disambiguate the effect of the two atmospheric parameters, it would be necessary to operate at a longer baseline.  However, since the DUNE FD will be located deep underground, it will also be able to collect a large sample of atmospheric neutrinos which probe a much larger range of $L/E$.  In addition, two of four DUNE FD modules are expected to be installed three years before the LBNF beam is commissioned.  Therefore, atmospheric neutrinos will play a complementary role to the beam neutrino samples.

We use the Honda flux prediction for Homestake mine~\cite{Honda:2015fha} and the cross section prediction from GENIE 2.12.2~\cite{Andreopoulos:2009rq} to determine that the DUNE FD should observe $\sim$1 $\nu_{\tau}$ CC event per kton-year.  This corresponds to $\sim$10 $\nu_{\tau}$ CC events per module each year. Inspired by Ref.~\cite{Conrad:2010mh}, we assume a 30\% signal efficiency and a 0.5\% NC background efficiency for selecting $\nu_{\tau}$ CC events where the $\tau$-lepton decays hadronically.  We assume that the calorimetric energy resolution for $\nu_{\tau}$ CC events is $\sim$17\% and the $\theta_{zen}$ resolution is $\sim$5$^{\circ}$.  Fig.~\ref{fig:DUNEAtmosSpec} shows the expected spectra for up-going $\nu_{\tau}$ candidates after smearing true energy and zenith angle.  For the most up-going neutrinos, the first and second oscillation maxima are visible. 

\begin{figure}
    \centering
    \includegraphics[width=0.48\textwidth]{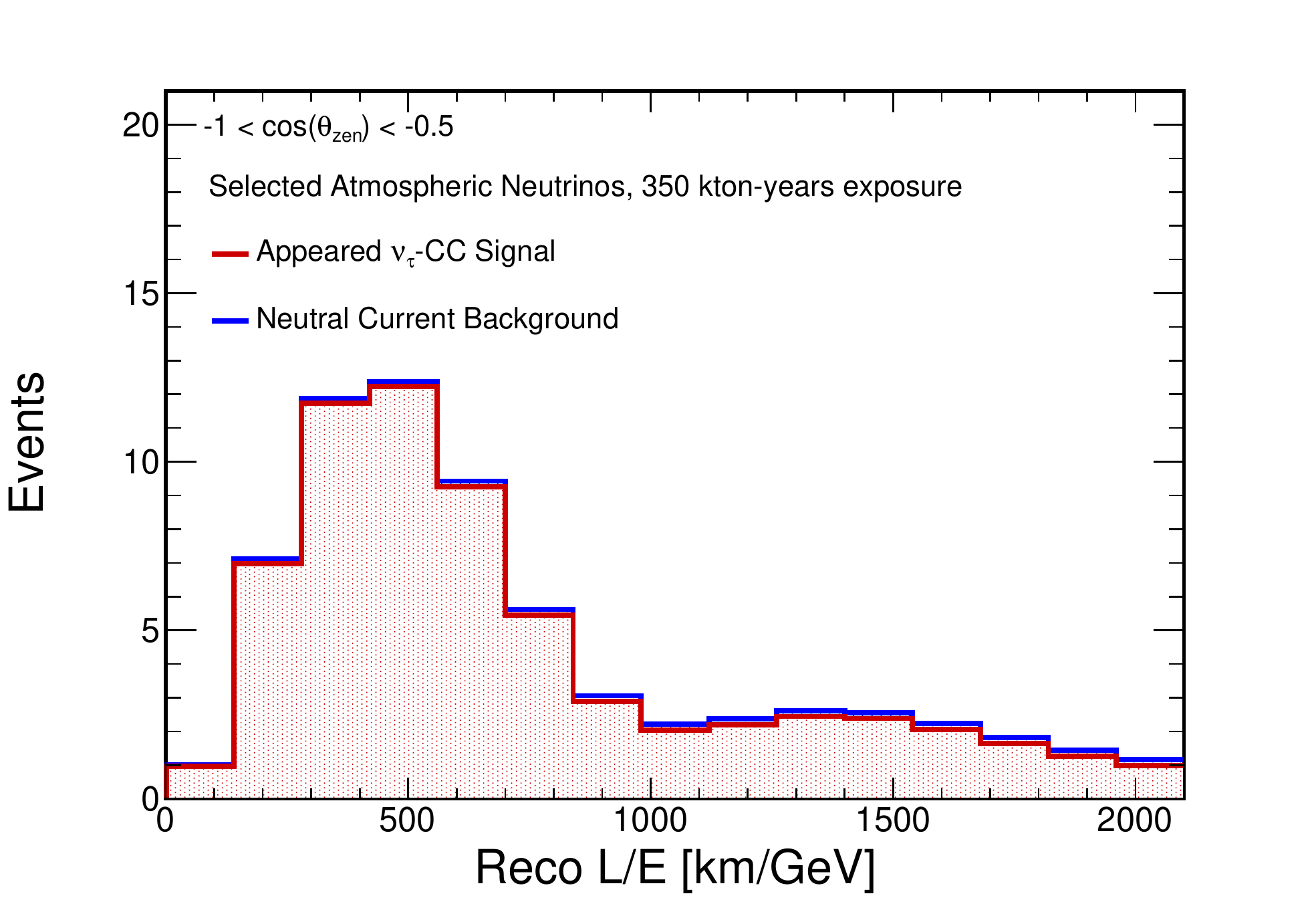}
    \includegraphics[width=0.48\textwidth]{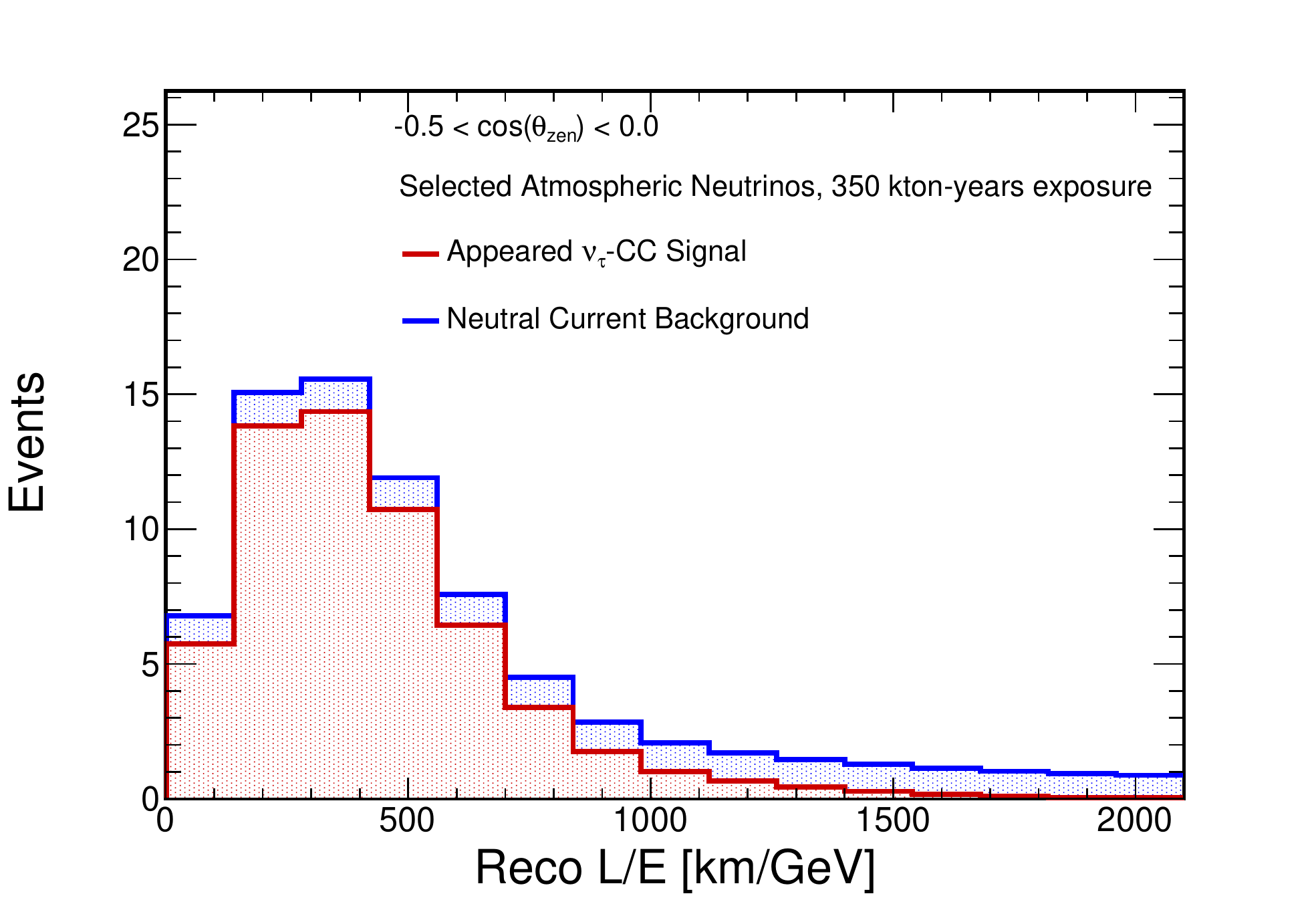}
    \caption{Selected atmospheric $\nu_{\tau}$ spectra for a 40 kton LAr detector at SURF.}
    \label{fig:DUNEAtmosSpec}
\end{figure}

To determine the physics reach of the atmospheric sample, we fit selected atmospheric neutrinos corresponding to a 350 kton-years exposure.  We split the data into 15 bins in reconstructed $L/E$ and 4 bins in reconstructed $\cos \theta_{zen}$.  Fig.~\ref{fig:DUNEAtmosSens}  shows the allowed region as a function of $\Delta m^2_{31}$ and $\sin^2\theta_{23}$ assuming a 25\% systematic normalization uncertainty.  Despite the lower statistics than expected from the beam sample, the atmospheric sample constrains the atmospheric parameters better than the beam sample.  This suggests that combining the beam and atmospheric samples will complement each other to constrain unitarity, sterile neutrinos, and NSI. 

\begin{figure}
    \centering
    \includegraphics[width=0.65\textwidth]{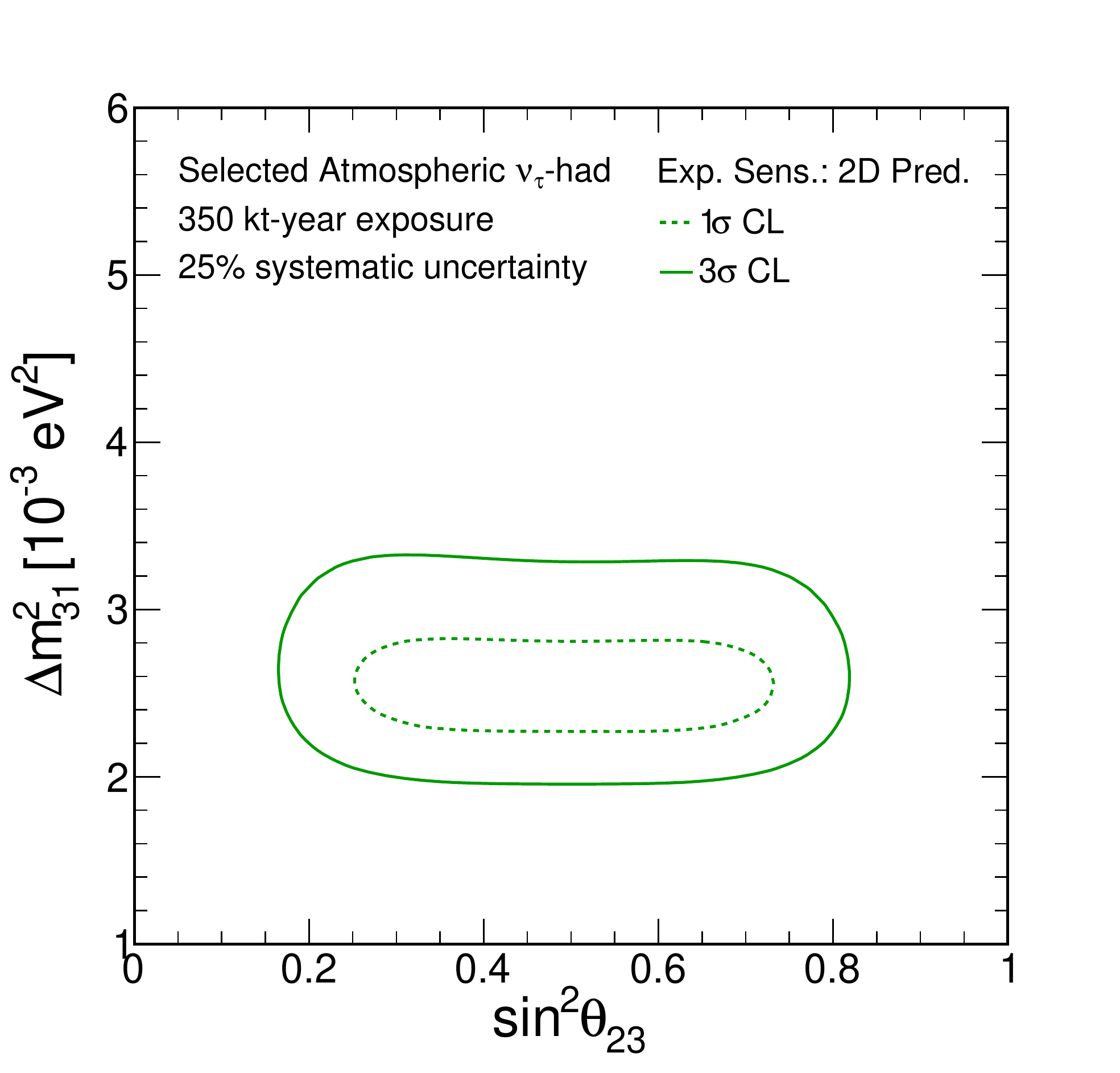}
    \caption{Sensitivity to three flavor parameters using 350 kt-years of exposure for a 40 kton LAr detector at SURF.}
    \label{fig:DUNEAtmosSens}
\end{figure}

\subsection{Neutrino Factory}
Another potential source of neutrinos is a muon storage ring in a racetrack configuration known as a neutrino factory \cite{Geer:1997iz,Cervera:2000kp,ISSAcceleratorWorkingGroup:2008laz,IDS-NF:2011swj,MICE:2012sli}.
While no concrete plans for such an experiment exist at the moment, such facilities have a number of advantages for neutrino oscillations in general and tau neutrino physics in particular.
First, the flux of neutrinos is reasonably expected to be very high and the decay spectrum of neutrinos is well understood, unlike in typical accelerator neutrino sources.
Second, there will be both $\nu_e$ and $\nu_\mu$ at the source provided an unprecedented capability to test many neutrino oscillation channels.
Third, due to the high energies typically considered, $\nu_\tau$ appearance is feasible.
In the past this channel was considered as a background to other channels \cite{Indumathi:2009hg,Donini:2010xk,Dutta:2011mc,Coloma:2012ji}, but depending on the detector technology, if the taus could be identified then they could be used as a separate measurement of the oscillation parameters.

In light of the recently renewed interest in muon colliders for electroweak precision observables \cite{AlAli:2021let}, it is definitely possible to perform $\nu_\tau$ appearance experiments at such a facility.

\pagebreak
\section{Experimental Probes at Intermediate Energies}
\label{sec:intermediate experiments}

For decades, hadron colliders have been the primary tool to explore and expand our understanding of nature at the energy frontier. The most energetic realization of this idea is the LHC, which collides two counter rotating proton beams with beam energies of up to 7~TeV. Its main objectives include the study of the Higgs boson and to search for signs of new physics at the TeV scale. However, as the accelerator experiment with the highest beam energy, the LHC is also the source of the most energetic human-made neutrinos. Indeed, the LHC produces an intense and strongly collimated beam of TeV-energy neutrinos along the direction of the proton beams. Notably, this neutrino beam includes a sizable fraction of tau neutrinos, mainly produced via the decay $D_s \to \tau \nu_\tau$ and subsequent tau decays, and hence provides a novel opportunity to study their properties. 

Already in 1984, De~Rujula and Rückl proposed to detect neutrinos from the LHC neutrino by placing a neutrino experiment in the far forward direction~\cite{DeRujula:1984pg}. This idea of detecting neutrinos from the LHC was revisited several times in the following decades~\cite{Vannucci:253670, DeRujula:1992sn, Park:2011gh, Buontempo:2018gta, Beni:2019gxv, FASER:2019dxq, XSEN:2019bel, FASER:2020gpr, Beni:2020yfy, SHiP:2020sos, Ahdida:2750060, Foldenauer:2021gkm},
but only recently concrete efforts started to build neutrino experiments at the LHC. In 2018, the FASER collaboration installed a suitcase size pilot detector employing emulsion films and reported the observation of the first neutrino interaction candidates at the LHC, demonstrating the feasibility of LHC neutrino experiments~\cite{FASER:2021mtu}. 

In upcoming Run~3 of the LHC, starting in the spring of 2022, two dedicated neutrino detectors at the LHC will start their operation: FASER$\nu$~\cite{FASER:2019dxq, FASER:2020gpr} and SND@LHC~\cite{SHiP:2020sos, Ahdida:2750060}. Both experiments are located around 480~m downstream from the ATLAS interaction point, in the previously unused side tunnels TI12 and TI18 respectively and will detect LHC neutrinos with high significance for the first time. Notably, tens of charged current tau neutrino interactions are expected in these detectors during LHC Run~3 with an integrated luminosity of 150~fb$^{-1}$. To further increase the event rate, larger experiments are considered for the HL-LHC in the context of a Forward Physics Facility (or FPF)~\cite{Anchordoqui:2021ghd,Feng:2022inv}. Three different dedicated neutrino detectors have been proposed to be housed in the FPF, FASER$\nu$2, FLArE and Advanced SND@LHC (AdvSND), with the opportunity to collect thousands of tau neutrino interactions for an integrated luminosity of 3~ab$^{-1}$. 

\begin{table}[tbp]
\setlength{\tabcolsep}{3.2pt}
    \centering
    \begin{tabular}{c|c|c|c||c|c|c}
    \hline\hline
      \multicolumn{4}{c||}{Detector} & 
      \multicolumn{3}{c}{Number of CC Interactions} \\
      \hline
      Name &  Mass & Coverage & Luminosity
      & $\nu_e\!\!+\!\bar{\nu}_e$ 
      & $\nu_\mu\!\!+\!\bar{\nu}_\mu$
      & $\nu_\tau\!\!+\!\bar{\nu}_\tau$
      \\
       \hline\hline
       FASER$\nu$  
       & 1 ton & $\eta \gtrsim 8.5$ & 150~fb$^{-1}$
       & 901 / 3.4k  & 4.7k / 7.1k & 15 / 97  \\
       \hline 
       SND@LHC  
       & 800kg & $7<\eta < 8.5$ & 150~fb$^{-1}$
       & 137 / 395  & 790 / 1.0k & 7.6 / 18.6  \\
       \hline\hline
       FASER$\nu$2  
       & 20 tons & $\eta \gtrsim 8$ & 3~ab$^{-1}$
       & 178k / 668k  & 943k / 1.4M  & 2.3k / 20k \\
       \hline
       FLArE 
       & 10 tons & $\eta \gtrsim 7.5$ & 3~ab$^{-1}$
       & 36k / 113k & 203k / 268k & 1.5k / 4k  \\
       \hline
       AdvSND  
       & 2 tons & $7.2 \lesssim \eta \lesssim 9.2$ & 3~ab$^{-1}$
       & 6.5k / 20k & 41k / 53k & 190 / 754 \\
       \hline\hline
    \end{tabular}
    \caption{Detectors and neutrino event rates: The left side we summarize the detector specifications in terms of the target mass, pseudorapidity coverage and assumed integrated luminosity for all LHC neutrino experiments. The right side shows the number of charged current neutrino interactions occurring the detector volume for all three neutrino flavors as obtained using two different event generators, \texttt{Sibyll~2.3d} and \texttt{DPMJET~3.2017}. The event rates for LHC and HL-LHC experiments were obtained in Ref.~\cite{Kling:2021gos} and Ref.~\cite{Anchordoqui:2021ghd}, respectively. 
    }
    \label{tab:lhc_eventrate}
\end{table}

In Table~\ref{tab:lhc_eventrate}, we present the detector specifications and estimated number of neutrino interactions for the different LHC neutrino experiments. Here we show the geometries and event rates as reported in Ref.~\cite{Kling:2021gos} for the LHC Run~3 and in Ref.~\cite{Anchordoqui:2021ghd} for the HL-LHC experiments, which use the fast neutrino flux simulation introduced in Ref.~\cite{Kling:2021gos} to propagate the SM hadrons through the LHC beam pipe and magnets and to simulate their decays into neutrinos. The primary interactions were simulated using two different event generators, \texttt{Sibyll~2.3d}~\cite{Fedynitch:2018cbl, Riehn:2019jet} and \texttt{DPMJET~3.2017}~\cite{Roesler:2000he, Fedynitch:2015kcn}, as implemented in the \texttt{CRMC} simulation package~\cite{CRMC_KIT}. Although we only present sum of the neutrino and anti-neutrino event rate for each flavor, we note that there is a similar number of neutrinos and anti-neutrinos in the LHC’s neutrino beam. Comparing the different experiments, we can see that the neutrino event rate drops towards smaller pseudorapidities, or equivalently when moving away from the beam axis. This is caused by a decreasing neutrino beam intensity away from the beam's center, which is most pronounced for muon neutrinos but also visible for tau neutrinos. 

Comparing the predictions of \texttt{Sibyll} and \texttt{DPMJET}, we note that there are large differences between the two event generator predictions, especially for tau neutrinos. These are mainly related to the modelling of the charm component. On the one hand, this imposes a challenge for neutrino physics measurements and new physics searches for which neutrino fluxes are an additional source of systematic uncertainties. Dedicated efforts to quantify and reduce the uncertainties of the tau neutrino flux are needed an indeed already ongoing: results of the tau neutrino flux via an perturbative NLO calculation are discussed in Sec.~\ref{sec:tools}. On the other hand, the currently large uncertainties also illustrate that a flux measurement is an interesting physics goal by itself, which will help us to better understand the underlying physics associated with forward charm production.

Before turning to the different experimental probes, we below summarize the physics potential and motivations of tau neutrino measurements at the LHC
\begin{enumerate}
    \item \textbf{Tau Neutrino Cross Section} The LHC neutrino experiments will perform the first measurement of neutrino interaction cross sections at TeV energies for all three neutrino flavors. This will provide a unique opportunity to probe lepton universality in neutrino scattering by comparing the interaction cross section of all three neutrino flavors. In addition, the magnetized detector components of the FPF neutrino experiments will also allow to differentiate tau neutrinos and tau antineutrinos for the first time.  
    \item \textbf{Tau Neutrino Production and QCD} Since tau neutrinos are mainly produced in $D_s$ meson decay, a measurement of their flux provides a novel probe of the currently poorly constrained forward charm production. This opens a window to the otherwise experimentally inaccessible kinematic regimes of QCD, resulting in many interesting opportunities as summarized in Ref.~\cite{Anchordoqui:2021ghd}: to validate the predictions of collinear factorization and BFKL-based approaches, to constrain gluon PDFs at low $x$; to probe gluon saturation effects; or to test models of intrinsic charm. In addition, constraints on forward charm production with LHC neutrinos would help to reduce systematic uncertainties associated with the prompt atmospheric neutrino flux and hence provide a valuable input for neutrino telescopes, such as IceCube, in their search for astrophysical neutrinos. 
    \item \textbf{Tau Neutrino Interactions and QCD} LHC neutrino experiments can be considered as a neutrino-ion collider with center-of-mass energies in the range of 10  to 50~GeV. This setup is complementary to the planned electron-ion collider (or EIC)~\cite{Accardi:2012qut}, which will operate at a similar center-of-mass energy, and will allow to address a variety of interesting hadronic effects associated with neutrino interactions. In particular, FPF neutrino experiments provide an opportunity to measure the strange quark PDF via the charm associated neutrino interactions $\nu s \to \ell c$ similar to CHORUS and NuTeV~\cite{KayisTopaksu:2011mx, Goncharov:2001qe}; constrain nuclear PDFs~\cite{Kovarik:2015cma, Eskola:2016oht, AbdulKhalek:2020yuc} of the target nuclei and independently measure shadowing, anti-shadowing and the EMC effect for neutrinos; probe the time dependence of hadronization and the prescriptions for the formation zone; test color transparency and final state interaction effects at the highest energies; and provide valuable input to tune neutrino generator tools used to simulate high-energy neutrino events~\cite{Buss:2011mx, Andreopoulos:2009rq, Golan:2012rfa, Hayato:2021heg, Battistoni:2009zza}.
    \item \textbf{Probes of New Physics:} The large intensity and energy of the tau neutrino beam at the LHC also allows for a variety of novel strategies to search for new physics. This includes the searches for anomalous electromagnetic properties of tau neutrinos, such as a tau neutrino magnetic dipole moment, as discussed in Refs.~\cite{Ismail:2021dyp, AbariTsaiAbraham} and presented in more detail in Sec.~\ref{sec:theory}; searches for sterile neutrinos with the multi-eV masses, leading to an excess of tau neutrinos at short baselines, as discussed for example in Ref.~\cite{FASER:2019dxq} and Ref.~\cite{Bai:2020ukz}; or searches for new tau neutrino philic mediators to dark matter, such as a $B-3L_\tau$ gauge boson discussed in Refs.~\cite{Kling:2020iar, Batell:2021snh} or a neutrino-philic scalar discussed in Ref.~\cite{Kelly:2021mcd}, which can enhance the tau neutrino flux or modify their interaction rates. 
\end{enumerate}

\subsection{FASER$\nu$}

FASER~\cite{FASER:2018bac,FASER:2018eoc} is the ForwArd Search ExpeRiment at the LHC. FASER$\nu$~\cite{FASER:2019dxq} is an extension of that experiment, and was designed to detect collider neutrinos for the first time and study neutrino properties at TeV energies. The FASER$\nu$ proposal~\cite{FASER:2020gpr} was approved in December 2019, and the detector is being installed 480 m downstream of the ATLAS interaction point in the tunnel TI12. Beam exposure and data collection will start in spring 2022. FASER$\nu$ is deployed on the beam collision axis to maximize the interaction rate of all three flavors of neutrinos, namely $\nu_e$, $\nu_{\mu}$, and $\nu_{\tau}$, and antineutrinos. This allows FASER$\nu$ to measure the interaction cross sections in the unexplored high-energy range. 

\begin{figure}[hbt!]
\centering
\includegraphics[width=0.95\textwidth]{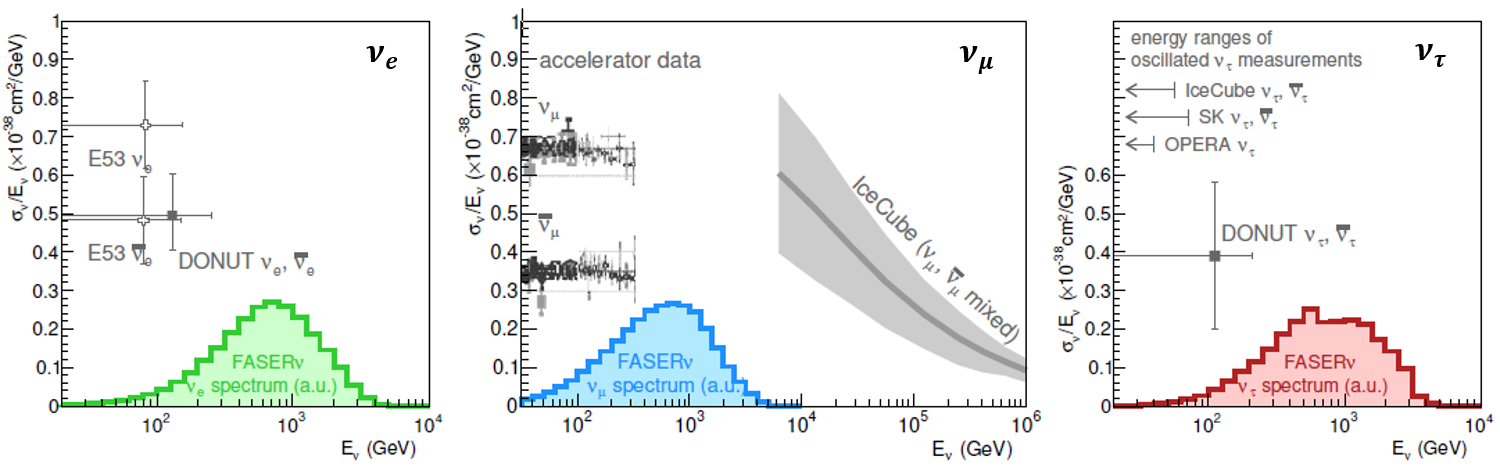}
\caption{Existing constraints on neutrino charged-current interaction cross sections, and the expected energy spectra of neutrinos interacting in FASER$\nu$~\cite{FASER:2019dxq}.}
\label{fig:xsec_and_spectra}
\end{figure}

Figure~\ref{fig:xsec_and_spectra} shows the existing constraints on neutrino charged-current interaction cross sections and the expected energy spectra of the neutrinos interacting in FASER$\nu$. 
For muon neutrinos, the FASER$\nu$ measurements can probe the gap between accelerator measurements ($E_\nu<360$ GeV)~\cite{PDG:2018} and the IceCube data ($E_\nu>6.3$ TeV)~\cite{Aartsen:2017kpd, Bustamante:2017xuy}. For electron and tau neutrinos, the cross section measurements can be extended to considerably higher energies. 
In addition to the measurements of charged-current interactions, neutral-current interactions can be measured. Such measurements can provide a new limit on nonstandard interactions of neutrinos to complement the existing limits~\cite{PhysRevD.103.056014}.

Furthermore, forward particle production, which is poorly constrained by the other LHC experiments at $\eta<5$, can be studied using FASER$\nu$ as it covers the range $\eta>8.5$. In particular, FASER$\nu$ measurements of high-energy electron and tau neutrinos, which mainly originate from charm decays, can provide the first data on forward charm production. 
Atmospheric neutrino production via charm decays, so-called ``prompt neutrinos'', has been the background for astrophysical neutrino observation by neutrino telescopes at the energy of 100 TeV or above, yet, a quantitative understanding hasn't been reached. A direct measurement using FASER$\nu$ with 13.6-TeV proton--proton collisions, corresponding to a 100-PeV proton interaction in the fixed-target mode, can provide basic data from a controlled environment. 

\begin{figure}[hbt!]
\centering
\includegraphics[width=0.85\textwidth]{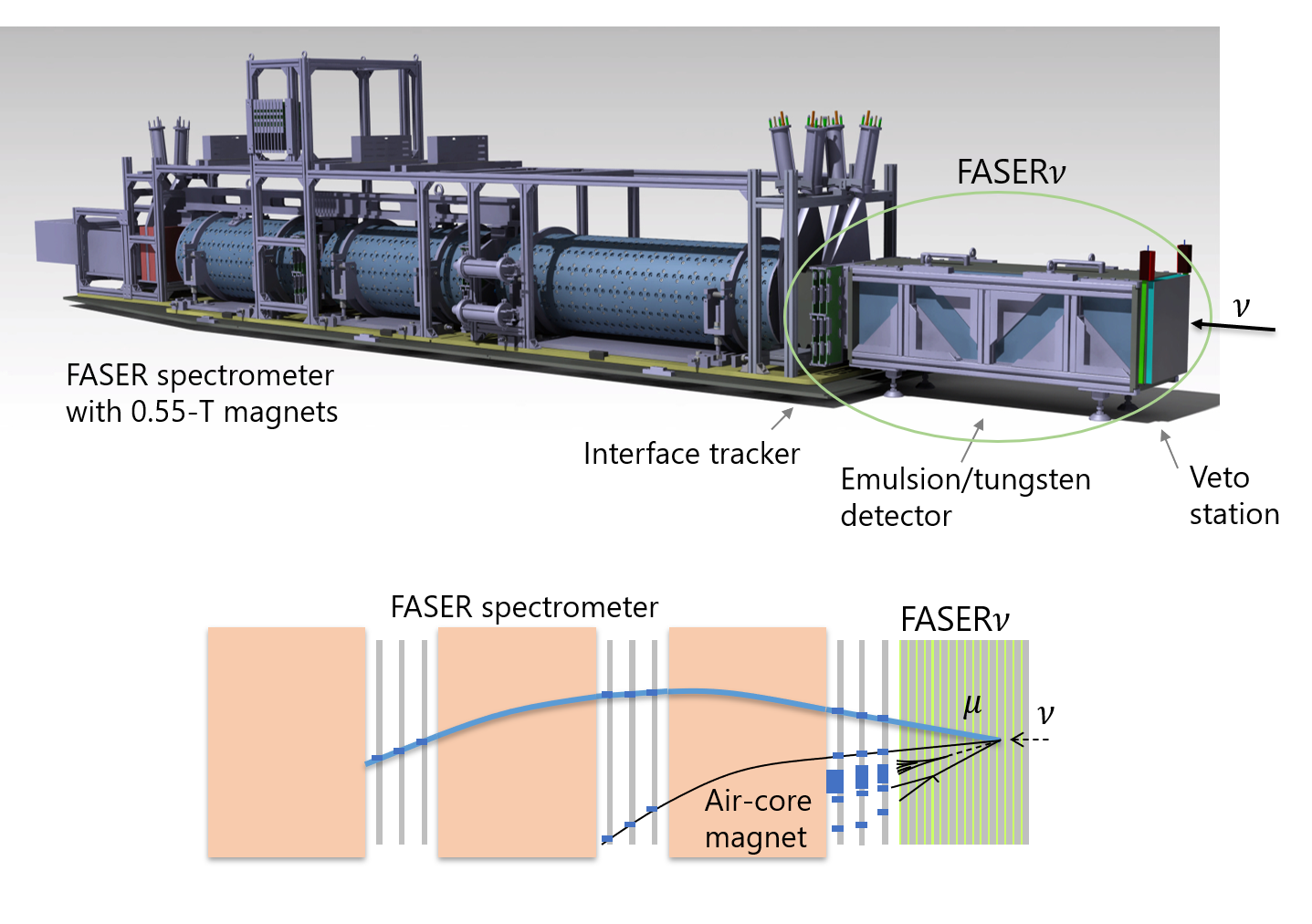}
\caption{Sketch of the FASER detector~\cite{Ariga:2022roc} (top) and the topology to be seen in the interface tracker and the FASER spectrometer~\cite{ariga:2021tau} (bottom).
}
\label{fig:FASERnu_detector}
\end{figure}

In LHC Run 3 starting in spring  2022, the FASER$\nu$ detector will be located in front of the main FASER detector~\cite{Ariga:2018pin} along the beam collision axis. Figure~\ref{fig:FASERnu_detector} (top) shows the FASER$\nu$ detector and the main FASER detector. The FASER$\nu$ detector includes a veto station, an emulsion/tungsten detector, and an interface tracker (IFT) coupled to the FASER magnetic spectrometer. 
The emulsion/tungsten detector is designed to identify different lepton flavors which will be produced in $\nu_e$, $\nu_{\mu}$, $\nu_{\tau}$ interactions. It has finely sampled detection layers (every 1 mm of tungsten) to identify electrons and to distinguish them from gamma rays, sufficient target material of 8 $\lambda_{int}$ to identify muons, and good position and angular resolutions to detect tau and charm decays. 
Additionally, the detector can measure muon and hadron momenta and energy of electromagnetic showers, which will be used to estimate energy of neutrinos. 
The IFT is located downstream of the emulsion/tungsten detector. Figure~\ref{fig:FASERnu_detector} (bottom) shows the topology of a neutrino event signal as it will be searched for in the IFT and the FASER spectrometer.
The emulsion/tungsten detector is consisted of a recurring structure of emulsion films interleaved with 1-mm-thick tungsten plates. The emulsion film has two emulsion layers, each 70-$\mu$m thick. These layers are added onto both sides of a 210-$\mu$m-thick polystyrene base. The emulsion detector contains a total of 770 emulsion films, each of dimensions 25 cm $\times$ 30 cm. The total tungsten mass is 1.1 tons. 

In 2018 during LHC Run 2, a pilot run was performed in the TI18 tunnel of the LHC to demonstrate neutrino detection at the LHC for the first time. The pilot detector was small and was not designed to identify muons. Its depth was only 0.6 $\lambda_{int}$, which is much shorter than the 8 $\lambda_{int}$ of the full FASER$\nu$ detector which is being prepared for LHC Run 3. The data from the pilot detector are used to demonstrate the feasibility of high-energy neutrino measurements in this experimental environment. 
Neutrino interactions were searched for by analyzing the data corresponding to 11 kg of the target mass. The first candidate events to be consistent with neutrino interactions at the LHC were observed~\cite{FASER:2021mtu}. A 2.7$\sigma$ excess of neutrino-like signals over muon-induced backgrounds was measured. These results demonstrate the ability of FASER$\nu$ to detect neutrinos at the LHC and pave the way for future collider neutrino experiments. 

\begin{figure}[hbt!]
\centering
\includegraphics[width=0.7\linewidth]{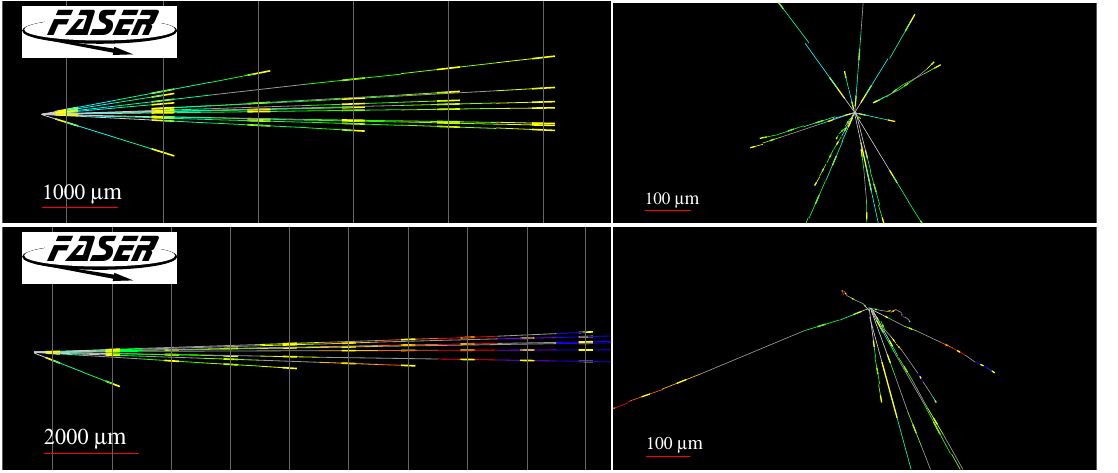}
\caption{Event displays of two neutrino interaction candidate vertices~\cite{FASER:2021mtu} in the $y$–$z$ projection longitudinal to the beam direction (left) and in the view transverse to the beam direction (right).
}
\label{fig:vtx_Pb_area14_and_area11}
\end{figure}

In March 2021, the main FASER detector was successfully installed into the TI12 tunnel. The FASER$\nu$ veto station and the IFT were also assembled and were installed in the tunnel in November 2021. The emulsion/tungsten detector will be installed just before the beam starts and will be replaced several times a year during planned technical stops of the LHC. The production of emulsion gel and films is scheduled few months before each replacement.

As shown in Table~\ref{tab:lhc_eventrate}, $\sim$10,000 flavor-tagged charged-current neutrino interactions, along with neutral-current interactions, will be collected in the FASER$\nu$ detector during LHC Run 3. About 
$\sim$20 ($\sim$100) $\nu_{\tau}$ charged-current neutrino interactions are expected based on the event generator \texttt{Sibyll~2.3d} (\texttt{DPMJET~3.2017}).

\subsection{SND@LHC}

SND@LHC (Scattering and Neutrino Detector @ LHC) 
is a compact and stand-alone experiment designed to perform measurements with neutrinos produced at the LHC in a hitherto unexplored pseudo-rapidity region of $7.2 < \eta < 8.4$, complementary to all the other experiments at the LHC, including FASER. The Collaboration submitted a Letter of Intent in August 2020~\cite{Collaboration:2729015}. Following investigations that confirmed the possibility of preparing the experimental area and installing the detector during 2021, with the LHC in cold operating conditions, the LHCC recommended the collaboration to proceed with the preparation of a Technical Proposal, submitted in January 2021~\cite{Ahdida:2750060}. Based on this document, the experiment was approved in March 2021 by the Research Board. 

\begin{figure}
\begin{center}
\includegraphics[width=1.0\linewidth]{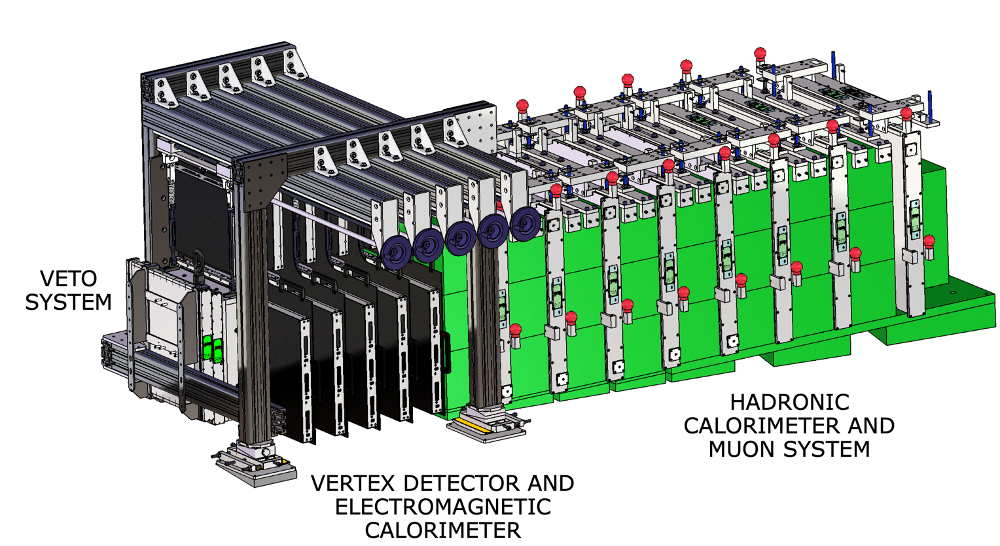}
\caption{SND Detector layout: the veto system is located upstream of the tungsten target, interleaved with emulsion and SciFi planes. Downstream of the target, 8 iron walls are followed by scintillating stations with the function of an hadronic calorimeter and a muon identification system.}
\label{fig:snddetector}
\end{center}
\end{figure}

The experiment is located 480\,m downstream of IP1 in the unused TI18 tunnel. The detector is composed of a hybrid system based on an 800\,kg target mass of tungsten plates, interleaved with emulsion and electronic trackers, followed downstream by an hadronic calorimeter and a muon identification system, as shown in Figure~\ref{fig:snddetector}. The configuration allows efficiently distinguishing between all three neutrino flavors, opening a unique opportunity to probe physics of heavy flavor production at the LHC in the region that is not accessible to ATLAS, CMS and LHCb. The detector concept is also well suited to searching for Feebly Interacting Particles via signatures of scattering in the detector target~\cite{boyarsky2021searches}. The first phase aims at operating the detector throughout Run~3 to collect more than 150\,fb$^{-1}$ overall.

The SND@LHC detector takes full advantage of the space available in the TI18 tunnel to cover the desired range in pseudo-rapidity. Figure~\ref{fig:sideview} shows the side and top views of the detector positioned inside the tunnel. It is worth noting that the tunnel floor is sloped, as can be seen from the side view, with the floor sloping down along the length of the detector. As shown in the top view, the nominal collision axis from IP1 comes out of the floor very close to the wall of the tunnel. The location is ideal to explore the off-axis region. Since no civil engineering work could have been done in time for the operation in Run 3, the tunnel geometry imposed several constraints. The following guidelines were adopted for the optimization of the detector design: a good calorimetric measurement of the energy requires about 10 $\lambda_{\mathrm{int}}$; a good muon identification efficiency requires enough material to absorb hadrons; for a given transverse size of the target region, the azimuthal angular acceptance decreases with distance from the beam axis. The energy measurement and the muon identification set a constraint on the minimum length of the detector. With the constraints from the tunnel, this requirement
competes with the azimuthal angular acceptance that determines the overall flux intercepted and therefore the total number of observed interactions. The combination of position and size of the proposed detector is an optimal compromise between these competing requirements. The geometrical constraints also restrict the detector to the first quadrant only around the nominal collision axis, as shown in Fig.~\ref{fig:sideview}.
\begin{figure}
\begin{center}
\includegraphics[width=0.6\linewidth]{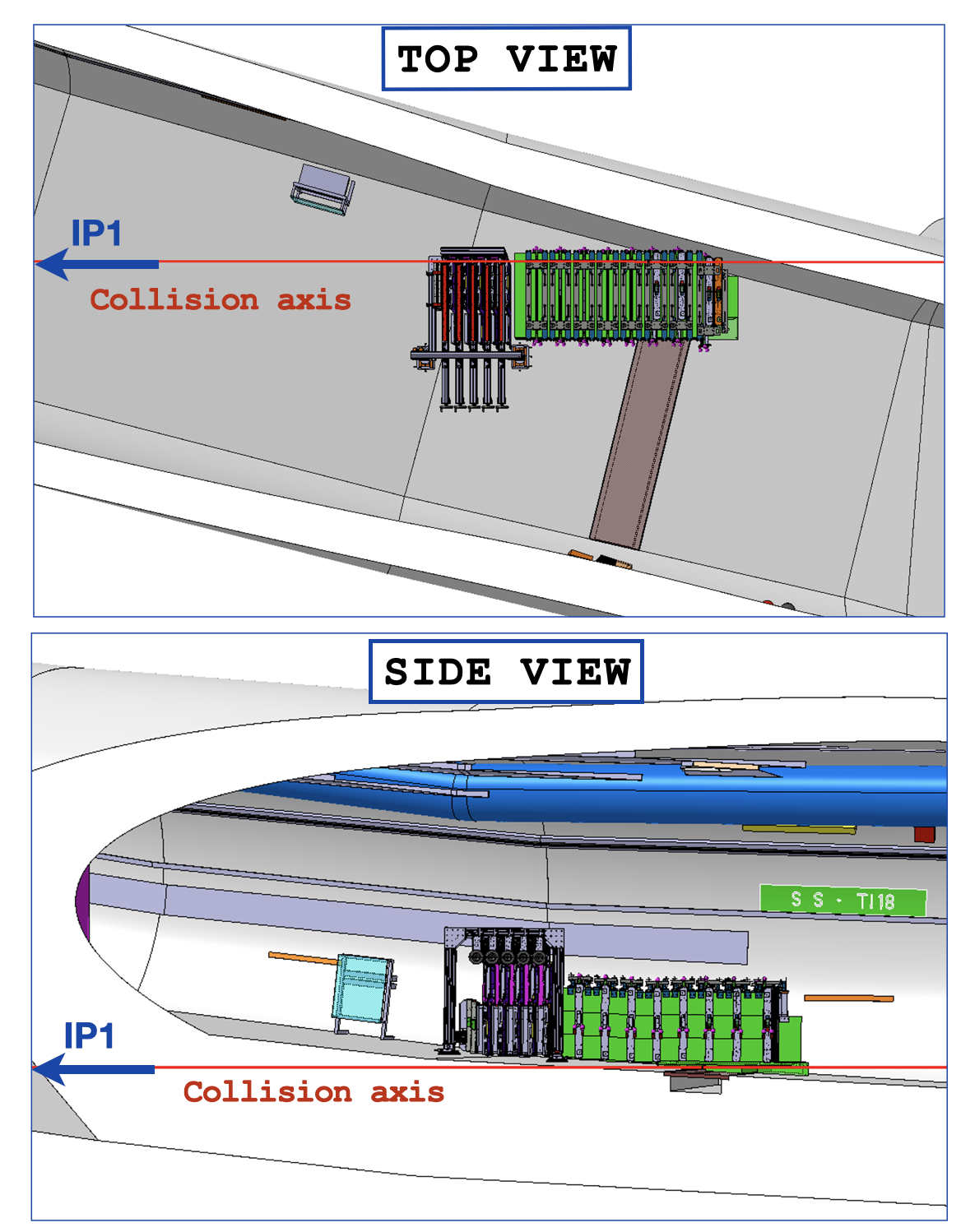}
\caption{Side and top views of the SND@LHC detector in the TI18 tunnel.
From \cite{Ahdida:2750060}.}
\label{fig:sideview}
\end{center}
\end{figure}
The result is a compact detector, 2.6 m in length. The energy measurement and the muon identification limit the target region to a length of about 80 cm. The transverse size downstream of about 80(H) × 60(V) cm$^2$ is limited by the constraint of the tunnel side wall. The transverse size of the target region is proportionally smaller in order to match the acceptance of the energy measurement and the muon identification for the vertices identified in the target volume. In order to maximise the number of neutrino interactions, tungsten has been selected as the passive material. The emulsion target will be replaced a few times per 
year during technical stops of the LHC.

With data from Run~3, SND@LHC will be able to study about two thousand high-energy neutrino interactions. Table~\ref{tab:NuFlux} reports the expected number of charged-current (CC) and neutral-current (NC) neutrino interactions in the detector target, assuming  150\,fb$^{-1}$ and an equal weight of upward and downward crossing-angle configurations. The average energies of the interactions are also reported. 

Performance studies show that the charmed-hadron production in the SND@LHC pseudo-rapidity range can be determined with a statistical and systematic accuracy of 5\% and 35\%, respectively. The result may be further used to constrain the gluon PDF~in the very-small-$x$ region~\cite{Ahdida:2750060}. Unique tests of lepton flavor universality with neutrino interactions can reach  10\% for both statistical and systematic uncertainties for $\nu_e$ and $\nu_{\mu}$ at high energy~\cite{Ahdida:2750060}. Here we discuss in some detail the measurements with tau neutrinos. 

\begin{table}[hbtp]
\centering
\caption{Number of CC and NC neutrino interactions in the SND@LHC acceptance.}
\begin{tabular}{c | c c | c c}
\hline
       &  \multicolumn{2}{c|}{CC neutrino interactions} & \multicolumn{2}{c}{NC neutrino interactions} \\
\hline
Flavor     &  $\langle\text{E}\rangle$ [GeV]  & Yield &  $\langle\text{E}\rangle$ [GeV]  & Yield \\
\hline
$\nu_\mu$       &  452  & 606 & 480 & 182 \\
$\bar{\nu}_\mu$ & 485  & 248 & 480 & 93\\
$\nu_e$         & 760  & 182 & 720 & 54\\
$\bar{\nu}_e$   & 680  &  97 & 720 & 35\\
$\nu_\tau$      & 740  &  14 & 740 & 4\\
$\bar{\nu}_\tau$& 740  &   6 & 740 & 2\\
\hline
TOT &   & 1153  & & 370\\
\hline
\end{tabular}
\label{tab:NuFlux}
\end{table}
 
{\bf Lepton flavor universality test with $\nu_\tau$ interactions}

In the pseudo-rapidity range of interest, tau neutrinos are essentially only produced in $D_s \rightarrow \tau \nu_\tau$ and the subsequent $\tau$ decays. According to the \textsc{Pythia} event generator, about 8\% of $\nu_\tau$s comes from beauty hadron decays. One can thus assume that the source of both $\nu_e$ and $\nu_\tau$ is essentially provided by semi-leptonic and fully leptonic decays of charmed hadrons. Unlike $\nu_\tau$s produced only in $D_s$ decays, $\nu_e$s are produced in the decay of all charmed hadrons, essentially $D^0$, $D$, $D_s$ and $\Lambda_c$. Therefore, the $\nu_e/\nu_\tau$ ratio depends only on the charm hadronization fractions and decay branching ratios. The systematic uncertainties due to the charm-quark production mechanism cancel out, and the ratio becomes sensitive to the $\nu$-nucleon interaction cross-section ratio of the two neutrino species. The measurement of this ratio can thus be considered a lepton flavor universality test in neutrino interactions. 
The $\nu_e$ to $\nu_\tau$ ratio ($R_{13}$) can be written as:
\begin{equation}
    R_{13} = \frac
    {N_{\nu_e+\overline{\nu}_e}}
    {N_{\nu_\tau+\overline{\nu}_\tau}}
    = \frac{\sum_i \tilde{f}_{c_i} \tilde{Br}(c_i \to \nu_e X) }
    {\tilde{f}_{D_s} \tilde{Br}(D_s \to \tau \nu_\tau) },
\end{equation}
where $\tilde{f}_{c_i}$ are the charmed hadron fractions
and $\tilde{Br}(c_i \to \nu_e X)$ are the branching ratios of each charm species. Notice that $\tilde{Br}(D_s \to \tau \nu_\tau)$ includes also the contribution from the subsequent $\tau$ decay, as estimated with the full simulation of the decay chain. The tilde symbol on the above mentioned quantities indicates that they refer to the expected values in the SND@LHC acceptance. 

The estimate of these ``weighted'' branching ratios is affected by a systematic uncertainty of about 22\%
while the statistical uncertainty is dominated by the low statistics of the $\nu_\tau$ sample, which corresponds to a 30\% accuracy. 
The systematic uncertainty was evaluated by studying the fluctuations of the ratio using different event generators, after having equalized the branching ratio $D_s \to \tau \nu_\tau$ to the PDG value~\cite{PDG:2018}.

{\bf Detector construction status}
All the detector systems were constructed in the labs by Summer 2021 and were assembled and tested at CERN. In October 2021, a test-beam was performed at the SPS with protons of different energies in order to calibrate the response of the hadronic calorimeter: seven scintillating bar stations interleaved by real-size iron blocks were used for the measurement. Moreover, the full detector was commissioned on the surface at CERN with penetrating muons in the H6 experimental hall. On November 1st, the installation underground started. Figure~\ref{fig:detconstr} shows the full detector installed in the middle of December 2021. A borated polyethylene  shielding box will be added to surround the target and absorb low-energy neutrons originated from beam-gas interactions. The detector installation is expected to be completed in February 2022, so as to be ready to take data as soon as the LHC operation resumes with the Run 3.  
\begin{figure}
\begin{center}
\includegraphics[width=0.7\linewidth]{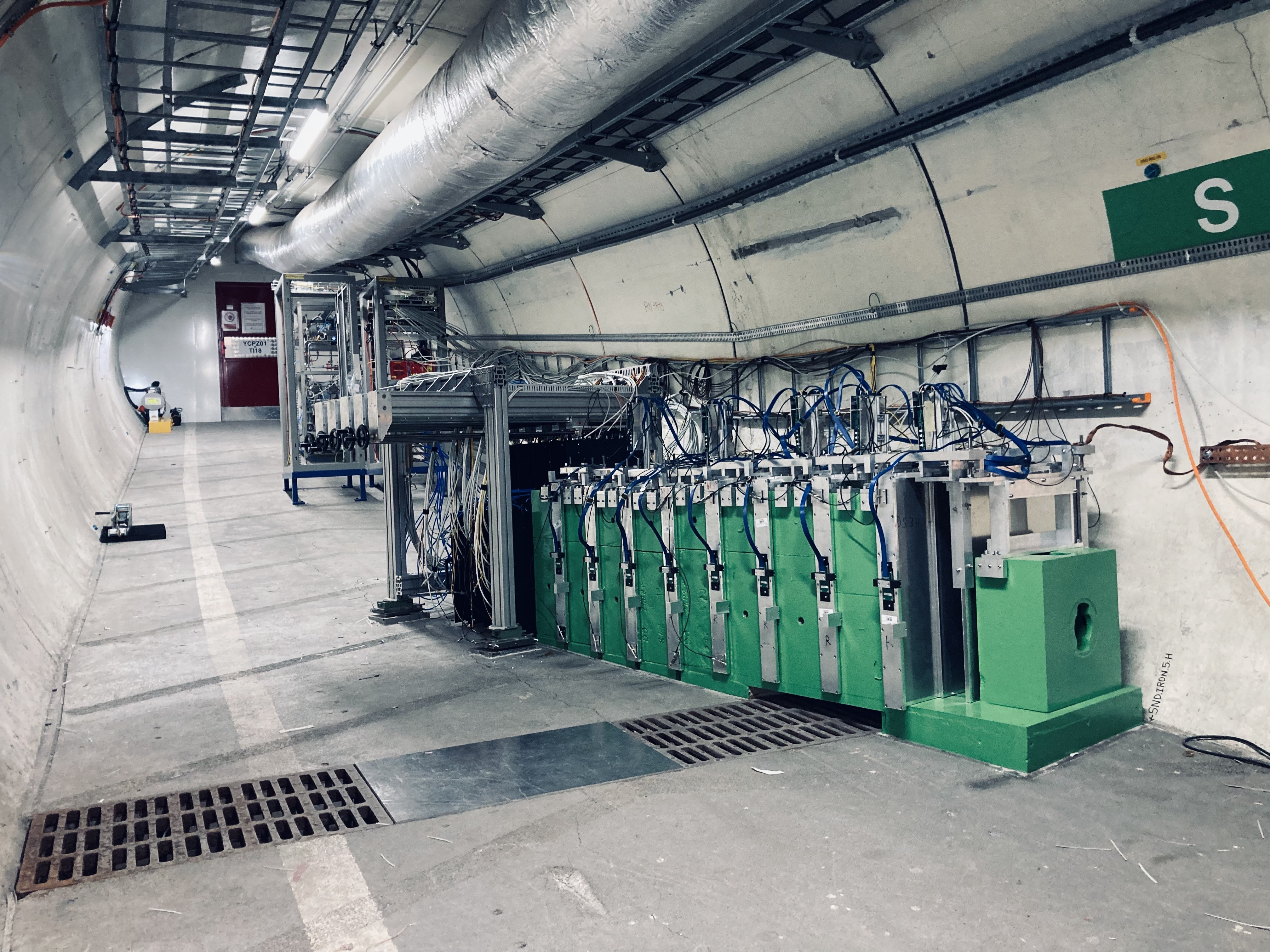}
\caption{Side and top views of the SND@LHC detector in the TI18 tunnel.}
\label{fig:detconstr}
\end{center}
\end{figure}

\subsection{The Forward Physics Facility Project}

With FASER$\nu$ and SND@LHC, the first two dedicated LHC neutrino detectors will start their operation in 2022. During the $3-4$ year long third run of the LHC, these experiments are expected to see tens of tau neutrino interactions.

Looking further in the future, these experiments also pave the way for a forward neutrino program during the HL-LHC era. After an upgrade of the accelerator infrastructure during the mid of this decade, the high luminosity runs of the LHC are expected to start around 2028 and to last about a decade. The nominal integrated luminosity of the HL-LHC is 3000~fb$^{-1}$, which provides a 20-fold increase compared to Run~3 of the LHC. In addition, the installation of improved detectors with larger target masses would further increase the event rate. Combining the effects of increased luminosity and larger target masses, neutrino experiments at the HL-LHC provide the opportunity to detect thousands of tau neutrino interactions.

\begin{figure}[t]
\centering
\includegraphics[width=0.99\textwidth]{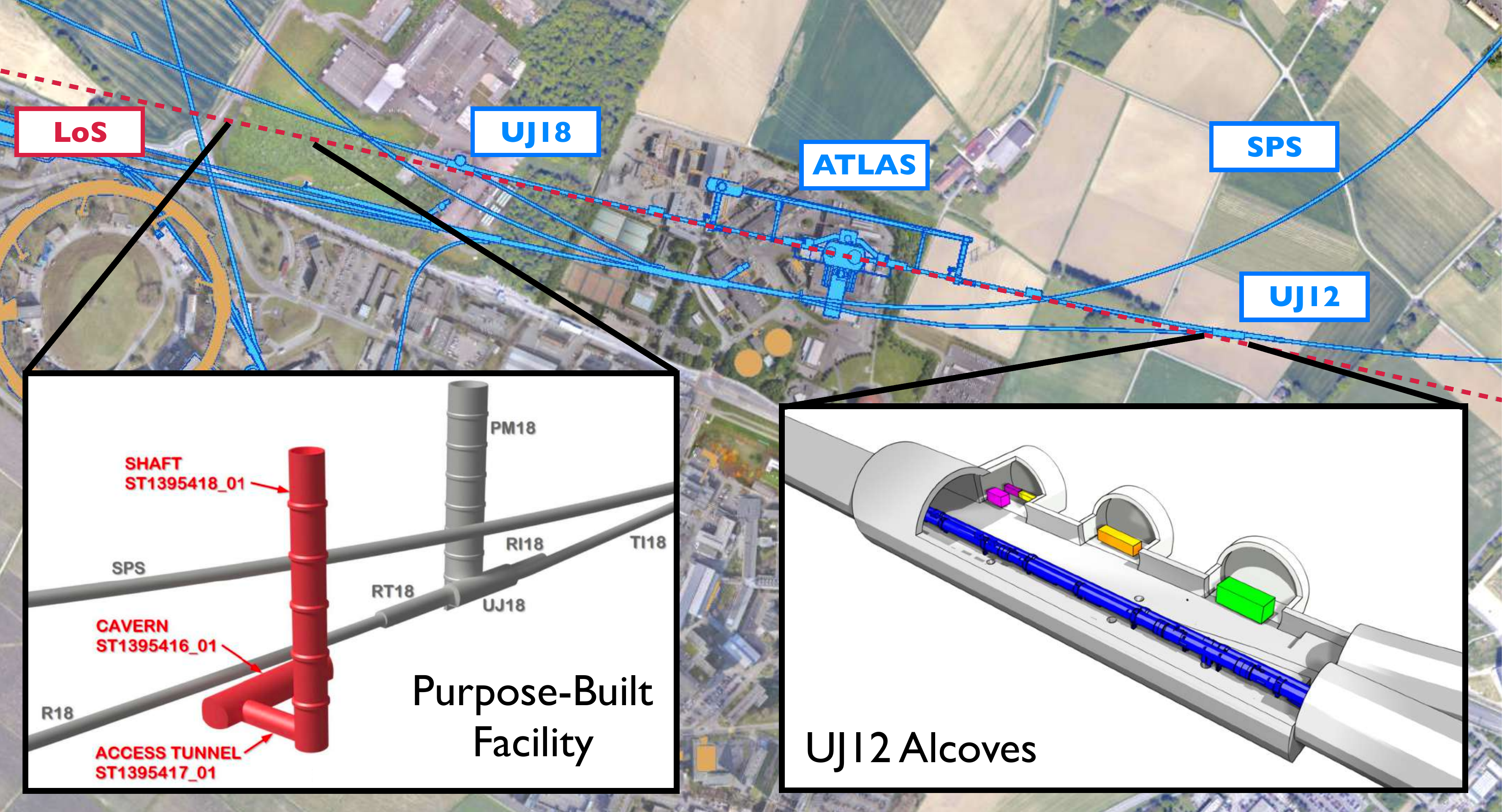}
\caption{Location of the two considered FPF sites: The upper part of the figure shows the CERN tunnel system near the ATLAS interaction point as well as the line of sight (LoS) which marks the center of the LHC neutrino beam. In the \textit{UJ12 alcoves} option, located about 500~m east of ATLAS, the UJ12 cavern is enlarged with alcoves to surround the LoS. For the \textit{purpose-built facility option}, a new experimental cavern and access shaft would be excavated around the LoS about 620~m meters west of ATLAS.}
\label{fig:fpf_location}
\end{figure}

The existing LHC neutrino experiments, FASER$\nu$ and SND@LHC, mentioned in the previous sections, are located at the intersection of the line of sight (LoS), which marks the center of the neutrino beam, and the tunnels TI12 and TI18. These tunnels originate from the LEP era, where they housed the injector that connected the SPS with LEP, but were never intended to host experiments. In particular, the available space and access and to LoS is severely constrained, which imposes strong limitations when considering larger detectors. To address this issue, the creation of the Forward Physics Facility (FPF), which would provide the space and infrastructure to support a suite of far-forward experiments at the LHC, has been proposed in Ref.~\cite{Anchordoqui:2021ghd}. Following a study by CERN's site and civil engineering department, two options for the FPF site are currently under consideration. The layout and locations of the two options with respect to the CERN tunnel system are shown in Fig.~\ref{fig:fpf_location}. 
\begin{description}
    \item [UJ12 Alcoves] One of the considered options is to enlarge the existing cavern UJ12, which is located roughly 500~m downstream from the ATLAS experiment near the current location of the FASER experiment. Currently, the LoS barely misses the UJ12 cavern at its southern end. The addition of three alcoves of about 6.4~m width would extend the cavern in this direction. This would allow to place experiments on the LoS and provide additional space for instrumentation around them. One disadvantage of this option is the limited access, especially when the LHC is in operation. 
    \item [Purpose-Build Facility] The other considered option would the  construction of a purpose build facility consisting of a new cavern which will be located roughly 620~m downstream from ATLAS on the French side. The cavern is envisioned to have a cylindrical shape with a length of approximately 65~m and a diameter of about 8.5~m. It would therefore provide space for significantly larger experiments. The cavern would be connected to the surface by an access shaft as well as to the LHC by a safety gallery. In this option, it would therefore be possible to access the cavern while the LHC is running, which would provide significantly more flexibility during both construction and operation. 
\end{description}

\begin{figure}[t]
\centering
\includegraphics[width=0.99\textwidth]{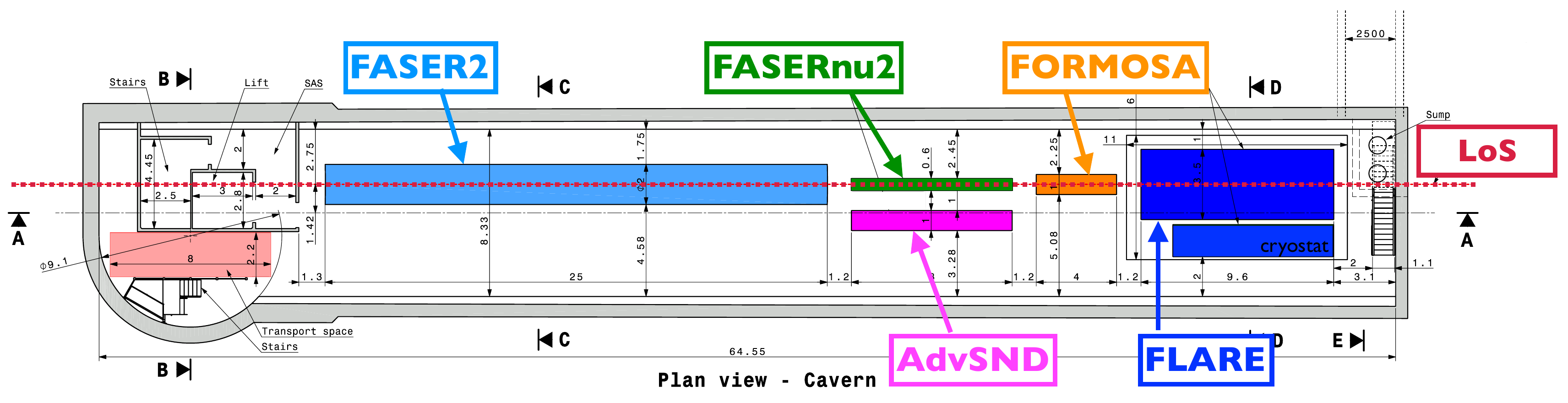}
\caption{Layout of the experimental cavern in the purpose-build facility option. The colored boxes indicate the location and dimensions of the proposed experiments: FASER2, FASER$\nu$2, AdvSND, FORMOSA and FLArE. The red line indicated the position of the line of sight (LoS).}
\label{fig:fpf_experiments}
\end{figure}

The FPF will house a suite of experiments, with the goal to detect the interaction of LHC neutrinos as well as to perform search for new physics. A an overview of the different experiments at the FPF for the purpose build facility option is shown in Fig.~\ref{fig:fpf_experiments}. The following experiments have been proposed: 

\begin{description}
    \item [FASER2:] An upgrade of the FASER experiment~\cite{Feng:2017uoz, FASER:2018ceo, FASER:2018bac}, consisting of a magnetized spectrometer and tracking detector, with the goal to search for decays of new long-lived particles predicted by models of new physics~\cite{FASER:2018eoc}. 
    \item [FORMOSA:] A plastic scintillator array, with a design similar to the milliQan experiment~\cite{Haas:2014dda, Ball:2016zrp, Ball:2020dnx}, which is dedicated to the search for millicharged particles~\cite{Foroughi-Abari:2020qar}. 
    \item [FASER$\nu$2:] A 10~ton emulsion based neutrino detector, centered around the LoS, to study neutrino interactions at the LHC. 
    \item [AdvSND:] A electronic neutrino detector with a multi-ton target mass, which is placed slightly off axis to extend the pseudorapidity coverage.
    \item [FLArE:] A liquid argon based detector designed for neutrino studies and dark matter searches~\cite{Batell:2021blf, Batell:2021aja, Batell:2021snh}. 
\end{description}
As we can see, three of the above-mentioned experiments (FASER$\nu$2, AdvSND and FLArE) are dedicated neutrino experiments and would have the ability to observe high energy tau neutrinos at the LHC. Below, each of these experiments is presented in detail.

\subsubsection{FASER$\nu$2 at the FPF}

The FASER$\nu$2 detector is designed as a much larger successor to FASER$\nu$ to greatly extend the physics potential for tau neutrino studies. It will be an emulsion-based detector able to identify heavy flavor particles produced in neutrino interactions, including $\tau$ leptons and charm and beauty particles. 
In the HL-LHC era, FASER$\nu$2 will be able to carry out precision tau neutrino measurements and heavy flavor physics studies, eventually testing lepton universality in neutrino scattering and new physics effects. FASER$\nu$2 will provide extraordinary opportunities for a broad range of neutrino studies, with additional and important implications for QCD and astroparticle physics as described in~\cite{Anchordoqui:2021ghd}.

Figure~\ref{fig:FASERnu2} shows a view of the FASER$\nu$2 detector. Its ideal location is in front of the FASER2 spectrometer along the beam collision axis to maximize the neutrino event rate per area for all three favors. The FASER$\nu$2 detector is envisioned to be composed of 3300 emulsion layers interleaved with 2 mm-thick tungsten plates. It will also include a veto detector and interface detectors to the FASER2 spectrometer, with one interface detector in the middle of the emulsion modules and the other detector downstream of the emulsion modules to make the global analysis and muon charge measurement possible. Both the emulsion modules and interface detectors will be put in a cooling system. The total volume of the tungsten target is 40 cm $\times$ 40 cm $\times$ 6.6 m, and the mass is 20 tonnes. The detector length, including the emulsion films and interface detectors, will be about 8 m.

The high muon background in the LHC tunnel might be an experimental limitation. The possibility of sweeping away such muons with a magnetic field placed upstream of the detector is currently being explored. Considering the expected performance, emulsion films will be replaced every year during the winter stops.

\begin{figure}[htb]
\centering
\includegraphics[width=0.95\linewidth]{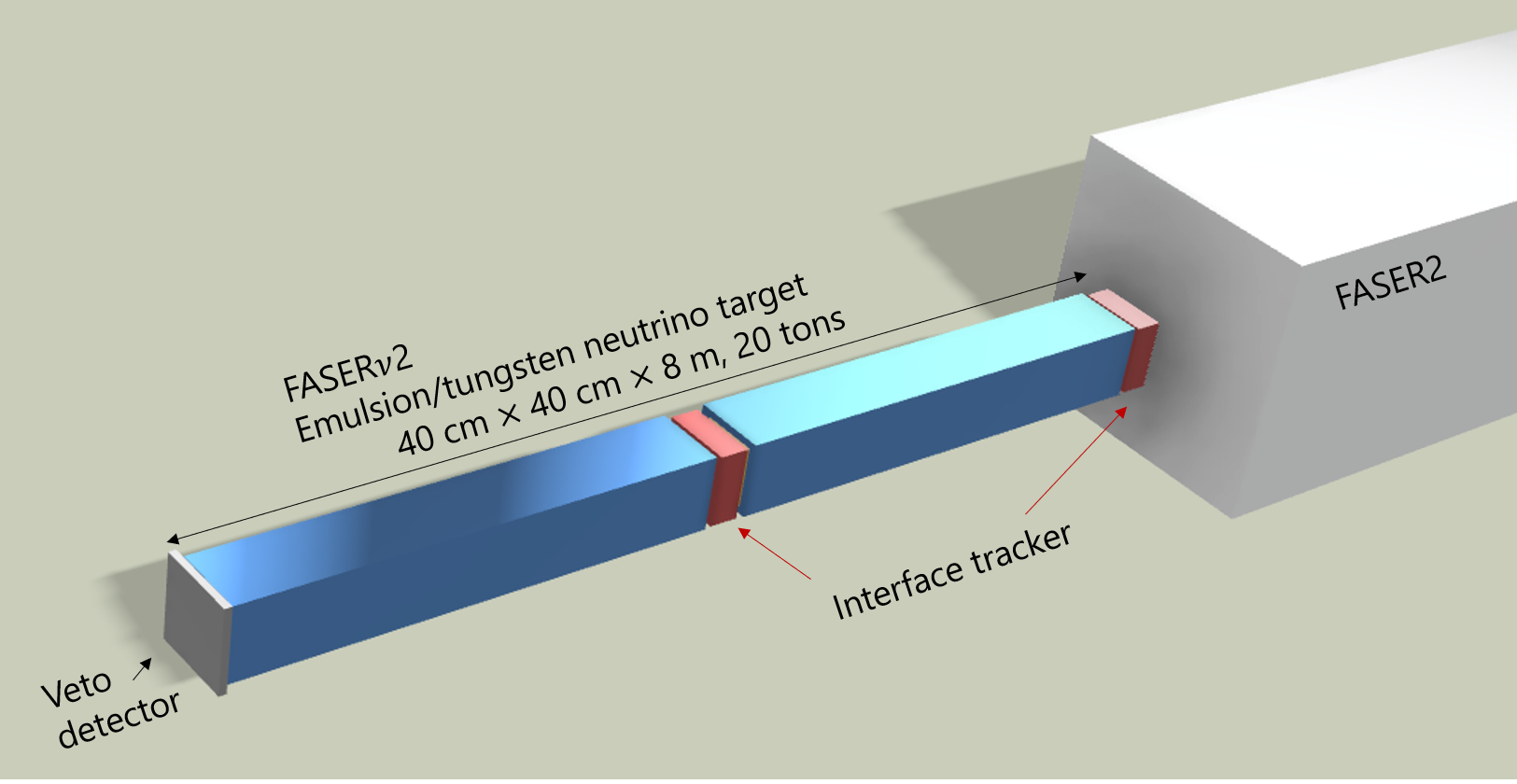}
\caption{Conceptual design of the FASER$\nu$2 detector~\cite{Anchordoqui:2021ghd}.}
\label{fig:FASERnu2}
\end{figure}

Analyses of the data collected in the emulsion modules will be based on readout of the full emulsion volume by the Hyper Track Selector (HTS) system~\cite{Yoshimoto:2017ufm}. The readout speed of the HTS system is 0.45~m$^2$/hour/layer. Currently, an upgraded system HTS2, which will be about 5 times faster, is under commissioning and a further upgraded system HTS3 with about 10~m$^2$/hour/layer is under development. 
The total emulsion film surface to be analyzed in FASER$\nu$2 is $\sim$530 m$^2$/year implying a readout time of $\sim$2400 hours/year with HTS or $\sim$420 hours/year with HTS2. It will be possible to finish analysing the data taken in each year within a year using either of the above systems.

Reconstruction of the emulsion data will make possible the localization of neutrino interaction vertices, the identification of muons, the measurement of muon and hadron momenta by the multiple Coulomb scattering method, and the energy measurement of electromagnetic showers. In addition, by conducting a global analysis that ties together information from FASER$\nu$2 with the FASER2 spectrometer via the interface detectors, the charges of muons will be identified. 
Figure~\ref{fig:FASERnu_sim_nutau} shows event displays of a simulated $\nu_{\tau}$ event in the FASERnu geometry. Similar events are expected in FASER$\nu$2.

\begin{figure}[tbhp]
\centering
\includegraphics[width=0.8\linewidth]{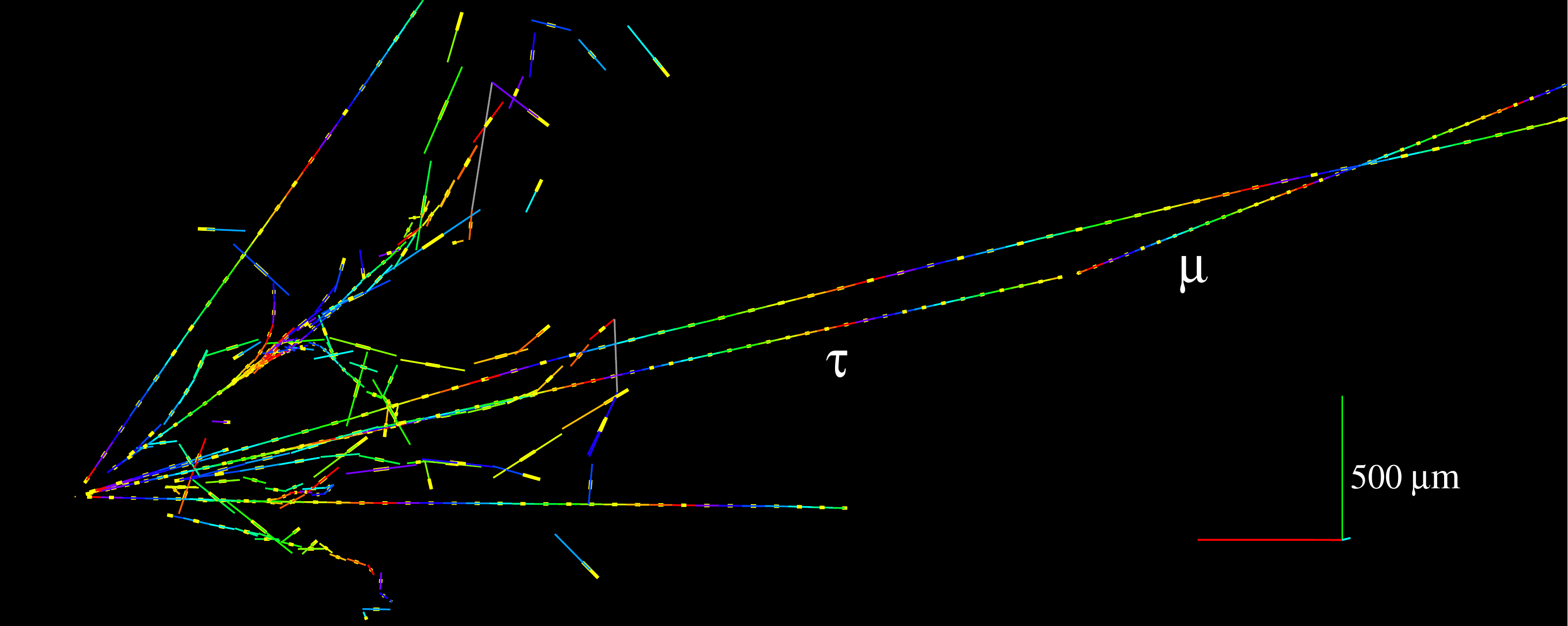}
\caption{Event display of a simulated $\nu_{\tau}$ event in the FASER$\nu$ emulsion detector~\cite{Feng:2022inv}.
}
\label{fig:FASERnu_sim_nutau}
\end{figure}

\begin{table}[ht]
\centering
\begin{tabular}{l|c|c|c}
\hline
\             & $\nu_e$+$\bar{\nu}_e$ CC  & $\nu_\mu$+$\bar{\nu}_\mu$ CC  & $\nu_\tau$+$\bar{\nu}_\tau$ CC  \\
\hline 
SIBYLL   & 178k &  943k & 2.3k \\ \hline
DPMJET   & 668k & 1400k & 20k  \\ \hline
\end{tabular}
\caption{The expected number of neutrino interactions in FASER$\nu$2~\cite{Anchordoqui:2021ghd} obtained using two different event generators, SIBYLL 2.3d and DPMJET 3.2017.
}
\label{table:event_rate}
\end{table}

In the HL-LHC, given the 20 times luminosity and 20 times target mass of FASER$\nu$, FASER$\nu$2 will collect two orders of magnitude higher statistics than FASER$\nu$, allowing precision measurements of neutrino properties for all three flavors. $\sim$2.3k ($\sim$20k) $\nu_{\tau}$ charged-current neutrino interactions are expected using the event generator \texttt{Sibyll~2.3d} (\texttt{DPMJET~3.2017}) as shown in Table~\ref{table:event_rate}. As for the uncertainty on tau neutrino flux, which is poorly constrained by the other experiments, it can be studied by measuring $\nu_e$ interactions in FASERnu2. $\sim$178k ($\sim$668k) $\nu_e$ charged-current neutrino interactions are expected in FASERnu2 using \texttt{Sibyll~2.3d} (\texttt{DPMJET~3.2017})~\cite{Anchordoqui:2021ghd}. Electron neutrinos at high energies above $\sim$500 GeV, which mainly originate from charm decays, can provide the data on forward charm production. The major remaining uncertainty could be a few 10\% level due to the dependency on the charm species, which might be constrained by the other experiments.

\subsubsection{AdvSND at the FPF}

The Advanced SND project is meant to extend the physics case of the SND@LHC experiment~\cite{Ahdida:2750060}. It will consist of two detectors: one placed in the same $\eta$ region as  SND@LHC, i.e.~$7.2 < \eta < 8.4$, hereafter called FAR, and the other one in the region $4 < \eta < 5$, hereafter denoted NEAR.  We review in the first part of this section the way the physics case would be extended, while in the second part we describe the detector design and layout. These two detectors are meant to operate during the Run 4 of the LHC. The FAR detector would nicely fit in  the Forward Physics Facility. The other one, given the higher average angle, would have to be placed more upstream to get a sizeable azimuth angle coverage. Note that the extension of the physics case covered here is related to neutrinos and, in particular, to tau neutrinos.   

\begin{table}[hbtp]
\centering
\begin{tabular}{|c | c c | c c|}
\hline
 \multicolumn{5}{|c|}{AdvSND - FAR} \\
\hline
       & \multicolumn{2}{c|}{$\nu$ in acceptance} & \multicolumn{2}{c|}{CC DIS} \\
Flavor &  hardQCD: $c\overline{c}$& hardQCD: $b\overline{b}$  &hardQCD: $c\overline{c}$& hardQCD: $b\overline{b}$  \\
\hline
$\nu_\mu$ +   $\bar{\nu}_\mu$   & $6.3 \times 10^{12}$ & $1.5 \times 10^{11}$ &  $1.2 \times 10^{4}$ &  200 \\
$\nu_e$  +  $\bar{\nu}_e$ & $6.7       \times 10^{12}$ & $1.7 \times 10^{11}$ & $1.2 \times 10^{4}$  & 220 \\
$\nu_\tau$  +  $\bar{\nu}_\tau$   & $7.1 \times 10^{11}$ & $4.7 \times 10^{10}$  &  880 & 40 \\
\hline
 Tot &\multicolumn{2}{c|}{ $1.4 \times 10^{13}$ }& \multicolumn{2}{c|}{ $2.5 \times 10^{4}$ } \\
\hline
\end{tabular}
\label{tab:NuFlux_Far}
\caption{Number of neutrinos passing through the far detector of Advanced SND@LHC and charged-current neutrino interactions in the detector target, assuming 3000\,fb$^{-1}$, as estimated with Pythia8 generator.}
 \end{table}

{\bf Neutrino cross-section measurements.} 

\begin{figure}[ht]
    \centering
    \includegraphics[width=0.98\textwidth]{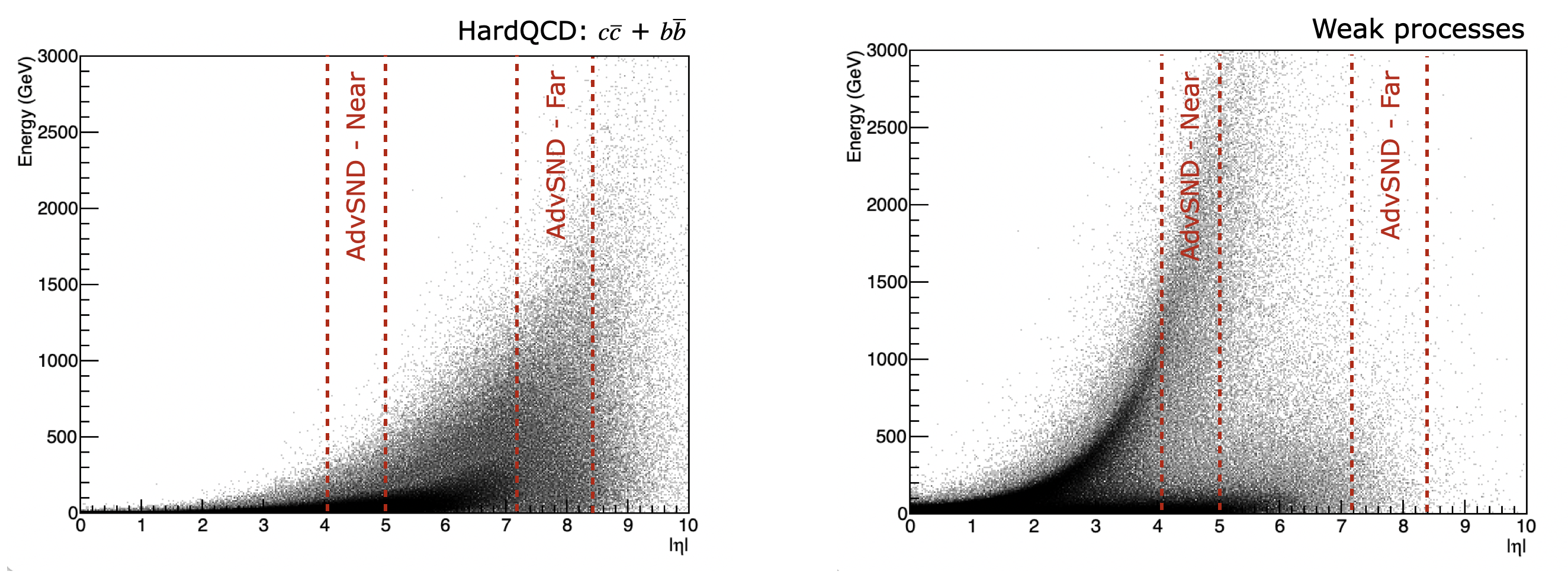}
    \caption{Scatter plot of neutrino energy versus pseudorapidity $\eta$ in $b$, $c$   (left) and $W$ (right) decays. All neutrino flavors are included~\cite{Beni:2019gxv}. The AdvSND regions are highlighted.}
    \label{fig:eta-energy}
\end{figure}

Figure~\ref{fig:eta-energy} shows the scatter plots of neutrino energy versus $\eta$ for neutrinos originated from $b$ and $c$ and from $W$ decays. Neutrinos from the leptonic W decays are seen to be kinematically well separated~\cite{Beni:2019gxv}. Note that LHCb has measured charm, beauty and $W$ production cross-sections in the $2 \div 5$~$\eta$ range: 1.5 nbarn for $W$, 144 $\mu$barn for beauty and 8.6 mbarn for charm. Note that the $W$ measurement was carried out at 7 TeV while the other two were done at 13 TeV. 
Accounting for all that, and considering the case of tau neutrinos which shows a low branching ratio in charm decays ($c \rightarrow \nu_\tau \sim 5 \times 10^{-3}$), we expect a factor $10^5$ more charm-induced than $W$ and $Z$-induced $\nu_\tau$s. The role of $W$ and $Z$ decays is therefore marginal in this context and we focus on charm and beauty in the following. 

Figure~\ref{fig:energy_all} shows the neutrino energy spectra for the two $\eta$ regions, separately for the different neutrino parents. It can be noticed that the energy spectrum of charm and beauty-induced neutrinos is much softer in the NEAR location, as expected. 
\begin{figure}[hbtp]
\centering
\includegraphics[width=0.98\columnwidth]{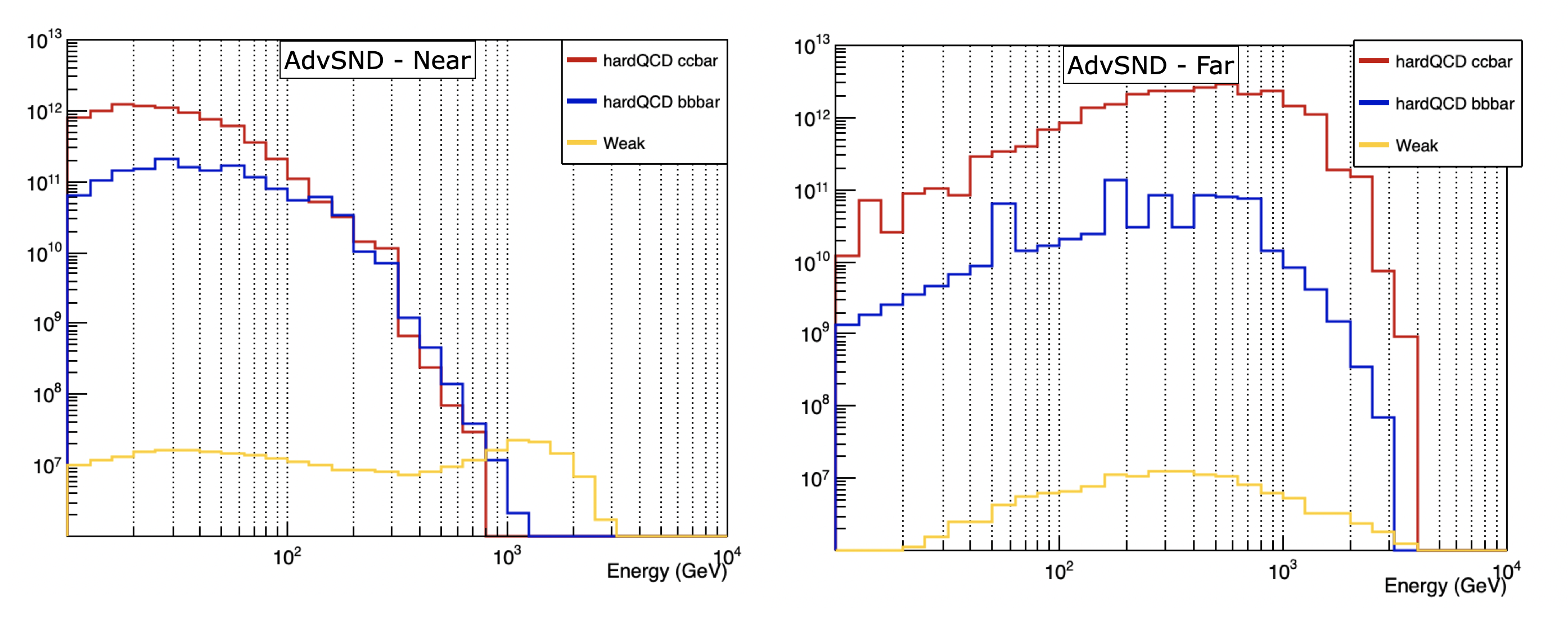}
\caption{Neutrino energy spectrum for three different sources: charm, beauty and $W$, $Z$ bosons. The Near (left) and Far (right) locations are considered. Figure from \cite{Feng:2022inv}.}
\label{fig:energy_all}
\end{figure}

The large uncertainty on the charm-induced neutrino flux in the large $\eta$ region prevents SND@LHC making a neutrino cross-section measurement. AdvSND will instead be able to perform this measurement with the NEAR detector, since the neutrino flux from charm and beauty in the $4.0 < \eta < 4.5$ region is very reliable, given the measurements performed by LHCb~\cite{Aaij:2015bpa}. This will lead to a neutrino cross-section measurement with very small systematic uncertainties of all three neutrino flavors, including tau neutrinos. The expected number of events in the NEAR detector is given in   Table~\ref{tab:NuFlux_Near}. Notice that, the lower average energy of neutrinos in the NEAR location results in a lower neutrino cross-section, which explains the differences between the neutrino yields in the two detectors, in spite of the similar flux. 

\begin{table}[hbtp]
\centering
\begin{tabular}{|c | c c | c c|}
\hline
 \multicolumn{5}{|c|}{AdvSND - NEAR} \\
\hline
       & \multicolumn{2}{c|}{$\nu$ in acceptance} & \multicolumn{2}{c|}{CC DIS} \\
Flavor &  hardQCD: $c\overline{c}$& hardQCD: $b\overline{b}$  &hardQCD: $c\overline{c}$& hardQCD: $b\overline{b}$  \\
\hline
$\nu_\mu$ +   $\bar{\nu}_\mu$   & $2.1 \times 10^{12}$ & $3.3 \times 10^{11}$ & 980 &  200 \\
$\nu_e$  +  $\bar{\nu}_e$ & $2.2       \times 10^{12}$ & $3.3 \times 10^{11}$ & 1000  & 200 \\
$\nu_\tau$  +  $\bar{\nu}_\tau$   & $2.7 \times 10^{11}$ & $1.4 \times 10^{11}$  &  80 & 50 \\
\hline
 Tot &\multicolumn{2}{c|}{ $5.4 \times 10^{12}$ }& \multicolumn{2}{c|}{ $2.5 \times 10^{3}$ } \\
\hline
 \end{tabular}
  \caption{Number of neutrinos passing through the near detector of Advanced SND@LHC and charged-current neutrino interactions in the detector target, assuming 3000\,fb$^{-1}$, as estimated with Pythia8 generator.}
  \label{tab:NuFlux_Near}
 \end{table}

Thus, one expects the leading uncertainty to be the statistical one: a few percent for electron and muon neutrinos and about 10\% for tau neutrinos as one can derive from Table~\ref{tab:NuFlux_Near}. Notice that the yield of muon neutrinos from $\pi$ and $K$ decays is not included in this table.

{\bf Lepton flavor universality with tau neutrino interactions.}
In the $7. 2 < \eta < 8.4$ region, electron and tau neutrinos come essentially only from charm decays. Therefore, the uncertainty on the flux cancels out in the ratio which can then be used to test lepton flavor universality with neutrino interactions. The corresponding measurement by SND@LHC is dominated by a 30\% statistical uncertainty due to the poor $\nu_\tau$ statistics. AdvSND will reduce the statistical uncertainty down to less than 5\%, see Table~\ref{tab:NuFlux_Far}. At this point, the systematic uncertainty due to the charm quark hadronization fraction into $D_s$ mesons, $f_{D_s}$, would be leading. This would turn into a measurement of the lepton flavor universality at the 20\% level.

More constraints on this ratio could come from the NEAR detector where all charmed hadron species, including $D_s$ have been identified by the LHCb Collaboration. Given the expected number of electron and tau neutrino interactions, as reported in Table~\ref{tab:NuFlux_Near}, lepton flavor universality with electron and tau neutrinos could be tested with an accuracy of 10\%. 

{\bf Detector layout.}
Both detectors will be made of three elements. The upstream one is the target region for  the vertex reconstruction and the electromagnetic energy measurement with a calorimetric approach. It will be followed downstream by a hadronic calorimeter and a muon identification system. The third and most downstream element will be a magnet for the muon charge and momentum measurement, thus allowing for neutrino/anti-neutrino separation for muon neutrinos and for tau neutrinos in the muonic decay channel of the $\tau$ lepton.

The target will be made of thin sensitive layers interleaved with tungsten plates, for a total mass of $\sim$ 5 tons. The use of nuclear emulsion at the HL-LHC is prohibitive due to the very high intensity that would make the replacement rate of the target incompatible with technical stops. The Collaboration is investigating the use of compact electronic trackers with high spatial resolution  fulfilling both tasks of  vertex reconstruction with micrometre accuracy and electromagnetic energy measurement. The hadronic calorimeter and the muon identification system will be  about 10~$\lambda$ which will bring the average length of the hadronic calorimeter to about 12~$\lambda$, thus improving the muon identification efficiency and energy resolution.
The magnetic field strength is assumed to be about 1 T  over about 2~m length. A schematic view of the detector is reported in Figure~\ref{fig:advsnd}.

 \begin{figure}[h]
    \centering
    \includegraphics[width=0.8\textwidth]{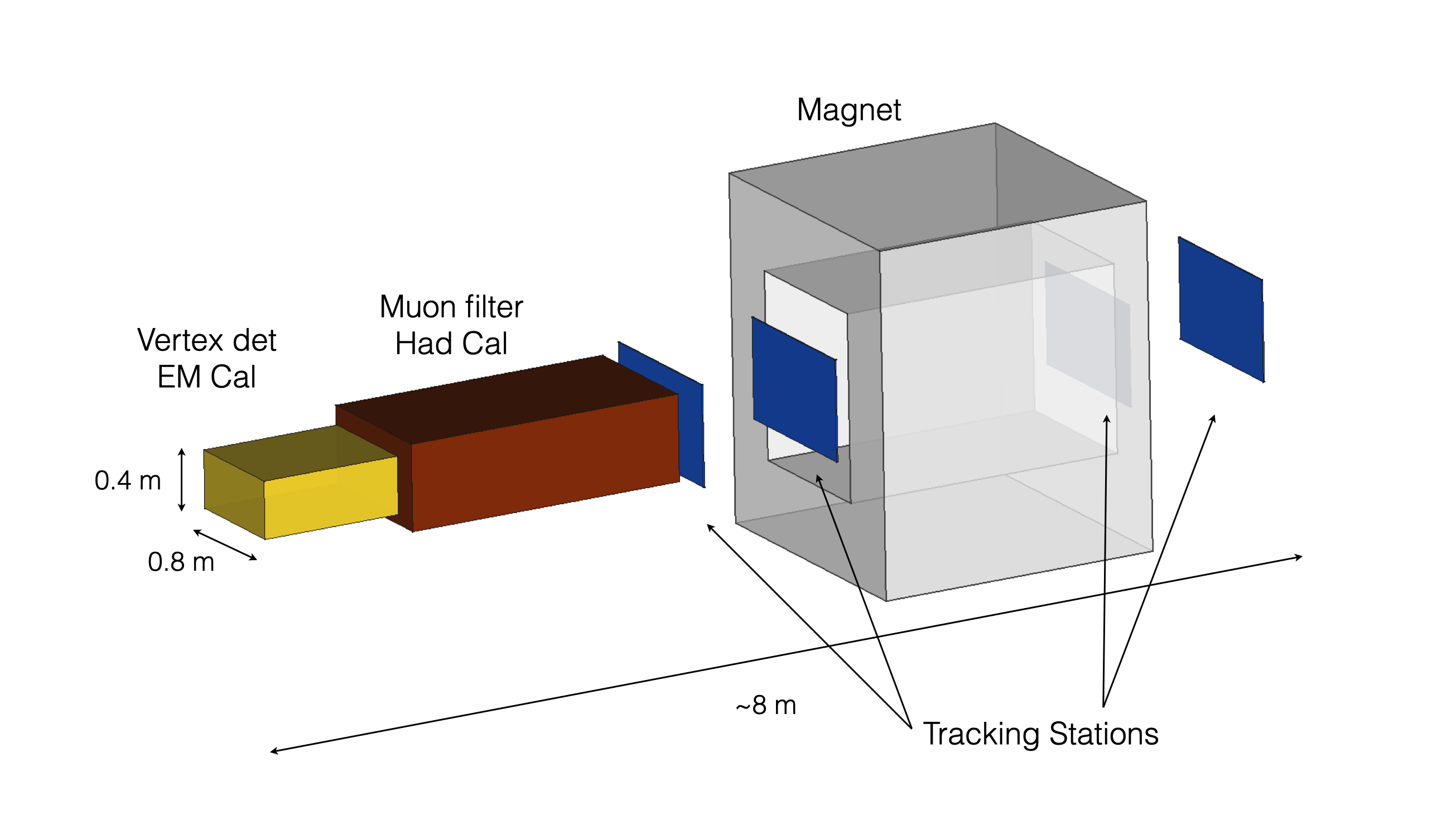}
    \caption{Layout of the AdvSND detector.}
    \label{fig:advsnd}
\end{figure}

The magnet is a key element in the detector design because it will allow to measure the leptonic number of muon neutrinos and of tau neutrinos when the tau lepton decays into a muon.  
The layout of a spectrometer measuring the bending angle of a track is shown in Figure~\ref{fig:magnet_scheme} with all the relevant parameters. We describe in the following the main parameters of the spectrometer for AdvSND. 

\begin{figure}[h]
    \centering
    \includegraphics[width=0.6\textwidth]{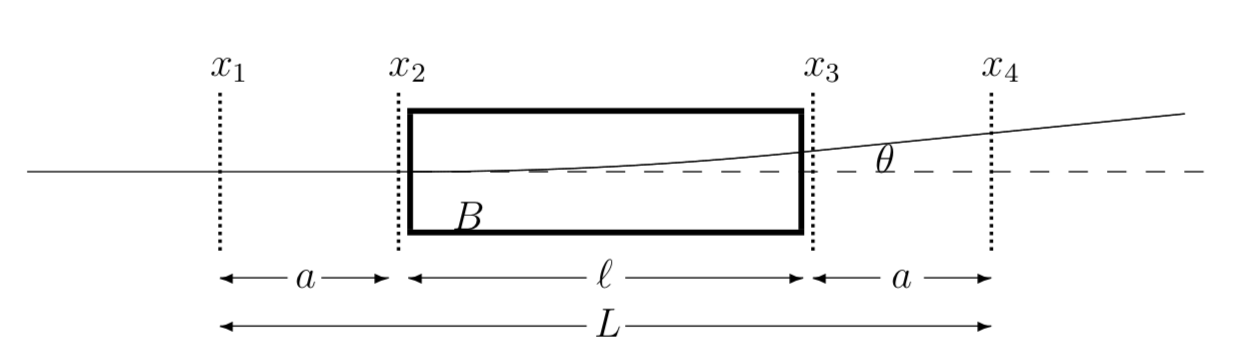}
    \caption{Schematic drawing of a magnetic spectrometer measuring the bending angle of a particle track.}
    \label{fig:magnet_scheme}
\end{figure}

For a given total length $L$ of the spectrometer the choice of the length $\ell$ of the magnet and of the lever arm $a$ between tracking stations, which results in the best momentum resolution, is defined by $a = \frac{L}{4} = \frac{\ell}{2}$.
In this configuration, the maximum momentum, up to which a muon charge assignment is possible, is obtained: 
\begin{equation} p_{max} = \frac{\mbox{e} B \ell a}{8 \epsilon}.
\end{equation}
If we use $B = 1$\,T for the magnetic field,
$\epsilon = 100\, \mu$m as the spatial resolution of the tracking chambers,
$\ell = 2\,$m and
$a = 1\,$m,  the spectrometer allows for a charge assignment up to 750 GeV/$c$,
thus covering 95\% of the momentum spectrum of CC $\nu_\mu$ interactions. 
The length of the spectrometer is 4~m. 

Table~\ref{tab:geometry} summarizes the main parameters of the two locations and the corresponding detectors. 
\begin{table}[hbtp]
\centering
\begin{tabular}{|l | c | c |}
\hline
&  AdvSND - NEAR & AdvSND - FAR \\
\hline
$\eta$ & [4.0, 5.0] &  [7.2, 8.4] \\ 
target mass (ton) & 5 & 5  \\
front surface (cm$^2$) & $120 \times 120$ & $100 \times 40$ \\
distance from IP (m) & 55 & 480 \\
\hline
\end{tabular}
\caption{Parameters of the two detectors in NEAR and FAR location.}
\label{tab:geometry}
\end{table}

\subsubsection{FLArE at the FPF}

A liquid argon time projection chamber (LArTPC) is considered for the suite of detectors for the FPF \cite{Anchordoqui:2021ghd}. For the following we will assume that this detector is located in a hall $\sim$ 600 meters from the interation point.   Such a detector offers the possibility to precisely determine particle identification, track angle, and kinetic energy over a large dynamic range in energies.  

A LArTPC is well motivated by the requirements of neutrino detection and the light DM search~\cite{Batell:2021blf,Batell:2021aja}. 
In particular the TPC is an excellent choice for detection and measurement of energetic electromagnetic showers.  Single muon tracks as well as showers of hadronic tracks also benefit from the superb spatial and charge resolution of this detector. The detector has no insensitive mass and therefore the energy loss and scattering can be measured along a long track. This capability leads to  superb particle identification at momentum of $\sim$ 1 GeV and also to excellent momentum resolution for high energy muons.  The kinematic resolutions in angle and momentum and how they affect various backgrounds for neutrino physics at the TeV scale needs further study. 

The detector is expected to measure millions of neutrino interactions, including tau neutrinos. The detector should have sufficient capability to measure these very high energy ($>100~ $GeV ) events, so that the cross section for each flavor can be measured. Identification of tau neutrinos with low backgrounds needs detailed simulations and reconstruction studies. As an approximate estimate we expect to see about 50 high energy neutrino events per ton per fb$^{-1}$ of collisions; this is approximately the daily luminosity during the high-luminosity running of the LHC.  The majority of this flux will be muon neutrinos with electron neutrinos forming about 1/5 of the event rate.  The tau neutrino rate is expected to be $\sim 0.1$ event/ton/fb$^{-1}$ with a very large uncertainty due to QCD modeling in the forward direction.  
The high energy electron neutrino and the tau neutrino flux come from charm meson decays in the forward region and therefore careful measurement of both of these event types has broad implications for particle physics as described in other parts of this report. 

Table \ref{lartab} summarizes the main parameters of a LArTPC for the FPF. A detector with a fiducial mass of approximately 10~tonnes of liquid argon is envisioned. We are also considering this same detector with a filling of liquid krypton.  
For 3 ab$^{-1} $, such a detector will collect hundreds of thousands of muon neutrino/antineutrino CC events, about a hundred thousand electron neutrino events, and thousands of tau neutrino events. These numbers have large uncertainties due to the poorly understood production cross section in the forward region~\cite{Bai:2020ukz}. It is also important to note that this flux of events will have the same time structure as the LHC accelerator with a bunch spacing of 25 ns. At the same time, muons from interactions at the IP will produce a background flux of about $\sim 1$~muon/cm$^2$/s at the nominal maximum luminosity of $5\times 10^{34}$~cm$^{-2}$~\text{s}$^{-1}$ at the HL-LHC.
If the TPC can be operated with liquid krypton several advantages are expected. The radiation length of LKr (4.7 mc) is much shorted than LAr (14 cm) leading to much more compact electromagnetic showers. This performance naturally leads to much higher event containment for neutrino events. The higher density of LKr should also yield a higher event rate. The overall increase of useful event rate is expected to be almost a factor of $2$ at energies above 1 TeV.  Detailed simulations of event reconstruction need to be performed, but the better resolution from LKr is expected to lead to much better performance for tau neutrinos.

\begin{table} 
\begin{center}
\begin{tabular}{ p{0.35\linewidth} || p{0.3\linewidth} | p{0.35\linewidth} }
 \hline
 \hline
 & Value & Remarks \\
 \hline 
 Detector length &  7~m & Not including cryostat \\ 
 TPC drift length &  $0.5$ ~m & 2 TPC volumes with HV cathode in center \\ 
 TPC height &  1.3~m &  \\ 
 Total LAr mass & $\sim 16$~tonnes & Volume in the cryostat \\ 
 Total LKr mass & $\sim 27.5$~tonnes &  As an option \\ 
 Fiducial mass LAr/LKr & $10/17$~tons &  \\ 
 Charge Readout & wires or pixels & Hybrid approach is possible \\ 
 Light readout & SiPM array & Needed for neutrino trigger \\ 
 Background muon rate & $\sim 1/$cm$^2/$s & Maximum luminosity of $5\times10^{34}/$cm$^2/$s \\
 Neutrino event rate & $\sim 50/$ton/fb$^{-1}$ & For all flavors of neutrinos \\ 
 Cryostat type & Membrane 0.5 m & Thickness of membrane \\
 Heat loss & $\sim$ 300 W &  \\  
 \hline
 \hline
\end{tabular}
\end{center}
\caption{Detector parameters for a LArTPC for the FPF. The top part of the table shows the nominal geometric parameters for a detector to be considered for the FPF, and the bottom part shows the basic properties of a LArTPC. \label{lartab}}  
\end{table} 

The nominal configuration for the LArTPC detector would include a central cathode operating at a large high voltage and two anode planes on two sides of the detector parallel to the beam from the ATLAS IP. The electric field between the cathode and the anode will be at $\sim500$~V/cm, providing a drift field for ionization electrons; the drift time for a 0.5~m-long drift will be about 0.3~ms. For a detector with approximate cross section of $1$~m$^2$, we therefore expect about 3 muon tracks to be within a single drift time. Neutrino and dark matter events must be selected out of these overlaying background particle trajectories. For the TPC, a readout using wires or pixels is possible\cite{Qian:2018qbv}.   A readout of the scintillation light is crucial to allow the measurement of the distance along the drift. It is also important for the selection of events that originate in the detector (such as a neutrino or a dark matter event), as well as generating the trigger necessary for acquiring the data.  Neutrino events need to be identified at the trigger level as events with tracks that originate from a common vertex within the detector volume. 

The LArTPC is expected to be installed in a membrane cryostat with passive insulation and with inner dimensions of 1.3~m 
$\times$ 1.2~m $\times$ 7.2~m. Following the example of ProtoDUNE~\cite{DUNE:2017pqt,DUNE:2021hwx}, the membrane cryostat technology allows the cryostat to be constructed underground. The insulation, being passive, ensures reliable and safe long-term performance. The cryogenic system must re-condense the boil-off, keeping the ullage absolute pressure stable to better than 1 mbar, and purify the LAr bath. A standard approach is to re-condense the argon with a heat exchanger with liquid nitrogen. A LAr flow of 500~kg/h through the purification circuit is considered sufficient to reach and maintain the required LAr purity. 

The total heat input due to the cryostat and the cryogenics system is estimated to be of the order of few kW.   A Turbo-Brayton ($\sim 8$m$ \times 1.6$m$ \times 2.7$m) TBF-80 unit from Air Liquid installed in the vicinity of the cryostat provides approximately 10~kW cooling power from $\approx$100~kW electrical power and 5 kg/s of water at ambient temperature. The design and technology for the cryosystem is well understood because of the experience from ProtoDUNE at CERN.  If Lkr is considered as a fill, then the requirements for the cryosystem need to be further examined in detail.  

A LAr detector could be an excellent choice for the detection of neutrino and dark matter events  at the 10-ton fiducial mass scale.  Further simulation work is needed to understand event reconstruction and background rejection, especially for tau neutrino events.  For detector design, in particular, simulation work is needed to understand neutrino event  containment and energy resolution in a 7 m-long detector.  Study of kinematic resolution in the case of wire readout versus pixel readout is needed. And finally, the design and performance of the photon detector system needs to be investigated and demonstrated by R\&D.   Lastly, we are considering liquid krypton as an option for the detector. Such a LkrTPC would have remarkable resolution for electromagnetic showers and the event containment is expected to excellent.  Looking further to the future, the addition of magnetic field and momentum measurement either with a downstream magnet or as part of the TPC needs to be explored.  
\pagebreak
\section{High Energy Astrophysical Experimental Probes}
\label{sec:he experiments}
With existing massive water Cherenkov and radio detectors, the era of high energy astrophysical neutrinos is upon us.
Due to the high neutrino energies that these experiments are sensitive to, they are excellent tau neutrino experiments and, in fact, many are \emph{only} sensitive to tau neutrinos.
Additionally, there are many proposed and planned astrophysical neutrino experiments looking to push to even higher energies. Astrophysical neutrinos have been observed in the TeV to PeV energy range and are expected to be observable at energies several orders of magnitude higher still. These energetic neutrinos act as probes of new physics models at an energy scale well beyond the center-of-mass energies of current terrestrial experiments. Moreover, because they are predominantly extragalactic, individually small perturbations can accumulate to yield collectively large effects. Flavor ratios, and in particular, observations of \textit{tau} neutrinos are an important observable for constraining new physics. The potential impacts of new physics models on important observables are summarized in Fig.~\ref{fig:HEnusNewPhysics}.

In this section we will first discuss the various tau neutrino physics concepts that can be probed in high-energy astrophysical neutrino experiments, and then review the existing and proposed plans.

\begin{figure}[tbh]
	    \centering
	    \includegraphics[width=120mm]{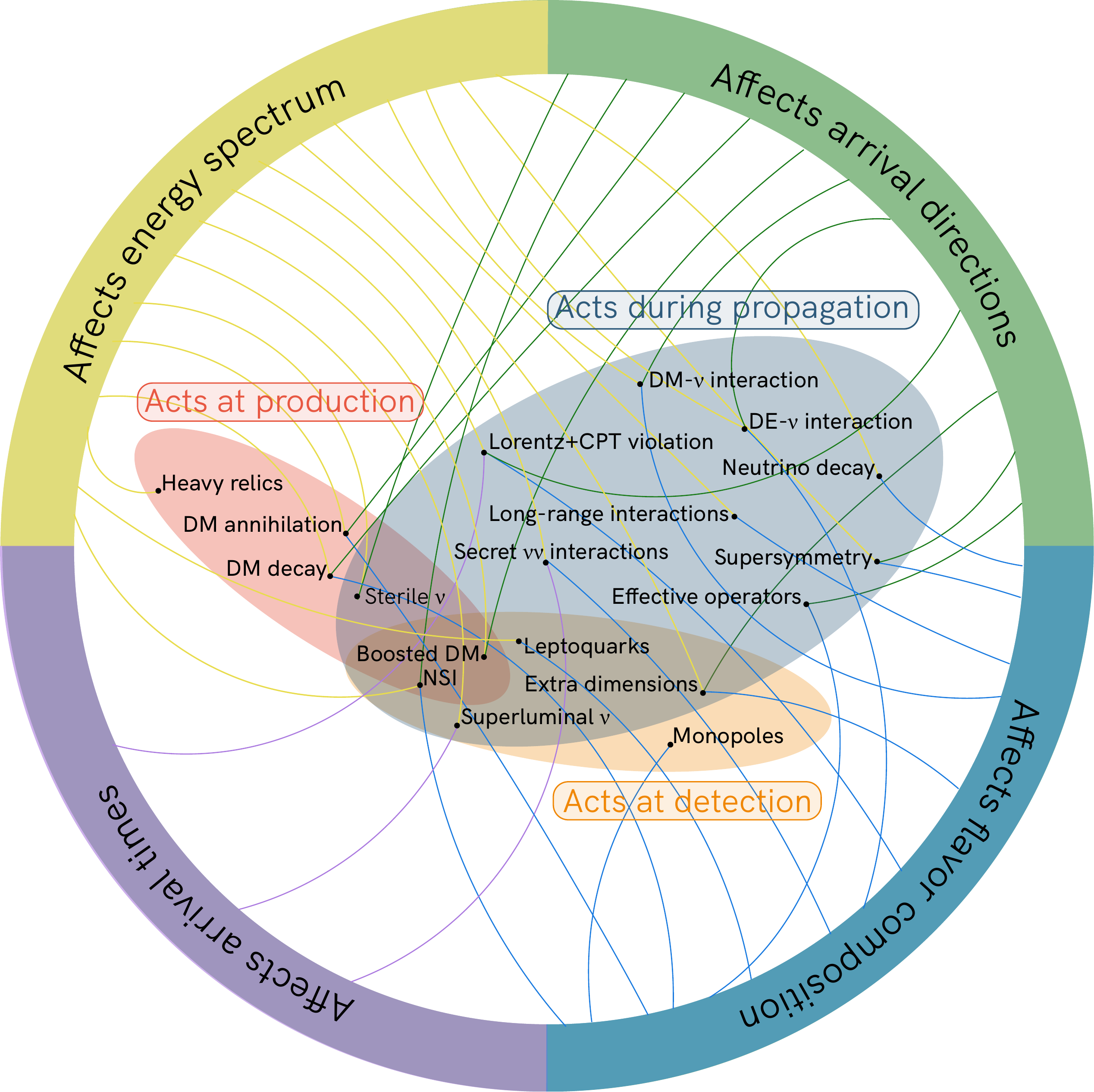}
	    \caption{Summary of models of new physics and their impact on the neutrino observables at energies above the TeV scale. Representative models are grouped by when they act (at production, propagation, or detection) and some of the observables they can impact like the energy spectrum, arrival times and directions, and the flavor ratios. Reproduced from \cite{Arguelles:2019rbn}.}
	    \label{fig:HEnusNewPhysics}
	    \end{figure}

\subsection{Cross Section} 

\begin{figure}[t]
  \centering
  \includegraphics[width=\textwidth]{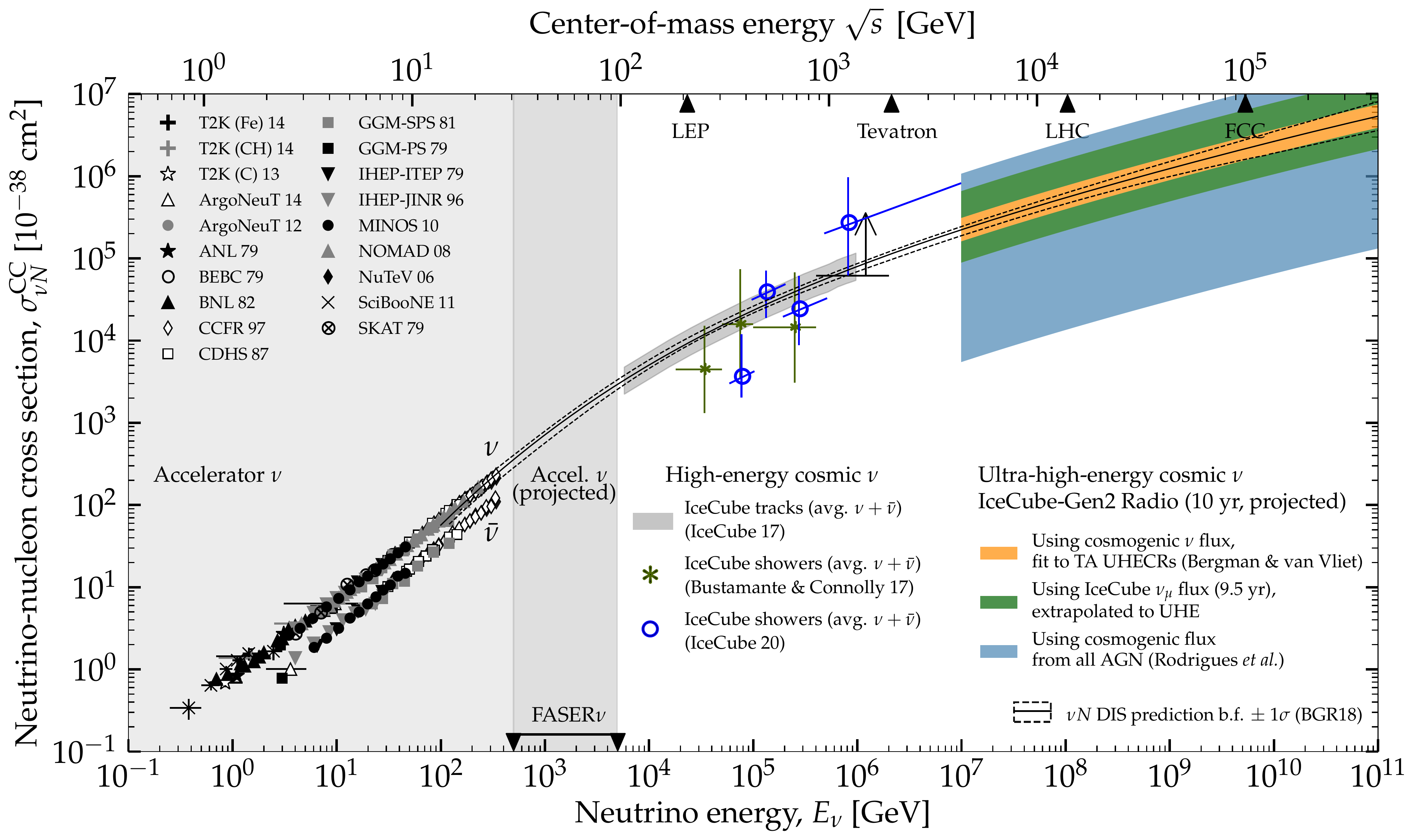}
  \caption{Neutrino-nucleon cross section measurements, compared to deep-inelastic-scattering (DIS) cross section prediction from Ref.~\cite{Bertone:2018dse} (BGR18).  Measurements in the TeV--PeV range are based on IceCube showers~\cite{Bustamante:2017xuy, IceCube:2020rnc} and tracks~\cite{IceCube:2017roe}.  Projected measurements at energies above 100~PeV~\cite{ValeraUHECrossSection} envision 10~years of operation of the radio component of IceCube-Gen2, assuming a resolution in energy of 10\% and a resolution in zenith angle of 2$^{\circ}$.  Since the flux at these energies remains undiscovered, projections for the measurement of the cross section are for different flux predictions; see also \cite{Denton:2020jft}.  Figure reproduced from Ref.~\cite{ValeraUHECrossSection}.}
  \label{fig:cross_section_uhe}
\end{figure}

The neutrino-nucleon interaction cross section, $\sigma_{\nu N}$, is an important observable, sensitive to both Standard Model physics (parton distributions) and beyond-the-Standard Model (BSM) scenarios; many BSM models predict increased new interactions which increase the cross-section, often drastically~\cite{Kusenko:2001gj,Anchordoqui:2001cg,Anchordoqui:2018qom,Klein:2019nbu}. Figure~\ref{fig:cross_section_uhe} shows the neutrino-nucleon interaction cross section, $\sigma_{\nu N}$, measured across GeV to PeV energies, its projected measurements at hundreds of PeV, and compared to a recent Standard Model prediction~\cite{Bertone:2018dse}. Neutrino interaction cross sections have been measured at accelerators, up to an energy of about 370 GeV, typically with a precision of $5 - 10\%$~\cite{Formaggio:2012cpf}.  Deep Inelastic Scattering (DIS) dominates the cross section at the higher-energy accelerator experiments, while quasi-elastic and resonant interactions become important at energies below about 30 GeV.  Unfortunately, these experiments were unable to detect $\nu_\tau$, so could only measure $\nu_\mu$ and $\nu_e$ cross-sections.  However, in the Standard Model, the high-energy cross sections for the three neutrino flavors are very similar, so it is not critical to measure each flavor separately.  This also holds true in most BSM models. One big exception are leptoquarks, where different leptoquarks might couple differently to different lepton generations, so the $\nu_\tau$ cross sections might be quite different from their $\nu_\mu$ and $\nu_e$ counterparts.

In the TeV range, looking forward, the FASER$\nu$ experiment at CERN is expected to measure $\sigma_{\nu N}$ for all three flavors~\cite{FASER:2020gpr}.  The expected precision is in the 10\% to 40\% range, depending on neutrino flavor and energy, with $\nu_\tau$ precision at the less-precise end of that range. The High-Luminosity upgrade of the Large Hadron Collider (HL-LHC) will dramatically increase the rate of collisions, and thus FASER$\nu$2 at the CERN Forward Physics Facility will significantly reduce $\sigma_{\nu N}$ uncertainties~\cite{Anchordoqui:2021ghd}.

At higher energies, studies use astrophysical and atmospheric neutrinos, in detectors like IceCube, KM3NeT, and Baikal-GVD. There has been considerable progress since the last Snowmass study \cite{Klein:2013xoa}. Two approaches can be used.  Either one can assume that the flux is known \cite{IceCube:2021jhz}, and infer the cross-section via the number of observed events, or one can measure absorption in the Earth.  The second method avoids uncertainties over the flux normalization, but only works at energies above 5-10 TeV, where absorption becomes significant.   So far, there have been three measurements, one using mostly $\nu_\mu$ \cite{IceCube:2017roe}, and the other involving starting events \cite{Bustamante:2017xuy, IceCube:2020rnc}.  

The former analysis~\cite{IceCube:2017roe} used about 10,000 through-going muons and achieved a 40\% precision, in the energy range from 6.3 to 980 TeV averaged over all flavors.  These events offer excellent angular resolution, but very poor energy resolution, due to the broad distributions in distance from the interaction to the detector (and consequent muon energy loss) and in the fraction of the $\nu$ energy transferred to the muon.

The sensitive energy range was determined by, at the lower end, the need for significant absorption, and, at the higher end, by the limited statistics.  An in-progress analysis of 8-years of through-going muons will have 10 times the statistical power of the published study, and will divide the energy spectrum into three bins \cite{IceCube:2021keu}.  

The first example of the second type used IceCube public data on 58 contained showers \cite{Bustamante:2017xuy}, and found cross sections (or, at the high-energy end, lower limits) in four energy bins from 18 TeV to 2 PeV. The second analysis, by IceCube, used 60 events with deposited energy above 60 TeV~\cite{IceCube:2020rnc}.  This analysis divided the events into four energy ranges, covering 60 TeV to 10 PeV. This sample included both cascades and starting tracks.   

These analyses did not explicitly consider $\nu_\tau$, although both of the event samples likely included some $\nu_\tau$ interactions, so would have been sensitive to a large increase in the $\nu_\tau$ cross-section. A direct way to test non-perturbative  physics is by measuring the ratio of the neutral-to-charge current cross sections~\cite{Kusenko:2001gj,Anchordoqui:2001cg,Anchordoqui:2018qom}. A first measurement of such a ratio using IceCube data shows no signs of new physics~\cite{Anchordoqui:2019ufu}.

Future optical Cherenkov measurements are likely to achieve increased precision, but will not expand their energy reach by very much.  The lower energy cutoff is determined by the systematic uncertainties, which will limit how well we can measure very small absorption.  The higher energy cutoff will be limited by the low number of detected events.   Because of the low selection efficiency, it seems unlikely that a separate $\nu_\tau$ cross-section will be measured, even at energies where astrophysical $\nu$ dominate.

Proposed future detection experiments will instrument much larger volumes.     In models where GZK neutrinos are mostly from protons, a detector volume of 100 km$^3$ of Antarctic ice is big enough to accumulate sizable samples of neutrinos with energies above $10^{17}$ eV, which are sensitive to quark distributions with Bjorken$-x$ below $10^{-4}$.  Similar active volumes are required for other detection techniques, including Earth-skimming and mountain-skimming experiments \cite{Denton:2020jft, Huang:2021mki}. 

Absorption lengths decrease with energy.  As it does, the zenith angle corresponding to an absorption-length through the Earth decreases, with a single absorption length corresponding to 10 degrees below the horizon, for $10^{16}$ eV $\nu$, decreasing to 3 degrees below at $10^{18}$ eV~\cite{Klein:2020nuk}.  So, at very high energies, most upgoing events are near the horizon \cite{Connolly:2011vc}.  These angular scales define the required acceptance. However, as these angles decrease, unbiased measurements become critical and systematic uncertainties on the zenith angle reconstruction become an important consideration; relatively small biases in zenith angle can alter the inferred cross-section.  These could come from a variety of sources, including estimates of the integrated thickness of mountains (for Earth-skimming experiments).  For air-based optical or ice-based radio-detection experiments respectively, uncertainties due to refraction in the atmosphere or the near-surface firn may be a significant concern. 

\subsection{Inelasticity}

Inelasticity is the fraction of the neutrino energy that is transferred to a hadronic target in a DIS interaction.  Inelasticity measurements nicely complement cross-section measurements.   If a new reaction contributes to the cross-section, it is unlikely to have the same inelasticity distribution as conventional CC DIS.  The inelasticity distribution is well-predicted in the Standard Model. Therefore, the inelasticity distribution of events detected at neutrino telescopes has long been envisioned as an important tool for revealing new physics~\cite{Anchordoqui:2006wc}. IceCube recently made the first measurement of $\nu_\mu$ inelasticity \cite{IceCube:2018pgc}, using 2650 starting tracks with energies above 1 TeV.  The cascade and track energy were determined separately, and the inelasticity calculated for each event.  The inelasticity distributions were in good agreement with the standard model.  The sample was also used for several different types of physics studies, including measurements of the astrophysical neutrino flux, spectral index and flavor composition, and of charm production in neutrino interactions.  The flavor sensitivity came from two sources.  One was a comparison with a companion 965 event cascade sample, which was enriched in $\nu_e$. The second was because the starting track sample include $\nu_\tau$ events with muons in the final state.  These muons carry a smaller fraction of the neutrino energy than $\mu$ from $\nu_\mu$.   The results of the fit are shown in Fig.~\ref{fig:probes_HESE-tau-flavor}.

Looking ahead, current and future optical Cherenkov detectors should be able to collect samples two to four times larger than were used by IceCube, and so should be able to make considerably more precise measurements.  The increased data will be helpful for flavor measurements, including the $\nu_\tau$.   The inelasticity approach complements the $\nu_\tau$ identification studies discussed below, with a different event sample, and a different set of systematic uncertainties. 

Inelasticity measurements are a challenging but important target for radio-detection experiments \cite{Klein:2019nbu}.  They require separate observation of the outgoing lepton and the hadronic cascade produced in neutrino interactions.  There are a couple of possible approaches, most of which rely on the fact that the Landau-Pomeranchuk-Migdal (LPM) effect lengthens electromagnetic showers, eventually, at high enough energies, leading to multiple subshowers from a single initial electron \cite{Gerhardt:2010bj}.   If one or more subshowers are observed separately from the initial hadronic shower, then it may be possible to separately infer the energy of the outgoing electron.
This measurement is most straightforward at extremely high energies ($10^{20}$~eV). However, at lower energies, where the LPM lengthening is lessened, it may be possible to separate the hadronic and electromagnetic cascades on the basis of their different radio-emission spectra, because the longer electromagnetic shower produces a radio pulse that is more concentrated near the Cherenkov angle, and, away from the Cherenkov angle, is cut off at lower frequencies than the hadronic component \cite{Alvarez-Muniz:1999qlp,Stjarnholm:2021xpj}.  At $10^{20}$ eV, the $\tau$ decay length is too large to be contained in a 100 km$^3$ detector, so the efficiency will be low, or significantly larger detectors are required.  Other approaches to these different topologies will be discussed in sec.~\ref{sec:tools}.

\subsection{Flavor Ratios}
        The flavor composition of astrophysical neutrinos carries information on the production mechanisms and environments at the cosmic sources of particle acceleration. While the standard production mechanism via pion decay produces a source flavor composition of $\nu_e:\nu_{\mu}:\nu_{\tau}=1/3:2/3:0$, the expected flavor composition on Earth assuming extragalactic sources is $\nu_e:\nu_{\mu}:\nu_{\tau}=0.30:0.36:0.34$ assuming best-fit oscillation parameters \cite{Esteban:2020cvm}. However, other source production mechanisms are possible, such as muon decay \cite{Hummer:2010ai}, muon damped \cite{Kashti:2005qa}, neutron decay \cite{Anchordoqui:2003vc} and charm decay \cite{Kachelriess:2006ksy}, each resulting in slightly differing flavor composition on Earth. Further, the flavor composition does not need to stay constant over the entire observable energy range: instead, a gradual shift from the pion decay to the muon-damped scenario has been proposed \cite{Kashti:2005qa, Lipari:2007su, Hummer:2010ai, Hummer:2011ms, Bustamante:2015waa}. A measurement of the flavor composition is thus complementary to searches for individual neutrino sources or source populations, which are typically carried out using track-like neutrino events observed with ice or water Cherenkov detectors, predominantly stemming from astrophysical and atmospheric muon neutrinos as well as atmospheric muons.

	    The flavor composition can also carry imprints of effects beyond the Standard Model affecting neutrino propagation (see sec.~\ref{sec:theory}). Due to the extremely large distances traveled by cosmic neutrinos, even very small effects can modify the flavor composition enough to be measurable \cite{Barenboim:2003jm, Keranen:2003xd, Bustamante:2010nq, Arguelles:2015dca, Bustamante:2015waa, Rasmussen:2017ert, Ahlers:2018yom, Ahlers:2020miq, Song:2020nfh}.
	    
	    To measure the flavor composition on Earth, a detector with direct sensitivity to each neutrino flavor is required. Currently, IceCube is the only complete detector capable of a flavor-composition measurement. It is accomplished by identifying flavor-specific or flavor-enhanced event signatures, such as shower-like (mainly from $\nu_e$), track-like (mainly from $\nu_\mu$) and double-shower-like (specific to $\nu_\tau$). At energies above tens of TeV, the flux of atmospheric tau neutrinos is extremely small (stemming only from a small fraction of the thus-far unobserved ``prompt'' atmospheric neutrino component), and any high-confidence tau-neutrino candidate is a high-confidence astrophysical neutrino candidate. Electron and especially muon neutrino fluxes do have sizable atmospheric fractions; these need to be carefully modeled, and / or suppressed. To date, one flavor-composition measurement with direct sensitivity to each neutrino flavor has been performed by IceCube \cite{IceCube:2020abv}. It uses an event selection of high-energy events starting within the detector \cite{IceCube:2013low}, and contains 60 events with deposited energies above 60~TeV in 7.5 years of data-taking \cite{IceCube:2020wum}. The resulting flavor composition is not constraining any production mechanisms at sources yet as shown in Figure \ref{fig:probes_HESE-tau-flavor}. However, some parameter space of the BSM models severely inhibiting neutrino flavor change during propagation can be disfavored at $>2 \sigma$.
	    
	    \begin{figure}[tbh]
	    \centering
	    \includegraphics[width=120mm]{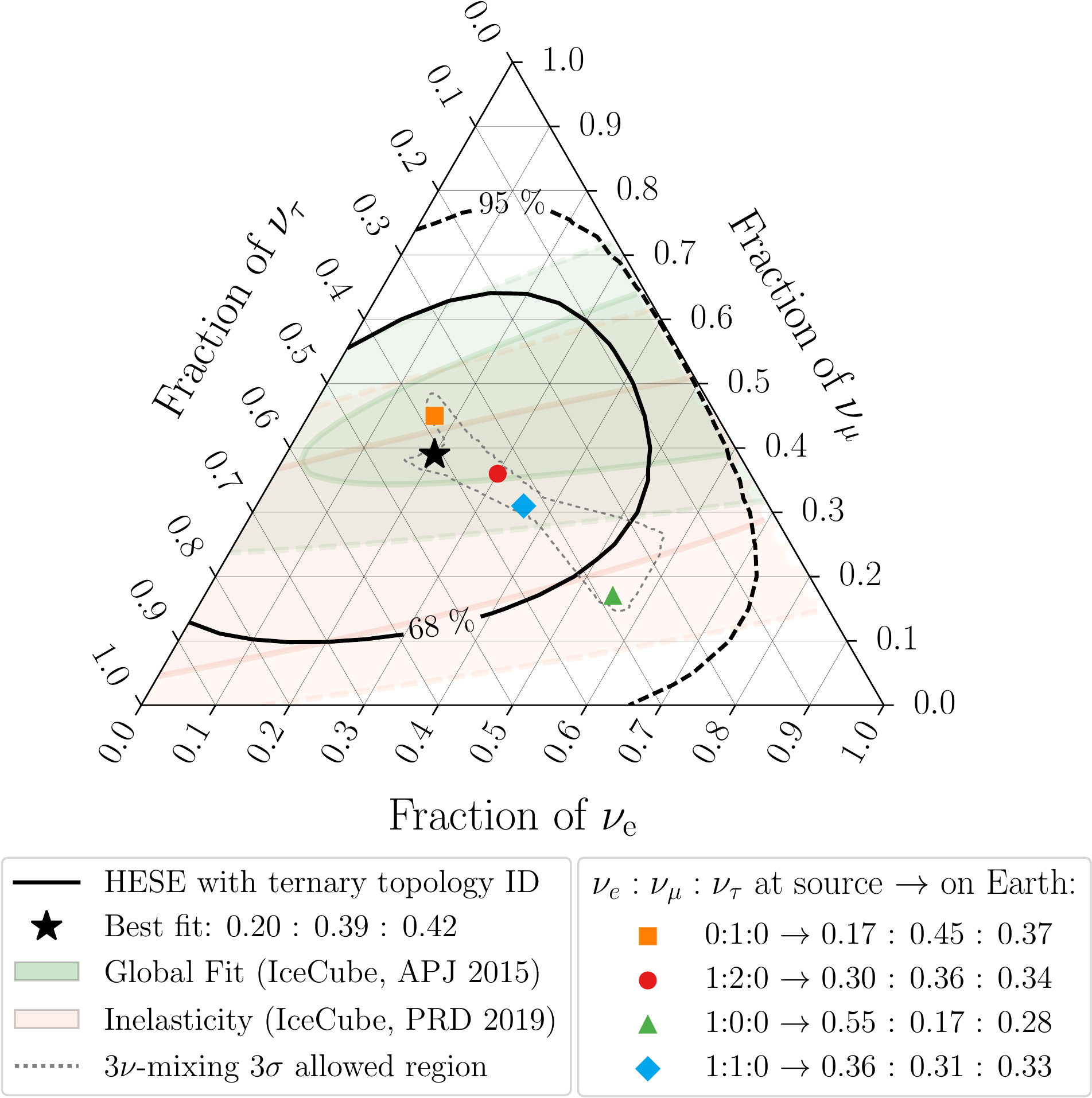}
	    \caption{First flavor composition measurement of astrophysical neutrinos with direct sensitivity to each flavor. Previously published results without direct tau sensitivity and expected flavor composition on Earth using $3 \sigma$ oscillation parameter uncertainties given in \cite{Esteban:2020cvm} are shown with various source production scenarios highlighted. Reproduced from \cite{IceCube:2020abv}.}
	    \label{fig:probes_HESE-tau-flavor}
	    \end{figure}
	    
	    In the next years, such a flavor-composition measurement can be performed on a combination of different event selections (pioneered using track-like and shower-like events \cite{IceCube:2015gsk}) and with the inclusion of other tau or flavor sensitive observables: inelasticity \cite{IceCube:2018pgc} and hadronicity \cite{Li:2016kra,Steuer:2017tca}, being able to tell the tau content in a track-like sample and the neutral-current to charged-current ratio in a shower-like sample, respectively.  Estimates show that the increased statistics and flavor identification power of upcoming neutrino telescopes should yield long-awaited high-precision measurements of the flavor composition, to within a few percent of uncertainty~\cite{Bustamante:2019sdb, Song:2020nfh}.
	    
	    Glashow resonance events \cite{Glashow:1960zz} are the only known signature, that can solely be made by (electron) anti-neutrinos. The identification of an event at the Glashow resonance \cite{IceCube:2021rpz} opens the possibility to  measure the neutrino-to-antineutrino ratio in the next decade, once more than one such intriguing event has been observed. 

	    In the 2040s, a truly global measurement of the neutrino flavor composition can be performed in the 100~TeV to 10~PeV regime~\cite{Song:2020nfh}, including data from the ice / water Cherenkov observatories IceCube(-Gen2), Baikal-GVD, KM3NeT, and P-ONE. All these data sets are complementary, with different sensitivities to each flavors, different systematic challenges, and the different energy ranges explored by the three observing techniques. It should be noted that IceCube-Gen2 alone may have the sensitivity to resolve a change in the source production mechanism with energy, provided the change happens at a favorable energy (e.g., 1~PeV) as shown in Figure \ref{fig:probes_Gen2Flavor} \cite{IceCube-Gen2:2020qha}.
	    
	    \begin{figure}[tbh]
	    \centering
	    \includegraphics[width=120mm]{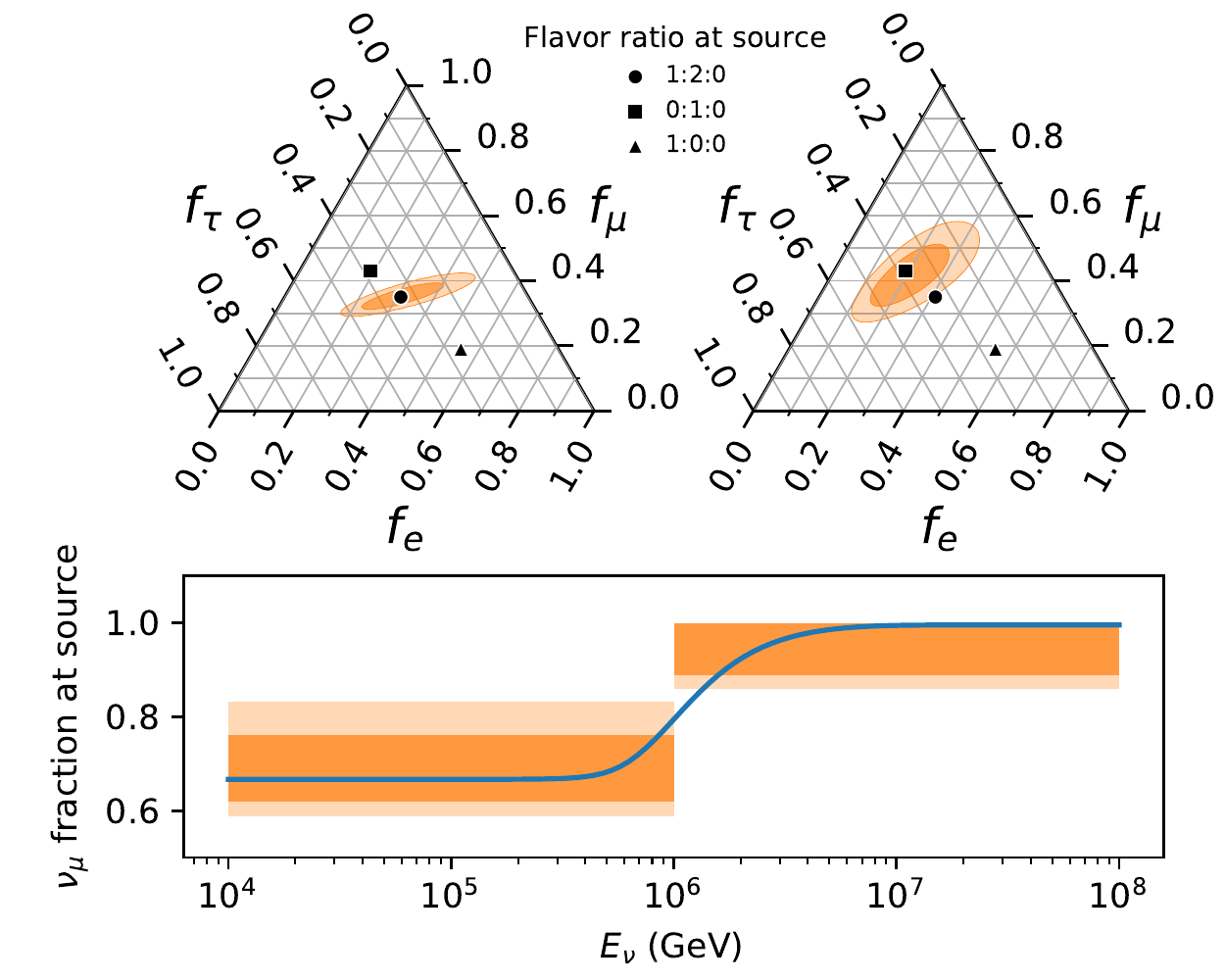}
	    \caption{IceCube-Gen2 sensitivity to the transition between pion-decay and a muon-cooling dominated neutrino production at sources assuming a muon critical energy of 2~PeV. Reproduced from \cite{IceCube-Gen2:2020qha}.}
	    \label{fig:probes_Gen2Flavor}
	    \end{figure}
	    
	    The effect of the combination of all present and near-future water / ice Cherenkov neutrino telescopes on the flavor composition of astrophysical neutrinos is shown in Figure \ref{fig:probes_NuFlavorFuture}. Note that systematic uncertainties are not included, and tau-neutrino identification is not assumed. The inclusion of tau-neutrino identification will predominantly reduce the $e-\tau$ degeneracy  and thus shrink the contours in the most elongated direction. Assessments of systematic limitations for flavor identification are difficult to make, especially for telescopes currently planned or under construction. IceCube-Gen2 can extrapolate the knowledge of known systematic uncertainties, such as due to the modeling of optical properties of the glacial ice at the South Pole from analyses performed with IceCube, IceCube DeepCore, and the expected performance of the IceCube-Upgrade. With the maturing of the Baikal-GVD and KM3NeT detectors, a proper assessment of the systematic uncertainties will be available within a decade.
	    
	    \begin{figure}[tbh]
	    \centering
	    \includegraphics[width=120mm]{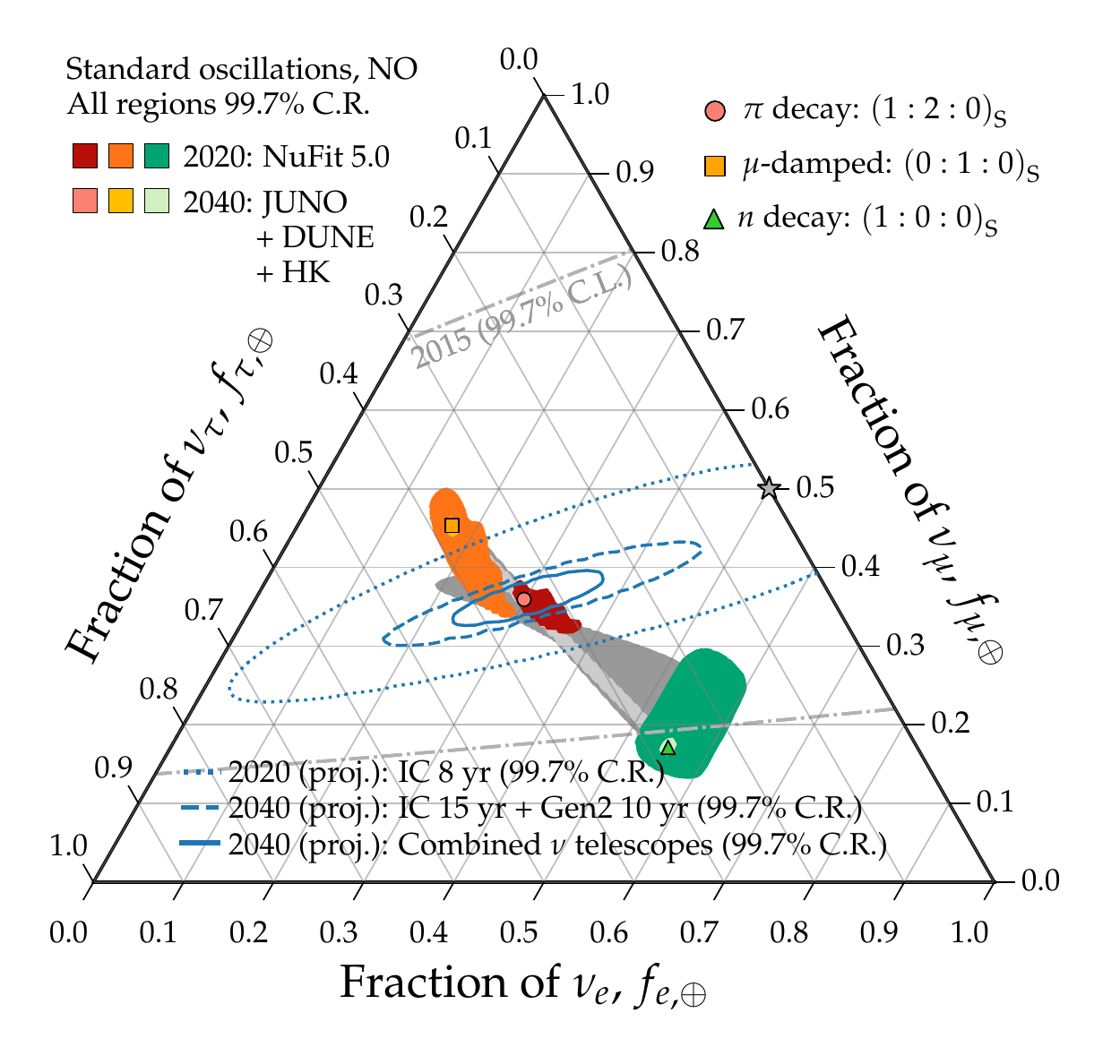}
	    \caption{Projected sensitivity to the flavor composition of astrophysical neutrinos for IceCube and IceCube-Gen2 data only and a combination of all water and ice Cherenkov telescopes operational in 2040. Allowed regions using a global fit to oscillation measurements and the expected knowledge of oscillation parameters in 2040 are also shown. Reproduced from \cite{Song:2020nfh}.}
	    \label{fig:probes_NuFlavorFuture}
	    \end{figure}
	    
	    The in-ice radio detection technique which is currently being used by ARA, ARIANNA, and RNO-G and will be employed by IceCube-Gen2-Radio is predominantly sensitive to showers in the ice. Tau neutrino interactions producing double-showers may be resolved and well-distinguished from single showers or multiple showers created by muon neutrinos. This distinction will extend flavor composition measurements into the energy region above 10~PeV. At the lower end, the energy range of in-ice radio experiments overlaps with the energy range of IceCube-Gen2, allowing for a continuous flavor composition measurement across several orders of magnitude.
	    
	    Complementary to the all-flavor sensitive neutrino observatories, there are several detection techniques targeting Earth-skimming tau neutrinos: \emph{Trinity}~\cite{Otte:2019aaf}, RET-N~\cite{Prohira:2019glh}, and shower detector TAMBO~\cite{Romero-Wolf:2020pzh} in an energy range overlapping with water and ice Cherenkov detectors; the in-atmosphere radio experiments under development (BEACON~\cite{Wissel:2020sec}, GRAND~\cite{GRAND:2018iaj}, and TAROGE-M \cite{Wang:2021ouz}) in an energy range overlapping with in-ice radio detectors; the balloon-borne PUEO experiment~\cite{PUEO:2020bnn} which will search for showers initiated both in the ice and in the atmosphere at the EeV scale; and the operational and proposed air shower detectors, Pierre Auger Observatory and POEMMA~\cite{POEMMA:2020ykm}, respectively, reaching into the EeV regime. These measurements will allow for a clean measurement of the astrophysical tau-neutrino spectrum, providing crucial knowledge of single-flavor spectral parameters that are difficult to obtain with all-flavor detectors due to flavor misidentification or low statistics for well-identified events. The single-flavor tau-neutrino measurements will be crucial inputs to global modeling of astrophysical flavor composition. 
	    
	    High-elevation observatories like POEMMA and PUEO will be able to extend our knowledge of astrophysical (tau) neutrinos into otherwise unobservable energy regimes by observing enormous fields-of-view. As tau neutrinos are not expected to be produced at cosmic sources in sizeable fractions, knowledge from lower energy flavor composition measurements will be combined at the higher energies to infer the total neutrino production at the highest energies.
	    
	    A combination of all flavor-sensitive identification techniques over all accessible energy ranges in the next decades will drastically expand our understanding of astrophysical neutrinos, their flavor composition, source production mechanisms, and possible variations in energy with respect to the dominant source populations and production mechanisms. It may also be the key to discover physics beyond the Standard Model affecting neutrino propagation.  

\subsection{Spectrum} The energy dependence of observed neutrino fluxes is an imprint of the source classes, propagation effects due to cosmological evolution, or new physics beyond the Standard Model of particle physics. At TeV to PeV energies, the observed energy spectrum is consistent with a power law, either one that continues to higher energies or has a cutoff. At the EeV scale, the expected energy spectrum has a predictable shape, tuned by observations of other messenger particles (cosmic rays and gamma rays). 
    
    In fact, a good example to illustrate the need for good spectral measurements is the astrophysical neutrino flux measured by IceCube. Several years after its detection and in spite of intense ongoing research, its origin is still not known. A more precise measurements of the spectral shape should narrow the number of potential source classes. In particular, as identifying individual sources remains a challenge, better spectral measurements may be the only handle we have to identify the origin of the bulk of high-energy astrophysical neutrinos.
    
    Spectral measurements can help in two ways. Firstly by better constraining the spectral index of what appears to be a power law. And secondly, by identifying deviations from a pure power law. Of particular interest would be to detect a break or a cut-off in the spectrum. Even if measurements at higher energies cannot directly resolve a break, a sensitive enough observation yielding a non-detection can constrain the existence of a cut-off. That, in itself, would already be able to narrow down the question of source classes.
    
    Moreover, new physics may imprint on the energy spectrum causing dips and spectral enhancements on top of the underlying astrophysical spectrum. Sharp dips or bumps may indicate new interactions between neutrinos and other neutrinos~\cite{Lykken:2007kp, Ioka:2014kca, Ng:2014pca, Blum:2014ewa, Ibe:2014pja, Kamada:2015era, DiFranzo:2015qea, Shoemaker:2015qul, Altmannshofer:2016brv, Kelly:2018tyg, Barenboim:2019tux, Murase:2019xqi, Bustamante:2020mep, Esteban:2021tub, Creque-Sarbinowski:2020qhz}, dark matter~\cite{Feldstein:2013kka, Esmaili:2013gha, Higaki:2014dwa, Rott:2014kfa, Dudas:2014bca, Ema:2013nda, Zavala:2014dla, Fong:2014bsa, Murase:2015gea, Anchordoqui:2015lqa, Boucenna:2015tra, Cohen:2016uyg, Dev:2016qbd, Arguelles:2017atb, Hiroshima:2017hmy, Chianese:2017nwe, Kelly:2018tyg, Sui:2018bbh, Arguelles:2019ouk}, and/or dark energy~\cite{Anchordoqui:2007iw, Klop:2017dim}. The reconstructed neutrino spectrum may be impacted by modifications of the neutrino-nucleon cross section. Fundamental symmetries of nature, such as lepton-number conservation, CPT, and Lorentz invariance, can further impact the observed energy spectrum~\cite{Amelino-Camelia:1997ieq, Hooper:2005jp, Gonzalez-Garcia:2005ryx, Anchordoqui:2005gj, Bazo:2009en, Bustamante:2010nq, Kostelecky:2011gq, Diaz:2013wia, Stecker:2014oxa, Stecker:2014xja, Tomar:2015fha, Ellis:2018ogq, Laha:2018hsh}.
    
\subsection{Timing} 
  In high-energy non-thermal astrophysical sources, the conditions that enable the production of high-energy neutrinos should enable the simultaneous production of high-energy gamma rays.  When these conditions are met only temporarily, and over a relatively short period, particle production might occur as a transient burst.  This could take place, e.g., in a gamma-ray burst or a flaring blazar.  In that case, if the region of particle production is transparent to gamma rays, then gamma rays and neutrinos should leave the source at roughly the same time.  Nominally, after leaving the source, neither will be delayed on their way to Earth---though gamma rays will be downgraded in energy.  As a result, simultaneous bursts of gamma rays and neutrinos should be detected roughly simultaneously at Earth.
  
  Hence, under the plausible assumption that gamma rays and neutrinos are emitted simultaneously from a transient source, the observation of significant differences between their arrival times could be evidence of either of them undergoing new interactions en route to Earth~\cite{Addazi:2021xuf}.  Often, these are studied as low-energy manifestations of high-energy quantum-gravity effects that introduce violations of Lorentz and CPT symmetries~\cite{Colladay:1996iz, Colladay:1998fq}.  Presently, there is no significant evidence for delays from such an origin, but there are important upper limits 
  
  For gamma rays, delays due to quantum-gravity effects could be energy-dependent and, thus, may occur between photons of different energies emitted by a transient source.  Upper limits on the energy scale of quantum gravity and on the strength of the new interactions come from, e.g., gamma-ray bursts~\cite{Amelino-Camelia:1997ieq, Ellis:2002in, Ellis:2005sjy, Bernardini:2017tzu, MAGIC:2020egb}, blazars~\cite{Amelino-Camelia:2002kwj}, the Crab nebula~\cite{Jacobson:2002ye, Ellis:2003sd}, and a combination of various types of high-energy sources~\cite{LHAASO:2021opi}.
  
  For high-energy neutrinos, limits on new interactions come instead by looking for delays between the arrival times of neutrinos and gamma rays emitted by the same source~\cite{Amelino-Camelia:1997ieq, Hooper:2005jp, Gonzalez-Garcia:2005ryx, Anchordoqui:2005gj, Bazo:2009en, Bustamante:2010nq, Kostelecky:2011gq, Diaz:2013wia, Stecker:2014oxa, Stecker:2014xja, Tomar:2015fha, Ellis:2018ogq, Laha:2018hsh, Wang:2020tej}.  Separately, high-energy neutrinos could also be delayed by repeatedly scattering off of the background of relic neutrinos or dark matter via new mediators, en route to Earth~\cite{Murase:2019xqi, Shalgar:2019rqe}.  In this case, for PeV-scale neutrinos, delays could range from tens of seconds to a handful of days, depending on the strength of the interaction and the distance to the source.  
  
  Presently, searches for relative delays between high-energy neutrinos and gamma rays using real observations~\cite{Ellis:2018ogq, Laha:2018hsh, Wang:2020tej} is largely limited to the single detection of a coincident high-energy neutrino with the flaring blazar TXS 0506+056.  Future observations of further sources of simultaneous neutrino and electromagnetic emission will improve on these searches.

\subsection{Arrival Direction}
    The arrival direction of neutrinos imprint both the distribution of neutrino sources in the Universe and the potential interactions of neutrinos with the cosmic environment~\cite{Olinto:2011ng, Guepin:2020lex}. Under the assumption that neutrino sources are isotropically distributed in the Universe, any observed anisotropy could result from new physics or dark matter interactions~\cite{Arguelles:2017atb,Cherry:2014xra}. These interactions would affect the propagation of neutrinos through the Universe, resulting into a neutrino horizon~\cite{Weiler:1997sh,Ioka:2014kca,Ng:2014pca}, as observed for other messengers such as cosmic rays and gamma rays~\cite{Berezinsky:1969erk}. To study such effects, combined analyses with these other messengers are needed in order to constrain the distribution of potential neutrino sources in the Universe, and correct for intrinsic anisotropies due to distribution of the sources~\cite{Fang:2020rvq}.
    
    To that end, accurate measurements of the neutrino arrival direction are of paramount importance in order to correctly identify neutrino sources~\cite{Fang:2016hop} or, at least, accurately reconstruct the neutrino sky to perform combined analyses with other messengers.  This is particularly important when performing tests of fundamental physics, such as of Lorentz invariance~\cite{Colladay:1998fq,Kostelecky:2011gq}, since otherwise the observation of neutrino anisotropies could be incorrectly attributed to new physics rather than to the underlying source distribution. Currently, the angular resolution achieved for tau neutrinos is within a few degrees~\cite{IceCube:2013dkx}, which is much larger than the resolution needed to reasonably identify a source in the sky. 
    
    The arrival direction also provides an excellent veto of tau neutrinos for experiments relying on Earth-skimming observations. Earth-skimming neutrinos are mostly tau neutrinos, which, unlike other neutrino flavors, are able to re-generate inside Earth and travel longer without being fully absorbed~\cite{Fargion:1999se}.  In this scenario, sub-degree angular resolution is enough to veto tau neutrinos in principle. However, for tau neutrino cross-section measurements, improvements in the angular resolution directly impact the experimental uncertainties (see dedicated section of this paper).

\subsection{Detection techniques} 

The detection of high energy (HE, few hundred TeV to PeV) and ultra-high energy (UHE, above PeV) neutrinos is particularly challenging because flux measurements, their extrapolations and existing limits constrain expectations from different production mechanisms to extremely low values, even in the most optimistic scenarios. To estimate event rates in given experiments these low fluxes are to be convolved with very reduced cross sections for neutrino interactions even at the highest energies. In order to obtain measurable event rates the demands on the detector are formidable, requiring over a gigaton of active target, equivalent to a 500 m height mountain. This forces the use of large natural target volumes to capture the effects of these elusive particles, leading to a diversity of proposals resulting from the combination of different detection techniques and different media. This can serve as a means of classifying the different proposals in a somewhat systematic way. The breadth of experiments operating or proposed are summarized in Fig.~\ref{fig:experiments_table} and their expected sensitivities are shown in Fig.~\ref{fig:UHE_diffuse_sensitivity}.

Astrophysical neutrinos have been detected at energies above few hundred TeV, the most energetic ones reaching out to just over 10 PeV. There are three possibilities to detect neutrinos in these energy ranges: showers, tracks, and Earth-skimming neutrinos.\\
{\noindent \bf Tracks:} It is possible to search for tracks of leptons produced in charged current interactions. While for electron neutrinos the electron starts a shower after order 10 cm of rock, both muons and particularly tau leptons can travel distances in rock well above a km before decaying, the taus reaching ranges up to order 50 km at the highest energies, because their electromagnetic interactions are suppressed by the ratio of the lepton to the electron mass squared. Muons and tau leptons have long straight tracks with reduced stochastic energy losses producing small showers along their track until they decay~\cite{Lipari:1991ut}. By instrumenting a given region to detect these tracks, the effective active volume is enhanced because of these long tracks that can start well outside the detector. The original neutrino direction can be accurately inferred from that of the recorded track. By searching for showers going upwards, the largest background due to muons produced by cosmic ray showers in the atmosphere is naturally shielded by the Earth. This was the approach inspiring the first projects to detect neutrinos instrumenting large volumes of either natural water or ice to measure the Cherenkov light produced by these long lepton tracks as they go through the detector. 

\clearpage
\begin{figure}[!th]
    \centering
    \includegraphics[width=\textwidth]{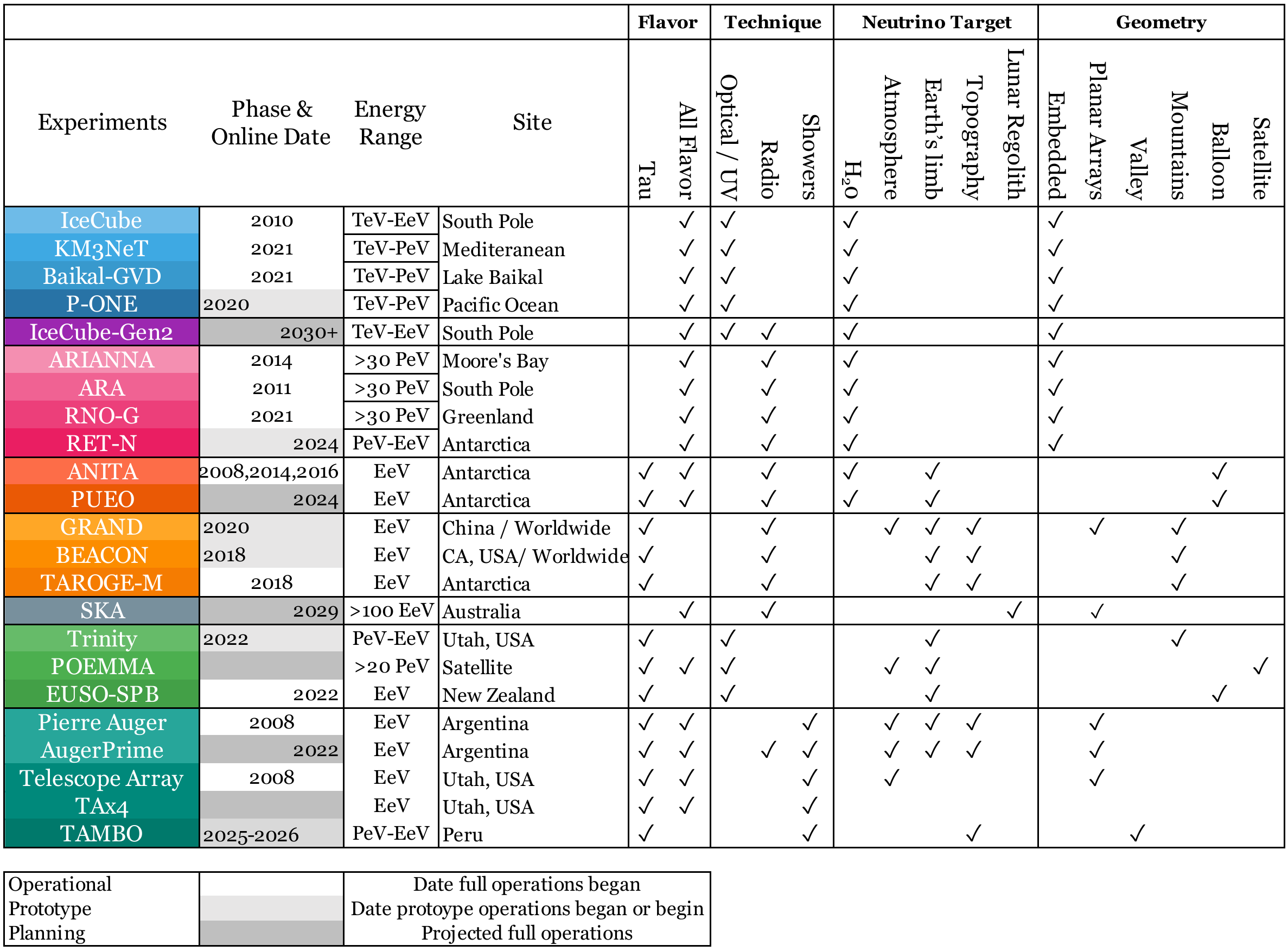}
    \caption{Landscape of operating and planned experiments~\cite{IceCube:2016zyt, Fermani:2020oxx, Rea:2021hjd, IceCube-Gen2:2020qha, Clark:2021fkg, Barwick:2006tg, Allison:2011wk, RNO-G:2020rmc, Prohira:2019glh, ANITA:2008mzi, PUEO:2020bnn, GRAND:2018iaj, Wissel:2020sec, Nam:2020hng, Bray:2015ita, Otte:2019aaf, POEMMA:2020ykm, Bacholle:2017dye,PierreAuger:2015eyc, PierreAuger:2016qzd,Tokuno:2011zz,Romero-Wolf:2020pzh} sensitivity to tau neutrinos at the highest energies showing their flavor sensitivity, detection technique, and geometries. The flavor sensitivity column indicates which experiments have observation channels sensitive only to tau neutrinos and those that are sensitive to all flavors but that need to tag tau neutrinos using event topology. Experiments in different stages (operational, prototype operation or development, or planned full operations) are indicated with the grey bands. Colors are the same as those in Fig.~\ref{fig:UHE_diffuse_sensitivity}. See text for more details about each experiment.}
    \label{fig:experiments_table}
\end{figure}
    
\noindent \textbf{Showers:} The second possibility is to search for the showers produced in both charged and neutral current neutrino interactions. The detection of showers opens many possibilities. The showers can develop in a dense medium such as water or ice but they can also develop in thinner ones like the atmosphere. Showers that develop in the atmosphere can be detected from very far distances if they are energetic enough, as will be discussed below. This channel is in principle sensitive to all flavors. Neutral currents produce showers of hadronic type with typically about $20\%$ of the neutrino energy at UHE according to the Standard Model predictions, with no difference between flavors. However in the case of charged currents electron neutrinos transfer the remaining energy ($80\%$) to an electromagnetic shower which is typically mixed with the hadronic one. As a result the sensitivity to electron neutrino showers is maximal relative to the other two flavors. Showers induced through stochastic energy losses of muons or tau leptons can increase their detection probability, particularly for very energetic leptons~\cite{Garcia-Fernandez:2020dhb}. More importantly, the tau with a short lifetime typically decays with high energy, inducing a shower in most decay channels that, on average, carries $\sim 50\%$ of the tau energy. This adds interesting detection and identification possibilities. Depending on the tau energy, a proxy for the neutrino energy, the decay shower can separate physically from that produced in the neutrino interaction, leading to a double shower that, when identified, is a signature of the tau neutrino~\cite{Learned:1994wg}. The muon has a much longer lifetime and it typically decays when it reaches GeV energies, after having lost most of its energy. The decay shower is thus very small and difficult to identify. 
    
The search for showers induced by neutrino interactions opens up a plethora of alternatives besides the optical Cherenkov in ice or water that serves for detecting muon and tau-lepton tracks. Showers that develop in ice, water, salt, the lunar regolith and the atmosphere can be detected with optical, radio, sound, radar techniques and using arrays of particle detectors. The optical and radio techniques are by now well established from the study of cosmic rays and gamma rays. 

\noindent \textbf{Earth-skimming tau neutrinos:} The lifetime and energy loss of the tau leptons open up a very interesting possibility for detection that combines a dense media -- in which tau neutrinos interact to produce a tau lepton -- with the atmosphere -- where the tau exits and eventually decays after traveling kilometers underground. The mechanism, first described at the turn of the  century~\cite{Bertou:2001vm,Fargion:2003kn,Feng:2001ue}, is especially effective for tau neutrinos so it naturally achieves flavor identification\footnote{Energetic muons exit the Earth tend to escape to space.}. A high energy tau suffers an $e$-fold energy reduction in about 5 km of rock which is the decay length of a 0.1~EeV tau. If an area of hundreds of square kilometers (common for UHECR detectors) can be monitored for exiting taus the effective target mass for neutrinos can be extremely large. 

The tau range in rock increases linearly as the energy rises while it is dominated by decay time but above $\sim 0.1$~EeV it turns to be dominated by energy loss and only increases logarithmically. Neutrino absorption in the Earth comes strongly into play limiting the solid angle at increasing neutrino energies. The trajectories of the neutrinos intersect the Earth over a chord distance that depends strongly on the local exit angle of the tau trajectory.Neutrinos at 1 EeV have a mean free path of about 500 km in rock, corresponding to the subtended chords for very shallow (3$^{\deg}$~\cite{Klein:2020nuk}) emergence angle relative to the horizontal. As the neutrino energy increases the available solid angle reduces. Moreover if very energetic taus emerge to the atmosphere they can decay so far away that the remaining depth of the atmosphere is insufficient for the shower to fully develop~\cite{Zas:2005zz}. The tau neutrino conversion to an air shower is most effective for a broad region about 0.1 EeV and thus we refer to these events as Earth-skimming neutrinos~\cite{Feng:2001ue}. One can visualize points on the Earth surface as exit points of tau leptons produced by a diffuse tau neutrino flux. As the emergence angle increases from order one degree to vertical, the flux drops and the mean tau energy decreases from the EeV range to an energy at which the neutrino mean free path roughly matches the column depth to the corresponding Earth's chord~\cite{Zas:2005zz}. 

\subsubsection{Optical}
    {\noindent \bf In-ice/water:}
    The optical technique in ice or water consists of instrumenting large volumes with photodetectors to sample the emitted Cherenkov light of secondaries from all neutrino flavors.
These detectors are, depending on the selected instrumentation density,  sensitive to neutrino interactions in the energy range 
from 10 GeV to $10^9$ GeV.  Muons from energetic muon neutrinos allow good directional reconstruction so that they are especially suitable for  neutrino astronomy.
Muons have a large effective volume as  the primary interaction of through-going muons is outside of the instrumented volume. 
That fact however, as well as their stochasticity in energy loss, limits the energy resolution for muon tracks somewhat to a level of a factor 2. The Cherenkov light emitted by the showers produced in small stochastic energy losses is useful to constrain the energy of the muon or tau-lepton track. We note however that, as the energy losses of muons and taus are significantly different, there can be in addition a large intrinsic uncertainty because once a track is identified it is impossible to know if it is due to a muon or a  tau-lepton unless its decay is also observed in the detector volume.
        
    The same technique can be also used to reconstruct the showers in a complementary way. The angular reconstruction is much more challenging, because the shower dimensions are typically of order 10 meters, while the energy determination of the shower can be made more precisely, provided the shower is contained in the detector. This was actually the way the first PeV neutrinos were  detected~\cite{IceCube:2013cdw}. 

    The technique started with the DUMAND project in the 1970's, planned for ocean water~\cite{Roberts:1992re} and was followed by lake Baikal in lake water~\cite{BAIKAL:1997iok} and AMANDA in Antarctic ice, that led to the first successful detection of  astrophysical neutrinos in 2013 with its follow-up observatory, IceCube~\cite{IceCube:2006tjp,IceCube:2013low}. The lake Baikal initiative together with ANTARES~\cite{ANTARES:2011hfw}, in the Mediterranean sea, have been developed in parallel. A new generation of detectors is planned or under construction both as follow-ups of these initiatives, IceCube Gen2~\cite{Clark:2021fkg} in ice, Baikal-GVD in lake Baikal~\cite{Avrorin:2020dre}
    KM3NeT/ARCA~\cite{Fermani:2020oxx} following ANTARES or as new initiatives in study such as P-ONE~\cite{Rea:2021hjd} in the Pacific. 

    {\noindent \bf In-atmosphere:}  Detecting optical emission from particle showers developing in the atmosphere (air showers) is a well-established technique. Air-shower particles radiate Cherenkov light and, at a much lower intensity, also generate fluorescence emission by collision, exciting nitrogen molecules. 
        
    For example, imaging Atmospheric Cherenkov telescopes (IACTs) detect the Cherenkov emission from gamma-ray-induced air showers. After more than 30 years since the detection of the first gamma-ray source with the air-shower imaging technique, IACTs are still the most sensitive instruments to detect gamma-rays above 25\,GeV \cite{Hinton:2008ka}.
        
    Imaging air-showers with fluorescence light, on the other hand, has proven to be a crucial technique to study the composition of UHECR above $~10^{16}$\,eV and is, for example, used by the Pierre Auger and TA observatories \cite{Song:1999wq}.
        
    IACTs and fluorescence telescopes designed to detect gamma-rays and UHECR, respectively, have also been used to search for tau-induced air showers and demonstrated the power of the imaging technique for neutrino searches. Air-shower imaging is sensitive to two different types of neutrino events. The first event type is due to neutrinos entering the atmosphere at a shallow angle, interacting in the atmosphere, and inducing an air shower, which develops quasi horizontally. The second event type is due to Earth-skimming tau neutrinos, which induce upward-going air showers when the tau exits the Earth and decays in the atmosphere.
        
    In the first case, the neutrino interacts with quasi equal probability anywhere along the trajectory through the atmosphere because, even for 1 ZeV (1000 EeV) neutrinos, the interaction length by far exceeds the maximum 72,000 g/cm$^2$ grammage of the atmosphere (a trajectory tangent to the Earth surface at sea level). Detecting these events and discriminating them from cosmic-ray-induced air showers requires a good reconstruction of the shower geometry, particularly the starting point of the air shower. 
        
    In the case of Earth-skimming neutrinos, the mere fact that air-showers develop up-ward into the atmosphere or emerge from mountains provides a powerful means of rejecting cosmic-ray events, which are down-ward developing.
 
     Extensive air shower (EAS) imaging was the technique originally proposed to detect Earth-skimming neutrinos, defined by modest slant depths through the Earth, but did not gain much traction, for several reasons. Firstly, EAS imaging telescopes can only observe around astronomical night, resulting in a duty cycle of $\sim$20\%, Secondly, space-based telescopes are costly. And, thirdly, imaging using the fluorescence technique led to energy thresholds above 10 EeV. However, advances in technologies, new optics concepts, photo-detectors with wider wavelength response, and new observing strategies of proposed EAS Cherenkov imaging-based neutrino detectors, have made optical EAS imaging competitive over the past years. This includes reducing the neutrino detection energy threshold down to $\sim 10$ PeV using the optical Cherenkov EAS signal  \cite{Krizmanic_2019,Venters_ToO, Cummings:2020ycz,Cummings:2021_CR}. Radio experiments have also gained complexity and scale during the same period, which dilutes the cost argument. Thus, the optical Cherenkov imaging technique can be realized to have good sensitivity at very-high-energies $<10^8$\,GeV significantly lower than radio, thus bridging the critical energy range between IceCube and radio.
     
     The 20\% duty cycle is mostly compensated by a high acceptance stemming from the ability of detecting EAS developing at large distances from the detector, thus monitoring an immense neutrino target volume. Furthermore, tau neutrino regeneration allows for a significant sensitivity for slant depths in the Earth that are larger than that for `Earth-skimming' $\nu_\tau$'s, significantly increasing the tau neutrino target volume. Thus this technique is denoted as 'Earth-emergent' neutrino detection. These realizations are why all recently proposed imaging systems point at the horizon from an elevated position, like a mountain, a high-altitude balloon, or Earth-orbiting spacecraft. In part, the ability to detect very distant showers can be attributed to the silicon photomultiplier, which offers a very-high photon detection efficiency from the near UV to the far red, which spans the wavelength band of the EAS generated Cherenkov light.  For ground-based measurements, due to atmospheric scattering, only the red components of the Cherenkov spectrum survive over $>100\,$km distances, which is why good red sensitivity is critically important to capture a sufficiently strong signal with a comparably, ground-based small telescope. However, for sub-orbital or space-based measurements, a significant portion of the EAS optical Cherenkov signal can occur at altitudes above the aerosol layer of the Earth and leads to Cherenkov spectra peaked in the near-UV, cutoff below $\sim$ 300 nm by the Earth's ozone layer \cite{Krizmanic:2020shl}. Thus, even though the distance to the EAS is larger for sub-orbital and space-based experiments, there is a significant gain in the Cherenkov intensity due to the reduced atmospheric attenuation of the signal.
     
     While smaller telescopes reduce the cost of the system, much more impactful have been new technologies to process and digitize photon detector signals at a much lower price than previously possible. So while both optical and radio arrays may have a similar number of channels, i.e.\ $\sim 10^4$, to achieve good neutrino flux sensitivity, an optical telescope's channel are located within the telescope itself, while the channles for the radio, e.g.\ antennae, are dispersed over a large area, $> 10^5$ km$^2$ for the largest, such as GRAND200k \cite{GRAND:2018iaj} and BEACON \cite{Wissel:2020sec}.
     
     As the distance to the horizon increases when the elevation of observation increases, the detection threshold becomes larger and the detection area increases too. One approach has been taken by locating the detectors on mountains in the case of Trinity which is designed specifically to search for Earth skimming tau neutrinos. Alternatively detectors can be located at much higher altitudes with ultra-long duration ($\sim 100$ day) balloon flights as in EUSO-SPB~\cite{Bacholle:2017dye} or in low Earth orbiting (LEO) satellites as in  POEMMA~\cite{POEMMA:2020ykm}. The later approach results in large effective areas that are both sensitive to Earth-emergent tau neutrinos as well as having unique sensitive to higher energy neutrinos using the EAS fluorescence signal as measured by the FT detectors used to measure Ultra High Energy Cosmic Rays. 
        
    \subsubsection{Radio}
        
    The radio technique is most promising for UHE neutrino detection and offers multiple possibilities. The technique has many advantages. There are large volumes of natural materials where radio pulses can propagate such as air, ice, salt or the Moon regolith; the pulses are large in intensity at the highest energies and can propagate large distances with little absorption, i.e.\ hundreds of km in the atmosphere or $\sim 1$ km in ice; the duty cycle can be $100\%$ since signals can stand over the background also during the day; the emission is coherent and its pattern gives information about the shower structure, because of coherence the pulse samples a large portion of the shower and its intensity scales with shower energy; pulse polarization is linked to the emission mechanisms and it can be used in event reconstruction; and the detectors are relatively simple (antennas). The main difficulty resides in removing noise, often dominated by man-made technologies and which is thus very dependent on the site selection.  
        
    {\noindent \bf In dense media:} Askary'an in the early 1960s~\cite{Askaryan:1961} proposed the detection of neutrinos both under the Earth and under the Moon surface by searching for the coherent pulses produced by the excess negative charge of the induced showers. First detailed calculations of radio pulses in the early 1990s~\cite{Zas:1991jv} confirmed the great potential of the technique in ice, particularly for the highest energies and the first R\&D initiatives came to light within the decade searching for pulses in the Moon with the Parkes telescope~\cite{Hankins:1996} and in ice with RICE~\cite{RICE:2001ayk} in synergy with the development of AMANDA, the forerunner of IceCube, in the South Pole~\cite{Andres:1999hm}. The RICE initiative was followed with arrays buried deep in ice such as ARA~\cite{Allison:2011wk} or at shallow depths such as ARIANNA~\cite{Barwick:2006tg} with which the technique matured. The successors of these initiatives IceCube-Gen2-Radio in Antarctica and RNO-G in Greenland combine both deep and shallow antennas over large surface areas.  The ANITA experiment~\cite{ANITA:2008mzi}, with antennas flown in a long-duration balloon to the stratosphere to search for neutrino induced pulses in the Antarctica ice bed, set flux bounds at the highest energies~\cite{ANITA:2005gdw} and unexpectedly detected extensive air showers in the 200 MHz-1.2 GHz~\cite{ANITA:2010ect}. 

    In a dense medium such as ice the excess charge or Askary'an effect dominates the radio pulse. The polarization pattern is determined by the shower direction projected onto a plane perpendicular to the line of sight, the  electric field vector points in the direction of the shower axis. The emission pattern is also characterized by a Cherenkov cone at $\sim 56^\circ$ in ice (with refraction index $n\simeq 1.78$) where the coherence and the amplitude of the field are maximized. Coherence is kept up to few GHz in the Cherenkov cone but, as the observation angle is moved away from it, diffraction effects reduce the amplitude and width of the pulse (suppressing the highest frequencies) due to the longitudinal shower spread (typically a few radiation lengths except when the showers are ``stretched" by the Landau-Pomeranchuck-Migdal (LPM) effect~\cite{Alvarez-Muniz:1997wxh}). The choice of the frequency band is a trade off between sensitivity maximized for the Cherenkov direction at the highest frequencies, and acceptance which is increased by using lower frequencies. Here frequency bands can extend to the GHz. 
    
    One interesting question is to what extent ultra-high energy detection experiments can perform flavor-specific measurements.  Askary'an radio-detectors are sensitive to cascades, which come primarily from charged-current $\nu_e$ interactions and the hadronic cascades from all interactions.  The latter typically carry only 20\% of the neutrino energy, though.  That said, there may be prospects for observing $\tau$ double-bang events as two well-separated cascades, as long as both cascades are energetic enough to be detected independently.   Unfortunately, there are also potential background from muons, if a muon undergoes two very energetic stochastic energy losses, or from electromagnetic cascades that are split by the LPM effect \cite{Gerhardt:2010bj}.  More investigation is needed to understand if flavor identification is practical in this energy range. 
         
    {\noindent \bf In atmosphere:} The possibility of using the radio technique to detect extensive air showers was already suggested in the 1950s~\cite{Jelley:1958} and it was developed in the 1960s and 1970s to study cosmic rays~\cite{Jelley:1965}, but it was then halted because of difficulties to correlate the detected pulses to the primary properties of the incident particles~\cite{Mandolesi:1974}. The field reignited in the 2000s' with new experimental initiatives~\cite{LOPES:2005ipv,Kelley:2011zz} but its large potential for neutrino astronomy was recognized afterwards particularly due to the enhanced sensitivity to Earth-skimming tau neutrinos which makes it a flavor sensitive technique. The detection of pulses consistent with air showers going in the upward direction with ANITA, the often called {\it anomalous events}, attracted a lot of attention stressing the fact that this experiment is uniquely sensitive to both showers developing in the ice and in the air. This mechanism has become an objective for PUEO~\cite{PUEO:2020bnn} the follow-up of ANITA, and many other dedicated projects instrumenting large areas with antennas such as GRAND~\cite{GRAND:2018iaj} or putting them in mountains such as BEACON~\cite{Wissel:2020sec} and TAROGE \cite{Wang:2021ouz}.
        
    The pulses emitted from air showers are dominated by the geomagnetic effect, that is the transverse current that develops as the shower develops and particles deviate in the direction of the cross product of the shower direction and the Earth's magnetic field. The polarization is parallel to the transverse current and its amplitude scales with the amplitude of the component of the B-field transverse to the shower direction. Provided the shower develops sufficiently far from the antennas the coherence properties of the emission define a Cherenkov cone, of order a degree opening angle~\footnote{Its precise value depends on the air density at the region where the shower is maximally developed.}, within which the emission is both largest and most coherent (extending to highest frequencies and being sharpest in time). 
    The width of the cone is of diffractive origin and it is thus different for each frequency component. For the highest GHz frequencies it is a fraction of a degree but it becomes increasingly wider as the frequency drops. In the time domain the pulses become wider as they are observed away from the Cherenkov angle. ANITA observations of cosmic-ray air showers demonstrated experimentally that coherence is reached to the GHz as was also shown with detailed simulations~\cite{Alvarez-Muniz:2011ref,Alvarez-Muniz:2012ytl}. The choice of frequency band for detection is partly related to this effect, low frequencies favor  increase in acceptance while high frequencies favor higher sensitivity in the Cherenkov direction. The chosen frequency bands are also constrained by background noise and therefore site dependent, a common band for air shower arrays being 30-80 MHz \cite{Schroder:2016hrv, Huege:2017khw}. 
    
{\noindent \bf Interferometry:} Progress in the past decade has been enormous, both for detecting pulses in ice \cite{Connolly:2016pqr} and in the atmosphere~\cite{Huege:2016veh}. The phased array technique based on interferometry has been particularly important. Earliest efforts with LOPES~\cite{LOPES:2005ipv} successfully obtained the first angular images of the pulses. The technique allows an important increase in sensitivity~\cite{Vieregg:2015baa}, it was used with ANITA data~\cite{Romero-Wolf:2014pua} and provides the basis for most current experiments and future projects. Moreover, a modification of the technique has been developed to account for the fact that showers are at a finite distance from the antennas which allows the reconstruction of both the shower direction and its position. Using simulations it has been shown to have an enormous potential allowing angular accuracy of arcminutes~\cite{Schoorlemmer:2020low}, well beyond the resolution of other neutrino experiments and can have an enormous impact on neutrino (and multimessenger) astronomy. The possibilities of applying this technique for flavor tagging are also quite attractive. Both the effect of a double cascade as expected for a tau neutrino interaction~\cite{Alvarez-Muniz:2020ary} or an LPM-elongated shower as expected for a charged current electron interaction~\cite{Alvarez-Muniz:1997wxh,Alvarez-Muniz:1998rgq,Alvarez-Muniz:1999qlp} would have pretty well defined patterns of emission that could in principle serve for flavor identification.
         
    {\noindent \bf Lunar:}
    The largest natural target volume considered in UHE neutrino detection with the radio technique is the Moon \cite{NOT-IN-INSPIRE-Dagkesamanskii_SovPhys:1989}. The lunar Askaryan technique involves using ground-based radio telescopes to search for the nanosecond radio flashes produced in showers initiated by an UHE cosmic-ray or a neutrino inside the regolith of the Moon \cite{Bray:2016xrn}. These pulses are refracted upon exiting the Moon and can reach Earth.  The radio signal should be searched for pulses in real time with candidate events stored for later analysis, pointing the movable instruments directly at the Moon or {\it beamforming} electronically, compensating for ionospheric dispersion and filtering out local radio interference. Observations with a large number of existing radio telescopes have shown that this is technically feasible (see \cite{Bray:2016xrn} for a comprehensive list and description), although no detection has been achieved so far. Large geometrical areas for neutrino interaction can be achieved with this technique, but due to the $\sim1$ second-light distance to the Moon the energy threshold of all experiments to-date is typically above $10^{20}$ eV. 
        
    {\noindent \bf Radar:}
    The radar echo method, which uses active radar sounding to detect cascades, rather than detecting a signal produced by the cascade itself. In the radar echo method, a transmitter broadcasts a radio signal into a detection volume (nominally ice, though salt has also been proposed~\cite{Chiba:2012zz}). When a neutrino-induced cascade is produced within the detection volume, nearly all of the energy within the cascade is lost to ionization of the medium, which results in a dense and short-lived cloud of charge within the volume. This cloud can reflect the incident radio signal to remote receivers monitoring the same detection volume (like bouncing an air-traffic-control radar signal off an airplane). The radar echo method is a potentially powerful probe for UHE neutrinos for two primary reasons. First, the radar echo method is {\it active}, meaning that the received signal is not produced by the cascade, but is a convolution of the transmitted signal and the geometry of the transmitter---cascade---receiver system, giving experimenters an extra handle on the received signal properties and allowing for robust reconstruction capabilities. Second, this cloud of ionization is detectable over much of the solid angle, increasing angular efficiency with respect to other radio methods dependent on aperture limited emission (Askaryan type). The radar echo method is under development, with laboratory tests~\cite{Prohira:2019glh} indicating that it could be a suitable detection strategy. Forthcoming in-nature tests will lend further insight, and will be discussed below. 
        
    \subsubsection{Shower particles} 
    The particles themselves that constitute the showers induced in the atmosphere either by downward-going neutrinos or Earth-Skimming tau neutrinos, can also be observed with the same arrays of conventional particle detectors used to study cosmic-rays. An efficient discrimination of neutrino-induced showers from background cascades initiated by cosmic-rays in the top layers of the atmosphere can be achieved for observation of extensive air showers starting deep in the atmosphere, where only weakly interacting neutrinos (or nearly horizontal taus) can initiate the showers. The success of this technique depends crucially on measuring shower properties related to the depth of the first interaction. Based on this general idea, neutrino identification can be done more efficiently at large zenith angles ($\theta$) w.r.t. the vertical at ground \cite{Berezinsky:1969erk, Berezinsky:1975zz}, where cosmic-ray showers are to a first approximation a muonic front because the electromagnetic component is absorbed in the intervening atmosphere to be traversed in the inclined directions before ground level, while penetrating neutrinos induce a larger electromagnetic component. Typically, downward-going showers can be most efficiently discriminated between $\theta\sim 60^\circ$ and $90^\circ$ \cite{Zas:2005zz, PierreAuger:2011cpc}. Upward-going neutrino detection with this technique is more efficient for tau neutrinos in the zenith angle range $\theta\in(90^\circ, \sim 95^\circ)$ \cite{Fargion:2000iz, Bertou:2001vm} producing taus emerging from the Earth and decaying at altitude less than $\sim 2~\mathrm{km}$ \cite{PierreAuger:2007vvh}, otherwise the atmospheric upward-going shower would not intercept the instrumented ground. Coincidentally, the detection of Earth-Skimming neutrinos at EeV energies is favored for quasi-horizontal showers as stated above where, for instance, at $\theta \sim 91.5^\circ$ the chord of the Earth $\sim 300$ km matches the interaction length of the neutrino. The large zenith angle of the showers calls for detectors with sensitivity to particles arriving in the inclined directions, favoring {\it volumetric} detectors such as water-Cherenkov stations, over flat ones oriented parallel to ground as is typically done with scintillators in cosmic-ray detector arrays. The technique also requires the capability to separate the electromagnetic from the muonic component of the shower, either directly using buried or shielded detectors more sensitive to the muonic component, or indirectly exploiting the different time structure of the electromagnetic and muonic components of the shower front \cite{PierreAuger:2007vvh, PierreAuger:2019ens}. For detection of UHE neutrinos, individual detectors are spread over surface areas of thousands of $\mathrm{km}^2$ in existing observatories \cite{PierreAuger:2019ens, TelescopeArray:2019mzl}, separated by distances of the order of $\mathrm{km}$. As mentioned earlier, the topography of the site around the array, in particular the presence of high mountains or deep valleys, can benefit the identification and efficient detection of Earth-Skimming tau neutrinos and slightly downward-going neutrinos with particle arrays, also at sub-EeV energies in this case requiring more compact arrays \cite{Romero-Wolf:2021ftg, HAWC:2021dhc}.

\clearpage
\begin{figure}[!ht]
    \centering
    \includegraphics[width=0.9\textwidth]{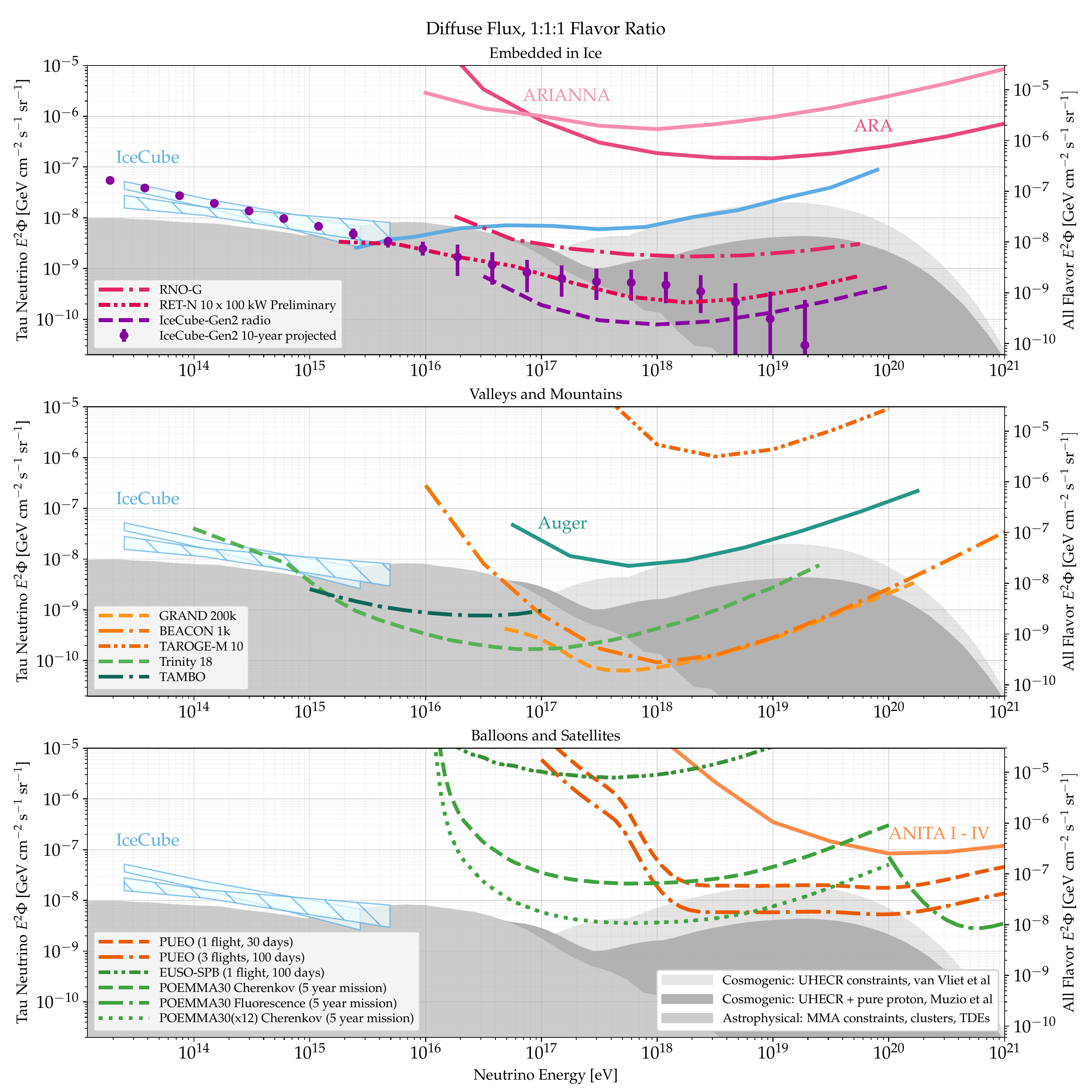}
    \caption{\small{ The expected differential 90\%~CL sensitivity to a diffuse neutrino flux, binned with full decade bins. The tau neutrino flux and experimental sensitivities are shown on the left, while the all-flavor flux is shown in the axis on the right, assuming even flavor ratios. We assume a ten-year exposure for planned experiments, unless otherwise noted. Colors and references for the experimental sensitivities are the same as in Fig.~\ref{fig:experiments_table}. The top panel compares experiments embedded in water and ice, which are sensitive to all flavors and use event topology to identify tau neutrinos. The middle panel shows experiments that observe air showers induced by neutrinos from valleys, mountains, and planar arrays with the number of stations for each experiment in the legend. The bottom panel compares balloon-borne experiments and satellites with the flight times and configurations listed in the legend. The blue bands show the astrophysical neutrino flux measured by IceCube ($\nu_\mu$~\cite{Stettner:2019tok} in solid band and high-energy starting events~\cite{IceCube:2015gsk} in hatch) and solid lines show experimental upper limits at higher energies~\cite{PierreAuger:2019ens, ARA:2019wcf, Anker:2019rzo,Gorham:2019guw}. The experimental sensitivities are compared to a range of model expectations for both cosmogenic neutrinos~\cite{vanVliet:2019nse, Muzio:2021zud} and astrophysical neutrinos~\cite{Fang:2017zjf,Biehl:2017hnb}.
    \label{fig:UHE_diffuse_sensitivity}}}
\end{figure}

\subsection{Experiments}
\subsubsection{In-ice or water optical experiments}\quad\\
\noindent \textbf{IceCube/IceCube-Gen2:}
The IceCube Neutrino Observatory~\cite{IceCube:2016zyt} is located near the geographic South Pole.
IceCube instruments one cubic kilometer of Antarctic glacial ice and is the largest instrumented volume neutrino detector currently operating. IceCube consists of 86 cables called ``strings" which are equipped with 60~Digital Optical Modules (DOMs) deployed between 1450~m and 2450~m deep in the ice. The typical horizontal distance between strings is 125~m, except in the more densely instrumented ``DeepCore'' region in the center of the detector~\cite{IceCube:2011ucd}. The DOM consists of a glass pressure vessel housing a single downward-facing 10~inch photomultiplier tube (PMT)~\cite{IceCube:2010dpc}, flasher LEDs for calibration~\cite{IceCube:2013llx} and digitizing electronics~\cite{IceCube:2008qbc}. The PMT signal is digitized with a sampling rate of 300 MSPS and the digitized signals are given a global time stamp accurate to 2~ns. IceCube completed construction in 2010 and operates with over 99\% uptime. IceCube uses the timing, location and amount of light deposited in the detector in order to reconstruct the arrival direction, time, energy, and flavor of neutrinos interacting in the ice. 
        
IceCube uses several methods to search for $\nu_{\tau}$. Above 1~PeV, $\nu_{\tau}$ can create a ``double cascade'' in the ice, however such high-energy events are rare. In order to search for $\nu_{\tau}$ at lower energies where the flux is higher, IceCube uses a likelihood reconstruction to determine whether cascade-like events are single-cascades or double cascades. Additionally, at energies between 100~TeV and 1 PeV, a $\nu_{\tau}$ interacting near a DOM can create a “double pulse” (DP), which is visible in the digitized signal of individual DOMs~\cite{IceCube:2015vkp}. Based on the diffuse flux of cosmic neutrinos measured by IceCube and the expectation of equipartition in flavors detected at Earth, IceCube expects to see 0.2 identifiable $\nu_{\tau}$ events per year and new techniques may increase this sensitivity. IceCube recently identified two tau neutrino candidates in 7.5~years of data~\cite{IceCube:2020abv}. One of these events passes both double cascade and double pulse search criteria~\cite{Wille:2019pub,Meier:2019ypu}. The largest systematic uncertainty for $\nu_{\tau}$ searches in IceCube is due to modeling of the optical properties of the ice, which remains an area of intensive study in IceCube~\cite{IceCube:2021xhi}. Updated ice modeling and other updated calibrations~\cite{IceCube:2020nwx} can be applied retroactively to the entire IceCube data-set, enabling re-analysis with higher sensitivity.
        
        The forthcoming IceCube Upgrade~\cite{Ishihara:2019aao} will deploy 7~new strings in the center of IceCube. Although these strings will not in themselves add to IceCube's instrumented volume, the Upgrade includes a suite of dedicated calibration light sources which will reveal the optical properties of the ice in unprecedented detail, which will be especially beneficial to $\nu_{\tau}$ searches. The Upgrade will use multi-PMT modules similar to KM3NET~\cite{IceCube:2019anq,IceCube:2021xht}, which will also serve as a testbed for a future next-generation neutrino detector.
        
        The envisioned expansion of IceCube is IceCube-Gen2~\cite{IceCube-Gen2:2020qha}, a broad-band neutrino observatory designed to resolve the sources of cosmic neutrinos. The detector includes a radio component (described below) and an expansion to the existing optical array, as well as a surface air shower array. The optical array will consist of 120~strings, drilled to a depth of 2.6~km, spaced 240~m apart. The array will be sensitive to 5~times fainter neutrino sources than the current IceCube detector, and the array is expected to see one $\nu_{\tau}$ event per year, a factor of 5 increase over the current rate. IceCube-Gen2 is expected to set very strong constraints on cosmic neutrino source models and BSM physics which affect flavor ratio after 10~years of operation.
        
\noindent \textbf{\emph{KM3NeT}:} 
        The KM3NeT research infrastructure \cite{KM3Net:2016zxf} is under construction in the deep Mediterranean Sea and hosts two underwater Cherenkov detectors: KM3NeT/ARCA and KM3NeT/ORCA  (Astro-particle / Oscillation Research with Cosmics in the Abyss). Both instruments use segmented sensors, called Digital Optical Modules (DOMs), housing 31 small 3-inch PMTs each. 18 vertically aligned DOMs are integrated on Detection Units anchored on the sea bed. The multi-PMT design provides a homogeneous $4\pi$ effective area while allowing for photon counting and signal arrival direction sensitivity. The spacing between DOMs is optimized to meet the different science goals of the two detectors: With a vertical spacing of 36\,m between DOMs and a horizontal spacing of $\sim\SI{90}{m}$ between Detection Units, KM3NeT/ARCA will instrument $\sim\SI{1}{km^{3}}$ of sea water with optimum sensitivity to TeV--PeV astrophysical neutrinos and good visibility towards the center of our Galaxy. KM3NeT/ORCA on the other hand is more densely instrumented (9.3\,m vertical, 20\,m horizontal) to detect
        neutrinos with a threshold of few-GeV and achieve resolutions close to the intrinsic limit of neutrino--nucleon scattering \cite{Adrian-Martinez:2016zzs}. 

        KM3NeT/ARCA is optimized to search for astrophysical neutrino sources up to $\sim$PeV energies. The Collaboration plans to instrument a volume of $\sim$1\,km$^{3}$ of sea water with 230 Detection Units. Analyses of data with the first deployed Detection Units start to see first neutrino candidates \cite{Sinopoulou:2021fom}.

        In KM3NeT, rather than storing the entire waveform for photon pulses, the analogue signals from the PMTs are digitized inside the DOMs and a start time, at which the signal surpasses a threshold, and the time-over-threshold is returned. Hence, double-pulse signatures as used in current IceCube analysis are therefore typically not observed on a single PMT-level. Due to the longer scattering length in sea water compared to ice however, more photons are registered that have not scattered before reaching the sensors. This allows to use the individual timing of the first photons seen in the different PMTs and the direction information for event reconstruction.  A maximum likelihood reconstruction method for double-cascades has been developed \cite{vanEeden:2021onr} which exploits these features. This reconstruction method targets double-cascades with a large separation between the neutrino interaction and the $\nu_\tau$ decay expected for high-energy $\nu_\tau$s. Based on simulated data, the angular resolution of the developed double-cascade reconstruction method achieves sub-degree level for tau lengths larger than 25\,m. Error spreads of 3.17\,m on the reconstructed $\tau$ length and 13\% on the visible energy can be reached \cite{vanEeden:2021onr}.

\noindent \textbf{\emph{Baikal-GVD}:}
        The Baikal Gigaton Volume Detector (Baikal-GVD)~\cite{Baikal-GVD:2021zsq} is a cubic-kilometer scale neutrino telescope under construction in Lake Baikal, the deepest freshwater lake in the world. The lake water has an optical absorption length of 22-24 m and a scattering length of 30-50 m. Baikal-GVD is installed approximately 3 - 4 km from shore at depths of 750 - 1275 m.  The Baikal-GVD optical module contains a single downward-facing 10-inch photomultiplier tube. The OMs are arranged in sections of 12, spaced at 15 m vertically. The detector is deployed in “clusters” of 8 strings spaced 60 m apart horizontally, each with 3 sections, for a total of 288 OMs per cluster. As of 2021, 8 clusters are deployed with an instrumented volume of about 0.4 km$^3$.  The goal is to deploy a total of 14 clusters by 2024. The Baikal-GVD collaboration is investigating both the double pulse and double cascade methods for $\nu_{\tau}$ detection~\cite{Baikal-GVD:2021ugo}. Double cascade events can occur in single clusters or as combinations of events from multiple clusters.

\noindent \textbf{\emph{P-ONE}:}
The Pacific Ocean Neutrino Experiment (P-ONE)~\cite{P-ONE:2020ljt} is a planned multi-cubic-kilometer neutrino telescope in the Pacific Ocean. P-ONE partners with the  Ocean Networks Canada (ONC), which operates the North East Pacific Time-series Underwater Networked Experiment (NEPTUNE) ocean observatory. NEPTUNE provides fiber-optic power and data infrastructure for the telescope. The NEPTUNE node in the Cascadia Basin, at a depth of 2600~m, has been selected as the site for P-ONE. The geometry is being optimized for horizontal muon neutrino induced tracks, where neutrino telescopes have the greatest sensitivity to high energy neutrinos, due to the lack of downgoing background from cosmic ray air showers and the attenuation of the Earth on upping high energy neutrinos. P-ONE is primarily designed for neutrino source searches, and the optimization for tracks also takes advantage of the superior angular resolution of the track channel. The detector design is segmented rather than evenly filled in, the final design calls for 7 segments, each with 10 strings.

An initial pathfinder called STtrings for Absorption length in Water (STRAW) has been deployed in the Cascadia basin to qualify the site~\cite{STRAW:2018osc}. STRAW-a, deployed in 2018, measured the. optical attenuation length at the site. The measured attenuation length is 28 metres at 450 nm, which qualifies the site for deployment of P-ONE~\cite{Bailly:2021dxn}. The STRAW lines include POCAMs (Precision Optical Calibration Modules)~\cite{Fruck:2019vam}, calibration devices which were designed for the IceCube Upgrade, as well as digital optical modules. STRAW-b, deployed in 2020, was designed to perform a full characterization of the water optical properties and light background of the Cascadia Basin site, with a focus on the bioluminescence emission spectrum~\cite{strawbref}. A prototype P-ONE line is under construction. The prototype will verify the design of the P-ONE optical module concept, a segmented multi-PMT design similar to that used for KM3NET and IceCube-Gen2. The prototype will also verify the mechanical structure and deployment procedure. The P-ONE Explorer, the first segment of 10 lines, is planned for deployment in
2023-2024 with the full detector planned for deployment by 2030~\cite{P-ONE:2020ljt}.

Although P-ONE is not optimized for cascade detection, the collaboration is exploring the possibility of detecting $\nu_{\tau}$ using the double pulse method~\cite{Katilwnppc:2021}. Initial studies with a double pulse algorithm indicate that a single 10-line segment of P-ONE can detect about 0.3 $\nu_{\tau}$ per year.

\subsubsection{In-air optical or fluorescence experiments}\quad\\
        
\noindent \textbf{\emph{Trinity}:} \emph{Trinity} is a proposed system of 18 air-shower Cherenkov telescopes optimized for detecting Earth-skimming neutrinos with energies between $10^6$\,GeV and $10^{9}$\,GeV \cite{Brown:2021lef,Wang:2021zkm}. \emph{Trinity} is an evolution of the original Earth-skimming concept, which proposed monitoring nearby mountains with Cherenkov telescopes \cite{Bertou:2001vm,Fargion:2003kn,Feng:2001ue}. The concept has been tested and validated by several groups. Two recent examples are NTA \cite{Ogawa:2021irx} and MAGIC \cite{MAGIC:2018gza}. A major benefit of air-shower imaging is the ability to reconstruct the arrival direction of the imaged air-shower with arcminute resolution and to estimate its energy with a few ten percent uncertainty. It is thus not surprising that air-shower imaging is widely used in very-high-energy gamma-ray and ultrahigh-energy cosmic ray observations since three decades \cite{Weekes:1989tc}.
        
\emph{Trinity}'s low-energy PeV threshold provides an overlapping energy range with IceCube. This unique feature allows studies of, for example, the astrophysical neutrino spectrum in regions of the sky (declinations $-75^\circ$ to $55^\circ$ \cite{Wang:2021zkm}) of which most is not accessible with in-ice or atmospheric radio experiments. That is because radio in atmosphere becomes sensitive only at $\sim10^{8}$\,GeV, and in-ice radio experiments, while sensitive down to $\sim10^{7}$\,GeV, have a limited sky acceptance due to their locations close to the poles. \emph{Trinity}, therefore, closes an important observational gap.
        
Trinity's telescopes are located on mountains in 2-3\,km altitude and point at the horizon. From there, they are sensitive to air-showers initiated by Earth-skimming tau neutrinos in distances of up to 200\,km \cite{Otte:2018uxj}. The ability to detect these very distant showers close to the threshold compensates for the 20\% duty cycle boosting \emph{Trinity}'s acceptance beyond radio below $10^8$\,GeV. 
        
\emph{Trinity}'s telescopes are optimized to deliver the best possible detection sensitivity per cost \cite{Otte:2019aaf}. A key feature is the extreme and unique 60-degree wide-field optics \cite{Cortina:2015xra}. The telescopes use several demonstrated key technologies, which lower costs and improve performance. For example, the mirrors are lightweight and fabricated in the same technology used by the Cherenkov Telescope Array. High efficient, mechanical, and optical robust silicon photomultipliers populate the focal plane outperforming classical photomultipliers. The digitizer system has been developed for high-energy physics experiments focusing on low cost and high-channel density. These technologies make it possible to build a high-performance system for a fraction of the cost of a conventional Cherenkov telescope and on par with the projected costs of proposed UHE-neutrino radio detectors.
        
\emph{Trinity} will be constructed on three different sites with six telescopes at each site. The location of the three sites at different longitudes will allow for continuous monitoring of a large fraction of the sky.
        
\emph{Trinity}'s sensitivity improves inverse proportional with the number of telescopes. One single telescope has the sensitivity to detect astrophysical neutrinos within five years if the spectrum extrapolates from IceCube energies without a cut-off.
        
\emph{Trinity} is currently in its demonstrator phase \cite{Brown:2021lef}. A $1\,$m$^2$ air-shower Cherenkov telescope will be deployed on Frisco Peak, UT in 2022. The demonstrator will validate the concept and camera technologies planned for a full \emph{Trinity} telescope.

\noindent \textbf{POEMMA:} (Probe Of Extreme Multi-Messenger Astrophysics) is a proposed NASA Astrophysics Probe-class, space-based mission that aims to represent the next generation of Ultra-High Energy Cosmic Ray (UHECR) and UHE neutrino detectors \cite{POEMMA:2020ykm}. Over a 5-year period, POEMMA will definitively measure high altitude backgrounds and cosmic rays, and either detect the first neutrino with $E_{\nu}>20$~PeV for tau neutrinos and $E_{\nu}>20$~EeV for all flavors of neutrinos or set stringent limits on both the diffuse flux and certain transient source models.
        
The POEMMA design consists of two identical telescopes that fly in a loose formation at 525~km altitude, separated by an average lateral distance of 300~km. Each POEMMA telescope utilizes a Schmidt optical system with a $6 \, \mathrm{m}^{2}$ collecting area and a $45^{\circ}$ field of view. The focal surfaces of these telescopes are divided into two sections, with each section dedicated to a different measurement scheme. The POEMMA Fluorescence Camera (PFC) occupies 80\% of the focal surface and is optimized to record the fluorescence light from EAS initiated by UHECR in the atmosphere. The POEMMA Cherenkov Camera (PCC) occupies the remaining 20\% of the focal surface and is oriented to observe near the Earth-limb optimized to measure the Cherenkov light produced by EAS sourced from Earth-skimming neutrino interactions. The POEMMA telescopes can slew in both azimuth ($90^{\circ}$ in $\sim 8$ minutes) and zenith, allowing for unprecedented follow-up on transient astrophysical events by tracking sources as they move across the sky \cite{Venters_ToO}. The separation of the POEMMA spacecraft can also be decreased to $\sim$25 km to put both telescopes in the upward-moving EAS light pool, thus reducing the neutrino detection energy threshold. The orbital period of the POEMMA telescopes is 95 minutes. Because of this fact, one of the main advantages of the POEMMA mission is being able to achieve full-sky coverage for both UHECR and UHE neutrino sources.
     
There are two science modes of POEMMA.  The first is a precision UHECR and UHE neutrino stereo mode where the telescope configuration is oriented to co-measure the EAS air fluorescence signal in a common volume corresponding to nearly $10^{13}$ tons of atmosphere. Due to the high accuracy of the EAS reconstruction from the stereo fluorescence technique and large field-of-view from LEO, POEMMA can accurately reconstruct the development of the EAS with $\lesssim 20^\circ$ angular resolution, $\lesssim 20\%$ energy resolution, $\lesssim 30$ g/cm$^2$ X$_{\rm Max}$ resolution \cite{Anchordoqui:2019omw}. This yields excellent sensitivity for all neutrino flavors for UHE EAS that begin deeper in the atmosphere and well separated from the dominant UHECR flux. 

At energies above a EeV, $\sim$80\% of the neutrino energy is put into the emergent lepton and $\sim 20$\% put into a hadronic cascade for both charged-current and neutral current interactions  \cite{Gandhi:1998ri}. Thus the properties of the composite EAS are determined by the emergent UHE lepton for both charged- and neutral-current interactions in the atmosphere. 
     
The second has the telescopes pointed to view slightly below the limb of the Earth to be sensitive to the beamed optical Cherenkov signal from upward-moving EAS sourced from tau neutrino interactions in the Earth. This $\nu_{\tau} \rightarrow \tau$ detection channel allows POEMMA to have with sensitivity to energies $E_{\nu}>10$~PeV. Using this observation channel, for an idealized $2 \pi$ azimuth telescope configuration with POEMMA optical Cherenkov performance, improves on the limits of the diffuse neutrino flux set by IceCube by roughly an order of magnitude for $E_{\nu}>$100~PeV \cite{Krizmanic_2019, Cummings:2020ycz}. In principle, POEMMA is also sensitive to Earth-emergent neutrinos through the $\nu_{\mu} \rightarrow \mu$ and $\nu_{\tau} \rightarrow \tau \rightarrow \mu$ interaction channels. These channels improve POEMMA's sensitivity for $E_{\nu}<10$~PeV due to the relatively long interaction lengths of the muon around 1 PeV, and in the case of the primary $\nu_{\mu}$, increased Earth emergence probabilities.\\
\indent While POEMMA has increased sensitivity to the diffuse neutrino flux with respect to current generation ground-based experiments, it is unlikely that it will observe diffuse neutrinos during the 5 year mission due to its $\sim 30^\circ$ azimuth coverage. However, due to its unique observational capabilities, including slewing the telescopes to the location of an astrophysical transient event, it is expected that, compared to ground-based experiments, POEMMA will improve upon the sensitivity to long-burst transient events (duration of $10^{5-6}$~s: e.g., binary neutron star mergers and tidal disruption events) for $E_{\nu}>100$~PeV by nearly an order of magnitude and to short-burst (duration of $\sim 10^{3}$~s: e.g, short gamma ray burst with extended emission) by at least an order of magnitude \cite{Venters_ToO}.
\\
\noindent \textbf{EUSO-SPB2:} The Extreme Universe Space Observatory aboard a Super Pressure Balloon 2 (EUSO-SPB2) is the follow-up mission to EUSO-Balloon (2014) and EUSO-SPB1 (2017). It is a pathfinder mission for future balloon and space-based observatories \cite{Wiencke:2019vke}. EUSO-SPB2 builds on the lessons and technologies developed during previous EUSO missions such as EUSO-Balloon, EUSO-SPB1, EUSO-TA, and Mini-EUSO and employees the technologies utilized for POEMMA in a near-space environment. EUSO-SPB2 will measure high altitude backgrounds and cosmic rays via both fluorescence and optical Cherenkov emission, thereby validating the detection method.\\
\indent EUSO-SPB2 is will fly two telescopes.  A Fluorescence Telescope (FT)  will point downwards and is optimized to measure the fluorescence light from EAS initiated by UHECR. A Cherenkov Telescope (CT) will point near the Earth limb to measure atmospheric backgrounds and the optical Cherenkov emission from cosmic rays. The Field of View (FOV) of the EUSO-SPB2 fluorescence camera is $11^{\circ} \times 35^{\circ}$, while that of the Cherenkov camera is $6.4^{\circ} \times 12.8^{\circ}$. The EUSO-SPB2 flight train will include an azimuth rotator configured for day and night pointing. Together with an elevation angle tilting mechanism on the CT, it will allow "target of opportunity" follow-up searches for neutrinos with the CT in response to selected alerts of astrophysical transients, for example from gamma ray bursts or gravitational wave events as the source direction crosses the Earth's limb. EUSO-SPB2 will fly at an altitude of 33~km for an expected 100 day duration, and is targeting a 2023 launch from Wanaka, NZ.\\
\indent EUSO-SPB2's observation of the background near the Earth limb is vital for the success of future missions, as there are currently no high altitude measurements of backgrounds on the timescales ($\mathcal{O}$(ns)) and wavelength range (300~nm to 1000~nm) relevant for neutrino detection. The Cherenkov camera of EUSO-SPB2 will observe partially above the Earth limb and is expected to measure hundreds of direct cosmic ray events per hour of detector live time for $E>1$~PeV \cite{Cummings:2021_CR}. These events share many characteristics with the signals generated from neutrino sourced events and are a prime candidate for event reconstruction and evaluation of detector optics, electronics, and triggering algorithms. EUSO-SPB2 will also search for possible background events, such as downward going showers reflected off of the ocean and atmospherically refracted cosmic ray events. The results of the EUSO-SPB2 mission will help constrain the optimal configuration of future balloon and space-based missions.

\subsubsection{In-ice radio experiments}\quad\\

\noindent \textbf{ARIANNA:} The ARIANNA experiment explores the usage of the in-ice radio technique for the detection of UHE neutrinos  \cite{COSPAR2019}. In a uniquely radio quiet area on the Ross-Ice-Shelf in Antarctic, a hexagonal array of pilot-stations has been taking data for several years. In addition, two detector stations have been installed at the South Pole. The radio detector stations consist of two parallel pairs of high-gain downward pointing log-periodic dipole antennas (LPDAs) a few meters below the ice surface. Additional upward pointing LPDAs were added for a cosmic-ray detection and vetoing of anthropogenic noise. ARIANNA derived a limit on the high-energy neutrino flux which demonstrates the feasibility of the in-ice radio detection technique \cite{Arianna:2021vcx}. The ARIANNA detector is also a test bench for future detector optimizations relevant for the future IceCube-Gen2, e.g., the detector was optimized through optimizations of the signal chain \cite{Glaser:2020pot} and trigger \cite{Arianna:2021vcx}. Furthermore, reconstruction algorithms for the neutrino energy, direction and flavor were developed, and probed with in-situ measurement using radio emitters that are lowered into the ice, as well as through the measurement of cosmic rays \cite{Anker:2019zcx, GaswintICRC2019, ARIANNA2020Polarization, Stjarnholm:2021xpj, NuRadioReco, GaswintPhD, ARIANNADirectionICRC2021, Barwick2017, NellesICRC2019, Welling2019, Arianna:2021lnr}. 

\noindent \textbf{ARA:} 
The Askaryan Radio Array (ARA) searches for radio emission from neutrino interactions in ice~\cite{Allison:2011wk}. ARA has been operating and expanding since 2011. Consisting of five stations deployed deep in the ice at the South Pole, it achieves a high effective volume for each station, with an array of 16 antennas buried 200 m below the surface of the ice. ARA serves as a testbed for deep radio instruments used in future experiments like IceCube-Gen2. The stations are separated by a large enough distance (2~km) that each station acts as its own independent interferometer. The ARA collaboration has recently analyzed a subset of its full dataset placing strong limits on the neutrino flux at EeV energies with the expectation of stronger limits to come in the near future~\cite{ARA:2019wcf}. One station, the Phased Array on ARA5, uses a novel phased array technique to lower the energy threshold of the instrument~\cite{Allison:2018ynt, Hughes:2021ndi}. Additionally, the effective volume of ARA may be enhanced by up to 25\% through the observation of secondary leptons (including tau leptons) from charged current interactions~\cite{Garcia-Fernandez:2020dhb}.

There are some indications that flavor may be tagged in ARA and other Askaryan detectors by separating electromagnetic and hadronic showers~\cite{Lai:2013kja, Garcia-Fernandez:2020dhb}. With ARA's large exposure, such studies warrant further scrutiny.  Similarly, such searches can be applied to RNO-G (which is currently being built) and the future IceCube-Gen2 radio array.

\noindent \textbf{RNO-G:} 
The Radio Neutrino Observatory in Greenland (RNO-G) is an in-ice detector that measures neutrinos through the Askaryan emission generated by in-ice showers~\cite{RNO-G:2020rmc}. The 3~km deep ice sheet above central Greenland with attenuation length of approx. 1~km at the relevant frequencies of 100MHz – 1~GHz provides a good target material for achieving large effective volumes. The ice will be instrumented with a sparse array of 35 autonomous radio detector stations with a separation of 1.5~km. The stations are solar powered with additional wind generators under development to power the stations during the dark winter months. The stations are connected through an LTE network to SUMMIT station. The first three stations have been installed in 2021, the remaining stations will be installed over the next three years. 

Each station is equipped with total of 24 antennas. An interferometric phased array provides a low-threshold trigger, consisting of 4 bicone antennas installed in close proximity vertically above each other at 100 m depth \cite{Allison:2018ynt}. Additional bicone (vertical signal polarization) and quad-slot (horizontal signal polarization) antennas above the phased array and horizontally displaced on two additional strings provide additional information to reconstruct the properties of the neutrino \cite{Aguilar:2021uzt}. LPDA antennas are installed close to the surface providing additional neutrino sensitivity to neutrinos with complementary uncertainties. Each station also has three upward facing LPDAs to veto and measure radio emission of air showers which provide in-situ calibration signals \cite{Anker:2021mcf}. 

Due to is relatively low latitude of 72\,deg, RNO-G observes the majority of the Northern sky within 24 hours adding UHE neutrino information to multi-messenger observations. Its diffuse flux sensitivity is large enough to start probing the parameter space of GZK neutrino production (see Fig.~\ref{fig:UHE_diffuse_sensitivity}). Furthermore, RNO-G will be a technical testbed and pave the way for the much larger radio detector array foreseen for IceCube-Gen2.
\\
\noindent \textbf{IceCubeGen2 Radio:}
To extend the energy reach to EeV energies, IceCube-Gen2 will comprise a sparse array of radio detector stations next to its optical component \cite{IceCube-Gen2:2020qha}. The radio technique allows for a cost-efficient instrumentation of the large volumes required to measure the low flux of UHE neutrinos. The ice at the South Pole provides an optimal target material for radio detection with attenuation lengths of more than \SI{2}{km} close to the surface where the ice is coldest. 
The radio array will cover an area of approximately \SI{500}{km\squared} with more than 300 radio detector stations. The proposed array consists of two types of radio detector stations that measure and reconstruct neutrino properties with complementary uncertainties to maximize the discovery potential by mitigating risks and adding multiple handles for rare background rejection \cite{IceCube-Gen2:2021rkf}. The radio component of IceCubeGen2 uses shallow stations with high-gain antennas near the surface of the ice and hybrid stations with additional deep, isotropic antennas deployed in boreholes up to 150~m deep, building on experience from RNO-G, ARA, and ARIANNA.
All shallow components are also equipped with upward facing LPDA antennas which provide sensitivity to cosmic rays to veto air-shower induced background \cite{Garcia-Fernandez:2020dhb, Glaser:2021hfi}, as well as to monitor the detector performance (see e.g.\ \cite{Arianna:2021lnr}).
\\
\indent The IceCube-Gen2 radio array will provide sufficient sensitivity to probe GZK neutrino production. The Gen2 sensitivity would reach the current best-fit models to CR data as measured by the Pierre Auger Observatory, assuming sources identical in CR luminosity, spectrum and composition, as well as a rigidity-dependent cut-off and thereby essentially no protons at the highest-energies \cite{AlvesBatista:2018zui, Heinze:2019jou}. In an only slightly more favorable scenario of 10\% protons, IceCube-Gen2 will detect at least 2 events per year above $\sim$100\,PeV. 
\\
\indent For an unbroken astrophysical neutrino spectrum that follows $E^{-2.28}$, as the one shown
in Figure~\ref{fig:UHE_diffuse_sensitivity}, the radio detector of Gen2 will measure close to ten neutrinos per year where most detected neutrinos will have energies between \SI{e17}{eV} and \SI{e18}{eV} \cite{IceCube-Gen2:2021rkf}.
Due to its location at the South Pole, the instrument continuously observes the same part of the sky with most sensitivity between $\delta \approx \SI{-40}{\degree}$ and $\delta \approx \SI{0}{\degree}$ as the Earth is opaque to neutrinos at ultra-high energies (UHEs). The instantaneous sensitivity will allow to explore neutrino production in transient events such as neutron-star mergers. 
\\
\indent The large sensitivity to neutrinos arriving from and slightly below the horizon enables the measurement of the neutrino-nucleon cross section at extremely high energies. Also a measurement of the inelasticity seems possible through the detection of high-quality electron neutrino charge-current interaction where the energy of the hadronic shower induced by the breakup of the nucleus can be measured separately from the electromagnetic shower induced by the electron. Furthermore, the production of high-energy muons in air showers with energy beyond PeV can be probed through a coincident measurement of the air shower via the in-air radio emission and a muon induced particle shower in the ice via the Askaryan emission. 
        
\subsubsection{In-air radio experiments}\quad\\
        
\noindent \textbf{ANITA}: The Antarctic Impulsive Transient Antenna (ANITA) was a
balloon-borne ultrahigh-energy neutrino observatory first proposed in the early
2000s that flew four successful Antarctic flights between 2006 and 2016
(ANITA-I, -II, -III and -IV, respectively)~\cite{ANITA:2008wdk, ANITA:2010hzc, ANITA:2019wyx,ANITA:2018vwl}. ANITA was originally designed to
detect the Askaryan radiation from neutrino-induced showers in the Antarctic ice
but was also, coincidentally, sensitive to multiple other detection channels for
both UHE neutrinos and UHECRs. In total, ANITA had four \textit{science}
channels: a) UHE neutrino detection via the Askaryan radiation produced by
neutrino-induced showers in the ice (primary channel); b) detection of UHECRs by
observing the geomagnetic radio emission \textit{after} it \textit{reflects} off
the Antarctic ice; c) detection of \textit{Earth-skimming} UHECRs, whose
trajectories never intercept the ground, via the geomagnetic emission created in
the atmosphere; and d) detection of UHE Earth-skimming tau neutrinos, $\nu_{\tau}$,
via the geomagnetic radiation produced by the in-air decay of the tau lepton.

While sub-orbital balloon instruments, like ANITA, have shorter livetimes
(typically 20-30 days) than ground based experiments like Auger and IceCube,
they compensate for this by having the largest instantaneous effective areas due
to their unique observing geometry: ANITA, when at its nominal float altitude of
${\sim}37$~km, can instantaneously instrument more than one million cubic
kilometers of ice. However, the large distance to the payload from a in-ice
neutrino interaction (in excess of tens or hundreds of kilometers) means
sub-orbital experiments have typically have a higher energy threshold for
detection. For example, ANITA-IV, turned on rapidly at neutrino energies of
${\sim}1$~EeV (for the Askaryan channel), exceeded the neutrino effective area of
Auger by ${\sim}10$~EeV, and continues to grow up to ${\sim}1$~ZeV. The only current
technique with larger instantaneous effective area to UHE neutrinos is the lunar
skimming techniques (Section Lunar) but these currently probe energies above
${\sim}1$~ZeV where few current models predict significant neutrino fluxes (Section
Lunar)~\cite{James:2008ed}

Each ANITA flight flew between 16 and 48 dual-polarization quad-ridged horn
antennas with a nominal bandwidth of 200 MHz to 1.2 GHz. While ANITA relied on a
combinatoric power-sum trigger to detect events during the flight, post-flight
analysis, including pulse-phase interferometry, or beamforming, of the received
signals, allowed ANITA to reconstruct the direction of the incoming
radio-frequency pulse to within $0.1^{\circ}$-$0.2^{\circ}$; the conversion of this high
angular resolution into a neutrino or UHECR direction depends on the detection
channel but can be as as good as ${\sim}1^{\circ}$ for the in-air $\tau$-decay channel.
In addition to the waveforms \& spectra of each event, ANITA uses the
\textit{polarization} and \textit{polarity} of the incident electric field to
reconstruct events. For a unipolar waveform, the polarity is determined by the
sign~($\pm$) of the impulse. For bipolar waveforms, the polarity is typically
indicated by the order of the two primary poles (i.e.\ $+,-$ or $-,+$). Due to
the Fresnel reflection coefficient at the air-ice boundary, reflected EAS
signals have a completely inverted polarity with respect to the signals observed
directly from an EAS without reflection~\cite{Schoorlemmer:2015afa}.
\textit{Polarity}, which is related to the \textit{sign} of the electric field
impulse, is distinct from \textit{polarization}, which describes the
\textit{geometric orientation of the electric field} and is used to separate EAS
events from in-ice Askaryan neutrino events. Over its four flights, ANITA has
observed seven direct events and 64 reflected UHECR
events~\cite{ANITA:2010ect,Schoorlemmer:2015afa,ANITA:2020gmv}

During the first and third flights (ANITA-I and ANITA-III, respectively), ANITA
observed two steeply upcoming~($\sim30^{\circ}$~below the radio horizon) events that
did \textit{not} show the characteristic polarity inversion upon reflection and
were observationally consistent with an \textit{upgoing} extensive air
shower~\cite{ANITA:2016vrp,ANITA:2018vwl}. While this $\nu_{\tau}-\tau$ EAS
was initially considered to be unlikely due to the attenuation of neutrinos
across the long chord lengths through Earth at these steep angles, several
analyses have studied the $\nu_{\tau}$-origin hypothesis for these
events~\cite{ANITA:2016vrp, IceCube:2020gbx}. A prior analysis by the ANITA
collaboration of these events under a diffuse $\nu_{\tau}$ flux
hypothesis~\cite{Romero-Wolf:2018zxt,ANITA:2021xxh} implied a
diffuse neutrino flux limit that is in strong tension with the limits imposed by
the IceCube~\cite{IceCube:2016uab} and Pierre
Auger~\cite{PierreAuger:2015ihf} observatories.

A follow-up analyses by the ANITA collaboration estimated the
sensitivity of ANITA to $\nu_\tau$ point sources to investigate the possibility that
a point-like neutrino source could be responsible for these steep  events~\cite{ANITA:2021fuf}. This
analysis determined that the \textit{instantaneous} effective area of ANITA to tau neutrino point sources is  significantly smaller than Auger's $\nu_\tau$ point source peak effective area at steep angles and that are still in strong tension with point-like neutrino limits set by Auger~\cite{PierreAuger:2019azx}.

A number of alternative hypothesis have been proposed to explain these
theoretically challenging events. These range from Beyond Standard Model~(BSM)
physics~\cite{Borah:2019ciw,Chauhan:2018lnq,Collins:2018jpg,Esmaili:2019pcy,Esteban:2019hcm,Fox:2018syq,Heurtier:2019git,Heurtier:2019rkz,Hooper:2019ytr} to more mundane effects such as transition radiation of cosmic ray air showers
piercing the Antarctic ice sheet~\cite{deVries:2019gzs} and subsurface
reflections due to anomalous ice features~\cite{Shoemaker:2019xlt} although
the latter has recently been experimentally constrained by the ANITA
collaboration~\cite{Smith:2020ecb}. As of this whitepaper, the origin of these
anomalous events remains unknown (see sec.~\ref{sec:anita} for a possible explanation).

The fourth flight of ANITA, ANITA-IV, also observed four extensive air
shower-like events that have the same polarity as above-horizon UHECRs (i.e.\
non-inverted implies a non-reflected radio signal), but reconstruct
\textit{below} the horizon where ANITA has significant exposure to UHE
$\tau$-induced EAS. These four events therefore appear to be upward-going air
showers emerging from the surface of the Earth, but unlike the ANITA-I and
ANITA-III anomalous events described previously, the ANITA-IV events reconstruct
very near to, but below the horizon ($\lesssim1^{\circ}$)~\cite{ANITA:2020gmv}. The
significance of finding 4 events with a polarity inconsistent with their
geometry out of the 27 air shower events with a well-determined polarity was
estimated to be greater than 3\,$\sigma$ by the ANITA collaboration, when considering
the possibilities that: the events could be an anthropogenic background; that
there might be an error in the reconstructed arrival direction; and that the
polarity might be misidentified~\cite{ANITA:2020gmv}. An analysis of these
events under a $\nu_{\tau}$ hypothesis found that while these events are consistent
with Monte Carlo simulations of $\nu_{\tau}$ detections by ANITA in terms of their
locations and spectral properties, the fluence implied by the detection of
${\sim}4$ of these events in ANITA-IV is in strong tension with Auger across all
energies and is also in tension with ANITA's own Askaryan limits above
$10^{19.3}$~eV~\cite{ANITA:2021xxh}.
\\

\noindent \textbf{PUEO}: The Payload for Ultrahigh Energy Observatory (PUEO) is the
direct successor of the ANITA instrument~\cite{PUEO:2020bnn, Deaconu:2019rdx}. The main
instrument follows the same overarching design of the ANITA instrument with
significant changes to the internal electronics and improvements to the signal
chain to \textit{significantly} increase the sensitivity in \textit{all} of
ANITA's four detection channels (Askaryan, above-horizon stratospheric UHECR,
reflected UHECR, and $\tau$-induced EAS). Like ANITA, PUEO will be sensitive to tau neutrinos through radio emission from both in-ice and in-air induced showers.

Unlike ANITA, PUEO is comprised of \textit{three} sub-instruments: a) the ``main
instrument'', the direct successor to ANITA is composed of 96 quad-ridged horn
antennas (compared to the 48 in ANITA-IV), now target the 300 MHz to 1200 MHz
band; b) a multi-channel dedicated ``low-frequency instrument'' targeting the 50
MHz to 300 MHz designed to detect EAS signals, from UHECRs, $\tau$-induced EAS, and
any new steeply upcoming anomalous events; and c) a ``nadir instrument'',
composed of the same quad-ridged horn antennas used in the main instrument, but
directed steeply down towards the ground in order to significantly increase
PUEO's sensitive to steeply upcoming events.

Despite deploying more than twice the number of antennas as the last flight of
ANITA, the biggest improvement in PUEO's sensitivity comes from
\textit{real-time interferometric beamforming} used in the trigger. This system,
which computes highly directional beams on the sky, in real-time, by coherently
summing waveforms with different time delays allows PUEO to have a 50\% trigger
efficiency level below a signal-to-noise ratio (SNR) of 1~\cite{PUEO:2020bnn}.
PUEO's beamforming trigger system is built on the Xilinx RFSoC (Radio-Frequency
System-on-Chip) platform which combine wide-bandwidth digitizers, large
field-programmable gate-arrays (FPGAs), and digital signal processing (DSP)
cores onto a single die.

\noindent \textbf{BEACON}: The Beamforming Elevated Array for COsmic Neutrinos is a detector concept that targets the detection of Earth-skimming tau neutrinos. BEACON can detect  radio-frequency radiation from air showers induced by these Earth-skimming events, and will be deployed at several sites around the world for a full sky coverage~\cite{Wissel:2020sec}.

To achieve this goal BEACON builds on two key factors: first it uses the radio-interferometer technique, an efficient radio technique for high sensitivity observations extensively used in the radio-astronomy community with tremendous results, and second it fully profits of its topography site. 
Each BEACON station is designed to be deployed on top of high elevation mountains and thus access a very large collecting area with a minimal number of antennas ($\sim10$). The instrument therefore achieves a large field of view with a high radio sensitivity which yields a high effective area towards UHE Earth-skimming tau neutrinos. The sensitivity estimates of BEACON reach down $\sim6\times 10^{-10}$\,GeV cm$^{-2}$ s$^{-1}$ sr$^{-1}$ after 3 years of integration and with 1000 stations~\cite{Wissel:2020sec}. The chosen frequency band is 30-80\,MHz but sensitivity studies have shown that a higher frequency band, for instance 200-1200\,MHz could also be suitable~\cite{Wissel:2020sec}. This leaves the possibility to deploy hybrid frequency-band arrays in order to combine them to improve the pointing and the reconstructions capabilities.

The BEACON experiment is currently in the demonstration phase, with a prototype deployed at the White Mountains Research Station in California. It is built with custom 4 crossed electrically-short-dipole antennas and a two stages of amplification. The trigger system is based on a low-power phased array trigger implemented on an FPGA. While the prototype electronics were originally  developed for the ARA experiment~\cite{Oberla:2019tht}, custom modular electronics are under design for future stations. With this instrumentation, the full waveforms are stored in the data set for interferometric reconstruction.

The prototype, too small to envision significant neutrinos detection for the moment, is used as a test-bench for calibration and data analysis on cosmic-ray observations~\cite{BEACON:2021fpe}, simulation study have shown that tens of events are already detectable with its current configuration~\cite{Zeolla:2021cbb}. Once the validation of this prototype is achieved the up-scaling of the experiment can be easily achieved by the deployment of a few hundreds of stations making it ultimately competitive to detect UHE neutrinos~\cite{AlvesBatista:2018zui}.
\\
        
        \noindent \textbf{GRAND}: The Giant Radio Array for Neutrino Detection (GRAND) is a planned large-scale radio observatory of UHE particles: cosmic ray, gamma ray, and neutrino-initiated extensive air showers in the atmosphere. GRAND will detect the geomagnetic radio emission from them.  In particular, GRAND aims to discover the long-sought UHE neutrinos, even if their flux is low $\sim10^{-10}$\,GeV cm$^{-2}$ s$^{-1}$ sr$^{-1}$, by targeting very inclined showers, i.e., showers coming from directions close to the horizon, expected from Earth-skimming UHE tau neutrinos~\cite{GRAND:2018iaj}. To reach this sensitivity, in its final configuration GRAND will consist of a large number of radio antennas - $20$ clusters of $10,000$ antennas each, totalling an instrumented area of $200,000$\,km$^2$. Within each cluster, a large portion of the field of view will be sensitive to the expected emerging directions of the tau-induced showers from Earth-skimming tau neutrinos. GRAND antennas will operate in the $50-200$\,MHz band, and will be installed in radio-quiet, and favorable locations from the point of view of the topography of the site, such as opposing mountain slopes~\cite{Decoene:2019izl}. Their relative elevation and the large number of antennas will allow for sub-degree angular resolution~\cite{Decoene:2021ncf,GRAND:2018iaj}. This will, in principle, allow GRAND to discriminate neutrinos, whose showers are horizontal or slightly upward-going from the large background of downward-going showers initiated mainly by cosmic rays. In addition, the large number of antennas and the large field of view make it a competitive instrument in the transient domain of radio astronomy. 
        GRAND builds on years of experience in the radio-detection of UHE particles demonstrated by previous experiments \cite{Ardouin:2010gz}. Yet, in order to validate key technical aspects, GRAND has a staged construction, with prototype stages focusing on the development and testing of the detector design, but also able to achieve scientific goals in themselves~\cite{GRAND:2018iaj}. The first prototype stage, GRANDProto300, is a $300$-antenna array, designed to develop and validate the autonomous radio detection and reconstruction of very inclined EAS. GRANDProto300 will also study cosmic-ray physics and astrophysics as well as gamma rays and radio astronomy~\cite{Decoene:2019sgx}. The second prototype stage, GRAND10k, will be a $10,000$-antenna array, with a design based on the results from the GRANDProto300 stage. It will be the first large scale cluster of the $20$ envisioned in total. It will demonstrate the feasibility of the radio detection with large-scale and sparse arrays. Finally, a sensitivity simulation study conducted on the southern rim of the Tian Shan mountains have shown that a GRAND10k array deployed there would lead to an integrated UHE neutrinos sensitivity of $8 \times 10^{-9}$\,GeV cm$^{-2}$ s$^{-1}$ sr$^{-1}$ after 3 years~\cite{GRAND:2018iaj}, allowing to detect cosmogenic UHE for optimistic fluxes scenario~\cite{AlvesBatista:2018zui}.
        \\
        
\noindent \textbf{TAROGE-M:} The Taiwan Astroparticle Observatory for Geo-synchrotron Emission at Mt. Melbourne is a synoptic radio array installed atop the high mountains, Mt. Melbourne, Antarctica. The detector is designed to observe radio emissions from EAS induced by the earth skimming tau neutrinos. One of the main goals of TAROGE-M is to probe the upward moving air-showers reported by ANITA ~\cite{ANITA:2016vrp}. This experiment utilizes the advantages of Antarctica, which has the quietest ambient RF noise and the strongest geo-magnetic field required for the radio emissions. Although the target volume in the field of view is smaller due to the lower observation altitude than the balloon-borne experiment, the shorter distance to the shower allows to lower the energy threshold. Furthermore, it is possible to maximize exposure through operation over multiple years with a high duty cycle. 

A station of TAROGE-M consists of a minimal antenna tower (6 to 8 receivers) and is designed for convenient transportation and installation in Antarctic mountain environments ~\cite{Nam:2020hng}. 7dBi log-periodic dipole array (LPDA) antennas are employed for the frequency band of 180 MHz – 450 MHz. Frequencies below 200 MHz are limited because the antenna size corresponding to that frequency is not suitable for transport by helicopter. 

In March 2019, a prototype station (2700m) was deployed near the summit of Mt. Melbourne and noise measurements were made. During the next austral summer season, December 2019-January 2020, the first station consisting of six LPDA antennas was installed and operated for about a month. In December 2019, the first station consisting of six LPDA antennas was installed and operated for about a month. An in-situ calibration performed post-installation using drone-mounted pulses showed angular resolutions of 0.2 degrees and 0.3 degrees for azimuth and elevation, respectively. TAROGE-M detected 7 UHECR candidate events during the operation of the first station, thereby validating the equipment and demonstrating the detection capability of EAS ~\cite{TAROGE-M:Wang}. TAROGE-M plans to deploy 10 stations within 5 years to provide a tau neutrino sensitivity of $ \sim 10^{-6}$\,GeV cm$^{-2}$ s$^{-1}$ sr$^{-1}$ in $ \sim 10^{18} ~ \rm {eV} $ and would confirm ANITA’s upward moving air showers.
\\

\noindent \textbf{Radar Echo Telescope:} The Radar Echo Telescope (RET) is an umbrella project for two distinct instruments, the Radar Echo Telescope for Cosmic Rays, RET-CR~\cite{RadarEchoTelescope:2021rca}, and the Radar Echo Telescope for Neutrinos, RET-N. RET-CR is a pathfinder detector to test the radar echo method in nature, using cosmic rays as a testbeam. A high energy cosmic ray air shower ($> 10^{16}$eV) deposits a fraction of its energy at the earth surface for zenith angles less than 30$^\circ$. The higher the elevation of the ground, the greater the deposited fraction, which at EeV energies can reach 25\%. As fraction, which is highly concentrated at the cascade axis, crosses the air-ice boundary, a dense cascade is produced in the medium, similar to that expected from a neutrino induced cascade. RET-CR will use charge particle detectors at the surface of a high elevation ice sheet, such as Taylor Dome Antarctica, to indicate when a cascade is present, and use these englacial cascades to test the radar echo method in nature via a buried radar system just beneath the surface. 

RET-N is a next-generation UHE neutrino observatory that would, contrary to RET-CR, be located deep in the ice. A central transmitter surrounded by numerous receiving antennas would comprise a station, and several stations will be deployed. RET-N has a projected sensitivity that can complement IceCube Gen-2 optical by having comparable sensitivity in the 10-100\,PeV range. Detailed sensitivity studies are underway at time of writing. 

\subsubsection{Lunar experiments}\quad\\

Several previous experiments using radio telescopes on Earth, at Parkes in Australia, Goldstone in the USA (GLUE), Kalyazin in Russia (RAMHAND), Westerbork in the Netherlands (NuMoon), Parkes and the Australian Telescope Compact Array (LUNASKA), the Extended Very Large Array in the USA (RESUN), the Lovell telescope in the UK (LaLUNA) and the Low Frequency Array in the Netherlands, have developed the necessary techniques to search for radio pulses produced in neutrino and cosmic-ray showers inside the regolith of the Moon (see \cite{Bray:2016xrn} for a review and further references). However, these instruments have lacked the necessary sensitivity to detect the known flux of UHE cosmic rays or neutrinos from such a large distance. This situation will change with the advent of the Square Kilometre Array (SKA) \cite{Carilli:2004nx}. 

\noindent \textbf{SKA:}
The Square Kilometer Array is the largest radio telescope ever conceived. It will consist of multiple components, with the low-frequency telescope SKA-Low covering the range 50 - 350 MHz. Located at the Murchison Radioastronomy Observatory in outback Western Australia, Phase~1 of SKA-Low (SKA1-Low), representing $\sim 10\%$ of the planned capability of the entire telescope array, will consist of an array of 131,072 log-periodic dipole antennas, grouped in 512 stations, each with 256 antennas. Half of these antenna stations will be placed in a dense core of less than \SI{1}{km} diameter at the central region and the rest will span out along three spiral arms stretching 65 kilometres end to end. 
Dual polarisation signals from all antennas in each station will be digitised and added in-phase (added coherently in the voltage domain) to form station beams with a beamwidth $>1.4^\circ$ that can be pointed at the Moon. This instrument is particularly appealing as a UHE particle detector, due to the increased angular width of the radio emission at low frequencies. When complete by 2023, it is estimated that SKA1-Low could detect of order 1 UHECR below $10^{20}$ eV per year of observing whenever the Moon is visible ($\sim$ 2500 hr) which may allow for a first detection \cite{James:2015cav, James:2017pvr}. The estimated rate for Phase 2 of SKA-Low (SKA2-Low) will be of order 100 UHECR per year with potentially sufficient angular resolution to study their origin. The sensitivity of SKA-Low in either phase to UHE$\nu$ will not be sufficient to detect the estimated flux of cosmogenic neutrinos, but both can place strong constraints and test models of UHE$\nu$ production in exotic scenarios beyond the Standard Model \cite{James:2017pvr} above $10^{20}$ eV.  The compact core of SKA1-Low also offers excellent conditions for the radio detection of extended atmospheric showers (EAS) at energies $\sim 10^{16} - 10^{18}$ eV in tandem with an array of particle detectors \cite{Buitink:2021pkz}. In order to enable these observations with a giant radio array, members of previous lunar experiments have joined forces to form the SKA High Energy Cosmic Particles focus group, together with an experiment to perform precision measurements of EAS.

\subsubsection{Shower-based experiments}\quad\\
        
\noindent \textbf{Pierre Auger Observatory:} The Pierre Auger Observatory \cite{PierreAuger:2015eyc} is the World's largest array for UHE cosmic-ray detection. Located in the province of Mendoza, Argentina, at a mean altitude of 1400 m above sea level ($\sim880~\mathrm{g~cm^{-2}}$ of vertical atmospheric column density), it consists of a surface detector (SD) to sample the front of shower particles at the ground level with an array of 1660 cylindrical water-Cherenkov detectors (WCD) of 1.2 m height and $\sim 3~\mathrm{m}^2$ top surface area, deployed over an area of $\sim 3000~\mathrm{km^2}$ arranged in a hexagonal pattern with 1.5 km spacing between detectors. The signals produced by the passage of shower particles through the SD stations are recorded as time traces in 25 ns intervals. The SD of the Pierre Auger Observatory can efficiently identify UHE neutrinos of all flavors in the background of showers initiated by UHECRs. The search strategy consists in selecting showers that arrive at the SD array in the inclined directions both downward $\theta\in (60^\circ,90^\circ)$ and upward-going $\theta\in (90^\circ,\sim 95^\circ)$, identifying those that exhibit a broad time structure in the signals induced in the SD stations indicative of an early stage of development of the shower, a signature of the shower developing close to the ground. The Pierre Auger Observatory is highly efficient as a neutrino detector, with its sensitivity mostly due to Earth-Skimming tau neutrinos that account for $\sim80\%$ of the expected event rate for a differential neutrino flux $dN_\nu/dE_\nu = k\times E_\nu^{-2}$. With the applied selection algorithms, the neutrino search is not limited by the background due to UHECR-induced showers but by exposure \cite{PierreAuger:2019ens}. A search for neutrino-induced showers was performed in the Observatory data from 1 January 2004 up to 31 August 2018. No neutrino candidates are identified and a restrictive upper limit on the neutrino flux was obtained \cite{PierreAuger:2019ens}. The single-flavor $90\%$ C.L.~integrated energy limit is: $k_{90} < 4.4 \times 10^{-9}~\mathrm{GeV~cm^{-2}~s^{-1}~sr^{-1}}$ or equivalently $k_{90} < 1.4~\mathrm{EeV~km^{-2}~yr^{-1}~sr^{-1}}$, with optimal sensitivity at $\sim$1 EeV. The upper bound constrains models of cosmogenic neutrino production, assuming the sources of UHECR produce mainly protons and evolve strongly with redshift. The SD array instantaneous sensitivity to UHE neutrinos extends to $\sim 30\%$ of the sky, with declinations ranging from close to the South Celestial Pole to $+60^\circ$, and with an unrivaled sensitivity to arrival directions in the Northern hemisphere and for $\lesssim 1$ hour-duration transient sources in the field-of-view of the Earth-Skimming channel \cite{PierreAuger:2019azx}. Follow-up in UHE neutrinos of sources of Gravitational Waves from compact binary mergers \cite{LIGOScientific:2017ync, ANTARES:2017bia} and of the blazar TXS0506+056 \cite{PierreAuger:2020llu} have been performed with the Observatory with no events identified, demonstrating the key role of the Pierre Auger Observatory in multimessenger astronomy at UHE \cite{PierreAuger:2019fdm}. 
        
        With the ongoing upgrade of the SD \cite{PierreAuger:2016qzd} dubbed AugerPrime, additional measurements of composition-sensitive observables will be possible with the data collected simultaneously with the WCD and the new flat scintillators (SSD) on top of them. The upgrade will enable to identify a possible proton component at the highest energies as small as $10\%$ \cite{PierreAuger:2016qzd}, a decisive ingredient for determining the role of cosmic-ray observations in astronomy, and for establishing the potential of present and future detectors to the observation of the cosmogenic neutrino flux \cite{Alvarez-Muniz:2017jxp}. The SD stations electronics will also be upgraded \cite{Nitz:2021kdx}. The new electronics has the resources to implement further trigger algorithms on top of the existing ones, targeted to searches for photons and neutrinos at sub-EeV energies, including the possibility of incorporating the WCD, SSD and RD to the trigger providing a wealth of information on a single event. An integral part of the AugerPrime upgrade is the Radio Detector (RD) \cite{Pont:2021pwd} with the goal to install an antenna on top of each WCD to detect the radio signal induced in air showers. The RD is most efficient for the detection of inclined showers \cite{PierreAuger:2018pmw}, and the shape of the radio footprint is known to contain information on the shower penetration in the atmosphere \cite{PierreAuger:2021rio}. These two facts open up the possibility of using the RD in the search for penetrating inclined showers induced by UHE neutrinos, complementing current searches.

\noindent {\bf Telescope Array:} 
        The Telescope Array (TA) experiment is the largest UHECR observatory in the Northern Hemisphere \cite{Tokuno:2011zz}. Located near Delta, Utah, USA, the surface detector of TA is an array of 507 flat plastic scintillator stations arranged on a square grid with 1.2 km spacing covering an approximate area of 700 $\mathrm{km}^2$. Each detector is composed of two layers of 1.2 cm thick extruded scintillator with a 3 $\mathrm{m}^2$ effective area. Data collected with this array between May 11, 2008 and May 10, 2017 (9 years of operation) has been used to search for UHE downward-going neutrinos in the zenith angle range $\theta\in[45^\circ,90^\circ]$ \cite{TelescopeArray:2019mzl}. For this purpose, a multivariate selection algorithm is applied combining 16 composition-sensitive observables. No neutrino candidate events have been identified. The flatness of the scintillator detectors reduces the detection efficiency of the most inclined showers for which the neutrino sensitivity is expected to be highest, including the most effective Earth-skimming tau neutrino channel. As a result, the upper bound on the integrated neutrino flux is $\sim 400$ less restrictive than that obtained with the Pierre Auger Observatory, with the smaller area of the TA array compared to the SD of Auger also contributing to the reduced sensitivity, a fact that will be improved at the highest energies with the extension of TA to TA$\times4$ a surface array with an area about 3 times that of the current TA \cite{TelescopeArray:2021dri} with the separation between the scintillators increased to 2.08 km for a total area of $\sim 2800~\mathrm{km^2}$.

\noindent {\bf TAMBO:} The Tau Air-Shower Mountain-Based Observatory (TAMBO) is a proposed high-energy astrophysical tau neutrino detector to be deployed in the Peruvian Andes~\cite{Romero-Wolf:2020pzh}. 
        It will measure 1-100 PeV Earth-skimming tau neutrinos in the Colca valley - a steep yet narrow valley uniquely able to maximize geometric acceptance of Earth-skimming neutrinos while narrow enough to accurately record tau lepton decay products. 
        The detector is planned to have 22,000 water Cherenkov tanks, each $1~{\rm m}^3$ in size, and positioned 150~m apart in a triangular grid.
        An alternative designs considers plastic scintillator panels instead of water tanks and is still under development. 
        TAMBO will have an effective area ten times larger than IceCube current tau neutrino effective area at $\sim$~3~PeV expecting a rate of 7 tau-neutrino events per year over when assuming an $E^{-2.5}$ spectra with the current IceCube best-fit normalization.
        Compared to on-going Earth-skimming experiments, such as the Pierre Auger experiment~\cite{Neto:2020zgu}, the deep-valley topography provides a larger geometric acceptance compared to a flat ground array.
        Given this increased acceptance and the lower energy threshold, due to the smaller tank-to-tank separation, TAMBO will characterize the high-energy astrophysical tau neutrino flux with significantly larger sample size and smaller backgrounds than other techniques. 
        In turn, this would allow a better understanding of high energy neutrino sources and to better constrain astrophysical neutrino flavor measurements. 
        TAMBO will also determine whether high-energy neutrino sources continue to accelerate particles above 10 PeV and therefore probe possible GZK neutrino scenarios. 
        Another consideration is that TAMBO will have a view of the galactic center and so makes detecting neutrinos from dark matter annihilation a possibility.

\pagebreak
\section{Tools}
\label{sec:tools}
Increasing our tau neutrino knowledge will help address many questions about neutrino oscillation physics and there is a rich experimental program spanning many decades of energies that will be detecting tau neutrinos.
Actually extracting and understanding this information requires care in several directions.
First, we need accurate flux simulations for accelerator produced neutrinos and atmospheric neutrinos.
Second, we need to understand neutrino-nucleus cross sections, in particular for lower energy experiments such as DUNE where the cross section is not fully DIS.
An intensive effort to improve our cross-section understanding is underway and tau neutrino physics can also benefit from progress there.
Third, the tau neutrino interaction needs to be identified in the detector.
Sometimes this is done with technology dedicated for $\nu_\tau$ detection such as emulsion, but more often the event must be reconstructed with the available technology designed for other physics goals.
Fourth, at ultra-high energies tau neutrinos experience regeneration during propagation in the Earth which depends on tau neutrino cross sections, tau lepton energy loss rates, and tau decays.
This section describes the state-of-the-art at the moment and looks to the needs of the future.

\subsection{Production Simulation}

Over the next decade, searches for tau neutrinos will be made with at least three accelerator-based sources: the Long-Baseline Neutrino Facility (LBNF), a beam dump at CERN's SPS, and the LHC.  Simulations for each of these sources have been developed and are described in more detail below.

\subsubsection{LBNF neutrino flux simulation}

Accelerator-based long-baseline neutrino experiments such as DUNE use neutrino beams which are generated through the decay of secondary hadrons predominantly charged pions and kaons. These secondary hadrons come from a high-energy primary proton beam hitting a target, and they eventually decay into a muon ($\mu$) and muon neutrinos ($\nu_{\mu}$).  The initial beam power of the LBNF optimized beamline is 1.2-MW and capability of up-gradable to 2.4-MW in the upcoming years. 

In the nominal LBNF design, the primary proton beam is 120-GeV in energy and impinges on a 1.5 m long, 16 mm diameter cylindrical Carbon (C) target. Charged particles produced in the target will be focused by 3 magnetic horns with 300 kA currents, with the target fully inside the first magnetic horn. The second (third) magnetic horn will be around 2.95(17.8) m downstream of the first horn.  Following the focusing horns, hadrons enter a 194 m long, 4 m diameter helium-filled decay pipe. This design is highly motivated by a genetic algorithm to optimize the CP-violation sensitivity at DUNE.

Even though the DUNE beam is mainly comprised of $\nu_{\mu}$'s, the intense neutrino beam, the massive long-distance LArTPC detector, and the versatile and high resolution Near Detector (ND) will ensure an opportunity to study the tau neutrino appearance because of $\nu_{\mu}\rightarrow \nu_{\tau}$ oscillations at the Far Detector (FD). Unlike the nominal LBNF design, which is explained above, in the tau-optimized beam design \cite{DUNE:2020ypp}, 120 GeV primary proton beam and two NuMI parabolic horns are used. The second horn will be at 17.5 m downstream of the first horn, and a 1.5 m long, 10 mm wide Carbon target located at 2 m upstream of the first magnetic horn. The reason for this, in the nominal LBNF design with the three CP-optimized horns, it is expected that approximately 130 $\nu_{\tau}$ charged-current (CC) neutrino interactions annually at the FD, supposing initial beam power is 1.2 MW and before detector efficiency. On the other hand, taking two NuMI-like parabolic horns into account, this number is significantly increasing to approximately 1000 $\nu_{\tau}$ yearly. The predicted neutrino fluxes at FD in the nominal and tau-optimized simulations can be seen in Fig.~\ref{tauoptim}. This figure shows the $\nu_{\mu}$ flux for standard and tau-optimized configurations. As it can be seen that the $\nu_{\mu}$ flux is important in the neutrino energy region between 0 GeV to 5 GeV, where the $\delta_{CP}$ is much smaller. However, the energy region is larger than 5 GeV, at which the cross-section of $\nu_{\tau}$ interactions is pivotal.

The LBNF beam simulation, known as G4LBNF, is a Geant4 based simulation using the QGSP\_BERT physics list. These models can be tuned to get more precise neutrino flux by using an external dataset provided by hadron production experiments, such as NA61/SHINE (CERN) and EMPHATIC (Fermilab) measurements.  The G4LBNF simulation is currently running for the version v3r5p8 tagged with the GEANT v4.10.3.p03b version and the default hadronic model is the QGSP\_BERT due to it runs importantly faster than FTFP\_BERT. Furthermore, the horn current is chosen to focus or to defocus of produced hadrons. For instance, if the horn current mode is set as a Forward Horn Current (FHC), it gives the $\nu_{\mu}$ flux, however, for the opposite one, it gives a $\bar{\nu}_{\mu}$ flux, called the Reverse Horn Current (RHC). 

\subsubsection{SHiP neutrino flux simulation}

With $2\times 10^{20}$ 400 GeV protons from the SPS at CERN incident on a molybdenum and tungsten target, many $\nu_\tau+\bar\nu_\tau$ events will be recorded in a SHiP detector in a 5 year period. 
The $\nu_\tau+\bar\nu_\tau$ flux incident on the 8 ton detector comes primarily from hadroproduction of $D_s^\pm$ since lighter charm hadrons have masses only just above the mass of the tau.
The prompt leptonic decays of the $D_s$ with a branching fraction of $B(D_s\to \tau\nu_\tau)=5.48\pm 0.23\%$ 
\cite{Zyla:2020zbs} and decays of $\tau$ leptons yield a flux of equal parts $\nu_\tau$ and $\bar\nu_\tau$ according to perturbative QCD production of $c\bar{c}$ pairs, absent intrinsic charm and nucleon spectator effects. At low energies, the flux is dominated by the direct $D_s\to \nu_\tau$ process. At high energies, the chain decay $D_s\to\tau\to \nu_\tau$ dominates \cite{Alekhin:2015byh}.

The SHiP detector cross sectional area of 0.8 m$\times 0.8$ m at a distance 46 m from the proton target means that neutrinos must arrive in the rapidity range of $\eta_\nu\gtrsim 5.8-6.1$. Evaluations of the flux using SHiP's FairShip software  based on Pythia 8 (Pythia 6 for heavy flavor, a leading order treatment) for the primary proton collisions, GEANT4 for propagation and GENIE for neutrino interactions yields of order 11,000 detectable $\nu_\tau+\bar\nu_\tau$ charged current events accounting for detection efficiencies \cite{SHIP:2021tpn}. Their simulations show an average energy in the $\nu_\tau\ (\bar\nu_\tau)$ CC events of 52 (70) GeV.

Evaluations of the $\nu_\tau+\bar\nu_\tau$ energy distributions for SHiP in an earlier configuration have been performed at next-to-leading order (NLO) in QCD \cite{Alekhin:2015byh,Bai:2018xum,Maciula:2019clb}. In the perturbative QCD evaluation, renormalization and factorization scale uncertainties lead to a range of between $\sim 0.5-1.5$ of the central result \cite{Bai:2018xum}. The inclusion of intrinsic transverse momenta through a Gaussian smearing factor impacts the number of events \cite{Alekhin:2015byh,Bai:2018xum}. Intrinsic charm that includes coalescence of heavy and light quarks to produce high energy charm hadrons can enhance the event rate and introduce charge asymmetries in the $\nu_\tau$ and $\bar\nu_\tau$ energy distributions \cite{Bai:2018xum}. This is also the case for non-standard sub-leading fragmentation in which, for example, $s\to D_s^-$ and $\bar{s}\to D_s^+$ \cite{Maciula:2019clb}. 

The DsTau experiment (NA65) will measure $D_s^\pm$ production and decay to $\tau \nu_\tau$ at the CERN SPS
\cite{DsTau:2019wjb}. They expect of order 1000 events for $2.3\times 10^8$ protons incident on a tungsten target. Already with data analyzed from a pilot run, more than 100 events have been observed with the characteristic double kink topology from $D_s\to \tau$ and $\tau$ decay, well above the background expectation \cite{Vasina:2021euh} for more than $3\times 10^7$ protons on target. Full scale runs are planned for 2021-2022. These measurement will serve to calibrate Monte Carlo simulations and NLO QCD evaluations of the energy distributions of $\nu_\tau+\bar\nu_\tau$ in the forward region in anticipation of future SHiP measurements.

\subsubsection{LHC neutrino flux simulation}
 
At the LHC in tunnels collinear with the beam near the ATLAS interaction point (IP), the neutrino experiments FASER$\nu$ 
\cite{FASER:2020gpr} and SND@LHC  \cite{SHiP:2020sos} have been approved for the LHC Run 3. FASER$\nu$ will cover a neutrino rapidity range of $\eta_\nu>8.9$ with a 1 ton neutrino detector, while SND@LHC will be off-axis, covering $7<\eta_\nu<8.5$ with a 800 kg detector. A Forward Physics Facility (FPF) \cite{Anchordoqui:2021ghd} is proposed to accommodate several neutrino detectors with a variety of detection techniques for running in the high-luminosity era of the LHC. Much work is underway to establish predictions of the forward flux of 
neutrinos 
\cite{Bai:2020ukz,Anchordoqui:2021ghd,Kling:2021gos,Maciula:2020dxv}.

Already in the SSC era, the idea to use the very forward region of high energy $pp$ collisions at a collider to collect neutrinos was proposed \cite{DeRujula:1984pg,DeRujula:1992sn,Vannucci:1993ud,Park:2011gh}. At the very highest neutrino energies, with configurations that include the magnetic fields and infrastructure near the IP and more than 100 m of concrete and rock between detector and IP, the dominant sources are charm hadron decays to neutrinos.
Tau neutrinos, again, come almost exclusively from the direct and chain decays of $D_s^\pm$.  Neutrino fluxes from a number of leading order Monte Carlo simulations \cite{Kling:2021gos} and from NLO QCD \cite{Bai:2020ukz,Bai:2021ira} show that thousands to tens of thousands of tau neutrino CC events can be accumulated in detectors at a FPF with ${\cal L}=3$ ab$^{-1}$.

Charm production at such forward rapidities has not been measured, but it is possible to anchor predictions with LHCb data for charm mesons in the range of $2<y<4.5$ \cite{Aaij:2016jht,LHCb:2013xam,Aaij:2015bpa}. Theoretical uncertainties include the parton distribution functions, renormalization and factorization scale dependence, and fragmentation functions. Transverse momentum smearing corrections to the collinear parton model can impact predictions at very forward rapidities where even a small $k_T$ can move a neutrino trajectory out of the detectable rapidity range \cite{Bai:2020ukz,Bai:2021ira}. An example of the flux of $\nu_\tau+\bar\nu_\tau$ from a NLO QCD evaluation with several parton distribution functions and two values for Gaussian smearing $\langle k_T\rangle$ is shown in figure \ref{fig:lhctaunu} \cite{Bai:2021ira}. A table showing corresponding predicted numbers of events for FASER$\nu$ and SND@LHC, and for a detector of 60 tons with $\eta>6.9$, is also shown. Already with a 29 kg pilot emulsion detector from spare parts of DsTau/NA65, a four-week run in 2018 with 12.2 fb$^{-1}$ at $\sqrt{s}=13$ TeV shows a neutrino signal above the expected background \cite{FASER:2021mtu}, ushering in a new era of neutrino physics in the forward region.

\begin{figure}
\centering
\includegraphics[width=.55\linewidth]{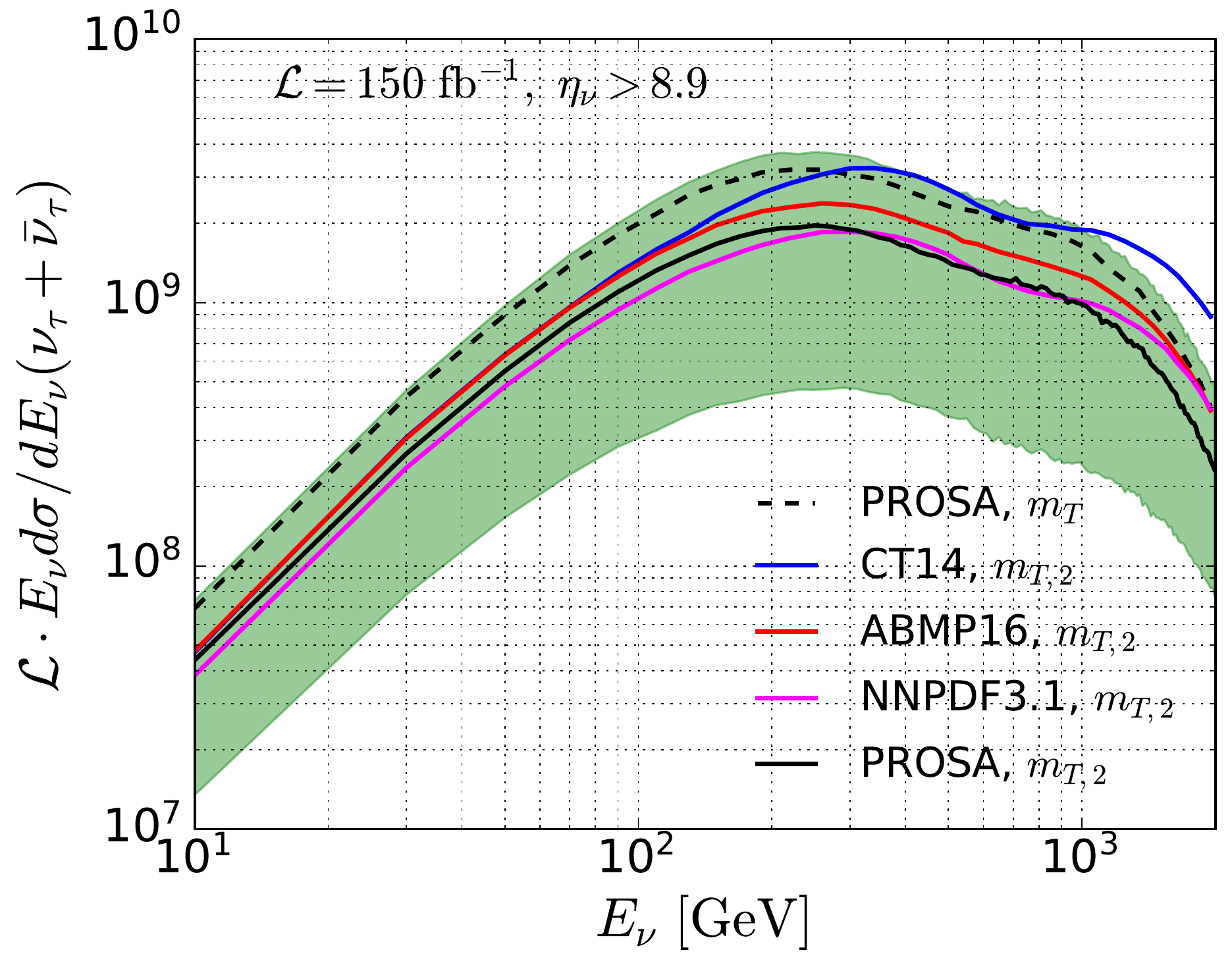}
\caption{\label{fig:lhctaunu} The number of $\nu_\tau+\bar\nu_\tau$ for $\eta_\nu>8.9$ given ${\cal L}=150$ fb$^{-1}$ for $\sqrt{s}=14$ GeV in $pp$ collisions at the LHC. 
The evaluations are done using the PROSA \cite{Zenaiev:2019ktw} PDFs with two different factorization and renomalization scale dependences and $\langle k_T\rangle$, and for CT14 \cite{Dulat:2015mca}, ABMP16 \cite{Alekhin:2018pai} and NNPDF3.1 \cite{Ball:2014uwa} with renormalization and factorization scales equal to $m_{T,2}=\sqrt{(2m_c)^2+(p_T)^2}$ and $\langle k_T\rangle=0.7$ GeV. The green band shows the scale uncertainties for the $m_{T,2}$ scale dependent result for PROSA PDFs. Ref.~\cite{Bai:2021ira} contains more details.} 
\end{figure}

\begin{table}[tbp]
    \centering
    \begin{tabular}{|c|c|}
    \hline\hline
      \multicolumn{2}{|c|}{PROSA, $\mu_R=\mu_F=m_{T,2},\ \langle k_T\rangle=0.7$ GeV} \\
     \hline
      \hline
       Detector, ${\cal L}$, $\eta$ range, Mass & N$_{\rm evt}(\nu_\tau+\bar\nu_\tau)$ CC \\
      \hline 
      \hline
      SND@LHC, 150 fb$^{-1}$, $7.2<\eta_\nu<8.6$, 830 kg & $4.2^{+3.8}_{-3.3}$ \\
      \hline
      FASER$\nu$, 150 fb$^{-1}$, $\eta>8.9$, 1.2 ton & $12.1 ^{11.6}_{-9.8}$ \\
      \hline
      3000 fb$^{-1}$, $\eta_\nu\gtrsim 6.9$, 60.63 ton &  $4775 ^{+4307}_{-3763}$\\
      \hline
   \end{tabular}
   \caption{\label{tab:prosanutau} Predictions for the number of $\nu_\tau+\bar\nu_\tau$ charged current events for $pp$ collisions at $\sqrt{s}=14$ TeV evaluated with NLO QCD with PROSA PDF sets \cite{Zenaiev:2019ktw}, renormalization and factorization scales set to $\sqrt{(2m_c)^2+p_T^2}$ and $k_T$ Gaussian smearing with $\langle k_T\rangle=0.7$ GeV \cite{Bai:2021ira}. Errors combine scale and PDF uncertainties.}
\end{table}

\subsubsection{Atmospheric neutrino simulations}

Atmospheric neutrinos produced in hadronic cascades from cosmic ray interactions with the atmosphere provide a natural source of $\nu_{\tau}$. There are four natural sources of $\nu_{\tau}$ at Earth: 

\begin{enumerate}
    \item The prompt flux from decays of ${\rm D}_{\rm s}$ and B mesons \cite{Pasquali:1998xf,Fedynitch:2018cbl}
    \item Tertiary tau lepton pair production from atmospheric muons traversing the atmosphere or the ground \cite{Bulmahn:2010pg}
    \item $\nu_{\tau}$ appearance due to oscillations of atmospheric $\nu_\mu$ and $\nu_{\rm e}$ \cite{IceCube:2019dqi}
    \item Astrophysical $\nu_{\tau}$ from oscillations of $\nu_\mu$ and $\nu_{\rm e}$ over astronomical distances \cite{IceCube:2020abv}.
\end{enumerate}

At energies $>100$ TeV, the dominant atmospheric $\nu_{\tau}$ contribution comes from the prompt flux whereas the tertiary production is negligible \cite{Bulmahn:2010pg}. The calculations of conventional and prompt atmospheric neutrinos are most conveniently performed with the tool {\sc MCEq} \cite{Fedynitch:2015zma}\footnote{\url{https://github.com/afedynitch/MCEq}}. {\sc MCEq} is distributed along with a variety of cosmic ray flux and hadronic interaction models thus simplifying estimates of flux uncertainties. For accurate angular distributions at 100 MeV -- few GeV energies, pre-computed flux tables are available for the two 3D neutrino flux calculations \cite{Honda:2015fha,Barr:2004br}. Estimates can be also computed semi-analytically, see {\it e.g.}~\cite{Gaisser:2016uoy}. The default prompt flux in {\sc MCEq} is computed using the charm production model from the {\sc Sibyll-2.3d} event generator \cite{Fedynitch:2018cbl}. Due to small cross-talk between conventional hadronic cascades and atmospheric charm production, prompt flux tables from the literature, such as \cite{Enberg:2008te,Bhattacharya:2015jpa,Benzke:2017yjn}, can be combined with the above flux calculation methods as long as the same cosmic ray flux model is used. The tertiary components can be computed by combining atmospheric muon fluxes with simulations using accurate lepton propagation tools, such as PROPOSAL \cite{Koehne:2013gpa}, or analytical estimates \cite{Bulmahn:2010pg}. However, both the prompt flux and the tertiary components are subdominant in comparison with the astrophysical diffuse $\nu_{\tau}$ flux, which is characterized by a simple power-law \cite{IceCube:2019dqi}.

At lower energies, the fluxes from $\nu_{\tau}$ appearance compete with the largely unknown low-energy astrophysical $\nu_{\tau}$ flux \cite{Athar:2004uk,Lee:2004zm}. The expected angular distributions and spectra are different. The astrophysical flux is expected to correlate with galactic diffuse gamma-ray emission and can be computed from Fermi LAT $\pi^0$ templates \cite{Fermi-LAT:2012edv} or numerical models, {\it e.g.}~\cite{Gaggero:2015xza}.

Cosmic ray interaction in the atmosphere is one of natural sources of tau neutrinos. Interactions of cosmic rays and air nuclei generate various hadrons, some of which decay to neutrinos, i.e. atmospheric neutrinos.
Atmospheric tau neutrinos can be produced directly from the decays of heavy flavor hadrons, mainly $D_s^\pm$ meson (prompt), and by oscillation of conventional neutrinos from the light hadron decays, predominantly $\nu_\mu/\bar{\nu}_\mu \to \nu_\tau/\bar{\nu}_\tau$ (conventional). 
The prompt tau neutrino flux has hard energy spectrum due to the extremely short decay length of the parent particles, while the flux of conventional neutrinos rapidly decreases with energy. 
Therefore, at high energies above a certain cross-energy near 1--10 TeV, the prompt neutrinos start to dominate and become the main component of atmospheric neutrinos.  

The flux of atmospheric tau neutrinos can be calculated using the semi-analytic $Z$-moment method to solve the coupled cascade equations, which describe the propagation of particles in the atmosphere.
In this method, the incident cosmic ray spectrum, the cross sections for hadron production and the decay distributions of hadrons to neutrinos are involved as primary input. 
(For the further details of the $Z$-moment method, see Refs.~\cite{Gaisser:2016uoy,Lipari:1993hd}.)
Alternative method to evaluate the atmospheric neutrinos is using a numerical tool to solve the cascade equations, MCEq (Matrix Casecade Eqations) \cite{Fedynitch:2015zma}, in which evaluation of the heavy flavor production is performed with the Sibyll-2.3 event generator.

\begin{figure}
\centering
\includegraphics[width=.55\linewidth]{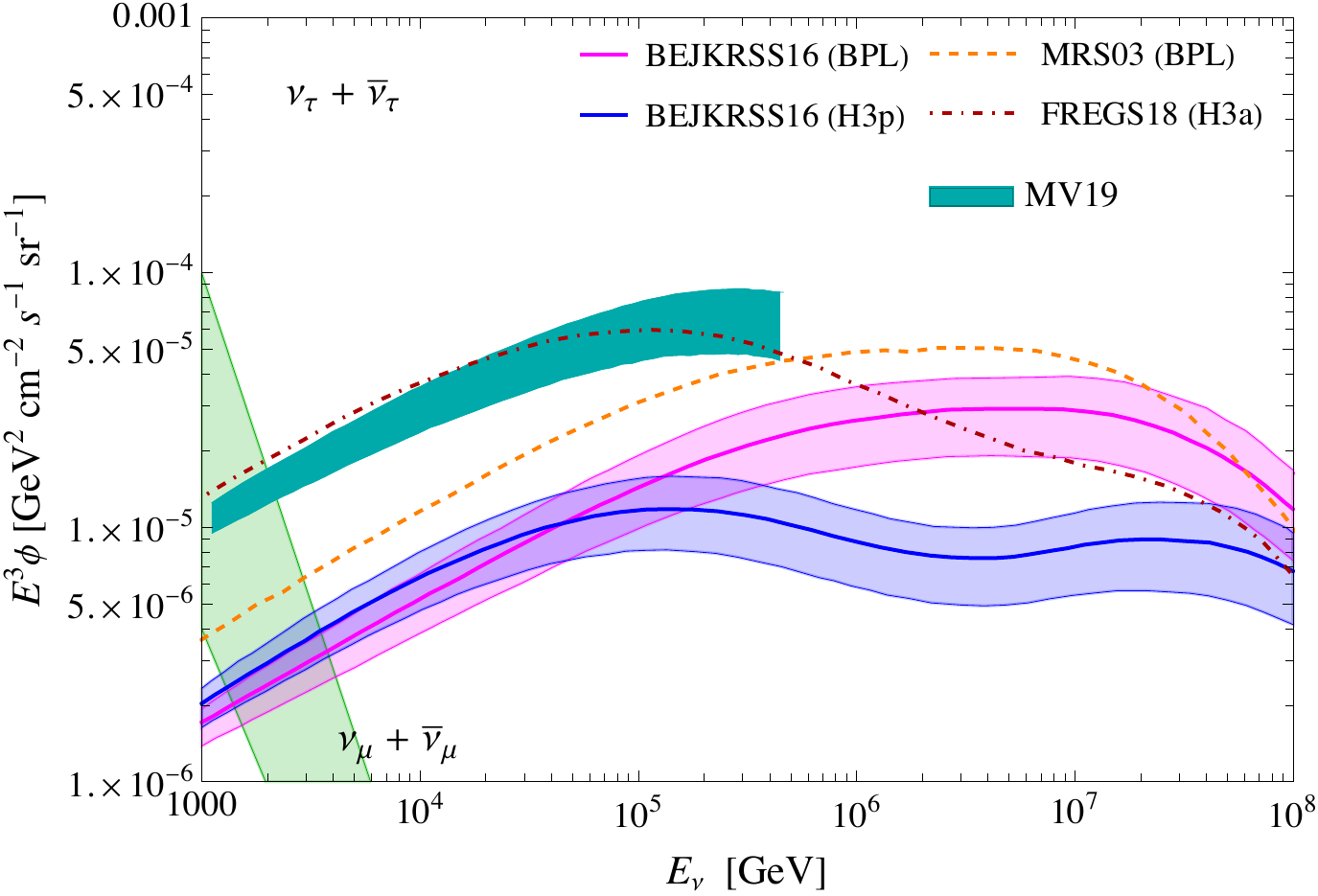}
\caption{\label{Fig:AtmNuTau} 
Comparison of the prompt atmospheric tau neutrino fluxes. The predictions are from Refs.~\cite{Bhattacharya:2016jce} (BEJKRSS16), \cite{Martin:2003us} (MRS03), \cite{Fedynitch:2015zma} (FREGS18) and \cite{Mascaretti:2019uqn} (MV19).}
\end{figure}
In the energy regime where the prompt atmospheric neutrinos are dominant, the cosmic ray spectrum and the cross sections for heavy flavor production have important impacts on estimating their fluxes. 
Figure \ref{Fig:AtmNuTau} shows several predictions of prompt atmospheric tau neutrino fluxes as well as the fluxes of the conventional tau neutrinos from oscillation. 
Prompt tau neutrino fluxes in the figure are evaluated with different cosmic ray spectra as indicated in the plot.
The cosmic ray spectrum has significant impact on the shape of the fluxes above  $\sim$ 100 TeV.
The overall large uncertainties are related with the interaction for the heavy flavor production and the evaluation method. 
In evaluating the cross sections for the heavy flavor production relevant for the atmospheric neutrinos, the involved partons have large momentum fraction $x$ from the cosmic ray side and very small $x$ from the air nucleus in the atmosphere, and the related $x$ ranges have not yet well constrained by the experiment. 
Depending on how to treat the parton distributions at the small-$x$ regime, there are several approaches for evaluating the heavy flavor production cross sections, and such different models bring about the large uncertainty in the theoretical prediction of atmospheric tau neutrino flux.
In the figure, the prediction by NLO in QCD \cite{Bhattacharya:2016jce} and the so-called dipole approach \cite{Martin:2003us} are compared as well as the evaluations by MCEq \cite{Fedynitch:2015zma,Mascaretti:2019uqn}.

\subsection{Neutrino Event Generators and Tau Neutrino Cross Section Code}

Experiments such as T2K \cite{DUNE:2015lol}, NOvA \cite{DUNE:2015lol}, MINERvA \cite{MINERvA:2013zvz} and others, report the calculation of the cross-section (neutrino interactions) as the major contributor of systematic uncertainties. Cross-section measurements depends on how accurate our interaction models are and also in how accurate we can determine the energy of neutrinos and neutrino-nucleus interactions via reconstruction techniques that are either based on kinematics (T2K/HK) or calorimetric methods (DUNE/NOvA/SBN).

While the theoretical physics sector have bring a wide variety of models that describe the interactions in different energy ranges, neutrino scattering measurements from MiniBooNE \cite{MiniBooNE:2008paa} MINERvA \cite{MINERvA:2013zvz}, MicroBooNE \cite{MicroBooNE:2019nio}, NOvA \cite{NOvA:2004blv}, and T2K \cite{T2K:2011qtm} points out that  these models needs improvements as they are miss-modeling real data. On the other hand, neutrino experiments have been running behind in adopting new models and corrections due to their high dependence on computation simulations and software development that must be implemented in the nuclear event generators. Having the theoretical sector moving faster than the experimental side provokes a deceleration in the field as a whole, testing and improving models of neutrino-nucleus in generators is a factor of fundamental importance. 

Monte Carlo (MC) simulations are the basis for any analysis, some well know neutrino generators are GiBBU \cite{Buss:2011mx}, NuWro \cite{Juszczak:2005zs}, NEUT \cite{Hayato:2009zz} , NUISANCE \cite{Stowell:2016jfr}, GENIE \cite{Andreopoulos:2009rq}, FLUKA \cite{Ferrari:2005zk} and TAUOLA \cite{JADACH1993361}; a recommended read about the particularities of these event generators can be found in this notes \cite{hugh_gallagher}. All of them take different approaches for different reasons, but as a general comment we can say that their task are: simulate neutrino interaction, simulate signals and backgrounds observed in the detector, be a bridge to compare real data and theories in order to extract neutrino oscillation parameters, reduce systematic uncertainties in measuring physics observations. Precision of neutrino event generators is required to better understand of neutrino interactions . In January 2020, it was held at Fermilab the last Generator Tools Workshop, bringing together neutrino instrumentalists, theorists, and event generator developers to agree on a plan for implementing several tools aimed at providing easier access by experiments to various generators and easing the process of getting new models into generators, see the summary here \cite{Barrow:2020gzb}.

While neutrino generators have evolved over the years and while much work must be settled, all of them have tuned their models to take into account only $\nu_{\mu}$ and $\nu_{e}$ and their interactions, being TAUOLA \cite{JADACH1993361} the only one that accounts for $\tau$-leptons measurements, TAUOLA decays the $\tau$ according to its branching ratios and accounts for $\tau$ polarization effects. Having remaining a third piece of this puzzle arise questions, for instance, how the effects in quasielastic (QE) scattering and $\Delta$ resonance production impact the cross-section, $Q^{2}$ distribution, and spin polarization of the produced $\tau^{\pm}$-leptons. These effects are pretty sensitive to the axial structure of the target. On the other hand, the nucleon pseudoscalar form factor neglected in QE interactions for $\nu_{\mu}$ and $\nu_{e}$ due to small mass of muon and electron compared to energies of experiment or mass of the proton can affect significantly interactions with $\nu_{\tau}$. In particular, polarization observables with polarized target or polarized knocked-out nucleon are sensitive to the pseudoscalar form factor slightly away from the tau production threshold when event rates are sizable \cite{Tomalak:2020zlv}. Future polarization measurements could provide independent access to the proton axial structure and allow the first extraction of the pseudoscalar form factor from neutrino data without commonly used, partially conserved axial current, ansatz and assumptions about the pion-pole dominance for this form factor, the latter is only expected to be approximately valid at low momentum transfer,\cite{LlewellynSmith:396544,Kaiser:2003dr,Chen:2020wuq}.

\subsubsection{GiBUU}
The Giessen Boltzmann-Uehling-Uhlenbeck (GiBUU) is the only generator that uses transport theory, i.e propagates phase-space distributions, not particles; all the others generators rely on intranuclear cascade Monte Carlos. Address for a semi-classical transport model in coupled channels \cite{Buss:2011mx} considering nuclear effects as local density approximation, mean-field, and Columb potentials, off-shell particle transport, 2p2h excitation, and in-medium spectral functions. GiBUU ensures consistency between nuclear effects in the initial state (Fermi motion, Pauli blocking, hadron self-energies, and medium-modified cross-sections) and the final state (particle reinteractions). Worth to mention, the final state interactions (FSI) part of GiBUU can be run or be checked separately from the specific initial interaction (ISI) by using special options \cite{Leitner:2008fg}, given in this way, a broader scope than most neutrino generators. Now, in real physics, ISI and FSI cannot be factorized, the nucleon potential links ISI and FSI, in order to factorize ISI and FSI it would require no potential at the point of first interaction. 

\subsubsection{GENIE}
GENIE is the most extensive international collaboration for a neutrino event generator; it emphasizes extensibility, modularity, and flexibility at the software level. Diverse experiments hosted in Fermilab (MINERvA, NOvA, SBN, and DUNE) have taken on many GENIE developments and, in this way, serve to support the large user base at Fermilab. Current versions of GENIE, starting with GENIE 2.10, offers new improvements to the intranuclear rescattering simulation to better characterize the A-dependence of rescattering processes, the inclusion of an effective spectral function model \cite{Bodek:2014pka} that combines a nuclear model based on super-scaling phenomenology with modeling of multi-nucleon scattering processes, the inclusion of a model for neutrino production of single kaons \cite{RafiAlam:2010kf}. 

The focus of neutrino interaction modelling in GENIE has been the few GeV neutrino energy range, which is relevant for atmospheric and accelerator-based neutrino experiments. A new package, called HEDIS, has been created implementing high-energy cross section calculation~\cite{GENIEv3}. This new module incorporates DIS off nucleons (modelled at NLO~\cite{Cooper-Sarkar:2011jtt,Bertone:2018dse}), Glashow scattering~\cite{Gauld:2019pgt} and sub-leading resonant effects due to neutrino interactions with the photon field of the nuclei~\cite{Zhou:2019vxt}. It can be used strictly for neutrino energies above 100 GeV, and has been validated up to 10 EeV.

\subsubsection{NuWRO}
Developed by a theory group at Wroclaw University \cite{Juszczak:2005zs}. Address for the problem of hadronization \cite{PhysRevC.86.015505} in nuclei and transitions between models, by example: the transition region between RES and DIS, or the transition between low mass hadronization and PYTHIA \cite{Sjostrand:2019zhc}. The final models are bench-marked against data, and good agreements are shown, \cite{Juszczak:2005zs}, \cite{Nowak:2006xv}. Includes the Berger-Sehgal model \cite{PhysRevD.79.053003} for coherent production in neutral current (NC) and charge current (CC) channels.

\subsubsection{NEUT}
The neutrino event generator (NEUT) was developed by the Kamiokande experiment \cite{Hayato:2009zz}, which translates into meets the needs of this experiment and other close experiments (K2K, SciBoone, T2K, Hyper-Kamiokande). NEUT address the Nieves model for multi-nucleon production \cite{Nieves:2011pp} and single pion cross-sections to the ANL/BNL data \cite{Wilkinson:2014yfa}. Notice that the RPA correction \cite{Nieves:2004wx} is included but as a reweight option and radiative CCQE is also an option. 

\subsubsection{FLUKA}
FLUKA \cite{Ferrari:2005zk} a simulation package for particle transport and interactions with matter in applications that include particle physics, cosmic ray physics, detector design and medical physics. It is distributed with different types of event generators. The FLUKA-specific generator PEANUT (Pre-Equilibrium Approach to Nuclear Thermalization) interaction model \cite{Battistoni:2006da} is used for hadronic and photon primaries up to energies of $E_\text{lab} \sim 10$ TeV. Above this energy FLUKA is linked to the DPMJET-III-19.2 \cite{Roesler:2000he,Fedynitch:2015kcn} generator. DPMJET is also eployed for nucleus-nucleus interactions above $\sqrt{s_\text{NN}} \sim 5$ GeV. Both models are based on the ideas of the Dual Parton Model \cite{Capella:1992yb}, and quark/parton string model.

Neutrino interactions are incorporated for QE, RES and DIS regimes \cite{Battistoni:2009zzb}. In QE and DIS scattering, the lepton mass dependent terms are included in the differential cross section. The Albright-Jarlskog relations are used for DIS, and the structure functions are extrapolated to $Q^2=0$ as in Ref.~\cite{Bertini:1995cr}. The RES cross section is evaluated with only $\Delta$ production using Rein-Sehgal model \cite{Rein:1980wg}.

\subsubsection{TAUOLA}
TAUOLA \cite{JADACH1993361} simulates tau decays for both the leptonic and hadronic decays modes. The hadronic currents implemented in TAUOLA are based on resonance dominance model (RDM), \cite{RDM}. Tauola offers an universal interface that reads information from the event record, allows it to be used with almost all MC generators, including Pythia 8, \cite{Sjostrand:2019zhc}. Tauola address for $\tau$ lepton(s) and spin states are calculated from kinematics configurations of hard processes.

\subsection{Special Considerations for High-Energy Neutrino Event Generators}

\begin{figure}
    \centering
    \includegraphics[width=\linewidth]{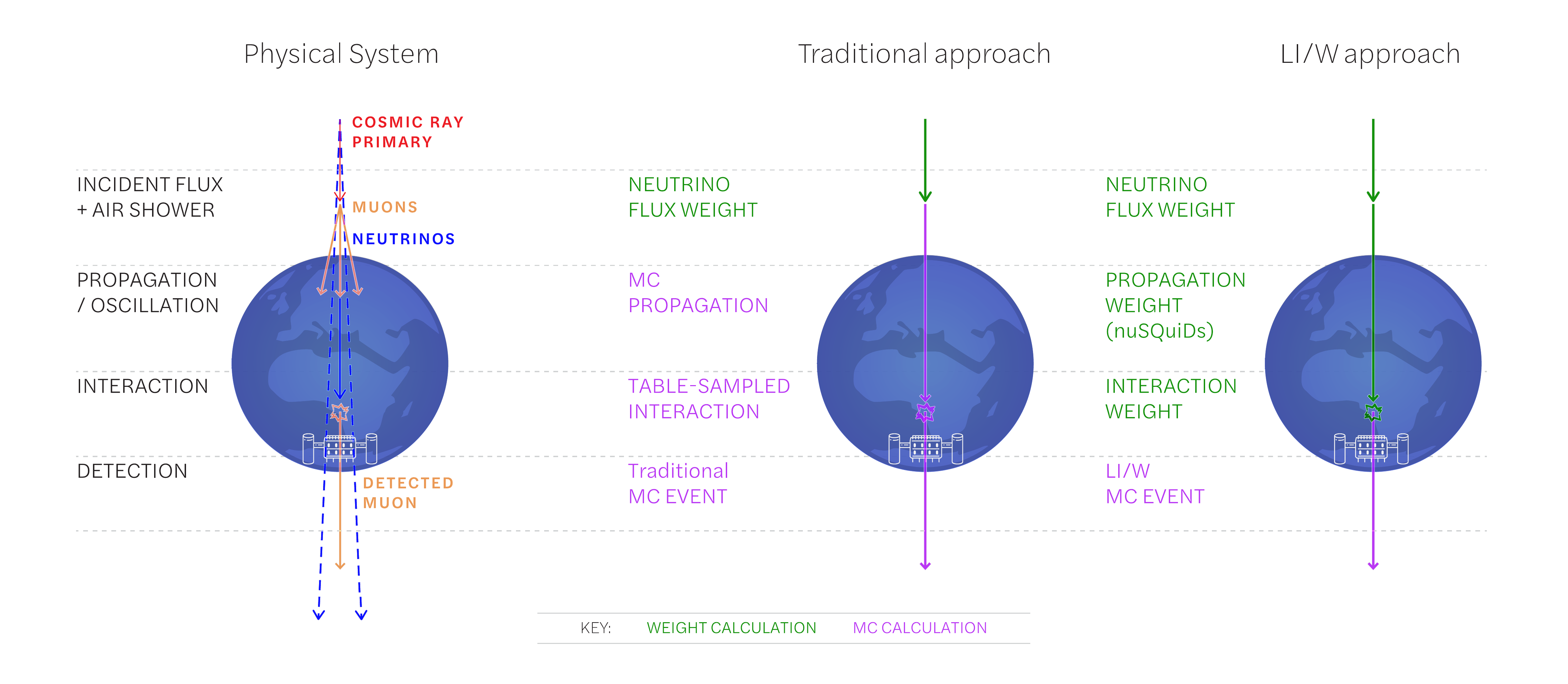}
    \caption{Reproduced from Ref.~\protect\cite{IceCube:2020tcq}. Diagram illustrates the different event generation stages for high-energy neutrino experiments. 
    Weighting steps for traditional methods compared with the LeptonInjector philosophy.}
    \label{fig:nufsgen}
\end{figure}

High-energy neutrino observatories have large detector spacing which does not allow them to resolve the neutrino interaction details.
This allows for simplification of the neutrino event generators, where the only relevant quantities are the out-going lepton energy and the light produced from the hadronic shower; see Ref.~\cite{Wiebusch:thesis,Radel:2012kw,Radel:2012ij} for recent parameterizations.
However, other problems need to be tackle in this high-energy regime, \textit{e.g.} the fact that the Earth is not transparent to neutrinos, and generators available separate the simulation in various stages as shown in Fig.~\ref{fig:nufsgen}.

An important consideration in these generators is the fact that muons produced either from the neutrino interactions or from the decay of tau leptons can travel several kilometers in rock or water.
Thus a significant problem in the high-energy neutrino event generators is to optimize the injection volume. 
Namely, to compute the volume such that most of the muons produced from neutrinos reach the detector.
This has been implemented in several generators, such as ANIS~\cite{Gazizov:2004va}, NuGen~\cite{DeYoung:865626}, and more recentlty LeptonInjector~\cite{IceCube:2020tcq}.

Another problem solved by these generators is the fact that in this regime the neutrino oscillation probability is not linear with the neutrino cross section and thus computing the weight of events requires the knowledge of all possible neutrino interaction processes; see Appendix D of Ref.~\cite{IceCube:2020tcq}.

\subsection{Special Considerations for Tau Neutrino Cross Sections}

In the Standard Model, $\nu_\ell\ (\bar\nu_\ell)$ charged current (CC) scattering is identical for all lepton flavors $\ell=e,\mu,\tau$ except for the impact of the charged lepton mass $m_\ell$. Such an impact requires a special consideration for the heaviest lepton $\tau$ and for the corresponding neutrino. The tau (anti-)neutrino energy threshold for CC interactions is $E_{\nu_\tau}^{\rm min}\simeq 3.5$ GeV. Large tau mass $m_\tau$ compared to the nucleon mass and masses of light hadrons limits the phase space for other particles produced in the CC interaction. For scattering of (anti-)neutrinos of a particular flavor $\nu_\ell\ (\bar\nu_\ell)$, the differential cross sections at leading order contains terms proportional to $m^2_\ell/M^2$ and $m^2_\ell/E^2_\nu$ that can be safely neglected for electron and muon $m_\ell=m_e,\, m_\mu$ but might change cross sections in case of $m_\ell=m_\tau$. 

In $\nu_\tau$ quasielastic scattering with nucleons, the contribution to the differential cross sections from the pseudoscalar form factor is proportional to $m_\tau^2/M^2$ and/or $m_\tau^2/E_\nu^2$ and therefore can be comparable to the contribution of vector and axial form factors. For unpolarized cross section, this contribution is much smaller than uncertainties from the axial form factor besides the region in the vicinity of the tau-production threshold where event rates are relatively low. However, the contribution from the pseudoscalar form factor is enhanced over the wide kinematic range for some polarization observables. The simplest polarization observables are asymmetries with one interacting particle polarized in a particular direction and the opposite one. Measurements of single-spin asymmetries for polarized target nucleons, recoil nucleons, and recoil charged lepton provide an independent probe of all form factors \cite{Graczyk:2019xwg,Tomalak:2020zlv,Fatima:2020pvv}. In particular, $\nu_\tau$ and $\bar{\nu}_\tau$ measurements of single-spin asymmetries in quasielastic scattering on polarized perpendicular to the beam direction protons inside the molecule would offer a way to measure the pseudoscalar form factor in neutrino scattering experiments \cite{Tomalak:2020zlv} while the charged lepton spin asymmetries would provide independent access to the axial form factor.

Deep-inelastic CC scattering with $\nu_\tau$ and $\bar{\nu}_\tau$ introduces a dependence on two structure functions, $F_4(x,Q^2)$ and $F_5(x,Q^2)$, with factors proportional to $m_\tau^2$. Albright and Jarlskog \cite{Albright:1974ts} showed that in the massless parton model at leading order, $F_4=0$ and $F_5$ is related to the structure function $F_2(x,Q^2)$ measured in $\nu_\mu$ and $\bar{\nu}_\mu$ scattering experiments. While suppressed by a factor of $m_\tau^2/(ME_\nu)$, the term in the differential cross section containing $F_5$ nevertheless affects the $\nu_\tau N$ cross section at a level of $\sim 10\%$ for $E_\nu=100$ GeV, and $\sim 30\%$ for $\bar\nu_\tau N$ for the same energy \cite{Jeong:2010nt,Ansari:2020xne}. Finally, in consideration of mass corrections, in addition to $m_\tau$, there are corrections due to the target mass $M$ and where applicable, heavy quark masses such as $m_c$, both comparable in scale to $m_\tau$ \cite{Kretzer:2002fr,Kretzer:2003iu,Ansari:2020xne}.

It is worth noting that Very large volume neutrino telescope (VLVnT) analyses including lower energy $\nu_\tau$ incorporate the contribution of $\nu_\tau$ in relation to both $\nu_\mu$ and $\nu_e$. Subsequently, an overall change in the neutrino cross section is unlikely to be a dominant source of uncertainty unless it impacts $\nu_\tau$ events differently than $\nu_\mu$ and/or $\nu_e$. For $\nu_\tau$ events in the DIS regime, the cross section dependence on the structure functions, $F_4(x,Q^2)$ and $F_5(x,Q^2)$, induces a difference between the cross sections for $\nu_{\tau}$ compared to $\nu_\mu$ and $\nu_e$. However, the uncertainty on these structure factors is not large enough to suggest that the cross section ratio ($\sigma^{CC_{DIS}}(\nu_\tau)/\sigma^{CC_{DIS}}(\nu_\mu)$) requires any additional treatment. This is because $F_4=0$ at leading order, and the $F_5$ structure function is proportional to both $F_1$ and $F_2$ in such a way that uncertainties from the parton distribution functions in $F_5$ propagate to $F_1$ and $F_2$ result in a relative cross section which remains mostly constant.

As VLVnTs come online with better detection thresholds at lower energies (notably KM3NeT-ORCA and the IceCube Upgrade) the fraction of non-DIS events in an analyzed sample will increase, and any \textit{relative} cross section uncertainties between $\nu_\tau$ and $\nu_\mu/\nu_e$ for QE and/or RES interactions will likely have to be examined, in addition to the absolute QE and RES uncertainties for $\nu_\tau/\nu_\mu/\nu_e$ which are already included in analyses. In this regard, the future of VLVnT-based $\nu_\tau$ analyses (as well as $\nu_\mu$ and any $\nu_e$ analyses) will be coupled to improvements in QE/RES cross section uncertainties that are relevant for neutrino accelerator experiments as well.

Muon and tau leptons decay after they travel some distance. 
The kinematics of the decayed products depends on the spin polarization of charged leptons as shown in Fig.~\ref{fig:polarization}.
Several works have investigated the spin polarization of tau leptons produced in charged current interactions \cite{Hagiwara:2003di,Kuzmin:2003ji,Levy:2004rk,Payet:2008yu}.
Currently, most of generators compute the polarization of final state leptons in neutrinos interactions assuming they are massless (FLUKA uses the approach of \cite{Hagiwara:2003di}).
Hence, fully polarized left(right)-handed leptons are generated if they are negatively (positively) charged.
This assumption is valid when $m/E\rightarrow0$.
Therefore, it is a good approximation for electrons and muons in the few GeV regime but not for taus.
At higher energies, the approximation is valid for any charged lepton. 

\begin{figure}[h!]
\centering
\includegraphics[width=0.49\textwidth]{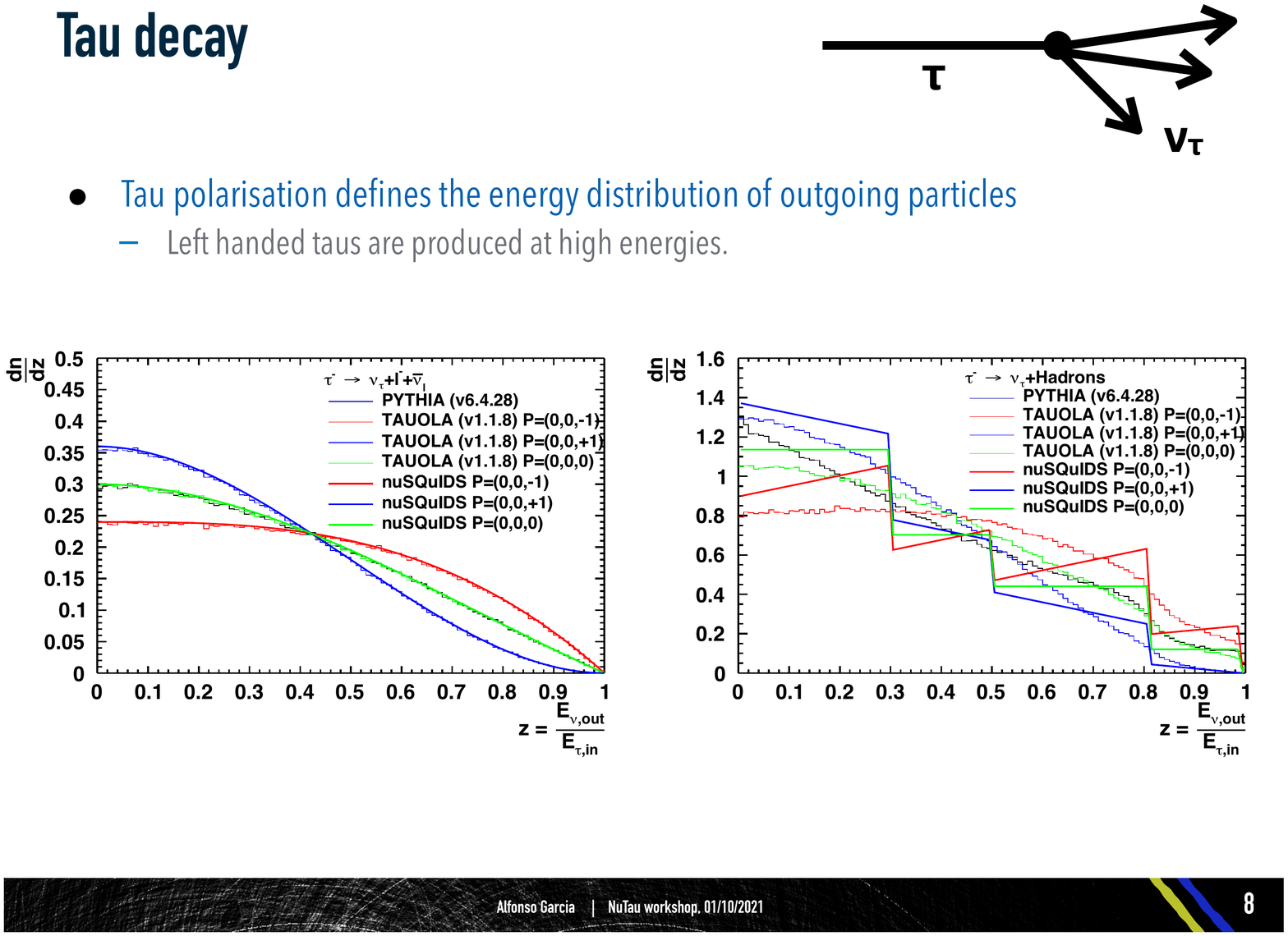}
\caption{Energy fraction of the outgoing $\nu_\tau$ for 1 million $\tau^{-}$ decaying leptonically. Different polarization has been assumed for the incoming tau (red shows the left-handed scenarios). \texttt{PYTHIA6} does not include polarization effects, so it assumes unpolarized taus. \texttt{nuSQuiDS} uses the parameterization described in~\cite{Dutta:2000jv}.}
\label{fig:polarization}
\end{figure}

\subsection{Reconstruction Techniques}

\subsubsection{Emulsion detectors: $\tau$-decay identification}

The identification of tau neutrinos requires the identification of the tau leptons at the neutrino interaction vertex. Since a tau decay always 
includes a tau neutrino which escapes measurement, the detection of the decay topology is a crucial issue (otherwise one could analyze the invariant mass). The required spatial resolution is a fraction of $c\tau$ of the tau lepton (not $\gamma c \tau$). For this purpose emulsion detectors, which have an excellent spatial resolution, provide an attractive and practical solution.

Nuclear emulsions \cite{Ariga:2020lbq} are made of 
a very large number of silver bromide (AgBr) crystals dispersed uniformly in a gelatin layer. Each crystal, with a typical diameter of 200~nm, acts as an independent detector channel. Hence, an emulsion layer has $O(10^{14})$ detection channels per cm$^3$. When a charged particle passes through nuclear emulsion, it ionizes the AgBr crystals, forming a latent image along its trajectory. After signal amplification by chemical methods such ionization sites induce filaments of metallic silver, called ``grains'', which can be observed as dark spots under a microscope, thus making the trajectory of the charged particle visible, as shown in Fig.~\ref{fig:emulsiongrain}. Details on the image formation can be found in~\cite{Ariga:2020lbq}.  

\begin{figure}[h]
\centering
\includegraphics[width=12cm]{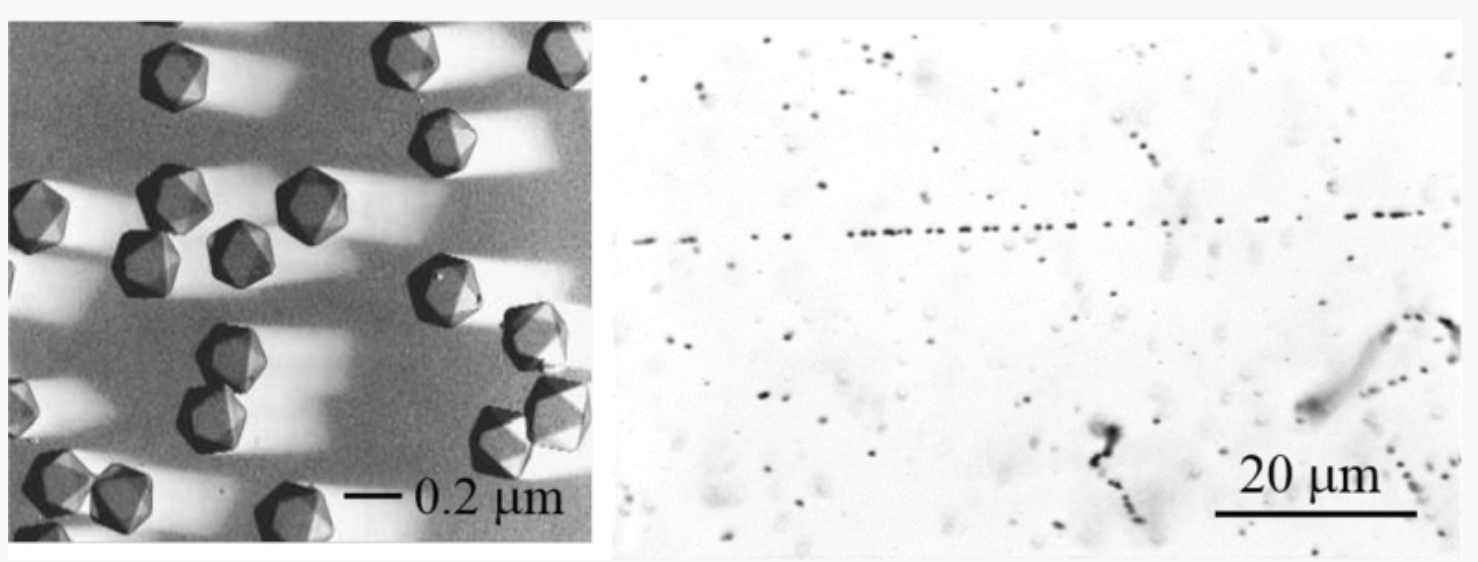}
\caption{\label{fig:emulsiongrain}Left: halide bromide crystals (0.2 $\mu m$ linear size) seen under 
an electron microscope. Right: the track left by a minimum ionizing particle (10~GeV $\pi^{-}$) in an emulsion film. Compton electrons are also visible on the right-bottom of the view.}
\end{figure}

A modern emulsion film has two sensitive layers of $\sim$ 50 $\mu$m thick on both sides of a plastic sheet of $\sim$ 200 $\mu$m, e.g. as for the OPERA Film~\cite{Nakamura:2005xov}. Such a structure minimizes effects of gelatin layer distortions, and provide 
3D-vector information with a high position resolution of effectively 0.2~$\mu m$ and angular resolution of 0.3~mrad. For the neutrino detection, the so-called ECC (Emulsion Cloud Chamber) technique is often employed. An ECC has a sandwich structure of emulsion films and massive target plates, to fulfill the detector requirements for tau neutrino detection: a``large mass'' and  a``high spatial resolution''. 
It provides:
\begin{description}
\setlength{\itemsep}{0pt}
\item [topology characterization] with a high spatial sub-micrometric resolution, allowing for a separation of secondary particles trajectories even in high multiplicity events in the deep inelastic scattering regime. Decay topology of short-lived particles such as $\tau$ leptons, charm and beauty hadrons can be detected.
\item [a sampling calorimeter] with the ability to count each single shower track for both electromagnetic and hadron shower measurements.
\item[a momentum estimator] by using the multiple Coulomb scattering. Through coupling with high-Z material, such as lead or tungsten, the scattering of particle can be measured thanks to the high spatial resolution, and can be translated into the particle momentum.
\item[muon identification] by the track range in the detector. Using a sufficiently large detector structure, muons can be separated from hadrons and electrons.
\item[$e$/$\gamma$ separation]; $\gamma$'s can be separated by the distance of their tracks from the interaction vertex. Furthermore, the energy deposition (darkness of the track) in emulsion layers can tell whether the track is caused by a single particle ($e$) or two particles ($\gamma\rightarrow e^+e^-$).
\end{description}
The trajectories in emulsion detector are measured by fully automated optical microscopes. The scanning speed, measured in terms of the amount of film surface processed per unit time, has increased significantly in recent years~\cite{Yoshimoto:2017ufm}. During the OPERA experiment film processing times a speed of 72~cm$^2$/h was reached; with the new generation of scanning 
systems, called hyper-track selector (HTS), the scanning speed exceeds 4700~cm$^2$/h. The advances in the scanning systems in terms of high efficiency and speed led in recent years to the re-appreciation of  emulsion detectors as an experimental technique.

Emulsion technology has efficiently contributed to the history of particle physics with fundamental discoveries: from the discovery of $\pi\rightarrow\mu$ decay in 1947, to the discovery of open charm particles in cosmic ray in 1971~\cite{Niu:2008zz}, and the first observation of $\nu_\tau$ in 2001~\cite{DONuT:2007bsg}. Moreover, emulsion technology was used to search for charmed particles in neutrino charged current interaction and to measure their lifetimes, to search for beauty productions in hadron interactions, to measure $\nu_\tau$ cross sections for the first time, and to study neutrino oscillations both in short baseline~\cite{CHORUS:2007wlo} and long baseline experiments~\cite{OPERA:2019kzo}. Last but not least, for the first time, the recent detection of neutrino candidates at the LHC has again demonstrated the effectiveness of the use of emulsion detectors~\cite{FASER:2021mtu}. With its rich history and unique capabilities, as mentioned above, emulsion based detectors will surely continue to serve as an important tool in high energy physics.

\subsubsection{Tracking calorimeters: transverse kinematic reconstruction}

It was first suggested by Albright and Shrock in 1979 that in principle one could reconstruct $\tau$ neutrinos in beam experiments using kinematic criteria~\cite{Albright:1979vd,Albright:1978ni}. The basic idea was to take advantage of the presence of two undetected neutrinos in the final state of the leptonically decayed tau lepton, and statistically search for a leptonic tau decay signal. This method had the potential to lead to the first observation of a $\tau$ neutrino, and was largely exploited by the NOMAD collaboration in the 90's. 

NOMAD was a short baseline (820~m) neutrino experiment which successfully operated with the CERN SPS wide neutrino beam (mean neutrino energy of 24~GeV). In its final analysis the collaboration covered more than 80\% of the tau decay branching ratio, including several hadronic decay modes and the electronic decay mode. The muonic decay mode was too drowned in the high level of $\nu_{\mu}$ charged current background but was used to generate data driven $\nu_{e}$ charged current events for the background study of the $\tau^{-} \rightarrow e^{-} \bar{\nu}_{e} \nu_{\tau}$ decay mode analysis.

The NOMAD detector~\cite{NOMAD:1997pcg}, which is best described as a low-density spectrometer, was composed of a set of subdetectors each assigned to a specific task. In particular the electromagnetic calorimeter aimed at precisely measuring the energy of the electron with a precision at the few percent level, and it was combined with a powerful electron identifier which reached an electron identification of more than 90\% with a charged pion rejection factor of more than $10^{3}$. The main subdetector was the drift chamber with a fiducial mass of 2.5 t and an excellent tracking performance allowing for a momentum resolution of charged particle, typically pions, at the level of few percents. 

The correct reconstruction of the kinematics in the transverse plane was indeed a key requirement to set up the test proposed by Albright and Shrock. Indeed, a powerful kinematic variable to look at for leptonic decays modes of the tau lepton is the transverse missing momentum. In such cases the decay of the charged lepton tau produces two final state undetected neutrinos which carry away a substantial fraction of the impinging neutrino energy, thus resulting in a large transverse momentum unbalance between the visible products (electronic/muonic and the hadronic system). The main associated background are the beam charged current electron/muon neutrino interactions but for which the transverse missing momentum is dominated by the Fermi momentum of the hit nucleon and the incorrect reconstruction of final state particles such as neutrons. Moreover, one can take advantage of the correlation, in the transverse plane of the interaction, between the visible leptonic momentum (here the electron/muon) and the hadronic momentum. For true electron/muon neutrino interactions, the two should indeed have a back to back configuration while for tau neutrinos interactions (with the tau lepton decaying into an electron/muon plus two neutrinos), many other configurations are possible because of the two undetected neutrinos.

\begin{figure}
    \centering
    \includegraphics[scale=0.30]{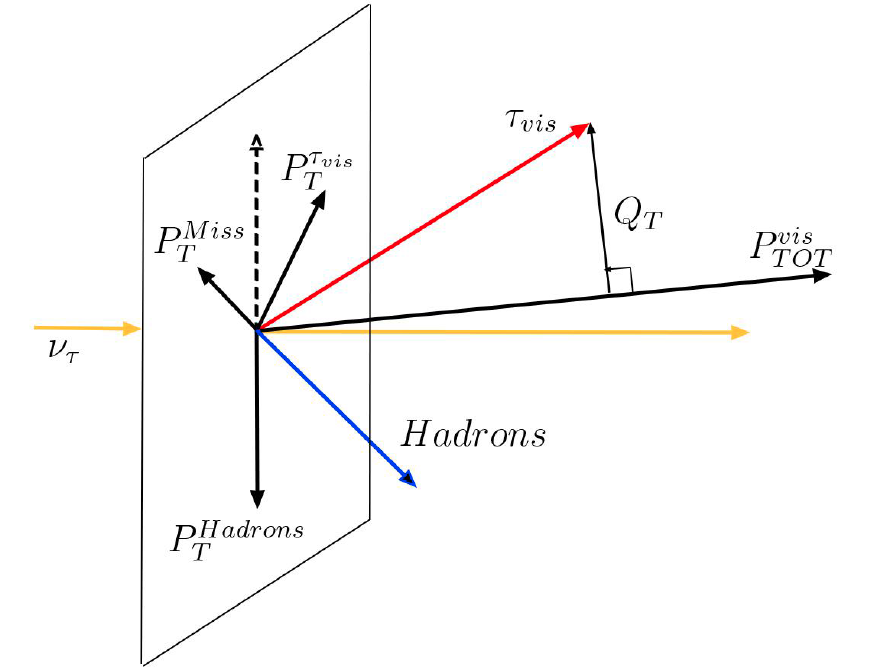}
    \caption{Schematic view of a tau neutrino charged current interaction, and the projection of the visible final state products (hadronic system and tau decay products) momenta in the transverse plane of the interaction.}
    \label{fig:tools:TransversePlaneKin_nutauCC}
\end{figure}

One important aspect for the reconstruction of tau neutrinos with tracking calorimeters is that the signal selection analysis strongly depends on the searched tau decay signature. Relevant kinematic variables for a given decay mode might be poorly sensitive in an other decay mode. As for illustration, one can think to the transverse missing momentum which is very discriminating between tau neutrinos (where $\tau^{-}\rightarrow e^{-}\bar{\nu}_{e}\nu_{\tau}$) and electron neutrino charged current interactions. However, for an hadronic decay mode ($\tau^{-}\rightarrow\pi^{-} \nu_{\tau}$, $\tau^{-}\rightarrow\rho^{-} \nu_{\tau}$...), for which the main background component are the neutral currents, both the signal and the background have one undetected final state neutrino. Thus it is likely that the transverse missing momentum be comparable in the tau neutrino signal and the neutral current background. One should stick to the philosophy of: one tau decay mode equals one dedicated analysis, as already suggested by the pioneering NOMAD experiment.

In the future the DUNE experiment will be operating gigantic liquid argon time projection chambers on a baseline favorable to the oscillation of muon neutrinos into tau neutrinos. It is expected to occur few tens of tau neutrino charged current interactions per year and per far detector module. DUNE will thus consist in a unique and unprecedented tau neutrino observatory with associated rich and opportunistic scientific perspectives~\cite{Ghoshal:2019pab, DeGouvea:2019kea}. Moreover DUNE may have the opportunity to run with an alternative configuration, resulting in a higher energy neutrino beam, which would foster the tau neutrino statistics by approximately a factor of 6. Even though liquid argon TPCs do not operate with a magnetic field, as opposed to NOMAD, it is nonetheless possible to achieve a reasonable particle identification using the mean ionization signal (dE/dx) along the particles trajectory combined with the range of the particle. Such possibilities were already explored in the protoDUNE single phase (horizontal drift) to distinguish among others, muons from protons~\cite{DUNE_FirstresultProtoDUNESP_2020}. In principle such methods can also be deployed to identify charged pions. Future results that go in that direction must be scrutinized, since identifying final state particles is a key requirement to search for tau neutrinos in tracking calorimeters. Early studies already suggest DUNE will be able to achieve a reasonable signal selection and background rejection~\cite{Machado:2020yxl}.

It should be recalled that DUNE is not primarily designed to study tau neutrinos, however it will offer a unique opportunity to probe the $\nu_{\mu} \rightarrow \nu_{\tau}$ oscillation channel. The selection and reconstruction of the tau neutrino sample will allow performing 3-flavor phenomenology. It is also expected to play a determining role in the PMNS matrix unitarity constraint, in particular for its last row (see sec.~\ref{sec:3fl_paradigm}). The possibility to run with an alternative higher energy beam is also a promising opportunity to perform cross section measurements.  

The quality of transverse kinematic reconstruction depends strongly on the quality of the underlying reconstruction of tracks and showers, including accurate particle identification and energy and momentum estimation. Therefore, modern machine learning techniques have the potential to significantly improve the physics reach of high resolution tracking calorimeters.  Two promising techniques currently under investigation are Graph Neural Networks (GNNs) and Panoptic Segmentation.

Recent work from the Exa.TrkX collaboration to develop GNN techniques for LArTPC event reconstruction provides promising prospects for reconstructing high-energy $\nu_{\tau}$ interactions with sufficient efficiency to reject background candidates based on event kinematics. This approach operates directly on detector hits, assigning each a semantic label based on patterns learned from simulated neutrino interactions. A GNN trained on charged-current quasielastic neutrino interactions achieved an 84\% efficiency in labeling detector hits from the charged lepton and hadronic systems \cite{ExaTrkX:2021}. Crucially, this approach does not rely on any intermediate reconstruction steps such as clustering or track/shower forming.

GNN approaches are natively sparse, which means they scale to events with a large spatial extent (such as high-energy $\nu_{\tau}$ interactions in the DUNE far detector) considerably more efficiently than a dense CNN-based method. Work is currently underway to extend this technique to a more comprehensive taxonomy of semantic particle categories -- in the context of $\nu_{\tau}$ interactions, this approach could be specifically targeted towards separating the hadronic and leptonic systems at the hit level for CC $\nu_{\tau}$ interactions in which the $\tau$ decays leptonically. This separation which would enable reconstruction of the transverse momentum for atmospheric $\nu_{\tau}$ interactions, which can be used to disambiguate leptonic CC $\nu_{\tau}$ interactions from CC $\nu_{\mu}$ and $\nu_{e}$ background events. If achieved, further in-progress developments such as instance labeling to group hits into individual particle instances would provide a detailed reconstruction of $\nu_{\tau}$ events, and improve kinematic reconstruction prospects even further.

Panoptic segmentation is a computer vision task that unifies two distinct techniques to segment images: semantic and instance segmentation. Semantic segmentation is the task to assign a class label to each pixel/voxel in 2D/3D images indicating what type of particle caused the energy deposit. Instance segmentation is the task to assign an instance label to pixels/voxels that belong to the same instance object in the image which is equivalent to high level object formation in many physics experiments. Therefore, panoptic segmentation assigns a unique value to every pixel/voxel that encodes both semantic and instance label \cite{panoptic_deeplab}.

Experiments such us MicroBooNE, ICARUS, ProtoDUNE and the future DUNE utilize Liquid Argon Time Projection Chambers (LArTPCs) technology that is able to capture high-resolution images  and fine details of hadronic and electromagnetic interactions. Despite the advantages of this type of detectors, the reconstruction and classification of interacting particles is still a challenge, therefore, a multi-task machine learning algorithm using a Sparse Convolutional Neural Network could be suitable tool for particle identification and event reconstruction.

\subsubsection{Water and ice Cherenkov at lower energies}
\label{sec:tools water and ice low energy}
Below $\sim$ 50~GeV oscillation-induced $\nu_{\tau}$ from the atmosphere are visible at water and ice Cherenkov detectors, such as Super-Kamiokande (Super-K), Hyper-Kamiokande (Hyper-K) and IceCube. 
As the Cherenkov technique at these detectors lacks the granularity to accurately reconstruct all of the particles produced in the charged current interaction of these neutrinos, their reconstruction relies on statistically separating them from other atmospheric neutrino backgrounds. 
Both water- and ice-based technologies have demonstrated the ability to identify $\nu_{\tau}$ above considerable backgrounds using coarse properties of the interaction topology. 

At Super-Kamiokande the $\nu_{\tau}$ search is designed around the hadronic decays of the $\tau$ lepton. 
Due to the large production threshold of the $\tau$, most $\nu_{\tau}$ interact via deep inelastic scattering, meaning there are several hadrons produced at the vertex in addition to the initial lepton. 
If the latter decays hadronically additional hadrons will be produced.
The result is a final state with many charged particles distributed more isotropically in the detector volume than a background event that lacks the contribution from the heavy $\tau$'s decay.
We note that the leptonic decays are also detectable in principle, though they indistinguishable from ordinary atmospheric $\nu_{l}$ charged current interactions and form a small component of those samples. 

Currently the Super-K neural network uses seven input variables to identify $\nu_{\tau}$~\cite{Super-Kamiokande:2017edb}. 
The variables are formed to take advantage of the fact that hadronic $\tau$ decays in general have more charged particles that are distributed more isotropically throughout the detector than backgrounds.  
Further, background events tend to have more charge deposition in just outside the Cherenkov ring of their most energetic particle, indicative of forward-going particle flow not seen in the $\tau$ signal.
For a typical operating point in the neural network output, this algorithm achieves 76\% efficiency for selecting $\nu_{\tau}$ (28\% for $\nu_{l}$) , but results in a sample that is only 4.7\% pure.  
Among the backgrounds, neutral current interactions are the most challenging; their feature variables often have similar shapes to those of the signal, which leads to many events populating the very $\tau$-like values of the neural network discriminant.

The challenge going forward, particularly at the next-generation experiment Hyper-K, is to identify new ways of reducing these backgrounds.
Since 2008 Super-K has had the ability to identify neutrons via $n(p,d)\gamma$ (2.2 MeV) and via $n(Gd,X)\gamma$ ($\sim 8$) MeV since 2020. 
Though highly uncertain, $\nu_{\tau}$ interactions appear to have more neutrons than backgrounds, including neutral current interactions. 
Making use of the number and distribution of those neutrons throughout the detector is another potential handle for $\nu_{\tau}$ discrimination. 
Further, new reconstruction efforts such as the improved likelihood fitter in ~\cite{Super-Kamiokande:2019gzr} or future machine-learning based approaches may be increase sensitivity.

Very large volume neutrino telescopes (VLVnTs) collect unprecedented amounts of $\nu_\tau$ events from oscillated atmospheric neutrinos\cite{IceCube:2019dqi}, but struggle and are often incapable of identifying any individual data event as $\nu_\tau$. At energies $<\approx100$\,GeV, a tau lepton will travel $\mathcal{O}(1)$\,mm before decaying, which makes it nearly impossible to resolve for detectors such as IceCube which have instrumentation distances of 7\,m to 17\,m vertically and 45\,m to 125\,m horizontally. Only at $\mathcal{O}(100)$\,TeV energies where the tau travels $\mathcal{O}(0.5)$\,m do charged current $\nu_\tau$ start to have features, e.g.\@ double-pulse or even double-bang at PeV energies\cite{Learned:1994wg}, that are unique for $\nu_\tau$ interactions and can be resolved with instrumentation that is tens of meters apart, but which also has $\mathcal{O}(3)$\,ns timing resolution. So any sub-TeV $\nu_\tau$ is largely indistinguishable from a CC $\nu_e$ or any NC event. As such, sub-TeV $\nu_\tau$ analyses in VLVnTs rely on the large statistics and quality reconstruction to determine the statistical impact of $\nu_\tau$ events on top of a large background, similar to `bump hunting' in collider experiments.

Reconstructing the event morphologies in VLVnTs such as IceCube DeepCore\cite{IceCube:2011ucd} is necessary to determine the important quantities of vertex position (x, y, z), time (t), azimuth and zenith direction ($\theta_{zen}, \phi_{azi}$), and energy of the neutrino $E_\nu$; where $E_\nu$ is equal to the cascade energy ($E_{cscd}$), except CC $\nu_\mu$ events which must also include the track-like signature from muon ($E_{trk}$). Despite an $\approx 17\%$ branching ratio of a tau lepton decay containing a muon, this particular decay channel is difficult to identify as track-like at lower energies because the out-going muon is too low in energy to extend far enough outside of the cascade to be identified. For example, the 3-year oscillated $\nu_\tau$ appearance analysis from IceCube DeepCore\cite{IceCube:2019dqi} only classified as track-like if the reconstructed track length was greater than $>50$\,m. With improved reconstructions algorithms and future detectors with better sensors and closer spacings between sensors, such as the IceCube Upgrade and KM3NeT-ORCA, the ability to classify $\nu_\tau$ interactions which produce a track-like muon as actual track-line events will improve; otherwise most $\nu_\tau$ events are classified as cascade-like regardless of the decay channel.

For any events in IceCube DeepCore which include a cascade-like signature, the current reconstruction method uses sensor-by-sensor tables which model the likelihood for a specific sensor to observe a photon as a function of time, and are varied during the minimization process to best match the observed photon distribution amongst all sensors in the events. The tables include the impact of photon propagation differences due to the varying photon scattering and absorption in the glacial ice at the South Pole, as well as the different photon characteristics depending on incident particle energy, particle type, direction, and neutrino interaction vertex. Whilst the table-based approach has been important for current lower energy analyses in IceCube DeepCore (both $\nu_\tau$ related and otherwise), the reconstructions are a significant computational bottleneck and will only become more onerous in the next generation of lower energy VLVnTs due to the increasing dimensionality of the segmented optical sensors to be deployed. Notably, the multi-PMT digital optical modules (mDOMs)--pioneered by the KM3NeT consortium\cite{KM3NeT:2013qjg} and modified for in-ice use in future IceCube extensions\cite{IceCube:2021eij}--consist of 24 PMTs with 3'' diameter, which in comparison to the single PMT in the existing IceCube DOMs would increase the reconstruction time beyond `bottleneck' category and into the `functionally unfeasible' category.

The future of reconstruction algorithms for large volume neutrino telescopes with multipixel sensors is being actively pursued with machine learning. Both KM3Net\cite{KM3NeT:2020zod} and IceCube\cite{IceCube:2021oxp} are developing convolution neural networks focused on energy regions relevant for atmospheric oscillated $\nu_\tau$ physics, e.g.\@ $<100$\,GeV, for both reconstruction as well as particle identification. Because current and upcoming VLVnTs have only semi-regular 3-dimensional sensor positions, there has also been a concerted effort to develop graph neural network reconstruction algorithms for lower energy events \cite{IceCube:2021dvc,Reck:2021zqw}. The data already resembles `point clouds', and the GNNs preserve and learn adjacency in point cloud data that can be irregular, whereas other methods (CNNs) require regular data formats for convolution kernels.

A significant advantage to ML reconstruction/classification tasks is the speed with which they can be executed. The IceCube DeepCore table-based reconstructions can take minutes per event to evaluate, whereas CNN and GNN algorithms take $\mathcal{O}(1-10)$\,ms and have comparable reconstruction resolutions that are improving with further development. The faster and less computational intensive reconstructions allow analyzers to include better reconstruction information at earlier stages of the event selection which will improve purity and efficiency, while also introducing the ability to run lower energy event reconstructions at the South Pole for the use in sending sub-TeV astrophysical alerts\cite{IceCube:2021oos}.

\subsubsection{Water and ice Cherenkov at high energies}

At higher energies, the longer tau decay length may allow for the tau decay shower and the neutrino interaction shower to be separated in reconstruction. Depending on the tau decay length and the detector geometry, the following reconstruction methods are currently available:
\begin{itemize}
	\item \textbf{Individual reconstruction of two well-separated showers:}  The light induced by the event is split and the  constituent showers reconstructed separately. Typically, the tau track is not reconstructed. This ``Double Bang'' method was the first one proposed to identify astrophysical tau neutrinos in an almost background-free way \cite{Learned:1994wg}.\\
	This method has been employed in Baikal-GVD in a search for tau neutrino events where the interaction happens in / close to one cluster, and the decay in another, as it is particularly suited for the Baikal detector geometry. Using a maximum likelihood shower reconstruction algorithm, the individual showers are reconstructed separately \cite{Baikal-GVD:2021ugo}. The tau neutrino induced event is built from two almost coincident showers with matching directions. Due to the cluster layout of Baikal-GVD with blind spots between the clusters, the method is prone to muon induced background. Spurious coincidences should be rare due to the good directional reconstruction in water.\\
	In IceCube, this method would become sensitive above tau decay lengths of 500~m, where in addition to the falling spectrum also the detector geometry limits the number of identifiable events with both showers contained within the detector volume, resulting in very low expected event rates. No algorithmic search using event splitting into the constituent cascades has been applied, however, shower reconstruction and event splitting codes both exist. 
	\item \textbf{Reconstruction of two connected showers:} The entire light deposited used to reconstruct the event, using the hypothesis of two causally connected showers and a maximum likelihood algorithm. As the separation between the showers can be arbitrarily small in principle, the detection threshold in terms of tau decay length achievable with this method depends on the goodness of the reconstruction, the confusion with single showers, and systematic uncertainties. In this method of a ``double cascade'' reconstruction the individual showers typically overlap, thus the (comparatively dim) tau track is easily hidden and not reconstructed.\\
	In Baikal-GVD the algorithm used splits the light into two subsets used for reconstruction of the individual showers. It has been applied to Monte Carlo generated events with a required minimum shower separation of 10~m, and achieved a position resolution of 5~m or better \cite{Baikal-GVD:2021ugo}.\\
	In IceCube, the algorithm used is an extension to the shower reconstruction algorithm by two degrees of freedom, the inter-shower separation (a proxy for the tau decay length) and the second shower's energy. Although the sensors are separated by $\sim$125~m horizontally and $\sim$17~m vertically, the very good time resolution of the collected data allows for a reconstruction resolution of $\sim$2~m. To keep a low misclassification fraction of single showers, the analysis threshold was set to 10~m. Two tau neutrino candidates have been observed using the double cascade method \cite{IceCube:2020abv}. As precise timing information is crucial, this method relies on precise modeling of photon propagation through the detector medium. In IceCube, particular care needs to be taken considering the anisotropy of photon propagation in the Antarctic ice \cite{Chirkin:2013lpu, IceCube:2021xhi}. If uncorrected, this anisotropy can lead to an elongation of single showers and thus a higher misclassification of single showers as double showers along the glacial flow direction. The IceCube-Upgrade \cite{Ishihara:2019aao} will contain new devices to calibrate the optical properties of the ice. This will result in a new, more precise model of ice optical properties, and thus improve the confidence in separating single and double showers down to lower shower separation lengths.\\
	In KM3NeT/ARCA, the algorithm is also an extension of a single shower maximum likelihood algorithm. The likelihood is maximized for the time pattern of when each sensor first detected light. The algorithm has thus far been tested on Monte Carlo generated events with a required energy above 100~TeV, with a resolution of $\sim$2~m on the tau decay length \cite{vanEeden:2021zzv}.\\
	The double cascade reconstruction algorithms can be used in a self-consistent way with shower and track reconstruction algorithms, for flavor identification and flavor composition measurements \cite{IceCube:2020abv}. 
	\item \textbf{Search for tau-specific features:} At even shorter tau decay lengths, a distinction between single and double shower by direct reconstruction becomes unfeasible. However, if the tau-neutrino interacts close to a light sensor, light from the neutrino interaction shower and the tau decay shower can be distinguished on that individual light sensor, resulting in a ``double pulse'' feature in the light collected on that sensor over time. While this feature can be caused by a muon having two large stochastic losses in close proximity, it can not be mimicked by a single shower, electron neutrino interaction. In absence of track signature, the presence of a double pulse points to a tau neutrino origin and the event's properties can be reconstructed using a single shower reconstruction algorithm. Timing is the most crucial aspect in the identification of double pulses: a time-binned readout of the light sensors' full waveforms is absolutely necessary, and the finer the binning (provided statistical fluctuations can be kept low), the lower the detection threshold can be pushed in tau decay length.\\
	This method was developed in IceCube \cite{IceCube:2015vkp}. The double pulse feature can also be incorporated into a tau neutrino identification scheme employing machine learning techniques \cite{Meier:2019ypu}. Two tau neutrino candidates have been observed using the double pulse method, one of which has also been observed with the double cascade method \cite{Meier:2019ypu, Wille:2019pub}.\\
	Baikal-GVD has studied the performance of a double pulse algorithm on Monte Carlo \cite{Baikal-GVD:2021ugo}.\\
	While the double pulse algorithms do not reconstruct the tau properties, they are a sensitive tool to identify tau neutrino candidates. In a sparsely instrumented, segmented detector, the expected event rates scale approximately with the number of light sensors that are not at the detector boundary.
	\item \textbf{Other features with indirect sensitivity to tau neutrinos:} Events containing a shower and a track (``starting track'') can be induced by either tau neutrinos or muon neutrinos. At very high energies, the tau may leave the detector, while at lower energies the tau may decay producing the muon (branching ratio of 17\%). The \textbf{inelasticity} has a different distribution for tau neutrino induced starting tracks than for muon neutrino induced starting tracks. This makes it possible to measure the tau contribution among starting tracks, given a large enough number of events. However, the method suffers from the low branching ratio of the muonic tau decay, as well as the large event-by-event variations in elasticity. This method has been employed in IceCube \cite{IceCube:2018pgc}.\\
	A \textbf{stopping track} is the result of a tau entering the instrumented volume and decaying within. In IceCube, a segmented spline track reconstruction \cite{IceCube:2021oqo}, initially developed for tracks, can be employed as it reconstructs large stochastic losses along the track. No such event has been observed.\\
	The first shower created when the tau neutrino interacts, is a hadronic shower, the second shower can be electromagnetic or hadronic. The \textbf{hadronicity} distribution of tau neutrino events is thus between fully hadronic neutral current events, and electron neutrino charged current events with large electromagnetic contributions. The hadronicity can be estimated by looking for the afterglow of high energy events, a ``neutron echo'' or ``muon echo'', in the medium \cite{Li:2016kra}. IceCube has searched for neutron echoes in events with reconstructed energies above 100~TeV, however, the understanding of the afterglow signal relies on a precise understanding of the medium and its intrinsic luminescence \cite{Steuer:2017tca,Pollmann:2021jlo}. 
\end{itemize}
Several tools have been established for tau neutrino searches at high energies in water and ice Cherenkov detectors, and applied to data in IceCube \cite{IceCube:2015vkp,Meier:2019ypu,Wille:2019pub,IceCube:2020abv} and Baikal-GVD \cite{Baikal-GVD:2021ugo}. Within this decade, a maturing of the techniques can be expected in the detectors under construction, Baikal-GVD and KM3NeT/ARCA. In IceCube, a combination of several of the established tools in an IceCube analysis is feasible, and could be applied to a combined sample of high energy neutrino induced events \cite{IceCube:2015gsk,IceCube:2021jmr}. Further, the IceCube-Upgrade will allow for the reduction of systematic uncertainties in the event classification due to better modeling of the optical properties of the South Pole Ice. New tools for fast classification of event topologies using machine learning techniques will likely be soon available. In the next decade, IceCube-Gen2 \cite{IceCube-Gen2:2020qha} will provide a much larger in-ice detection volume, yielding higher numbers of identifiable tau neutrinos per year, especially at the highest energies above 1~PeV. As both KM3NeT/ARCA and IceCube-Gen2 use segmented optical sensors with several PMTs per sensor, new reconstruction algorithms making use of the segmentation or tau-neutrino-sensitive signatures on multiple PMTs of an individual sensor will likely be developed. Such a signature could be a combination of a double pulse and a double cascade: light from the two vertices could be distinguished due to the photon arrival times at different PMTs of the same sensor, and provide higher identification confidence and better directional resolution simultaneously.\\
When designing new instruments for high-energy neutrino detection, the geometry and data collection choices should be chosen with care. While a sparse, clustered detector layout \cite{Baikal-GVD:2021zsq,P-ONE:2020ljt} enhances the sensitivity to neutrino sources, the inter-cluster blind spots can lead to lower muon-rejecting performance critical to tau-neutrino identification. Readout and time-resolved digitization of the full waveform on at least the brightest sensors should be pursued to enable tau-neutrino identification via double pulses. In KM3NeT/ARCA, only time of first light and total time over threshold is recorded, thus lowering the sensitivity to double pulses to either very well separated showers, or the hybrid signatures accessible due to sensor segmentation.

\subsubsection{In-ice radio detection at UHE energies}
Obtaining sensitivity the neutrino flavor at UHE energies with in-ice radio detectors is challenging. Neutral current interactions of all flavors will produce the same event signature but through charge-current interactions sensitivity to tau neutrinos can be obtained via three different channels
\begin{itemize}
    \item \textbf{Radio emission from secondary tau leptons} A tau lepton generated in $\nu_\tau$-CC interactions will induce several high-energy particle showers through stochastic energy losses during its propagation through the detector volume. The particle showers generate radio emission through the Askaryan effect which provides the measurable signature. The tau decay is only relevant at the low-energy threshold of in-ice radio detection because the tau decay-length quickly exceeds the detection volume. A detection of a radio signal from the initial neutrino interaction and a secondary interaction will provide a signature of a tau. The secondary tau channel will increase the detection rate of tau neutrinos by 20\% at \SI{e18}{eV} and  by up to 40\% at energies above \SI{e19}{eV} \cite{Garcia-Fernandez:2020dhb, Glaser:2021hfi}. At high energies, often both the initial and secondary interaction is detected in neighboring detector stations. 
    The largest background for this detection channel are muons generated in $\nu_\mu$-CC interactions which generate a very similar signature as taus. 
    
    \item \textbf{Identification of LPM elongated showers} Due to the LPM effect, $\nu_e$-CC interactions can be distinguished from NC interactions or CC interactions of muon and tau neutrinos \cite{Klein:2020nuk,Gerhardt:2010bj,Stjarnholm:2021xpj}. The cross-section of the electron generated in $\nu_e$-CC is reduced through the LPM effect which leads to a delayed and stochastic shower development with several sub-showers compared to the compact hadronic shower generated in other types of interactions \cite{NuRadioMC2019}. This enables a estimation of the combined muon and tau neutrino flux vs. the electron neutrino flux at UHE energies.
    
    \item \textbf{Tau regeneration through Earth} Nominally the Earth is opaque to UHE neutrinos but tau neutrinos can propagate larger distances through continuous tau regeneration \cite{Ritz:1987mh, Halzen:1998be}. The short lifetime of the charged tau lepton causes it to decay before losing a large fraction of its initial energy. As a result, EeV tau neutrinos propagating through Earth will escape with energies that peak between O(100) TeV to O(10) PeV depending on the arrival angle with respect to the horizon \cite{Safa:2019ege, Safa:2021ghs}. This provides a unique signal for the detection of tau neutrinos; namely, the search for Earth-traversing PeV events \cite{IceCube:2021pue}.
    
\end{itemize}

\subsubsection{Skimming techniques}
The final reconstruction technique leverages the unique properties of tau neutrinos.
When an ultra-high energy tau neutrino passes through the Earth or a mountain, its mean free path is relatively short and it will interact quickly.
If the interaction is CC it will produce a tau lepton which could live long enough to escape to the atmosphere depending on the geometry.
If the tau decays hadronically or to an electron which happens $\sim83\%$ of the time then it will create a shower in the atmosphere.
Since this is the only SM process that leads to an extensive airshower coming out of the Earth, any detection must be of a tau neutrino \cite{Feng:2001ue,Kusenko:2001gj}.
This technology is leveraged by several current and upcoming experiments as described in section \ref{sec:he experiments}.
This would lead to a scenario where we may have information, such as flux and cross section, about tau neutrinos but neither of the other two flavors, see e.g.~\cite{Denton:2020jft} on UHE neutrino cross sections.

\subsection{Ultra-High Energy Tau Neutrino and Tau Lepton Propagation Codes}
Several experiments aim to detect tau neutrinos with energies greater than a PeV benefit from tau neutrino regeneration where tau neutrinos propagate through the Earth with minimal energy loss such that a UHE neutrino detector on the other side has some chance to detect them.
The propagation of tau neutrinos depends on several factors including the medium through which the tau neutrinos and leptons propagate, neutrino cross sections, tau energy losses, and tau decay processes.

Recently several groups have emerged to accurately model tau propagation, including NuPropEarth, NuTauSim, TauRunner, nuPyProp in $\nu$SpaceSim, and DANTON. Table~\ref{tab:software_Tau_neutrino_propagation} outlines the main features of these codes, and we briefly describe the benefits of these modeling tools here. Fig.~\ref{fig:pexit-compare} and Fig.~\ref{fig:energy-compare} show a comparison of the tau exit probability and energy distribution of outgoing leptons for different packages. 

\begin{table}[!h]
\footnotesize
\centering 
\begin{tabular}{l|l|l|l|l|l}
\hline
\hline
 Software & Medium & Cross-Section & Energy Loss & Decay & Secondaries \\
\hline
\hline
 \texttt{NuPropEarth} & PREM$^{*}$ & DIS+Others (GENIE) & PROPO./TAUSIC & TAUOLA & $\nu$(all), $\tau$ \\
\texttt{TauRunner}   &  PREM, Sun$^{*}$ &    DIS (Table)       & PROPOSAL & Param.  & $\nu$(all), $\tau$, $\mu$\\
\texttt{nuPyProp}   & PREM$^{*}$ &    DIS (Table)       & Table    & Param. & 
$\tau$\\   
\texttt{NuTauSim}     & PREM     &    DIS  (Param.) & Param.$^{**}$ & Table & $\nu_\tau$, $\tau$ \\
\texttt{Danton}       & PREM*   & DIS+GLRES (ENT)    & PUMAS   & TAUOLA (Alouette) & $\nu$(all), $\tau$ \\
\hline
\hline
 \end{tabular}
 \caption[Tau Neutrino Propagation]{Tau Neutrino Propagation: the propagation of neutrinos through matter is fundamental to quantify the sensitivity to detect high-energy tau neutrinos. 
* Other geometries can be imported, **Not stochastic}
\label{tab:software_Tau_neutrino_propagation}
\end{table}

\begin{figure}[!h]
\centering
\includegraphics[width=.4\linewidth]{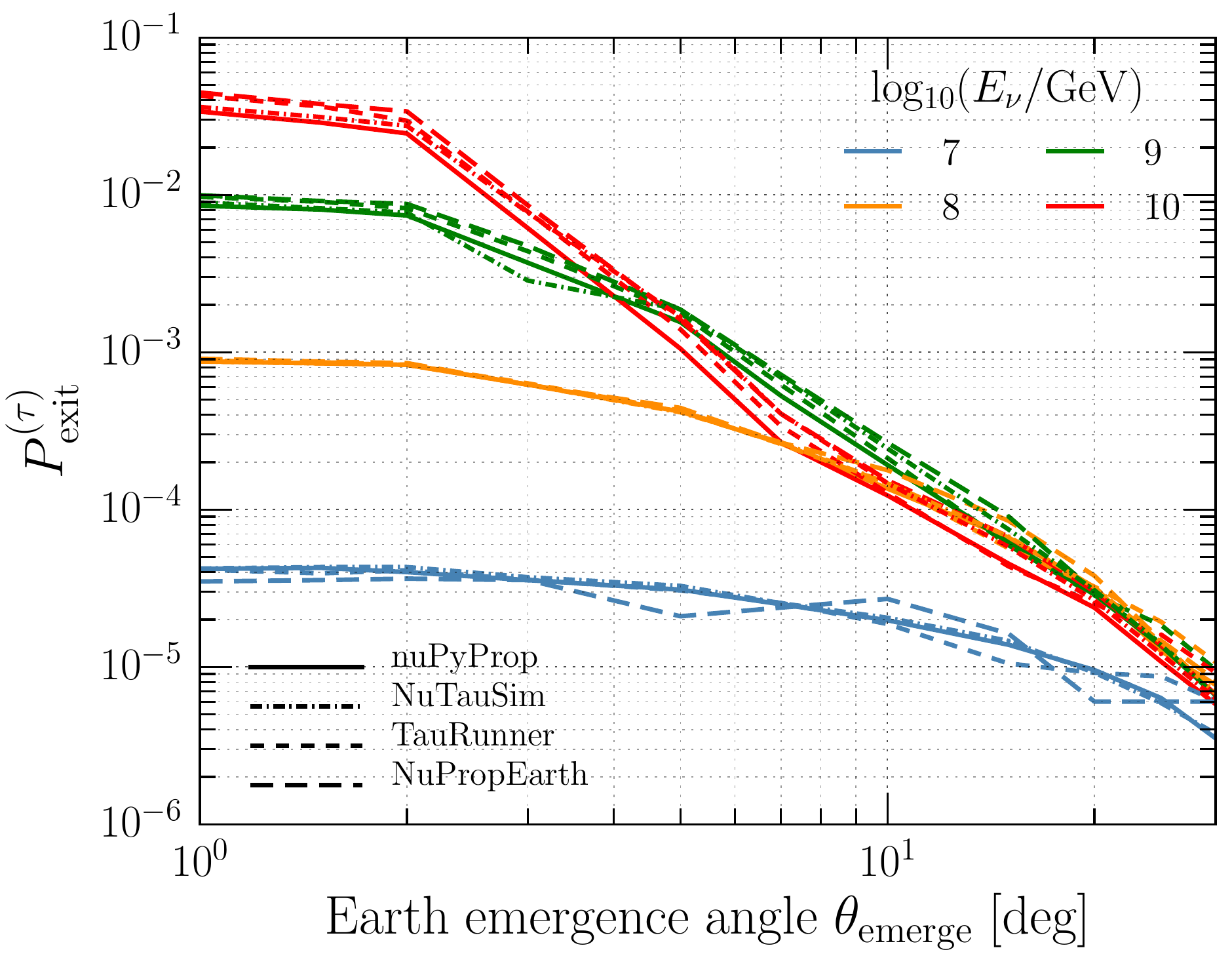}
\caption{\label{fig:pexit-compare} The tau exit probability as a function of Earth emergence angle for $E_\nu=10^7$, $10^8$, $10^9$ and $10^{10}$ GeV from nuPyProp \cite{NuSpaceSim:2021hgs}, NuTauSim \cite{Alvarez-Muniz:2018}, TauRunner \cite{Safa:2021ghs} and NuPropEarth \cite{Garcia:2020jwr}, all using the PREM Earth model \cite{Dziewonski:1981xy} modified for a water layer of 4 km and the ALLM parameterization \cite{Abramowicz:1997ms} of the structure function for tau photonuclear electromagnetic loss. }
\end{figure}

\begin{figure}[!h]
\centering
\includegraphics[width=1.\linewidth]{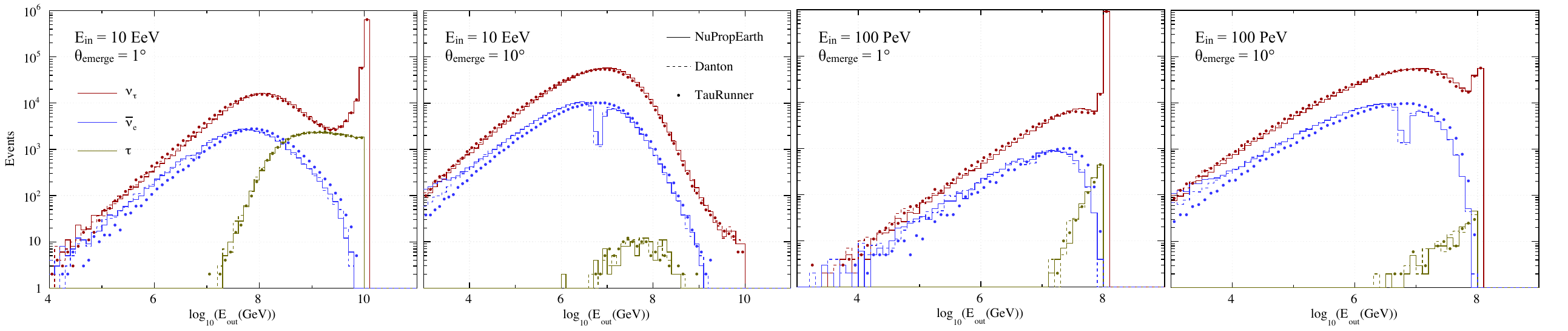}
\caption{\label{fig:energy-compare} A comparison of the energy distribution of neutrinos and taus exiting the Earth. Distributions have been generated assuming two monochromatic fluxes (100 PeV and 10 EeV) of 1 million $\nu_\tau$ with different emergence angles. The following configuration was assumed: PREM model with 4km of pure water; CSMS cross section~\protect\cite{Cooper-Sarkar:2011jtt}; photo-nuclear cross sections using ALLM97 parametrization and shadow effects~\protect\cite{Dutta:2000hh}; bremsstrahlung and electron pair production without LLPM effect~\protect\cite{Soedingrekso:2019qlr}; left-handed taus.}
\end{figure}

\subsubsection{\texttt{nuPyProp}}
The nuPyProp \cite{NuSpaceSim:2021hgs} open source program is designed to provide look-up tables for charged lepton exit probabilities and energies that are input to nuSpaceSim \cite{Krizmanic:2020bdm}, a package for the end-to-end modeling of optical and radio signals of extensive air showers that originate with tau neutrino and muon neutrino incident on the Earth. The intended use of nuSpaceSim is for space-based and suborbital detector design and analysis. The nuPyProp code is a stand-alone python package and command line utility that is available on GitHub\footnote{\url{https://github.com/NuSpaceSim/nupyprop}} 
and through pip and conda installations. Sub-modules written in fortran, wrapped with F2PY, and the use of OpenMP make this code relatively fast. Neutrino interactions and charged lepton interactions are implemented stochastically. Several neutrino cross section models and charged lepton photonuclear energy loss models are included along with standard bremsstrahlung, pair production and ionization energy loss inputs. Templates are provided for the user to input other neutrino and charged lepton interaction models, for example, to allow for BSM physics. Approximate density dependent correction factors to the nuclear charge $Z$ and atomic number $A$ are incorporated in the charged lepton electromagnetic energy loss evaluation. The tau neutrino energy distribution from tau decays in the regeneration process is parameterized. The PREM Earth density is the default Earth model, with the option to set the outer water layer depth from 0-10 km (in 1 km units), thus allowing its adaptation to neutrino and muon or tau propagation through mountains, for example. 

\subsubsection{\texttt{TauRunner}}

\texttt{TauRunner}~\cite{Safa:2021ghs} is an open-source, \texttt{Python}-based package for propagating neutral and charged leptons.
The standalone version of the code can be installed via \texttt{pip}.
Once installed this may be imported into a \texttt{Python} script or a \texttt{Jupyter} notebook.
Additionally, the user may download the source code from GitHub\footnote{\url{https://github.com/icecube/TauRunner}} and use the command-line interface.
\texttt{TauRunner} uses a Monte Carlo approach to propagate leptons, including stochastic energy losses for all charged leptons via the \texttt{PROPOSAL}~\cite{Koehne:2013gpa} package.
The program propagates all leptons that can travel macroscopic distances---\textit{i.e.} all except $e^{\pm}$---and returns the initial and final particle energies, particle type, parent particle, final position, and number of interactions.
This allows the user to track the full physics of propagation. Furthermore, \texttt{TauRunner} can simulate an Earth-traversing EeV $\nu_{\tau}$ in ${\sim}10~\mu\rm{s}$, allowing millions of events to be simulated on a single core in ${\sim}\rm{hours}$.
Notably, stochastic loss treatments allow for this software to function as a Monte Carlo event generator for next-generation experiments as it can track individual events and record their interactions.
\texttt{TauRunner} is also quite flexible, allowing the user to provide their own cross-section models, flux predictions, and propagation media.
By default, the software provides two representative cross-section models, one based on NLO QCD calculations, and the other based on a dipole model of the nucleus.

\subsubsection{\texttt{NuPropEarth}}

\texttt{NuPropEarth} \cite{Garcia:2020jwr} is a open-source, \text{C++}-based package available on GitHub~\footnote{\url{https://github.com/pochoarus/NuPropEarth}}.
It has the structure of a general-purpose Monte Carlo event generator, and therefore allows following the path and interactions of individual neutrinos and tau leptons as they travel through Earth on an event-by-event basis.
Neutrino interactions are simulated with differential cross sections in \texttt{GENIE}, with the possibility of using customized DIS structure functions (SF). 
Several precomputed SF tables (as function of $x,Q^{2}$) based on NLO models are available~\cite{Cooper-Sarkar:2011jtt,Bertone:2018dse}. 
\texttt{GENIE} also accounts for other interaction channels like Glashow resonance~\cite{Gauld:2019pgt} or W boson production~\cite{Zhou:2019vxt}.
These subdominant interactions can play a significant role in the detection of tau neutrinos from cosmic origin~\cite{Soto:2021vdc}. 
Energy losses due to electromagnetic interactions of tau leptons are modeled with \texttt{TAUSIC}~\cite{Kudryavtsev:2008qh} or \texttt{PROPOSAL}~\cite{Koehne:2013gpa}, while its decays are computed with \texttt{TAUOLA}~\cite{Davidson:2010rw}.
Customized geometries can be imported using the \texttt{ROOT} geometry package. 
A driver to construct a geometry based on the PREM model is already available in the code.
The outputs stores the four-momentum and position of all incoming and outgoing leptons in the geometry. 
Other information like number of interactions, probe and target types is also accessible. 
All the steps in the simulation are done assuming 3D space, so deflection of particles is also taken into account. 

\subsubsection{\texttt{NuTauSim}}
\texttt{NuTauSim} \cite{Alvarez-Muniz:2018} is an open source \text{C++}-based Monte Carlo code which simulates the propagation of $\tau$ neutrinos through the Earth, taking into account neutrino interactions and $\tau$-lepton energy losses. Compared to other neutrino propagation codes, \texttt{NuTauSim} is fast ($\sim1$~ms average propagation time for a 1~EeV $\nu_{\tau}$ for Earth emergence angles $0^{\circ}$--$90^{\circ}$) largely because it considers only the average losses of the $\tau$-lepton, as opposed to modeling the losses stochastically. Including the stochastic losses of the $\tau$-lepton becomes important for $E_{\tau}>10^{19}$~eV and has been shown to provide a $\sim$10\% difference between the results of other propagation codes. The neutrino interactions in \texttt{NuTauSim} are calculated considering different high-energy extrapolations of the neutrino cross section \cite{Connolly:2011vc}, which are selected via user input. The photonuclear energy losses of the $\tau$-lepton are similarly calculated using the extrapolations to high energies given in \cite{Abramowicz:1997ms} and \cite{Armesto:2004hh} and selected via user input. The inelasticity of a neutrino interaction is sampled using standard results calculated from  \texttt{CTEQ5} parton distributions \cite{Lai_2000} and the decay products of the $\tau$-lepton are generated using the \texttt{PYTHIA8} Monte Carlo code, assuming a fixed (negative) polarization for the $\tau$-lepton. The grammage profile used to propagate events is calculated using the \texttt{PREM} Earth density model with an added outer layer with a user input depth and density to model more local topographies. \texttt{NuTauSim} is a highly modular code and can easily be modified to include different models not currently involved in the base implementation to quickly evaluate novel scenarios. \texttt{NuTauSim} is available for download on GitHub~\footnote{\url{https://github.com/harmscho/NuTauSim}}. Recent updates to \texttt{NuTauSim} have been made to allow for the propagation of muon neutrinos and muons from $\tau$-lepton decay, as well as the consideration of events from Glashow Resonance. This updated version, called \texttt{NuLeptonSim} will be available for public release in 2022. 

\subsubsection{\texttt{Danton}}

\texttt{Danton} is a detailed Monte~Carlo engine dedicated to the coupled $\nu_\tau$-$\tau$ transport problem. A peculiarity of \texttt{Danton} is that it can operate in forward or in Backward Monte~Carlo (BMC), using the method discussed in~\cite{Niess2018bmc}. The BMC technique allows to achieve CPU performances comparable to parametrized solvers, while keeping the accuracy of a full Monte~Carlo, as illustrated e.g. in~\cite{niess2018danton}.

\texttt{Danton} by itself is only a glue layer, relying on external BMC libraries. Neutrino interactions are simulated with ENT~\footnote{\url{https://github.com/niess/ent}}. DIS is randomized from the DDCS in $x$ and $Q^{2}$, using LO expressions with a configurable PDF. Since Danton~\texttt{v0.4}, the total DIS cross-section can however be rescaled to more detailed computations, e.g. the CSMS~\protect\cite{Cooper-Sarkar:2011jtt} or BGR18~\protect\cite{Bertone:2018dse} cross-section. Electronic interaction channels are also considered, like Glashow resonance~\cite{Gauld:2019pgt}. The $\tau$ transport is performed by \texttt{PUMAS}~\cite{Niess2018bmc}, a Monte~Carlo engine initially developed for precision muography applications. Decays are delegated to \texttt{Alouette}~\footnote{\url{https://github.com/niess/alouette}}, a BMC wrapper inverting TAUOLA~\cite{Davidson:2010rw}. In addition, \texttt{Danton} allows to use detailed Earth topography data through the \texttt{TURTLE}~\cite{Niess2019turtle} library.

The \texttt{Danton} package is hosted on GitHub~\footnote{\url{https://github.com/niess/danton}} under the terms of the LGPL-3.0 license. It can be used as a \texttt{C} library (\texttt{libdanton}) or as an executable (\texttt{danton}). The \texttt{danton} executable is steered by data cards, in JSON format. On Linux systems, it is available as a standalone AppImage.

\subsubsection{\texttt{$\nu$SpaceSim}}

The simulation package \texttt{$\nu$SpaceSim} \cite{Krizmanic:2021eyu, Krizmanic:2020bdm} provides a vectorized Python framework that used the Earth-emergent tau flux predicted by tau neutrino generators, baselined with nuPyProp, to then generate the extensive air showers (EAS) from the tau decays, generate the optical Cherenkov and geomagnetic radio EAS signals, and record these using simulated detector responses. The unique nature of the tau decays, including effects of depolarization due to energy losses in the Earth, induces variability in the EAS particle profiles, as shown in Fig.~\ref{TauDecay}. Furthermore, the nature of the upward-moving EAS development in the rarified upper atmosphere and the different atmospheric attenuation yields to a much different Cherenkov light and radio signatures at high-altitude detectors as compared to downward-moving EAS \cite{Cummings:2020ycz,Romero-Wolf:2021kxe} measured by ground-based detectors, which requires a distinctive modeling methodology.  The \texttt{$\nu$SpaceSim} framework allows for inclusion of any tau exit probability and energy distribution package, and thus allows for the quantification of modeling systematic errors in the determination of neutrino-induced optical Cherenkov and radio EAS signals for sub-orbital and space-based instruments.

\begin{figure}[ht]
\centering
    \includegraphics[width=.4\linewidth]{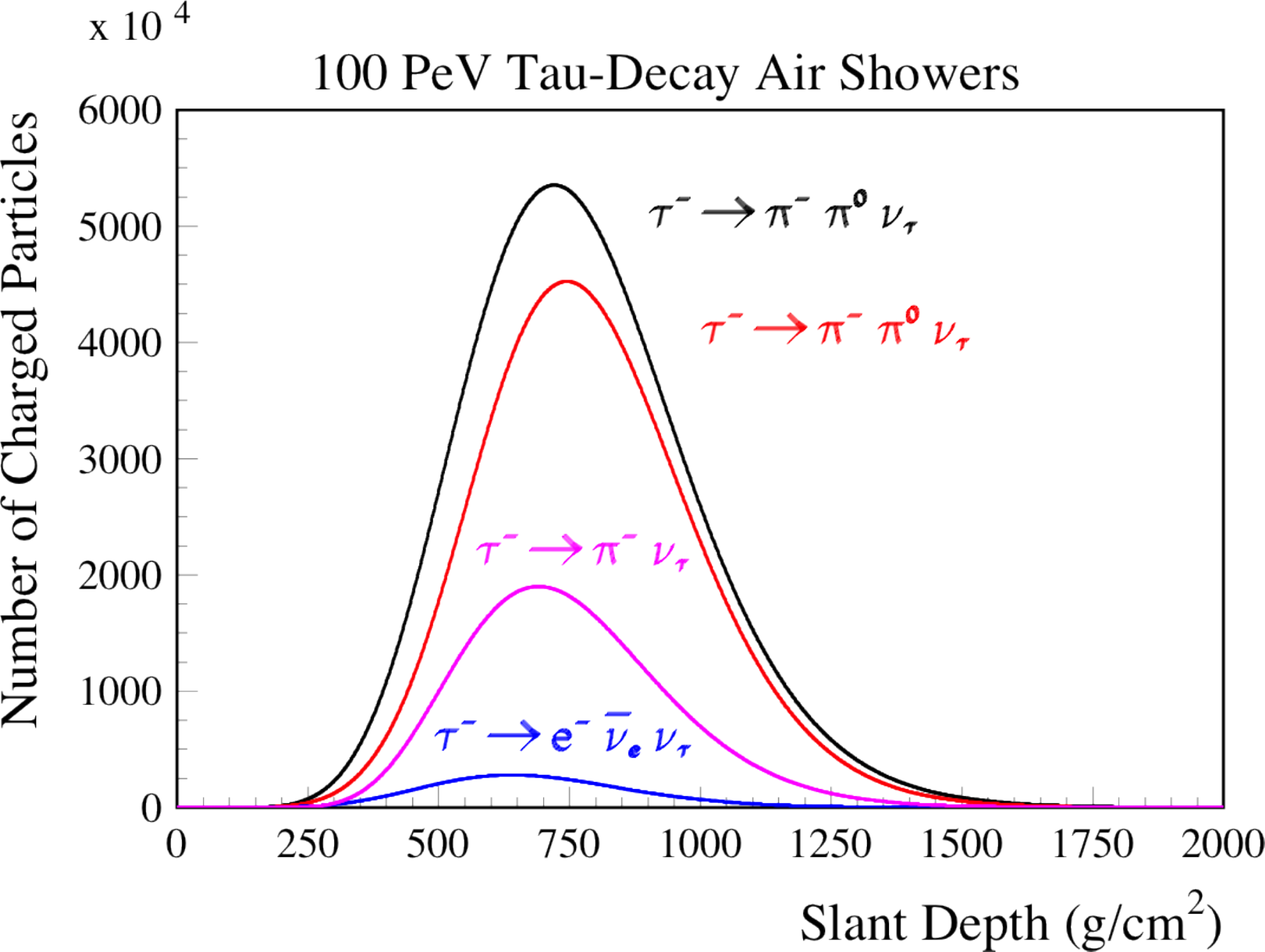}
 \caption{The longitudinal profiles of four different random, simulated extensive air showers (EAS) from 100 PeV $\tau$-lepton decays. PYTHIA provided the decays and the fractional energy of each decay product. From top down, the 1$^{\rm st}$ decay has 82\% of the initial $\tau$-lepton energy in the hadronic system ($f_{\pi^-}$=49\%; $f_{\pi^0}$=33\% ) the 2$^{\rm nd}$ has 62\% ($f_{\pi^-}$=22\%; $f_{\pi^0}$=40\% ), the 3$^{\rm rd}$ has $f_{\pi^-}$=62\%, while the 4$^{\rm th}$ has $f_{e}$=3\%. Starting point fluctuations for the $\pi^-$ EASs vary from 12 to 180 g/cm$^2$. From Ref.~\cite{Krizmanic:2020bdm}.
 \label{TauDecay}}
\end{figure}

\subsection{In-ice Radio Detection Simulation Codes}

To simulate the prospects of in-ice neutrino detectors, interpret its data, as well as to develop reconstruction algorithms, a precise and accurate MC code is required. The in-ice radio community is jointly developing the NuRadioMC simulation code \cite{NuRadioMC2019} which is available open-source through github\footnote{\url{https://github.com/nu-radio/NuRadioMC}} building upon the codes ARASim \cite{ARASIM,ARA:2014fyf}, ShelfMC \cite{PersichilliPhD} and PyREX\footnote{\url{https://github.com/bhokansonfasig/pyrex}} \cite{Hokanson-Fasig:2020vpw}.
The codes were thoroughly checked against each other and yield the same results within a few percent for the same physics settings.  NuRadioMC simulates the neutrino interaction in the ice, the generation of the radio signal, the signal propagation to the antennas, and a full detector and trigger simulation. Several milestones have already been achieved which are relevant for tau detection. A precise calculation of the Askaryan emission including a realistic modeling of the LPM effect for high-energy $\nu_e$-CC interactions \cite{Alvarez-Muniz:2020ary} was implemented which allows to study the identification of $\nu_e$-CC interactions vs. other types of neutrino interactions, and thereby enables a estimation of the combined muon and tau neutrino flux vs. the electron neutrino flux at UHE energies \cite{Stjarnholm:2021xpj}. In addition, the PROPOSAL code \cite{Koehne:2013gpa} was integrated into NuRadioMC to simulate the radio emission generated by secondary muons and taus produced in the initial neutrino interactions \cite{Garcia-Fernandez:2020dhb,Glaser:2021hfi}. These features allow to study the prospects for tau and tau neutrino detection in in-ice radio detectors. The simulation of tau regeneration through the Earth is currently being integrated into NuRadioMC which will provide another handle on tau detection with in-ice radio detectors. 

The next step in precision will be achieved with CORSIKA8 \cite{Engel:2018akg,Karastathis:2021akf} that will allow a microscopic simulation of particle cascades and the resulting radio emission in complex media. Current simulation codes decouple the calculation of the radio emission in particle cascades and the propagation of the radio wave in a medium with varying index-of-refraction. With CORSIKA8, both effects will be combined to calculate the radio emission generated by the movement of every shower particle at an arbitrary observer position in a medium with varying index-of-refraction. Furthermore, also complex scenarios such as a particle cascade transitioning from air into ice can be simulated with CORSIKA8. \pagebreak
\section{Conclusions}
\label{sec:conclusions}
The discovery of the tau neutrino in 2000 by DONuT completed the search for known fermions.
Several decades later, our measurements of tau neutrinos have improved with additional detections of tau neutrinos from oscillations at OPERA, SuperK, and IceCube.
As shown in this whitepaper, the field is poised for an explosion of new opportunities to detect tau neutrinos in the next decade and beyond.
In particular, DUNE can detect tau neutrinos from oscillations with very good identification, especially in a high-energy beam tune.
Atmospheric neutrino detection at SuperK/HyperK, IceCube/DeepCore, and KM3NeT/ORCA will continue to improve not only in statistics but also in systematics with hardware and analysis developments.
The LHC as a tool for neutrino physics is finally being realized with FASER$\nu$ and SND@LHC in the immediate future and additional plans down the road within the Forward Physics Facility framework.
Excellent precision can be reached with beam dump experiments such as SHiP as well as the DUNE near detector facility.
Finally, a plethora of high energy astrophysical experiments are being designed and constructed with a goal of studying the cosmos, but with tau neutrino detection as an added bonus.

Opportunities for tau neutrinos with a high-energy beam run at DUNE, which is not currently a part of the DUNE program, can bring qualitatively new physics understanding and should not be overlooked.
In addition, it is important to ensure that large volume detectors are well suited to identify tau neutrinos.
As the various collider and fixed target experiments shape up, ensuring that technology to identify tau neutrinos exists will be a crucial aspect of their experiments to reach their goals.
Moreover, as various next-generation collider experiments such as the FCC, CLIC, muon colliders, and others are being discussed, it is important to plan for a forward physics program there and ensure that tau neutrino identification is possible.
On the highest energy side, it is important to connect the astrophysics and particle physics communities to further study the particle physics benefits of these primarily astrophysical experiments.
Moreover, with the detection of ultra-high energy tau neutrinos we will enter a unique era where in some cases we will know more about tau neutrinos than the other flavors.

Detecting, identifying, and truly understanding any tau neutrino data set requires fairly advanced tools.
For collider and beam dump experiments, advanced flux predictions are required to match the expected statistical precision.
Identifying tau neutrinos in LArTPCs or water Cherenkov detectors requires an accurate knowledge of tau properties as well as clever reconstruction tools.
High energy astrophysical experiments need to account for tau neutrino regeneration through the Earth.
In addition, having an understanding of the sources of high energy neutrinos is also important for characterizing the flux.

To ensure that the tau neutrino goals for the broad experimental program are met, these tools require further work.
In particular, the intrinsic flux predictions need to be improved.
Additional work on reconstructing tau neutrinos in LArTPCs and water Cherenkov detectors is also required to realize the full potential of those experiments to identify tau neutrinos.
Finally, to ensure that particle physics information can be extracted from high energy astrophysical neutrino experiments, the propagation codes need to be cross-checked and validated and then integrated with the experimental analysis pipelines.

The theoretical interests of tau neutrino physics are also broad.
Measurements of tau neutrinos will tell us about three-flavor oscillation parameters.
They are crucial for testing the unitarity of the lepton mixing matrix and the general robustness of the three-flavor oscillation paradigm.
They also provide key tests of many well-motivated new physics scenarios such as sterile neutrinos, unitarity violation, non-standard interactions, secret interactions, neutrino decay, magnetic moments, and connections to dark matter and other aspects of cosmology.
There are many connections not only among the various probes presented in this whitepaper, but also other oscillation experiments, the LHC, and dark matter searches both direct and indirect.
There are also connections with the above mentioned tools such as flux predictions which are crucial for probing the zero-distance effect coming from steriles, unitarity violation, or non-standard neutrino interactions.

Improving our understanding of the three-flavor oscillation picture by testing for unitarity violation requires a careful global analysis of a range of experiments.
There is also an underdeveloped model building connection between the $b$-physics flavor anomalies and tau neutrinos which are involved in many of these anomalies.
As the forward physics program at the LHC ramps up it is important to continue the theoretical support to interpret their results in the broader new physics scenarios.
Finally, while the high energy astrophysical neutrino experiments are primarily designed to understand the physics of extreme accelerators, they are also potentially powerful probes of particle physics topics such as parton distribution functions and new physics scenarios, but it is essential to build more connections between the astrophysics and particle physics communities to ensure that they achieve their full physics potential.
\pagebreak

\section*{Acknowledgments}
\addcontentsline{toc}{section}{Acknowledgments}
PBD and JG acknowledge support from the US Department of Energy under Grant Contract DE-SC0012704.
SAW acknowledges support from the National Science Foundation through Awards 2111232 and CAREER 2033500 and NASA through Awards 80NSSC20K1288, 80NSSC20K0925, and 80NSSC20K0925.
MHR acknowledges support from US DOE grant DE-SC-0010113.
SRK acknowledges support from the National Science Foundation under grant number PHY-1307472 and the US Department of Energy under contract numbers DE-AC-76SF00098.
AA acknowledges support from the US Department of Energy, Office of Science, Office of High Energy Physics, under Award Number DE-SC0011784.
CAA is supported by the Faculty of Arts and Sciences of Harvard University and the Alfred P.~Sloan Foundation.
AG acknowledges support from European Union’s H2020-MSCA Grant Agreement No.~101025085.
JLF is supported in part by US NSF Grants PHY-1915005 and PHY-2111427, Simons Investigator Award \#376204, Simons Foundation Grant 623683, and Heising-Simons Foundation Grants 2019-1179 and 2020-1840.
ST is supported by the grant ``AstroCeNT: Particle Astrophysics Science and Technology Centre'' carried out within the International Research Agendas programme of the Foundation for Polish Science financed by the European Union under the European Regional Development Fund and by the Polish Ministry of Science and Higher Education through its scholarship for young and outstanding scientists (decision no 1190/E-78/STYP/14/2019). 
YSJ acknowledges support from the National Research Foundation of Korea (NRF) grant funded by the Korea government through Ministry of Science and ICT Grant 2021R1A2C1009296.
OT acknowledges support from LANL's Laboratory Directed Research and Development (LDRD/PRD) program under project number 20210968PRD4. LANL is operated by Triad National Security, LLC, for the National Nuclear Security Administration of U.S.~Department of Energy (Contract No.~89233218CNA000001).
FK is supported by the Deutsche Forschungsgemeinschaft (DFG, German Research Foundation) under Germany's Excellence Strategy – EXC 2121 Quantum Universe – 390833306.
YF acknowledges support from  European Union's Horizon 2020 research and innovation programme under the Marie Skłodowska -Curie grant agreement No 860881-HIDDeN and from Saramadan under contract No.~ISEF/M/400279.
JAM and EZ acknowledge support from Xunta de Galicia (Centro singular de investigaci\'on de Galicia accreditation 2019-2022), from European Union ERDF, from ''Mar\'\i a de Maeztu'' Units of Excellence program MDM-2016-0692, from the Spanish Research State Agency, and from Ministerio de Ciencia e Innovaci\'on PID2019-105544GB-I00 and RED2018-102661-T (RENATA).
JFK acknowledges support from NASA grants NNX17AJ82 and 80NSSC19K0626.
MB acknowledges support from the Villum Fonden under project no.~29388.
JYZZ acknowledges support from the China and Germany Postdoctoral Exchange Program from the Office of China Postdoctoral Council and the Helmholtz Centre (Grant No.~2020031).
\pagebreak


\section*{Endorsers}
\addcontentsline{toc}{section}{Endorsers}
\begin{multicols}{2}
\begin{tabular}{ll}
S.~Agarwalla & Bhubaneswar\\
L.~Anchordoqui & CUNY\\
R.~Alves Batista & IFT\\
J.~Barrow & MIT\\
S.~Barwick & Irvine\\
J.~Beacom & Ohio State\\
J.~Boyd & CERN\\
N.~Charitonidis & CERN\\
X.~Chen & Tsinghua\\
B.~Clark & Michigan State\\
J.~Coelho & CNRS Paris\\
D.~Cowen & Penn State\\
S.~De Kockere & Vrije Brussel\\
K.~de Vries & Vrije Brussel\\
C.~Deaconu & Chicago\\
R.~Dobre & ISS Romania\\
M.~DuVernois & Wisconsin\\
J.~Eser & Chicago\\
I.~Esteban & Ohio State\\
G.~Farrar & NYU\\
E.~Firu & ISS Romania\\
A.~Giarnetti & Rome\\
C.~Giunti & INFN Torino\\
A.~Guler & METU Turkey\\
F.~Halzen & Wisconsin\\
J.~Hanson & Whittier\\
A.~Haungs & KIT\\
D.~Hayakawa & Chiba\\
Z.~Hu & Tsinghua\\
K.~Hughes & Chicago\\
J.~Hyde & Moravian University\\
K.~Kampert & Wuppertal\\
A.~Karle & Wisconsin\\
T.~Katori & KCL\\
L.~Koerner & Houston\\
\end{tabular}

\begin{tabular}{ll}
M.~Komatsu & Nagoya\\
I.~Kravchenko & Nebraska\\
S.~Li & Fermilab\\
G.~Marsella & INFN Catania\\
F.~McBride & Penn State\\
M.~Miloi & JINR Bucharest\\
H.~Minakata & Tokyo\\
M.~Mostafa & Penn State\\
K.~Murase & Penn State\\
M.~Muzio & Penn State\\
A.~Neagu & ISS Romania\\
A.~Nelles & DESY\\
K.~Okui & Chiba\\
I.~Padilla-Gay & Niels Bohr Institute\\
S.~Parke & Fermilab\\
Y.~Perez-Gonzalez & Durham\\
T.~Preda & ISS Romania\\
X.~Qian & Brookhaven\\
M.~Reininghaus & KIT\\
A.~Romero-Wolf & JPL/Caltech\\
O.~Sato & Nagoya\\
M.~Schimp & Wuppertal\\
I.~Shoemaker & Virginia Tech\\
D.~Southall & Chicago\\
A.~Suliga & Berkeley\\
T.~Stuttard & Niels Bohr Institute\\
I.~Taboada & Georgia Tech\\
Z.~Tabrizi & Northwestern\\
Y.~Takubo & KEK\\
T.~Thakore & Cincinnati\\
O.~Toranosuke & Chiba\\
Y.~Tsai & Irvine\\
D.~Veberic & KIT\\
J.~Zupan & Cincinnati
\end{tabular}
\end{multicols}\pagebreak

\section*{References}
\addcontentsline{toc}{section}{References}
\bibliographystyle{iopart-num-mod}
\bibliography{references}

\end{document}